\documentclass[10pt]{article}
\usepackage{graphicx, yk}
\usepackage[margin=2.5cm]{geometry}
\usepackage{setspace}
\usepackage{float}
\usepackage{amsmath, amsfonts, framed}
\usepackage{amssymb}
\usepackage{booktabs}
\usepackage{multirow}
\usepackage{longtable}
\usepackage{caption}
\usepackage{wrapfig}
\usepackage{afterpage}
\usepackage{placeins}
\usepackage{multicol}
\usepackage{hyperref}    
\usepackage{textcmds}
\usepackage{adjustbox}
\usepackage{circuitikz}
\usepackage{subcaption}
\usepackage[normalem]{ulem}
\usepackage{algpseudocode}
\usepackage{algorithm}
\usepackage{cancel} 
\usepackage{tikz}
\usepackage{xcolor}
\usepackage{titlesec}
\usepackage{enumerate}
\usepackage{natbib}
\usepackage[most]{tcolorbox}

\definecolor{d1fill}{RGB}{222,235,247}   
\definecolor{d1line}{RGB}{ 66,114,161}
\definecolor{d2fill}{RGB}{219,238,235}   
\definecolor{d2line}{RGB}{ 45,132,120}
\definecolor{estfill}{RGB}{252,240,220}  
\definecolor{estline}{RGB}{190,120, 40}
\definecolor{outfill}{RGB}{237,237,237}
\definecolor{outline}{RGB}{ 90, 90, 90}
\definecolor{lane}{RGB}{247,247,249} 
\definecolor{lanelbl}{RGB}{140,140,148}
\definecolor{thmblue}{RGB}{181,222,232}
\definecolor{lemmagreen}{RGB}{145,235,145}
\definecolor{mainyellow}{RGB}{255,221,51}
\definecolor{grpline}{RGB}{205,175,120}
\usetikzlibrary{arrows.meta,positioning,calc,fit,backgrounds,shapes.geometric}

\newcommand\blfootnote[1]{
  \begingroup
  \renewcommand\thefootnote{}\footnote{#1}
  \addtocounter{footnote}{-1}
  \endgroup
}

\tikzset{
    proofthm/.style={
        rectangle, rounded corners=3pt, draw=black!70,
        fill=thmblue, very thick,
        minimum width=2.7cm, minimum height=0.75cm,
        align=center, font=\small
    },
    prooflemma/.style={
        rectangle, rounded corners=3pt, draw=black!60,
        fill=lemmagreen, thick,
        minimum width=2.65cm, minimum height=0.68cm,
        align=center, font=\scriptsize
    },
    proofmain/.style={
        rectangle, rounded corners=3pt, draw=black!70,
        fill=mainyellow, thick,
        minimum width=3.0cm, minimum height=0.72cm,
        align=center, font=\scriptsize
    },
    proofarrow/.style={->, thick, black!80}
}

\floatstyle{ruled}
\restylefloat{algorithm}

\titleclass{\subsubsubsection}{straight}[\subsubsection]

\newcounter{subsubsubsection}[subsubsection]
\renewcommand\thesubsubsubsection{\thesubsubsection.\arabic{subsubsubsection}}
\titleformat{\subsubsubsection}[block]
  {\normalfont\normalsize\bfseries}
  {\thesubsubsubsection}{1em}{}
\titlespacing*{\subsubsubsection}
  {0pt}{3.25ex plus 1ex minus .2ex}{1ex}
\newtcolorbox{algobox}[1][]{colback=gray!5!white, colframe=black,
  fonttitle=\bfseries, title=Estimating mediation functional, #1}
\tcbuselibrary{listingsutf8}

\newtcolorbox{myalgorithm}[1][]{enhanced,
  sharp corners,        
  colback=white,        
  colframe=black,      
  boxrule=0.5pt,         
  title=Estimating mediation functional,
  fonttitle=\bfseries,
  #1
}
\hypersetup{colorlinks, linkcolor={red}, citecolor={blue}, urlcolor={red}}  
\tikzset{every picture/.style={line width=0.75pt}} 
\setcitestyle{authoryear,open={(},close={)}} 

\newcommand{\independent}{\perp\!\!\!\perp} 

\providecommand{\keywords}[1]
{ \small	
  \textbf{\textit{Keywords---}} #1}

\title{Causal Mediation Analysis with High-Dimensional Surrogates: Identification, Estimation and Inference under Latent Factor Models}
 \author{Himani Yadav, Shi Bo, AmirEmad Ghassami$^\dagger$, Debarghya Mukherjee$^\dagger$}
\date{}

\begin{document}

\maketitle

\begin{abstract}
Causal mediation analysis is an important tool for understanding how a treatment affects an outcome through intermediate variables referred to as mediators. However, in many applications, these mediators and pre-treatment covariates are not directly observed and are instead measured through noisy, high-dimensional surrogates. Mediation analysis when such latent variables are observed only through these surrogates remains less explored.  In this paper, we propose Factor-Adjusted Regression for Mediation Analysis (FARMA), a factor-based framework for estimating direct and indirect effects when either one or both of the mediators and relevant pre-treatment covariates are latent. FARMA uses the surrogate measurements to recover information about the underlying latent variables. Under a latent factor structure, we show that the direct and indirect effects are approximately identified, with an approximation error that vanishes as the surrogate dimensions grow. We further show that our estimator is $\sqrt{n}$-consistent and asymptotically normal. We combine the theory with extensive simulation studies and apply it to data from the Alzheimer's Disease Neuroimaging Initiative (ADNI) study, where we investigate whether the effect of depressive symptoms on cognitive impairment is mediated through the latent process underlying high-dimensional DNA methylation measurements.
\end{abstract}

\keywords{Mediation Analysis; Latent Factor Models; Ultra-High-Dimensionality; Identification}

\blfootnote{$^\dagger$ Corresponding authors.}

\newpage

\begingroup
\hypersetup{linkcolor=black}
\tableofcontents
\endgroup

\newpage

\section{Introduction}\label{section:introduction}

Causal mediation analysis provides a principled framework for understanding the mechanisms through which a treatment affects an outcome \citep{baron1986moderator,imai2010identification,vanderweele2014mediation,pearl2022direct}. It moves beyond testing whether an overall effect exists to ask how that effect operates by separating the total effect into a direct effect, representing the effect through other pathways, and an indirect effect, representing the effect through an intermediate variable (referred to as the \textit{mediator}). The statistical literature on mediation analysis has developed from its early regression-based formulations. The product-of-coefficients method of \citet{baron1986moderator} offered an intuitive way to study mediated pathways. Later, the potential-outcomes framework of \citet{rubin1974estimating} allowed mediation to be defined in formal causal terms through natural direct and indirect effects \citep{robins1992identifiability,pearl2022direct}, along with the conditions for their identification \citep{imai2010identification}. Subsequent work developed semiparametric, weighting, and multiply robust methods for their estimation and inference \citep{tchetgen2012semiparametric,vanderweele2014mediation,vanderweele2015explanation}. More recent work has addressed high-dimensional mediation problems, even when the number of mediators is much larger than the sample size through variable selection, sparse regression, and related inferential techniques \citep{zhang2016estimating,huang2016hypothesis,chen2018high,zhang2018distance,zhao2021pathway,perera2022hima2,guo2022high,guo2023statistical,guo2024estimations,rakshit2024statistical,bo2024debiased}. 
However, in many real-world applications, the true mediators are not directly observed; instead, researchers may rely on a potentially large number of surrogate measurements that provide indirect information about the underlying mediator. Our motivating application uses DNA methylation profiles from the Alzheimer's Disease Neuroimaging Initiative (ADNI) study to investigate a latent biological pathway between depressive symptoms and cognitive impairment. Oxidative stress has been implicated in both depression and Alzheimer's disease (AD) \citep{black2015depression,buccellato2021role}. Direct quantification of reactive oxygen species (ROS) is difficult because of their short lifespan and high reactivity \citep{katerji2019approaches}. Researchers therefore rely on more stable molecular signatures that reflect oxidative activity. Among them is DNA methylation, which can be altered in response to oxidative stress and can be measured at scale \citep{miranda2000role,pandya2013antioxidants,black2015depression}. While DNA methylation provides a useful surrogate for investigating the mediating role of oxidative stress, its measurement across hundreds of thousands of CpG sites using high-throughput assays produces a high-dimensional data structure in which individual sites capture only partial aspects of the underlying biological process. When the pre-treatment covariates needed for confounding adjustment are also measured indirectly, both the mediator and the covariates must be related to their observed surrogates. One way to describe this measurement structure is through a latent factor model, in which a small number of unobserved variables generate the common variation across many noisy measurements \citep{bai2003inferential,fan2013large}.

Mediation analysis with indirectly measured variables has received attention in fixed-dimensional settings \citep{albert2016causal}, with fewer developments for high-dimensional measurements \citep{derkach2019high}. Proxy-based methods establish identification of mediation effects when the mediator is hidden and suitable proxy measurements are available \citep{ghassami2025causal}. Related factor-based approaches account for dependence among observed high-dimensional exposures and mediators \citep{zhang2026statistical} or accommodate individualized and time-varying mediation effects \citep{zhang2026individualized}. 
Despite these developments, existing results do not directly characterize how the dimensions of noisy surrogate measurements determine the error in recovering the latent mediation functional when both the mediator and the pre-treatment covariates are unobserved. In particular, it remains to be established when mediation regressions based on low-dimensional projections of these surrogates approximate the latent target closely enough to permit inference with the same first-order precision as if the latent factors were observed. This question becomes more delicate when the working factor dimensions exceed the true dimensions, because some directions then contain only measurement noise. For example, if there are three true factors but we estimate five, two linear combinations of the estimated factors contain no signal from the true factors. Their variances shrink as the number of surrogates grows, which can make the regression matrices nearly singular and the estimated regression coefficients unstable.

To address these questions, we propose Factor-Adjusted Regression for Mediation Analysis (FARMA), a factor-based framework for identification, estimation, and inference with high-dimensional surrogates of latent mediators and pre-treatment covariates. Building on factor-model approaches for high-dimensional data \citep{bai2002determining,bai2003inferential,fan2013large,fan2020factor,fan2022learning}, FARMA constructs low-dimensional projections of the surrogate covariates and mediators and incorporates these projected factors into cross-fitted mediation regressions. Under some standard assumptions, these projections retain information about the underlying latent factor spaces while reducing the contribution of measurement noise. Our analysis allows the fixed working dimensions to exceed these true factor dimensions and accounts for the additional directions that contain only measurement noise. The main contributions of this paper are threefold:

\medskip\noindent\textbf{Identification.}
We establish asymptotic identification of the causal natural direct and indirect effects when both the mediator and the relevant covariate factors are latent. With a fixed number of surrogates, the target effects are generally not point identified from the observed data alone without strong restrictions on the data-generating process. We show that suitable projections of the high-dimensional surrogates yield an observed-data parameter that approaches the latent target as the number of measurements increases, leading to asymptotic point identification.

\medskip\noindent\textbf{Estimation.}
Building on the identification result, we develop a cross-fitted estimator of the causal mediation effects using projected surrogate factors. The method does not require exact knowledge of the true factor dimensions and remains valid when fixed working dimensions are chosen as upper bounds on the true dimensions. Under suitable regularity and growth conditions on the surrogate dimensions, we show that the resulting estimator is $\sqrt{n}$-consistent and asymptotically normal for the parameter of interest. This result provides the basis for statistical inference on the natural direct and indirect effects even when mediators are not observed directly. 

\medskip\noindent\textbf{Application.}
We apply FARMA to investigate the effect of depression on Alzheimer's disease (AD) mediated by oxidative stress, where the mediating process is not directly observed and is instead represented by high-dimensional DNA methylation measurements. We illustrate the approach using data from the Alzheimer's Disease Neuroimaging Initiative (ADNI) Study. 

\medskip\noindent The remainder of the paper is organized as follows. Section~\ref{section:2model} presents the problem setting and model. Section~\ref{section:3identification} establishes identification of the target parameter. Section~\ref{section:4algorithm} introduces the FARMA. Section~\ref{section:5theory} develops the asymptotic theory. Section~\ref{section:6simulation} reports the simulation results. Section~\ref{section:7realdata} presents the application to the ADNI dataset. Section~\ref{section:8conclusion} concludes with a discussion of future work.
\section{Model Setting} \label{section:2model}
We consider a binary treatment variable $A \in \{0,1\}$ (also called \textit{exposure} or \textit{action} variable), and a real-valued outcome $Y \in \mathbb{R}$. For each individual, we consider mediators ${\mathbf f_M} = [{f}_{M_1}, \ldots, { f}_{M_{\bar q}}] \in \mathbb{R}^{\bar q}$ through which treatment may influence the outcome, and pre-treatment covariates ${\mathbf f_X} = [{f}_{X_1}, \ldots, { f}_{X_{\bar p}}] \in \mathbb{R}^{\bar p}$, which capture baseline characteristics that may confound the relationship among treatment, mediators and outcome. We use lowercase notation, such as $f_{M_i}$ and $f_{X_j}$, to denote the individual coordinates within these factors, respectively. The treatment can influence the outcome in two ways, indirectly through the mediators, or directly without passing through them. Formally, the variables are assumed to be generated from the following structural equations:
\begin{equation}\label{eq:factor}
\begin{aligned}
\mathbf f_M & = (1-A)\Big(\delta_{0}+\mathbf B_{0}\mathbf f_X+\mathbf u_{XM,0}\Big) + A\Big(\delta_{1}+\mathbf B_{1}\mathbf f_X+\mathbf u_{XM,1}\Big), \\
Y  & = (1-A)\Big(\alpha_{0}+\beta_{0}^\top\mathbf f_X + \gamma_{0}^\top\mathbf f_M + \epsilon_0\Big) + A\Big(\alpha_{1}+\beta_{1}^\top\mathbf f_X + \gamma_{1}^\top\mathbf f_M + \epsilon_1\Big),
\end{aligned}
\end{equation}
where $\epsilon_0$ and $\epsilon_1$ are centered outcome errors on the control and treated arms, $\mathbf u_{XM,0}\in\mathbb{R}^{\bar q}$ and $\mathbf u_{XM,1}\in\mathbb{R}^{\bar q}$ are the corresponding centered mediator structural noises, $\mathbf{B}_0, \mathbf{B}_1 \in \mathbb{R}^{\bar q \times \bar p}$, $\beta_0, \beta_1 \in \mathbb{R}^{\bar p}$, and $\gamma_0, \gamma_1 \in \mathbb{R}^{\bar q}$ are unknown coefficients, and $\delta_0, \delta_1 \in \mathbb{R}^{\bar q}$ and $\alpha_0, \alpha_1 \in \mathbb{R}$ are intercepts. Note that we allow for an $A-\mathbf{f}_X$ interaction in the generation of $\mathbf{f}_M$, and $A-\mathbf{f}_X$ and $A-\mathbf{f}_M$ interactions in the generation of the outcome $Y$.

\begin{figure}[t]
\centering
\scalebox{0.85}{%
\begin{tikzpicture}[
    >={Stealth[length=2.2mm]},
    latent/.style={
        rounded corners=2pt,
        draw=black,
        line width=0.75pt,
        minimum width=22mm,
        minimum height=9mm,
        align=center
    },
    obsX/.style={
        rounded corners=2pt,
        draw=red,
        line width=0.75pt,
        minimum width=22mm,
        minimum height=9mm,
        align=center
    },
    obsM/.style={
        rounded corners=2pt,
        draw=red,
        line width=0.75pt,
        minimum width=22mm,
        minimum height=9mm,
        align=center
    },
    term/.style={
        rounded corners=2pt,
        draw=black,
        line width=0.75pt,
        minimum width=12mm,
        minimum height=9mm,
        align=center
    },
    edge/.style={->, draw=black, line width=0.8pt},
    meas/.style={->, draw=black!55, line width=0.65pt}
]

\node[latent] (fX) at (0,0) {$\mathbf f_X $};
\node[term]   (A)  at (2.7,0) {$A$};
\node[latent] (fM) at (5.4,0) {$\mathbf f_M$};
\node[term]   (Y)  at (8.1,0) {$Y$};

\node[obsX] (X) at (0,3.0) {$\mathbf X$};
\node[obsM] (M) at (5.4,3.0) {$\mathbf M$};

\draw[meas] (fX) -- (X);
\draw[meas] (fM) -- (M);
\draw[edge] (fX) -- (A);
\draw[edge] (A) -- (fM);
\draw[edge] (fM) -- (Y);

\draw[edge] (fX) to[bend left=28] (fM);
\draw[edge] (fX) to[bend left=38] (Y);
\draw[edge] (A)  to[bend right=35] (Y);

\end{tikzpicture}
}
\caption{\footnotesize Causal diagram for the factor-augmented mediation model. The observed high-dimensional variables $\mathbf X$ and $\mathbf M$ are generated from the latent factors $\mathbf f_X$ and $\mathbf f_M$, respectively.}
\label{fig:DAG}
\end{figure}
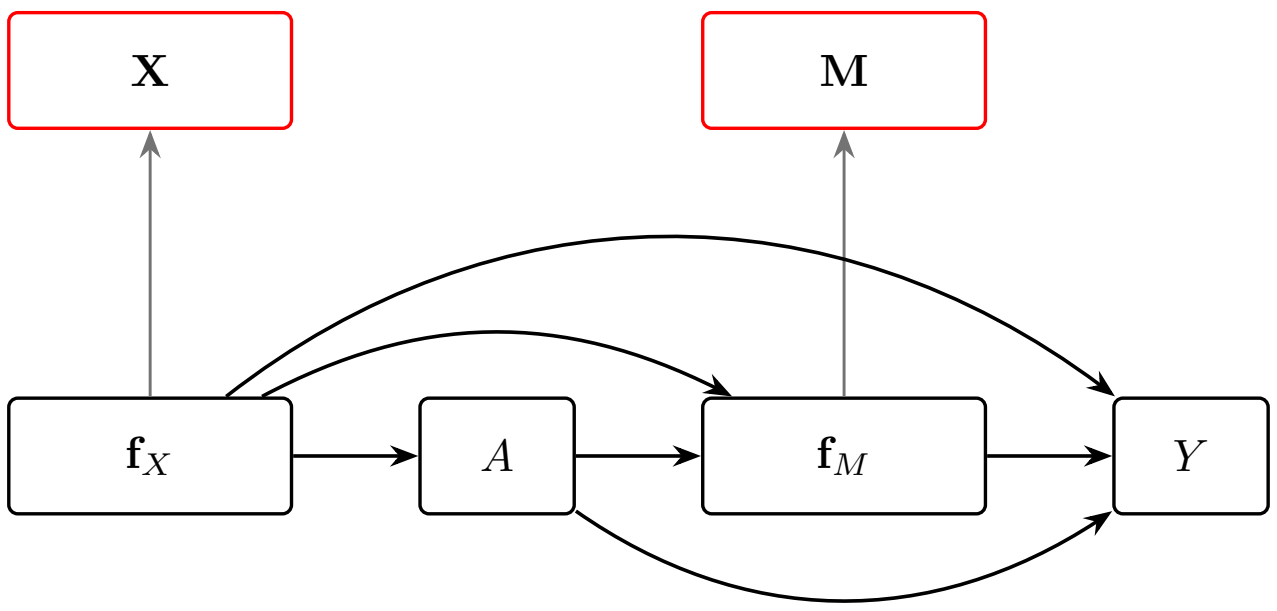

In many real-world applications, $\mathbf{f}_X$ and $\mathbf{f}_M$ are \emph{latent} (i.e., unobserved). Instead, we observe  measurements $X \in \mathbb{R}^p$ and $M \in \mathbb{R}^q$ that serve as noisy surrogates for these latent variables. We focus on high-dimensional measurements with $p \gg \bar p$ and $q \gg \bar q$. We adopt the following factor model for the observed measurements:
\begin{equation}\label{eq:factors}
X = \mathbf{\Gamma}_X \mathbf{f}_X + \mathbf{u}_X, \qquad M = \mathbf{\Gamma}_M \mathbf{f}_M + \mathbf{u}_M,
\end{equation}
where $\mathbf{\Gamma}_X \in \mathbb{R}^{p \times \bar p}$ and $\mathbf{\Gamma}_M \in \mathbb{R}^{q \times \bar q}$ are unknown loading matrices, and $\mathbf{u}_X$ and $\mathbf{u}_M$ are mean-zero errors. Henceforth, we refer to $\mathbf{f}_X$ and $\mathbf{f}_M$ as latent factors. Figure~\ref{fig:DAG} summarizes the causal relationships among the latent factors, treatment, and outcome specified in \eqref{eq:factor} with the measurement relationships in \eqref{eq:factors}. The standard mediation model with directly observed covariates and mediators is recovered as a special case by setting $p=\bar p$, $q=\bar q$, $\mathbf{\Gamma}_X=I_p$, and $\mathbf{\Gamma}_M=I_q$, with vanishing errors. In this case, $X=\mathbf{f}_X$ and $M=\mathbf{f}_M$, so the structural equations in \eqref{eq:factor} can be written directly in terms of the observed variables $X$ and $M$.

We first consider the setting in which $\mathbf{f}_X$ and $\mathbf{f}_M$ are directly observed, and define the causal estimands using the potential-outcomes framework \citep{rubin1974estimating, robins1992identifiability}. For each individual, let $Y^{(a,f_M)}$ denote the potential outcome under treatment $A=a$ and mediator value $\mathbf{f}_M=f_M$, and let $\mathbf{f}_M^{(a)}$ denote the potential mediator value under treatment $A=a$, where $a\in\{0,1\}$. The potential outcome under treatment $a$ is then $Y^{(a)}=Y^{(a,\mathbf{f}_M^{(a)})}$. The average treatment effect (ATE) admits the following decomposition:
\begin{equation}\label{eq:NDENIE}
\begin{aligned}
\underbrace{\mathbb{E}\left[Y^{(1)}-Y^{(0)}\right]}_{\text{Total Effect}}
&=\underbrace{\mathbb{E}\left[Y^{(1,\mathbf{f}_M^{(1)})}-Y^{(1,\mathbf{f}_M^{(0)})}\right]}_{\text{Indirect Effect}}+\underbrace{\mathbb{E}\left[Y^{(1,\mathbf{f}_M^{(0)})}-Y^{(0,\mathbf{f}_M^{(0)})}\right]}_{\text{Direct Effect}}.
\end{aligned}
\end{equation}
The two terms on the right-hand side are the total indirect effect and the pure direct effect, respectively, in the terminology of \citet{robins1992identifiability}. Following \citet{pearl2022direct}, we refer to them as the natural indirect effect (NIE) and the natural direct effect (NDE). In our setting, these effects are defined with respect to $\mathbf{f}_M$. The NIE measures the effect of changing the mediator from its potential value under control to its potential value under treatment, while holding treatment fixed at $A=1$. The NDE measures the effect of changing treatment from $A=0$ to $A=1$, while holding the mediator at its potential value under control.

The decomposition in \eqref{eq:NDENIE} depends on three counterfactual means, $\mathbb{E}[Y^{(1)}]$ ($\equiv \mathbb E[Y^{(1,\mathbf f_M^{(1)})}]$), $\mathbb{E}[Y^{(0)}]$ ($\equiv \mathbb E[Y^{(0,\mathbf f_M^{(0)})}]$), and $\mathbb{E}[Y^{(1,\mathbf{f}_M^{(0)})}\}$. Under consistency, positivity and sequential ignorability, the first two can be identified using some standard covariate adjustment. $\mathbb{E}[Y^{(1,\mathbf{f}_M^{(0)})}\}$ is a cross-world mean that combines the outcome under treatment with the mediator value that would arise under control. Since its identification requires additional assumptions that links the outcome and mediator across treatment regimes, we consider this as our parameter of interest,
\begin{equation}\label{theta_0}
\theta_0:=\mathbb{E}\big[Y^{(1,\mathbf{f}_M^{(0)})}\big].
\end{equation}
To identify $\theta_0$, we impose the following standard assumptions.

\begin{assumption}[Consistency]\label{ass:consistency}
\(Y^{(a, {{f}}_{M})}=Y ~\text{ if } A=a, ~ \mathbf{f}_M = f_M; ~~~~
     \mathbf{f}_M^{(a)} = \mathbf{f}_M \text{ if } A=a ~~~~ \forall ~~ a, f_M.\)
\end{assumption}

\begin{assumption}[Positivity]\label{ass:positivity}
    For all strata of $f_X$, the probability of being in the treatment and control group is positive, i.e., $0<P(A=a \mid  \mathbf f_X={f}_X)$, and the conditional density of the mediator is positive, i.e., $0 < P({f}_M\mid A=a, \mathbf{f}_X={f}_X)$ for all $A, f_X$ and $f_M$.
\end{assumption}

In addition to positivity and consistency, we operate under the assumption of sequential exchangeability \citep{imai2010identification}, which implies that there are no unobserved confounders for the treatment-outcome, mediator-outcome, and treatment-mediator relationships:
\begin{assumption}[Sequential exchangeability]\label{assumption:Sequential} 
Let $X_1 \independent X_2 \mid X_3$ indicate that the random variables $X_1$ and $X_2$ are conditionally independent given the random variable $X_3$. 
\begin{enumerate}
    \item $Y^{(a,{f}_M)} \independent \{A, {\bf f}_M\} \mid  {\bf f}_X$, $\forall$ $a, f_M$, 
    \item ${\bf f}_M^{(a)} \independent A \mid  {\bf f}_X$, $\forall$ $a$, 
    \item $Y^{(a,{ f}_M)} \independent  {\bf f}_M^{(a')}\mid  {\bf f}_X$, $\forall$ $a,a', f_M$. 
\end{enumerate}
\end{assumption}
These assumptions concern the latent covariates $\mathbf{f}_X$ and latent mediators $\mathbf{f}_M$. Under consistency, positivity, and sequential exchangeability, the mediation formula of \citet{imai2010identification} expresses $\theta_0$ as a functional of the joint distribution of $(Y,A,\mathbf{f}_X,\mathbf{f}_M)$:
\begin{equation}\label{theta:functional}
\begin{aligned}
\theta_0&=\iint\mathbb{E}\left[Y \mid \mathbf{f}_X=f_X,\,\mathbf{f}_M=f_M,\,A=1\right]dP_{\mathbf{f}_M\mid\mathbf{f}_X,A}(f_M\mid f_X,0)\,dP_{\mathbf{f}_X}(f_X).
\end{aligned}
\end{equation}
Here, the mediation functional combines three components: the outcome regression under treatment, the conditional distribution of the mediator factors under control, and the marginal distribution of the covariate factors. This representation identifies $\theta_0$ from the distribution involving the latent factors. In our setting, however, $\mathbf{f}_X$ and $\mathbf{f}_M$ are unobserved, so these components cannot be estimated directly from the available measurements $X$ and $M$.
\noindent
Under the linear structural model, \eqref{theta:functional} yields the closed form
\begin{equation}\label{eq:identification_1}
\theta_0=\mathbb{E}\big[Y^{(1,\mathbf{f}_M^{(0)})}\big]=\alpha_1+\gamma_1^\top\delta_0,
\end{equation}
where the covariate contribution vanishes under $\mathbb{E}[\mathbf{f}_X]=\mathbf{0}$. Building on this, we can more generally define the mediation functional for any $a,a'\in\{0,1\}$ as $\theta_{aa'}=\mathbb{E}[Y^{(a,\mathbf{f}_M^{(a')})}]$. In particular, $\theta_{10}=\theta_0$, where the two baseline potential outcomes are as follows:
\begin{equation}\label{eq:farma-ordinary-means}
\textstyle
\theta_{00}=\mathbb{E}[Y^{(0)}]=\alpha_0+\gamma_0^\top\delta_0,\quad\theta_{11}=\mathbb{E}[Y^{(1)}]=\alpha_1+\gamma_1^\top\delta_1.
\end{equation}
Following the decomposition in \eqref{eq:NDENIE}, the natural direct and indirect effects are given by
\begin{equation}\label{eq:farma-natural-effects}
\textstyle\theta_{\mathrm{NDE}}=\theta_{10}-\theta_{00},\quad\theta_{\mathrm{NIE}}=\theta_{11}-\theta_{10},
\end{equation}
so that the total effect is $\theta_{11}-\theta_{00}=\theta_{\mathrm{NDE}}+\theta_{\mathrm{NIE}}$. Hence, evaluating these effects requires recovering $\theta_{10}$, $\theta_{00}$, and $\theta_{11}$ from the observed data. However, a central challenge that still remains is that their coefficients are defined through models involving unobserved latent factors rather than direct measurements. In Section~\ref{section:3identification}, we show how suitable projections of the high-dimensional surrogates allow us to overcome this and identify the parameters from observed data under appropriate conditions as the surrogate dimensions increase.
\\\\
\noindent
\textbf{Notations.} Throughout the paper, let $n$ denote the sample size, and let $p$ and $q$ denote the dimensions of the surrogate vectors $X$ and $M$, respectively. For any random variable, vector, or matrix $\mathbf{X}$, let $\mathbb{E}[\mathbf{X}]$ denote its expectation and $\mathbb{E}[\mathbf{X} \mid \cdot]$ its conditional expectation. Let $P(\cdot)$ denote probability and $P(\cdot \mid \cdot)$ conditional probability. Let $\operatorname{Cov}(\mathbf{X})$ (or $\Sigma_{\mathbf{X}}$) denote its covariance matrix, and let $\operatorname{Cov}(\mathbf{X}\mid \cdot)$ represent its conditional covariance matrix. Let $\mathcal{N}(\mu, \Sigma)$ denote the multivariate Gaussian distribution with mean $\mu$ and covariance $\Sigma$. For any vector $\mathbf{x}$, let $\|\mathbf{x}\|_2$ denote its Euclidean norm. For any matrix $\mathbf{X}$, let $\mathbf{X}^\top$ denote its transpose, and let $\|\mathbf{X}\|_{\mathrm{op}}$ and $\|\mathbf{X}\|_F$ denote its operator norm and Frobenius norm, respectively. Let $\mathbf{1}_d$ and $\mathbf{0}_d$ denote vectors of all ones and all zeros in dimension $d$ (such as $\mathbf{1}_p$ or $\mathbf{1}_q$), and let $I_d$ denote the $d \times d$ identity matrix. For real numbers $a$ and $b$, define $a\vee b = \max\{a,b\}$ and $a \wedge b = \min\{a,b\}$. We denote the smallest and largest eigenvalues of a symmetric matrix $\mathbf{X}$ by $\lambda_{\text{min}}(\mathbf{X})$ and $\lambda_{\text{max}}(\mathbf{X})$, and the smallest and largest singular values of a rectangular matrix $A$ by $\nu_{\min}(A)$ and $\nu_{\max}(A)$. For two possibly random sequences $a_n$ and $b_n$, write $a_n \ll b_n$ (equivalently $b_n \gg a_n$) if $a_n = o_p(b_n)$, meaning $a_n/b_n \overset{p}{\to}0$ whenever the ratio is well defined.
\section{Identification}\label{section:3identification}

In this section, we show that high-dimensional surrogate measurements provide the information needed to identify the mediation functional when both the covariates and mediators of interest are latent. Recall from Section~\ref{section:2model}, under Assumptions~\ref{ass:consistency}--\ref{assumption:Sequential}, the structural models, and $\mathbb{E}[\mathbf{f}_X]=\mathbf{0}$, the causal estimand reduces to $\theta_0 = \mathbb{E}[Y^{(1,\mathbf{f}_M^{(0)})}] = \alpha_1 + \gamma_1^\top \delta_0$. When the surrogate dimensions $p$ and $q$ are fixed and the latent factors are unobserved, $\theta_0$ is not generally identifiable. Since the surrogates are noisy measurements, their idiosyncratic errors do not vanish at fixed dimension, and the map from observed moments to structural coefficients is confounded by the error covariances $\Sigma_{\mathbf{u}_X}$ and $\Sigma_{\mathbf{u}_M}$.

To overcome this, we establish approximate identification results for $\theta_0$, where the target is recovered up to an error that vanishes as $p,q\to\infty$. Since the latent factors are pervasive (Assumption~\ref{Assumption:pervasiveness}(ii)), information about each factor is shared across many surrogate measurements, which allows low-dimensional projections to preserve common factor information while filtering out idiosyncratic noise. We show that, under the structural models in \eqref{eq:factor}--\eqref{eq:factors}, $\theta_0$ becomes point identifiable as the dimensions $p,q\rightarrow\infty$. We begin by stating the assumptions needed for identification of $\theta_0$. These assumptions ensure that the latent factors and the structural and proxy errors are well-behaved.

\begin{assumption}[Sampling, latent factors, and errors]
\label{Assumption:covariances}
There exist constants $0<c<C<\infty$, independent of $n,p,q$, such that the following conditions hold for each $a\in\{0,1\}$.
\begin{enumerate}
\setlength{\itemsep}{2pt}
\setlength{\parsep}{0pt}
\item[(i)] The sequence $\{(\mathbf f_{X,i},A_i,\mathbf u_{X,i},\mathbf u_{M,i},\mathbf u_{XM,0,i},\mathbf u_{XM,1,i},\epsilon_{0,i},\epsilon_{1,i})\}_{i=1}^n$ is independent and identically distributed.
\item[(ii)] For $i=1,\ldots,n$, the errors $\epsilon_{a,i}$, $\mathbf u_{X,i}$, $\mathbf u_{M,i}$, and $\mathbf u_{XM,a,i}$ are mean zero and mutually independent, and $\bigl(\epsilon_{a,i},\mathbf u_{X,i}, \mathbf u_{M,i},\mathbf u_{XM,a,i}\bigr) \independent (\mathbf f_{X,i},A_i).$
\item[(iii)] The factor dimensions $\bar p$ and $\bar q$ are fixed, and $\mathbb E[\mathbf f_X]=\mathbf 0$. For $\mathbf v\in\{\mathbf f_X,\mathbf f_M\}$, $\|\mu_{\mathbf v\mid a}\|_2\le C$.
\item[(iv)] Let $\mathbf f=(\mathbf f_X^\top,\mathbf f_M^\top)^\top$. The conditional covariance matrices satisfy $c\le\lambda_{\min}\bigl(\Sigma_{\mathbf f\mid A=a}\bigr)
\le\lambda_{\max}\bigl(\Sigma_{\mathbf f\mid A=a}\bigr)\le C.$
\item[(v)] For $\mathbf u\in\{\mathbf u_X,\mathbf u_M,\mathbf u_{XM,a}\}$. The error covariance matrices satisfy $c\le\lambda_{\min}(\Sigma_{\mathbf u})
\le\lambda_{\max}(\Sigma_{\mathbf u})\le C.$
\end{enumerate}
\end{assumption}

Assumption~\ref{Assumption:covariances} provides baseline conditions which are common in the factor model literature and ensure the data-generating process is well-behaved. It provides the independence and moment conditions that are used in the identification argument. Assumption~\ref{Assumption:covariances}(ii) yields the orthogonality relations between the latent factors and idiosyncratic errors, while Assumption~\ref{Assumption:covariances}(iii)--(v) control the conditional means and keep the relevant covariance matrices uniformly bounded and nondegenerate. Moreover, the independence from $A$ also implies that the proxy-error covariance matrices are the same in both treatment groups.

\begin{assumption}[Model Parameters]\label{Assumption:pervasiveness}
There exist universal constants $0<c<C<\infty$ such that, for $a\in\{0,1\}$, the following conditions hold.
\begin{itemize}
\item[(i)] The structural parameters are uniformly bounded, $\max\{|\alpha_a|,\|\delta_a\|_2,\|\beta_a\|_2,\|\gamma_a\|_2,\|\mathbf B_a\|_{\mathrm{op}}\}\le C.$
\item[(ii)] The loading matrices satisfy $$c\le\lambda_{\min}\left(\frac{\mathbf\Gamma_X^\top\mathbf\Gamma_X}{p}\right)\le\lambda_{\max}\left(\frac{\mathbf\Gamma_X^\top\mathbf\Gamma_X}{p}\right)\le C,$$
and $$c\le\lambda_{\min}\left(\frac{\mathbf\Gamma_M^\top\mathbf\Gamma_M}{q}\right)\le\lambda_{\max}\left(\frac{\mathbf\Gamma_M^\top\mathbf\Gamma_M}{q}\right)\le C.$$
\end{itemize}
\end{assumption}

Assumption~\ref{Assumption:pervasiveness} (i) requires the structural parameters to remain uniformly bounded, which rules out models in which the strength of the structural relationships diverges with the dimensions of the observed surrogates and allows the approximation error to be controlled uniformly in $p$ and $q$. Assumption~\ref{Assumption:pervasiveness} (ii) requires the latent factors to be pervasive, in the sense that their loadings do not become asymptotically sparse as $p,q\to\infty$. This ensures that each factor contributes a non-vanishing amount of variation to $X$ and $M$. A sufficient condition for Assumption~\ref{Assumption:pervasiveness}(ii) is that the rows of $\Gamma_X$ and $\Gamma_M$ are independently generated from common distributions with finite second moments and positive definite second-moment matrices. Then, by SLLN, $\mathbf\Gamma_X^\top\mathbf\Gamma_X/p$ and $\mathbf\Gamma_M^\top\mathbf\Gamma_M/q$ converge to fixed matrices, so that the eigenvalue bounds in Assumption~\ref{Assumption:pervasiveness}(ii) hold. This condition is widely used in literature \citep{stock2002forecasting,bai2003inferential,fan2013large} and see Example 2.1 in \cite{fan2022learning} for more discussion.

To exploit the high-dimensional structure, we introduce projection matrices that map the observed variables onto a lower-dimensional space while preserving information about the latent factors.

\begin{definition}[Full-Projection Matrices]\label{definition:H}
The matrices $\mathbf W_X\in\mathbb R^{p\times\tilde p}$ and $\mathbf{W}_M\in\mathbb R^{q\times\tilde q}$ are called \textit{full-projection matrices} with $\mathbf{W}_X =\sqrt{p} V_X$ and $\mathbf{W}_M = \sqrt{q}V_M$ where $V_X\in\mathbb R^{p\times\tilde p}$ and $V_M\in\mathbb R^{q\times\tilde q}$ have minimum and maximum bounded eigenvalues, i.e., $c\le \lambda_{\min}(V_X^\top V_X) \le\lambda_{\max}(V_X^\top V_X)\le C$ and $c \le \lambda_{\min}(V_M^\top V_M)\le \lambda_{\max}(V_M^\top V_M)\le C$ for some constant $C$, if the resulting matrices
\begin{equation}\label{eq:fullproj}
\textstyle
    \mathbf{H}_X = \frac{1}{p}\mathbf{W}_X^\top \Gamma_X; \quad \mathbf{H}_M = \frac{1}{q} \mathbf{W}_M^\top \Gamma_M
\end{equation}
have full column rank.
\end{definition} 
In Section \ref{section:4algorithm}, we will provide a specific choice of full-projection matrices. We next show by leveraging full-projection matrices, the parameter $\theta_0$ is (approximately) identifiable with fixed $p$ and $q$. Define
\begin{align*}\label{eq:theta0p}
\textstyle
\theta_{0,p,q}&=e_1^\top \mathbb{E}[L L^\top \mid A = 1]^{-1}\mathbb{E}[L Y \mid A = 1]  \\
&\quad+ \left(e_1^\top \mathbb{E}[Q Q^\top \mid A = 0]^{-1}\mathbb{E}[Q N^\top \mid A = 0]\right)\left(\mathbf S_{\tilde q} \mathbb{E}[L L^\top \mid A = 1]^{-1}\mathbb{E}[L Y \mid A = 1]\right).
\end{align*}
where $L=\bigl(1,\,p^{-1}X^\top\mathbf{W}_X,\,q^{-1}M^\top\mathbf{W}_M\bigr)^\top$, $Q=\bigl(1,\,p^{-1}X^\top\mathbf{W}_X\bigr)^\top$ and $N=q^{-1}\mathbf{W}_M^\top M$. Here, \(\mathbf S_{\tilde q}=\begin{bmatrix}0_{\tilde q\times(1+\tilde p)} & I_{\tilde q} \end{bmatrix} \in\mathbb R^{\tilde q\times(1+\tilde p+\tilde q)}\) be the selection matrix that selects last $\tilde q$ elements. 
\begin{theorem}
\label{thm:id}
Suppose Assumptions \ref{ass:consistency}--\ref{assumption:Sequential} and \ref{Assumption:covariances}--\ref{Assumption:pervasiveness} hold. Let $\mathbf{W}_X$ and $\mathbf{W}_M$ be any full-projection matrices, and $\theta_0$ be defined as in Equation \eqref{eq:identification_1}. Then, there exists a universal constant $C > 0$ independent of $p$ and $q$, such that
\begin{equation*}
|\theta_0 - \theta_{0,p,q}|\;\le\;C \left(\frac{1}{p \, \nu_{\min}^2(\mathbf H_X)}+\frac{1}{q \, \nu_{\min}^2(\mathbf H_M)}\right)\left(1+ \frac{1}{p \, \nu_{\min}^2(\mathbf H_X)}\right).
\end{equation*}
If, in addition, $p \nu_{\min}^2(\mathbf H_X) \uparrow \infty$ and $q \nu_{\min}^2(\mathbf H_M) \uparrow \infty$ as $p, q \to \infty$, then $\theta_{0,p,q} \to \theta_0$, i.e., $\theta_0$ is point identifiable as $p,q \to \infty$. 
\end{theorem}

\begin{remark}
    The theorem assumes that both $X$ and $M$ are high-dimensional. A similar argument can be developed when either $X$ or $M$ is treated as fixed (i.e., no factor structure is imposed), we can replace $\mathbf f_X$ with $X$ in the model and identification can still be achieved through the remaining high-dimensional component. For instance, if $X$ is fixed, then \(|\theta_0 - \theta_{0,q}| \le C \left(q^{-1}\nu_{\min}^{-2}(\mathbf H_M)\right).\) An analogous result holds when $M$ is fixed.
\end{remark}

Theorem~\ref{thm:id} provides an upper bound for the approximation error. As a consequence, it yields identification of $\theta_0$ when $p$ and $q$ grow large. To the best of our knowledge, identification results of this type, have not been explicitly established in this form in the literature, including parameters such as the average treatment effect. We formalize this in the Corollary \ref{cor:ate} where we define $\theta_{\text{ATE},p,q}$ as follows:
\begin{equation*}
\begin{aligned}
\textstyle
        \theta_{\text{ATE},p,q}&= \bigg\{e_1^\top \mathbb{E}[L L^\top \mid A = 1]^{-1}\mathbb{E}[L Y \mid A = 1]  \\&\quad+ \left(e_1^\top \mathbb{E}[Q Q^\top \mid A = 1]^{-1}\mathbb{E}[Q N^\top \mid A = 1]\right)\left(\mathbf S_{\tilde q} \mathbb{E}[L L^\top \mid A = 1]^{-1}\mathbb{E}[L Y \mid A = 1]\right)\bigg\} \\
        & ~~~ - \bigg\{e_1^\top \mathbb{E}[L L^\top \mid A = 0]^{-1}\mathbb{E}[L Y \mid A = 0]  \\&\quad+ \left(e_1^\top \mathbb{E}[Q Q^\top \mid A = 0]^{-1}\mathbb{E}[Q N^\top \mid A = 0]\right)\left(\mathbf S_{\tilde q}\mathbb{E}[L L^\top \mid A = 0]^{-1}\mathbb{E}[L Y \mid A = 0]\right)\bigg\}
    \end{aligned}
\end{equation*}

\begin{corollary}\label{cor:ate}
Consider the structural models in \eqref{eq:factor}--\eqref{eq:factors}. Suppose Assumptions \ref{ass:consistency}--\ref{assumption:Sequential} and \ref{Assumption:covariances}--\ref{Assumption:pervasiveness} hold. Define \(\theta_{\mathrm{ATE}} := (\alpha_1+\gamma_1^\top\delta_1)-(\alpha_0+\gamma_0^\top \delta_0). \) Then, $\theta_{\mathrm{ATE}}$ satisfies the conditions of Theorem~\ref{thm:id}, there exists a universal constant $C>0$, independent of $p$ and $q$, such that
\begin{equation*}
    \textstyle    \bigl|\theta_{\mathrm{ATE}}-\theta_{\mathrm{ATE},p,q}\bigr| \le C\left(\frac{1}{p\,\nu_{\min}^2(\mathbf H_X)}+ \frac{1}{q\,\nu_{\min}^2(\mathbf H_M)}\right)\left(1+\frac{1}{p\,\nu_{\min}^2(\mathbf H_X)}\right).
\end{equation*}
Moreover, if \(p\,\nu_{\min}^2(\mathbf H_X)\to\infty \text{ and } q\,\nu_{\min}^2(\mathbf H_M)\to\infty,\) then \(\theta_{\mathrm{ATE},p,q}\to \theta_{\mathrm{ATE}},\) and hence $\theta_{\mathrm{ATE}}$ is point identifiable as $p,q\to\infty$.
\end{corollary}
\section{FARMA Algorithm for Mediation Analysis}
\label{section:4algorithm}

In this section, we present FARMA for estimating the mediation functional $\theta_0=\alpha_1+\gamma_1^\top\delta_0$. The procedure constructs projection directions on one subsample and fits two low-dimensional regressions on the other, then interchanges the subsamples and averages the resulting estimates. Let $\tilde p$ and $\tilde q$ denote the working dimensions of the covariate and mediator projections, respectively.

\begin{algorithm}[H]
\caption{Cross-fitted FARMA estimator of the mediation functional}
\label{algo:FARMA}
\small
\begin{algorithmic}
\setlength{\itemsep}{0.5em}
\State \textbf{Step 0:} Randomly partition the sample indices into two approximately equal folds, $\mathcal D_1$ and $\mathcal D_2$.

\State \textbf{Step 1:} Let $\hat\Sigma_X$ and $\hat\Sigma_M$ be the sample covariance matrices of $X_i$ and $M_i$ computed on $\mathcal D_1$. Form $\hat{\mathbf V}_X\in\mathbb R^{p\times\tilde p}$ and $\hat{\mathbf V}_M\in\mathbb R^{q\times\tilde q}$ from their leading orthonormal eigenvectors, and set $\mathbf W_X=\sqrt p\,\hat{\mathbf V}_X$ and $\mathbf W_M=\sqrt q\,\hat{\mathbf V}_M$.

\State \textbf{Step 2:} For each $i\in\mathcal D_2$, compute the projected factors
\[\textstyle\hat{\mathbf f}_{X,i}=p^{-1}\mathbf W_X^\top X_i,\quad\hat{\mathbf f}_{M,i}=q^{-1}\mathbf W_M^\top M_i,\]
and define $\hat Z_i=(1,\hat{\mathbf f}_{X,i}^\top,\hat{\mathbf f}_{M,i}^\top)^\top$ and $\hat R_i=(1,\hat{\mathbf f}_{X,i}^\top)^\top$.

\State \textbf{Step 3:} Let $\mathcal I_a=\{i\in\mathcal D_2:A_i=a\}$ for $a\in\{0,1\}$. Regress $Y_i$ on $\hat Z_i$ by ordinary least squares using $i\in\mathcal I_1$. Denote the fitted intercept and mediator-score coefficient by $\hat\alpha_1$ and $\hat\gamma_1\in\mathbb R^{\tilde q}$, respectively.

\State \textbf{Step 4:} Let $e_1=(1,0,\ldots,0)^\top\in\mathbb R^{1+\tilde p}$. Regress $\hat{\mathbf f}_{M,i}^\top\hat\gamma_1$ on $\hat R_i$ using $i\in\mathcal I_0$, and define the fitted intercept as
\[\textstyle\widehat{\gamma_1^\top\delta_0}=e_1^\top\bigl(\sum_{i\in\mathcal I_0}\hat R_i\hat R_i^\top\bigr)^{-1}\sum_{i\in\mathcal I_0}\hat R_i\bigl(\hat{\mathbf f}_{M,i}^\top\hat\gamma_1\bigr).\]

\State \textbf{Step 5:} Set $\hat\theta_0^{(1)}=\hat\alpha_1+\widehat{\gamma_1^\top\delta_0}$. Interchange $\mathcal D_1$ and $\mathcal D_2$ and repeat Steps 1--4 to obtain $\hat\theta_0^{(2)}$.

\State \textbf{Return:} $\hat\theta_0^{\mathrm{cf}}=\tfrac12\bigl(\hat\theta_0^{(1)}+\hat\theta_0^{(2)}\bigr)$.
\end{algorithmic}
\end{algorithm}

In Step 1, we use $\mathcal D_1$ to construct the projection matrices from the leading eigenvectors of the sample covariance matrices. In Step 2, we apply these directions to the surrogates in $\mathcal D_2$ to obtain the projected covariate and mediator scores. This separation allows the regression observations to be analyzed conditional on the projection directions, while retaining the factor-space information needed for estimation. 

Steps 3 and 4 implement the two regression stages underlying the mediation functional, a treated outcome regression and a control bridge regression. In Step 3, the treated outcome regression yields $\hat\alpha_1$ and $\hat\gamma_1$.

\begin{figure}[http]
\centering
\scalebox{0.84}{%
\begin{tikzpicture}[
  >={Stealth[length=2.2mm]},
  font=\small,
  d1/.style={rectangle, rounded corners=6pt, draw=d1line, line width=0.7pt, align=center, inner sep=6pt, minimum height=1.7cm, text width=3.5cm},
  d2/.style={rectangle, rounded corners=6pt, draw=red, line width=0.7pt,
     fill=white, align=center, inner sep=6pt, minimum height=1.7cm, text width=3.5cm},
  eqbox/.style={rectangle, draw=d1line, line width=0.6pt, fill=white,
     align=center, inner sep=4pt, minimum height=0.7cm, text width=2.3cm},
  ellip/.style={ellipse, draw=red, line width=0.7pt, fill=white,
     align=center, inner sep=2pt, minimum width=4.6cm, minimum height=1.5cm},
  est/.style={circle, draw=black, line width=0.8pt, fill=white,
     text=estline, inner sep=1pt, minimum size=0.95cm},
  outbox/.style={rectangle, rounded corners=6pt, draw=outline, fill=white, line width=0.7pt,
      align=center, inner sep=6pt, minimum height=1.4cm, text width=3.7cm},
  arr/.style={->, thick, rounded corners=3pt, draw=outline},
  line/.style={-, thick, rounded corners=3pt, draw=outline},
]

\node[d1] (s1) at (0,7.4)
  {\textit{Step 1: Compute} $\widehat{\Sigma}_X,\ \widehat{\Sigma}_M$ \textit{from} $\mathfrak{D}_1$; \textit{obtain} $\widehat{\mathbf V}_X,\ \widehat{\mathbf V}_M$};

\node[eqbox] (wm) at (-1.35,5.4) {$\mathbf W_X = p^{1/2}\widehat{\mathbf V}_X$};
\node[eqbox] (wx) at ( 1.35,5.4) {$\mathbf W_M = q^{1/2}\widehat{\mathbf V}_M$};

\node[d2] (s2) at (0,3.5)
  {\textit{Step 2: Estimate factors via diversified projection} $(\mathfrak{D}_2)$};

\node[ellip] (fh) at (0,1.3)
  {$\widehat{\mathbf f}_{X,i} = p^{-1}\mathbf W_X^{\top}X_i$\\[3pt] $\widehat{\mathbf f}_{M,i} = q^{-1}\mathbf W_M^{\top}M_i$};

\node[d2] (s3) at (6.2,1.3)
  {\textit{Step 3: Regress} $Y_i$ \textit{on} $(1,\widehat{\mathbf f}_{X,i},\widehat{\mathbf f}_{M,i})$, \textit{treated} $(\mathfrak{D}_2)$};

\node[d2] (s4) at (6.2,7.4)
  {\textit{Step 4: Regress} $\widehat{\mathbf f}_{M,i}^{\top}\widehat{\gamma}_1$ \textit{on} $(1,\widehat{\mathbf f}_{X,i})$, \textit{control} $(\mathfrak{D}_2)$};

\node[est] (g1) at (5.3,3.4) {\textcolor{black}{$\widehat{\gamma}_1$}};
\node[est] (a1) at (7.1,3.4) {\textcolor{black}{$\widehat{\alpha}_1$}};
\node[est] (gd) at (6.2,4.9) {\textcolor{black}{$\widehat{\gamma_1^{\top}\delta_0}$}};

\node[outbox] (s5) at (11.7,6.4)
  {\textit{Step 5: Set}\\[2pt] $\widehat{\theta}_0^{(1)} = \widehat{\alpha}_1 + \widehat{\gamma_1^{\top}\delta_0}$};

\node[outbox] (rep) at (11.7,4.2)
  {\textit{Repeat with} $\mathfrak{D}_1 \leftrightarrow \mathfrak{D}_2$\\[2pt] \textit{to obtain} $\widehat{\theta}_0^{(2)}$};

\node[outbox, line width=0.9pt] (cf) at (11.7,2.0)
  {\textit{Final Estimator} \\[2 pt] $\widehat{\theta}_0^{\,\text{cf}} = \tfrac{1}{2}\bigl(\widehat{\theta}_0^{(1)} + \widehat{\theta}_0^{(2)}\bigr)$};

\draw[arr] (s1.south -| wm) -- (wm.north);
\draw[arr] (s1.south -| wx) -- (wx.north);
\draw[arr] (wm.south) -- (wm.south |- s2.north);
\draw[arr] (wx.south) -- (wx.south |- s2.north);
\draw[arr] (s2.south) -- (fh.north);

\draw[arr] (fh.east) -- (fh.east -| s3.west);

\coordinate (b) at (3.6,1.3);
\draw[line] (fh.east) -- (b);
\draw[line] (b) -- (3.6,7.4);
\draw[arr]  (3.6,7.4) -- (s4.west);

\draw[arr] (s3.north -| g1) -- (g1.south);
\draw[arr] (s3.north -| a1) -- (a1.south);
\draw[arr] (g1.north) -- (s4.south -| g1);
\draw[arr] (s4.south) -- (gd.north);
\coordinate (join) at (9,5.6);   
\draw[line] (gd.east) -- (gd.east -| join) -- (join);
\draw[line] (a1.east) -- (a1.east -| join) -- (join);
\draw[arr] (join) |- ($(s5.south west)!0.7!(s5.north west)$);
\draw[arr] (s5.south) -- (rep.north);
\draw[arr] (rep.south) -- (cf.north);

\end{tikzpicture}
}
\caption{\footnotesize Schematic overview of FARMA. The estimates $\widehat{\alpha}_1$ and $\widehat{\gamma_1^{\top}\delta_0}$ combine into $\widehat{\theta}_0$, and interchanging the subsets gives the cross-fitted $\widehat{\theta}_0^{\,\text{cf}}$.}
\end{figure}
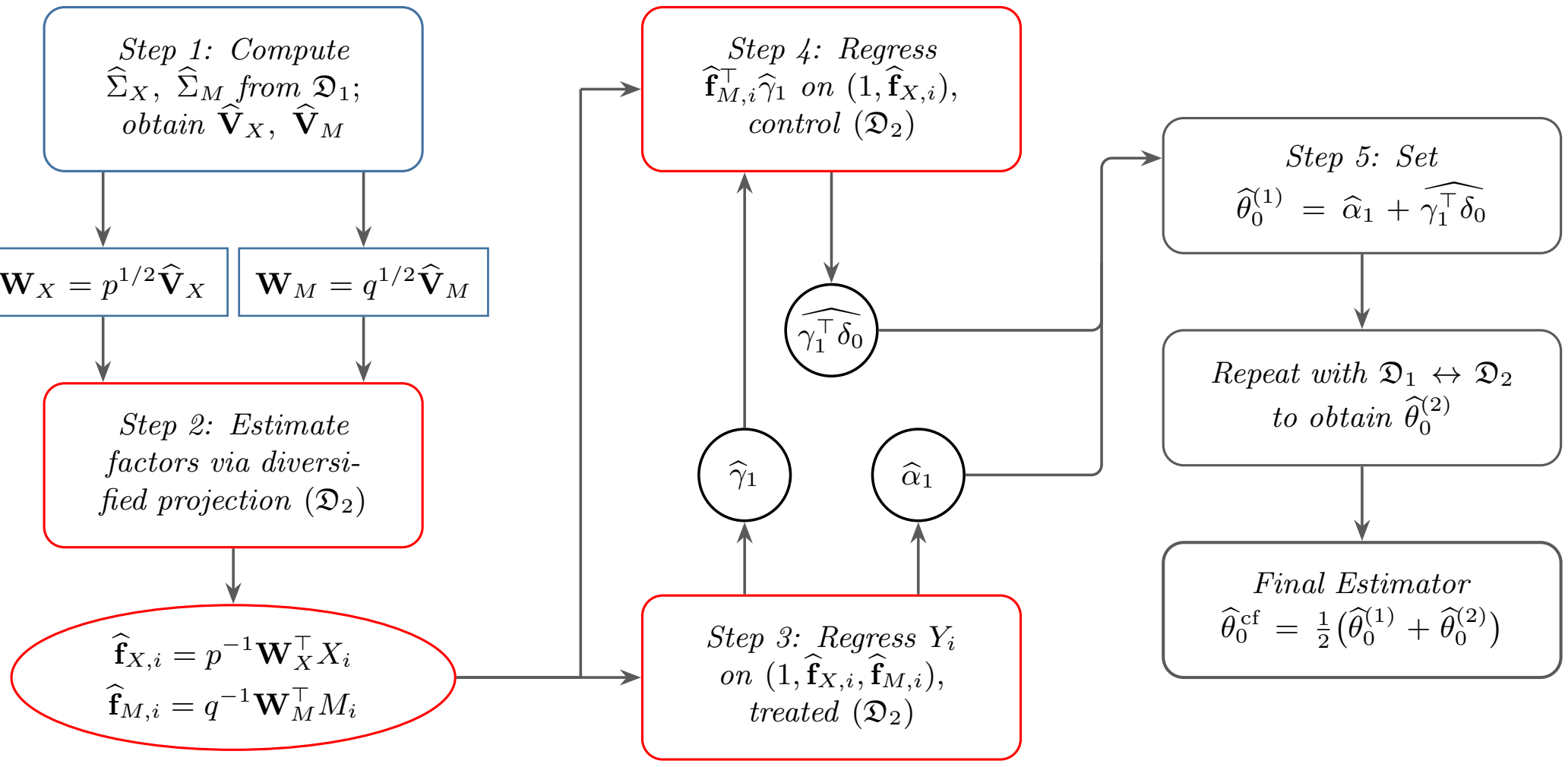

In Step 4, the bridge regression uses $\hat\gamma_1$ to form the scalar response $\hat{\mathbf f}_{M,i}^\top\hat\gamma_1$ and regresses it on the projected covariates among control observations. The fitted intercept, denoted by $\widehat{\gamma_1^\top\delta_0}$, estimates the contribution of the control mediator relation to the treated outcome.

In Step 5, we add the two intercepts to obtain $\hat\theta_0^{(1)}$, repeat the construction with the folds interchanged, and average the two estimates. Thus, every observation contributes to the regression steps in one split, and the procedure only requires recovery of the relevant factor spaces.

\medskip
\noindent\textbf{Natural direct and indirect effects.}
Under the centering convention $\mathbb E(\mathbf f_X)=0$, the potential outcomes in \eqref{eq:farma-ordinary-means} satisfy $\theta_{aa}=\alpha_a+\gamma_a^\top\delta_a$ for $a\in\{0,1\}$. Thus, the effects defined in \eqref{eq:farma-natural-effects} have the structural representations as follows:
\[\begin{aligned}
\theta_{\mathrm{NDE}}&=\textstyle\alpha_1-\alpha_0+(\gamma_1-\gamma_0)^\top\delta_0,\\\theta_{\mathrm{NIE}}&=\textstyle\gamma_1^\top(\delta_1-\delta_0).
\end{aligned}\]
Algorithm~\ref{algo:FARMA-effects} estimates these expressions using the same projections and scalar-regression construction as Algorithm~\ref{algo:FARMA}. The projected covariates remain in each regression, since the marginal centering condition does not imply $\mathbb E[\mathbf f_X\mid A=a]=0$. In Step 1 of Algorithm~\ref{algo:FARMA-effects}, we reuse the treated outcome fit and estimate the corresponding control outcome coefficients. In Step 2, the two within-arm mediator regressions provide $\widehat{\gamma_0^\top\delta_0}$ and $\widehat{\gamma_1^\top\delta_1}$. Step 3 combines these with $\widehat{\gamma_1^\top\delta_0}$ from Algorithm~\ref{algo:FARMA} to estimate the two effects. Finally, Step 4 interchanges the folds, and the two estimates of each effect are averaged.

\begin{algorithm}[t]
\caption{Cross-fitted FARMA estimators of the NDE and NIE}
\label{algo:FARMA-effects}
\small
\begin{algorithmic}
\setlength{\itemsep}{0.5em}

\State \textbf{Step 1:} Reuse the projected factors, design vectors, treatment groups, and fitted quantities from the first split of Algorithm~\ref{algo:FARMA}. Regress $Y_i$ on $\hat Z_i$ using $i\in\mathcal I_0$, and denote the fitted intercept and mediator-score coefficient by $\hat\alpha_0$ and $\hat\gamma_0\in\mathbb R^{\tilde q}$, respectively.

\State \textbf{Step 2:} For each $a\in\{0,1\}$, regress $\hat{\mathbf f}_{M,i}^\top\hat\gamma_a$ on $\hat R_i$ using $i\in\mathcal I_a$, and define the fitted intercept as
\[\textstyle\widehat{\gamma_a^\top\delta_a}=e_1^\top\bigl(\sum_{i\in\mathcal I_a}\hat R_i\hat R_i^\top\bigr)^{-1}
\sum_{i\in\mathcal I_a}\hat R_i\bigl(\hat{\mathbf f}_{M,i}^\top\hat\gamma_a\bigr).\]

\State \textbf{Step 3:} Form the fold-specific estimates
\[\begin{aligned}
\hat\theta_{\mathrm{NDE}}^{(1)}
&=\textstyle\hat\alpha_1-\hat\alpha_0+\widehat{\gamma_1^\top\delta_0}-\widehat{\gamma_0^\top\delta_0},\\\hat\theta_{\mathrm{NIE}}^{(1)}
&=\textstyle\widehat{\gamma_1^\top\delta_1}-\widehat{\gamma_1^\top\delta_0}.
\end{aligned}\]

\State \textbf{Step 4:} Interchange $\mathcal D_1$ and $\mathcal D_2$ and repeat Steps 1--3 using the corresponding fitted quantities from Algorithm~\ref{algo:FARMA}, yielding $\hat\theta_{\mathrm{NDE}}^{(2)}$ and $\hat\theta_{\mathrm{NIE}}^{(2)}$.

\State \textbf{Return:}
\[\textstyle\hat\theta_{\mathrm{NDE}}^{\mathrm{cf}}=\tfrac12\bigl(\hat\theta_{\mathrm{NDE}}^{(1)}+\hat\theta_{\mathrm{NDE}}^{(2)}\bigr),\quad\hat\theta_{\mathrm{NIE}}^{\mathrm{cf}}=\tfrac12\bigl(\hat\theta_{\mathrm{NIE}}^{(1)}+\hat\theta_{\mathrm{NIE}}^{(2)}\bigr).\]
\end{algorithmic}
\end{algorithm}

\begin{remark}[Cross-fitting]
The two-fold construction extends to any fixed number of folds $K\ge2$. For each fold, construct the projection directions from the remaining observations, run the regressions on that fold, and average the resulting $K$ fold-specific estimates of each target.
\end{remark}
\section{Theoretical Analysis}\label{section:5theory}

In this section, we study the asymptotic behavior of the cross-fitted FARMA estimator $\widehat\theta_0^{\mathrm{cf}}$ of $\theta_0=\alpha_1+\gamma_1^\top\delta_0$ and establish its $\sqrt{n}$-consistency and asymptotic normality. Furthermore, we also present a consistent estimator of the limiting variance, which aids in inference. Recall that, we allow the working factor dimensions to be anything larger than the true dimension, i.e., $\tilde p \ge \bar p$ and $\tilde q \ge \bar q$. The limiting variance does not depend on $(\tilde p, \tilde q)$, so the length of asymptotic confidence interval is independent of the choice of the working dimension as soon as it is larger than the true dimension. 

The main challenge is to control the contribution of the projected measurement errors to the two regression stages, including the additional directions introduced by overestimating the factor dimensions.

Let $\pi_1=\mathbb{P}(A_i=1)$ and $\pi_0=1-\pi_1$ denote the fixed treatment probabilities, with $0<\pi_1,\pi_0<1$ (Assumption \ref{assump:subgaussian}(iii)). We consider two approximately equal folds, so the treated and control sample sizes in each fold are of order $n$. For a given regression fold, the projection matrices are estimated using the other fold; we suppress this fold dependence in the notation. Under the measurement model \eqref{eq:factors}, the projected factors satisfy
\begin{equation}
\label{eq:estimation}
\begin{aligned}
\widehat{\mathbf f}_{X,i} &= \frac1p \mathbf  W_X^\top X_i = \underbrace{\left(\frac1p \mathbf W_X^\top \mathbf \Gamma_X\right)}_{=:\mathbf H_X}\mathbf f_{X,i} + \underbrace{\frac1p \mathbf  W_X^\top \mathbf u_{X,i}}_{=:\mathbf e_{X,i}},
\\
\widehat{\mathbf f}_{M,i}  & = \frac1q \mathbf W_M^\top M_i = \underbrace{\left(\frac1q \mathbf W_M^\top \mathbf \Gamma_M\right)}_{=:\mathbf H_M}\mathbf f_{M,i} + \underbrace{\frac1q \mathbf  W_M^\top \mathbf u_{M,i}}_{=:\mathbf e_{M,i}}.
\end{aligned}
\end{equation}
Here, $\mathbf  H_X\in\mathbb R^{\widetilde p\times\bar p}$ and $\mathbf H_M\in\mathbb R^{\widetilde q\times\bar q}$ map the latent factors into the space of working dimension, while \(\mathbf e_{X,i}\) and \(\mathbf e_{M,i}\) collect the projected idiosyncratic errors. We first state the regularity conditions used in the asymptotic analysis.

\begin{assumption}[Sub-Gaussian tails and nondegeneracy]\label{assump:subgaussian}
There exist constants $C<\infty$ and $\underline{\sigma}^2>0$, independent of $(n,p,q)$, such that, for $a\in\{0,1\}$ and $i=1,\ldots,n$, the following hold.
\begin{enumerate}
    \item[(i)] The random vectors $\mathbf f_{X,i}$, $\mathbf u_{X,i}$, $\mathbf u_{M,i}$, and $\mathbf u_{XM,a,i}$, and the scalar error $\epsilon_{a,i}$, are sub-Gaussian, with sub-Gaussian norms uniformly bounded by $C$.
    \item[(ii)] The outcome error variance is uniformly bounded away from zero: $\operatorname{Var}(\epsilon_{a,i})\geq\underline{\sigma}^2>0$.
    \item[(iii)] $\pi_1$ is uniformly bounded away from 0 and 1.
\end{enumerate}
\end{assumption}

Assumption \ref{assump:subgaussian}(i) is a commonly used tail condition in high-dimensional analysis and simplifies the concentration arguments for sample covariances and regression terms. It includes Gaussian and many non-Gaussian distributions with sufficiently light tails. This condition could be relaxed under suitable moment and growth conditions using an appropriate truncation-based argument, but we do not pursue this extension here. 
Assumption \ref{assump:subgaussian}(ii) is a mild nondegeneracy condition that prevents the outcome error variance from vanishing as $n,p,q$ grow. Together with the remaining assumptions, it ensures that the limiting variance is positive, so the usual normal approximation can be used for inference. Assumption~\ref{assump:subgaussian}(iii) ensures that both treatment groups contain a nonvanishing proportion of the sample.

\begin{assumption}[Dimensions and loadings]\label{assump:growth}
There exists a constant $C<\infty$, independent of $n,p,q$, such that the following hold.
\begin{enumerate}
    \item[(i)] The loading rows are uniformly bounded, $\max_{j\leq p}\|\mathbf\Gamma_{X,j\cdot}\|_2\leq C$ and $\max_{j\leq q}\|\mathbf \Gamma_{M,j\cdot}\|_2\leq C$.
    \item[(ii)] The working dimensions $\widetilde p\geq\bar p$ and $\widetilde q\geq\bar q$ are fixed, and
    $$
    \frac{\log(p\vee q)}{n}\longrightarrow0, \qquad \frac{p\wedge q}{n}\longrightarrow\infty \quad\text{as }n,p,q\longrightarrow\infty.
    $$
\end{enumerate}
\end{assumption}

Assumption \ref{assump:growth}(i) is a standard bounded-loading condition in factor models. It prevents any individual surrogate from having an increasingly large loading as $p$ and $q$ grow. Together with the pervasiveness condition in Assumption~\ref{Assumption:pervasiveness}(ii), it ensures that information about each latent factor is spread across many surrogate measurements. Assumption \ref{assump:growth}(ii) allows the fixed working dimensions to exceed the true dimensions, so the true factor dimensions need not be known exactly. 
The condition $\log(p\vee q)/n\to0$ is mild and allows $p$ and $q$ to be much larger than $n$, including any fixed polynomial growth. The additional condition $(p\wedge q)/n\to\infty$ requires the number of surrogates to grow faster than the sample size. This is sufficient condition for drawing inference at $\sqrt{n}$ rate. This condition can potentially be relaxed to $(p \wedge q)/\sqrt{n} \to \infty$, but we leave this for future research. We are now in a position to state our main results. First, we show that the projection matrices constructed in Algorithm~1 satisfy the full-projection condition in Definition~3.1 with probability tending to one: 

\begin{proposition}\label{prop:loading-cond}
Let $\widehat {\mathbf V}_X\in\mathbb{R}^{p\times\widetilde p}$ and $\widehat{\mathbf V}_M\in\mathbb{R}^{q\times\widetilde q}$ have orthonormal columns corresponding to the leading $\widetilde p$ and $\widetilde q$ eigenvectors of the sample covariance matrices $\widehat\Sigma_X$ and $\widehat\Sigma_M$, respectively, computed on the training fold. Take $\mathbf W_X=\sqrt{p}\,\widehat V_X$ and $\mathbf W_M=\sqrt{q}\,\widehat V_M$, and let $\mathbf H_X$ and $\mathbf H_M$ be defined as in Definition~\ref{definition:H}. Suppose Assumptions~\ref{Assumption:covariances}--\ref{Assumption:pervasiveness} and \ref{assump:subgaussian}--\ref{assump:growth} hold. Then there exist constants $0<c\leq C<\infty$ and $K_1>0$, independent of $n,p,q$, such that, for all sufficiently large $n$, with probability at least $1-n^{-K_1}$,
$$
c\leq\nu_{\min}(\mathbf H_X)\leq\nu_{\max}(\mathbf H_X)\leq C, \qquad c\leq\nu_{\min}(\mathbf H_M)\leq\nu_{\max}(\mathbf H_M)\leq C.
$$
\end{proposition}Proposition~\ref{prop:loading-cond} ensures that the projection preserves every true latent factor direction and that the projected factors remain well-conditioned with probability tending to one. When $\widetilde p>\bar p$, there are $\widetilde p-\bar p$ linearly independent combinations of $\widehat {\mathbf f}_X$ that contain no signal from the true latent factors $\mathbf f_X$. Similarly, when $\widetilde q>\bar q$,
there are $\widetilde q-\bar q$ such combinations of $\widehat {\mathbf f}_M$ containing no signal from $\mathbf f_M$. The variances of these combinations can be shown to be of order $p^{-1}$ and $q^{-1}$, respectively. Consequently, the Gram matrices constructed from the projected factors in the outcome and bridge regressions need not remain uniformly well conditioned. A key part of our analysis is controlling the contributions of these additional directions and showing that they are negligible in the first-order asymptotic expansion of $\widehat\theta_0^{\mathrm{cf}}$. The proof of Proposition~\ref{prop:loading-cond} is given in Appendix~\ref{app:loading-cond}.

Our next result is the asymptotic normality of the proposed estimator. To state the limiting variance, let $\mathbf f_{1,i}=(1,\mathbf f_{X,i}^\top,\mathbf f_{M,i}^\top)^\top$ and $\mathbf f_{X1,i}=(1,\mathbf f_{X,i}^\top)^\top$. Define the conditional second-moment matrices
$$
\Omega_1=\mathbb{E}[\mathbf f_{1,i}\mathbf f_{1,i}^\top\mid A_i=1], \qquad \Omega_{X0}=\mathbb{E}[\mathbf f_{X1,i}\mathbf f_{X1,i}^\top\mid A_i=0].
$$
These matrices correspond to the treated outcome regression and the control bridge regression, respectively. Let $a_1=(1,0_{\bar p}^\top,\delta_0^\top)^\top$, $e_{1,1+\bar p}=(1,0_{\bar p}^\top)^\top$, and $\eta_i=\gamma_1^\top u_{XM,0,i}$. Define
$$
\xi_{1,i}=a_1^\top\Omega_1^{-1}\mathbf f_{1,i}\epsilon_{1,i}, \qquad \xi_{2,i}=e_{\mathbf 1,1+\bar p}^\top\Omega_{X0}^{-1}\mathbf f_{X1,i}\eta_i.
$$
Here, $\xi_{1,i}$ and $\xi_{2,i}$ are the first-order terms from the treated outcome regression and the control bridge regression, respectively. 
\begin{theorem}[Asymptotic normality of $\widehat\theta_0^{\mathrm{cf}}$]\label{thm:clt}
Suppose Assumptions~\ref{ass:consistency}--\ref{assumption:Sequential}, \ref{Assumption:covariances}--\ref{Assumption:pervasiveness}, and \ref{assump:subgaussian}--\ref{assump:growth} hold, and let $\widehat\theta_0^{\mathrm{cf}}$ be the cross-fitted FARMA estimator defined in Algorithm~\ref{algo:FARMA}. Then
$$
\sqrt{n}\left(\widehat\theta_0^{\mathrm{cf}}-\theta_0\right) =\frac{1}{\sqrt{n}}\sum_{i=1}^n\left(\frac{A_i}{\pi_1}\xi_{1,i} +\frac{1-A_i}{\pi_0}\xi_{2,i}\right)+o_p(1).
$$
Consequently,
$$
\sqrt{n}\left(\widehat\theta_0^{\mathrm{cf}}-\theta_0\right) \xrightarrow{d}N(0,\sigma^2), \qquad \sigma^2=\frac{\mathbb{E}[\xi_{1,i}^2\mid A_i=1]}{\pi_1}+\frac{\mathbb{E}[\xi_{2,i}^2\mid A_i=0]}{\pi_0}\,.
$$
\end{theorem}
Theorem \ref{thm:clt} shows that using the projected factors does not affect the first-order limiting distribution of $\widehat\theta_0^{\text{cf}}$. The working dimensions $\widetilde p$ and $\widetilde q$ do not affect the limiting distribution as long as they do not grow with $n$. The estimator is $\sqrt{n}$-consistent, and its limiting variance remains \emph{oracle}, in the sense that it is the same as the asymptotic variance of the corresponding two-stage regression estimator had the true latent factors $(\mathbf f_X,\mathbf f_M)$ been observed exactly. The proof is given in Appendix~\ref{sec:main}.

For inference, we further require a consistent estimator of the limiting variance $\sigma^2$, which depends on unobserved latent factors and structural errors. We estimate it using the projected factors and regression residuals from Algorithm~1. Let $\widehat\pi_1=n^{-1}\sum_{i=1}^n\mathbf{1}(A_i=1)$ and $\widehat\pi_0=1-\widehat\pi_1$ denote the proportions of treated and control units in the full sample. For each observation, the estimated first-order terms are computed using the fitted quantities from its regression fold. For notational simplicity, define $\sigma_1^2 = \mathbb{E}[\xi_{1,i}^2\mid A_i=1]$ and $\sigma_2^2 = \mathbb{E}[\xi_{2,i}^2\mid A_i=0]$.  To estimate $\sigma_1^2$ and $\sigma_2^2$, we first replace $\epsilon_{1,i}$ and $\eta_i$ by the residuals $\widehat\epsilon_{1,i}$ and $\widehat\eta_i$ from the treated outcome regression in Step~3 and the control bridge regression in Step 4 of Algorithm 1, respectively.  Within each regression fold, the weights involving $\Omega_1^{-1}$ and $\Omega_{X0}^{-1}$ are estimated using the inverses of the empirical second-moment matrices of $\widehat{\mathbf f}_{1,i}$ among treated observations and $\widehat{\mathbf f}_{X1,i}$ among control observations. Overestimating the factor dimensions can make these matrices ill-conditioned, so we carefully control for the contribution of the additional directions. We then square the estimated first-order terms and average them separately over treated and control observations across both folds to obtain $\widehat\sigma_1^2$ and
$\widehat\sigma_2^2$, respectively. Appendix~\ref{sec:appendix-variance} gives the details of this construction and its justification.

\begin{lemma}[Consistency of the variance estimator]\label{lemma:ve}
Suppose the assumptions of Theorem~\ref{thm:clt} hold, and let
\[\widehat\sigma^2 =\frac{\widehat\sigma_1^2}{\widehat\pi_1}+\frac{\widehat\sigma_2^2}{\widehat\pi_0}.\]
Then $\widehat\sigma^2\xrightarrow{p}\sigma^2$.
\end{lemma}

Lemma \ref{lemma:ve} establishes consistency of the variance estimator even when the working dimensions exceed the true factor dimensions. The proof is given in Appendix~\ref{sec:appendix-variance}. Combining this lemma with Theorem \ref{thm:clt} yields
$$
\frac{\sqrt{n}\left(\widehat\theta_0^{\mathrm{cf}}-\theta_0\right)}{\widehat\sigma} \xrightarrow{d}N(0,1),
$$
where $\widehat\sigma=\sqrt{\widehat\sigma^2}$. Thus, an asymptotically valid $100(1-\alpha)\%$ confidence interval for $\theta_0$ is
$$
\left[\widehat\theta_0^{\mathrm{cf}} -z_{1-\alpha/2}\frac{\widehat\sigma}{\sqrt{n}}, \quad \widehat\theta_0^{\mathrm{cf}} +z_{1-\alpha/2}\frac{\widehat\sigma}{\sqrt{n}}\right],
$$
where $z_{1-\alpha/2}$ is the $(1-\alpha/2)$-quantile of the standard normal distribution.
\section{Simulation Studies}
\label{section:6simulation}

In this section, we design simulation experiments to evaluate the performance of our estimator ($\hat\theta_0^{\mathrm{cf}}$ as defined in Section~\ref{section:4algorithm}) under high-dimensional surrogates for the covariates and mediators, in terms of root mean squared error (RMSE), standard deviation (SD), confidence interval length, and coverage. We consider a data-generating process where both covariates and mediators are low-dimensional latent factors as stated in Section~\ref{section:2model}. The data is generated as follows:
\begin{enumerate}
    \item We generate a latent factor $\mathbf f_X \in \mathbb{R}^{\bar{p}}$ for each sample, with $\mathbf f_X \sim \mathcal{N}(0, \Sigma_X)$. The covariance matrix is constructed as $\Sigma_X = U \Lambda U^\top$, where $U \in \mathbb{R}^{\bar{p} \times \bar{p}}$ is an orthonormal matrix, and $\Lambda = \mathrm{diag}(\lambda_1,\dots,\lambda_{\bar{p}})$ is diagonal with $\lambda_j \sim \mathcal{U}(1, 2)$ for $1\leq j \leq \bar p$. Given covariates $\mathbf f_X$, the observed surrogates $X \in \mathbb{R}^p$ are generated as:
    \[
    X = \mathbf \Gamma_X \mathbf f_X + \mathbf u_X, \quad \mathbf u_X \overset{i.i.d.}{\sim} \mathcal{N}(0, \sigma^2 \mathbf I_p),
    \]
    where $\mathbf \Gamma_X \in \mathbb{R}^{p \times \bar{p}}$ has entries sampled from $\mathcal{U}(0.5, 1.5)$.
    
    \item The treatment assignment follows a logistic model:
    \[
    \mathbb{P}(A = 1 \mid \mathbf f_X) = \text{logit}^{-1}(\mathbf f_X^\top \xi), \quad \text{where } \xi \in \mathbb{R}^{\bar{p}} \text{ with } \xi_j \sim \mathcal{U}(0, 2).
    \]
    Then, $A \sim \mathrm{Bernoulli}(\mathbb{P}(A=1 \mid \mathbf f_X))$.
    
    \item The latent mediator $\mathbf f_M \in \mathbb{R}^{\bar{q}}$ is generated according to:
    \[ \mathbf f_M = (1-A)(\delta_0 + \mathbf B_0 \mathbf f_X + \mathbf u_{XM,0}) + A(\delta_1 + \mathbf B_1\mathbf f_X + \mathbf u_{XM,1}), \quad \mathbf u_{XM,0},\mathbf u_{XM,1} \sim \mathcal{N}(0, \Sigma_U),\]
    where $\Sigma_U = R_U \Lambda_U R_U^\top$, $R_U \in \mathbb{R}^{\bar q \times \bar q}$ is an orthonormal matrix, $\Lambda_U = \mathrm{diag}(\lambda_1,\ldots,\lambda_{\bar q})$ with $\lambda_j \overset{i.i.d.}{\sim} \mathcal U(1, 2)$, $\delta_0, \delta_1 \in \mathbb{R}^{\bar{q}}$ have entries drawn from $\mathcal{U}(0.5, 1.5)$, and $\mathbf B_0, \mathbf B_1 \in \mathbb{R}^{\bar{q} \times \bar{p}}$ have entries sampled from $\mathcal{U}(0.5, 1.5)$. Given mediators $\mathbf f_M$, we generate the observed surrogates $M \in \mathbb{R}^q$ by:
    \[ M = \mathbf \Gamma_M \mathbf f_M + \mathbf u_M, \quad \mathbf u_M \overset{i.i.d.}{\sim} \mathcal{N}(0, \sigma^2\mathbf I_q), \]
    where $\mathbf \Gamma_M \in \mathbb{R}^{q \times \bar{q}}$ has entries from $\mathcal{U}(0.5, 1.5)$.
    
    \item We generate response variable $Y$ from the model:
    \[Y = (1-A)(\alpha_0 + \beta_0^\top \mathbf f_X +\gamma_0^\top \mathbf f_M + \epsilon_0)+ A(\alpha_1 + \beta_1^\top \mathbf f_X + \gamma_1^\top \mathbf f_M + \epsilon_1), \quad \epsilon_0, \epsilon_1 \overset{i.i.d.}{\sim} \mathcal{N}(0, \sigma^2),\]
    where we set $\alpha_0 = \alpha_1 = 1$, and vectors $\beta_0, \beta_1 \in \mathbb{R}^{\bar{p}}$, $\gamma_0, \gamma_1 \in \mathbb{R}^{\bar{q}}$ are sampled from $\mathcal{U}(0.5, 1.5)$.
\end{enumerate}

We vary that the number of samples $n$ over $\{500, 1000, 1500, 2000, 2500\}$ and the dimensionality $p = q \in \{500, 1000, 2500, 5000\}$ to reflect different high-dimensional scenarios. For all simulations, we vary $\sigma \in \{0.1, 1\}$, and set $\bar{p} = \bar{q} = 3$ with $\tilde{p} = \tilde{q} = 7$. This choice of $\tilde p$ (resp.\ $\tilde q$) comes from the lowest-RMSE rule in large samples according to our sensitivity analysis (Appendix~\ref{sensitivity}). Our estimator ($\hat \theta_0^{\mathrm{cf}} = \hat \alpha_1 + \widehat{\gamma_1^\top \delta_0}$ as defined in Section~\ref{section:4algorithm}) is computed as in Algorithm~\ref{algo:FARMA}. The weight matrices in Step~1 (i.e., $\mathbf W_X$, $\mathbf W_M$) from principal component analysis are obtained via the R package \texttt{irlba}, and the OLS estimators in Steps~$3$--$4$ via the \texttt{lm} function in R. All results are averaged over $500$ replications.

\begin{table}[t]
\centering
\small
\begin{tabular}{cc cccc cccc}
\toprule
 &  & \multicolumn{4}{c}{$\sigma = 0.1$} & \multicolumn{4}{c}{$\sigma = 1$} \\
\cmidrule(lr){3-6} \cmidrule(lr){7-10}
$p = q$ & $n$ & RMSE & SD & CI Length & Coverage & RMSE & SD & CI Length & Coverage \\
\midrule
\multirow{5}{*}{500} & 500 & 0.2068 & 0.2065 & 0.8094 & 0.950 & 0.2291 & 0.2294 & 0.8991 & 0.958 \\
 & 1000 & 0.1451 & 0.1452 & 0.5692 & 0.944 & 0.1461 & 0.1462 & 0.5732 & 0.956 \\
 & 1500 & 0.1154 & 0.1153 & 0.4520 & 0.952 & 0.1239 & 0.1236 & 0.4846 & 0.960 \\
 & 2000 & 0.1009 & 0.1001 & 0.3925 & 0.964 & 0.1018 & 0.1017 & 0.3986 & 0.942 \\
 & 2500 & 0.0830 & 0.0831 & 0.3256 & 0.956 & 0.0927 & 0.0918 & 0.3597 & 0.946 \\
\midrule
\multirow{5}{*}{1000} & 500 & 0.1642 & 0.1639 & 0.6424 & 0.952 & 0.1348 & 0.1346 & 0.5278 & 0.958 \\
 & 1000 & 0.1109 & 0.1110 & 0.4350 & 0.954 & 0.0902 & 0.0903 & 0.3538 & 0.946 \\
 & 1500 & 0.0937 & 0.0938 & 0.3677 & 0.946 & 0.0783 & 0.0784 & 0.3072 & 0.954 \\
 & 2000 & 0.0793 & 0.0793 & 0.3110 & 0.950 & 0.0632 & 0.0632 & 0.2476 & 0.952 \\
 & 2500 & 0.0703 & 0.0703 & 0.2755 & 0.948 & 0.0574 & 0.0572 & 0.2241 & 0.946 \\
\midrule
\multirow{5}{*}{2500} & 500 & 0.1754 & 0.1751 & 0.6865 & 0.954 & 0.2526 & 0.2527 & 0.9907 & 0.954 \\
 & 1000 & 0.1200 & 0.1201 & 0.4709 & 0.956 & 0.1895 & 0.1893 & 0.7420 & 0.952 \\
 & 1500 & 0.1039 & 0.1040 & 0.4077 & 0.944 & 0.1462 & 0.1463 & 0.5736 & 0.956 \\
 & 2000 & 0.0810 & 0.0811 & 0.3179 & 0.954 & 0.1242 & 0.1244 & 0.4875 & 0.952 \\
 & 2500 & 0.0764 & 0.0765 & 0.2997 & 0.938 & 0.1159 & 0.1161 & 0.4549 & 0.956 \\
\midrule
\multirow{5}{*}{5000} & 500 & 0.1700 & 0.1701 & 0.6670 & 0.934 & 0.2072 & 0.2074 & 0.8129 & 0.940 \\
 & 1000 & 0.1223 & 0.1223 & 0.4793 & 0.956 & 0.1350 & 0.1352 & 0.5298 & 0.952 \\
 & 1500 & 0.0989 & 0.0989 & 0.3878 & 0.954 & 0.1095 & 0.1089 & 0.4269 & 0.946 \\
 & 2000 & 0.0856 & 0.0856 & 0.3356 & 0.946 & 0.0945 & 0.0946 & 0.3706 & 0.952 \\
 & 2500 & 0.0794 & 0.0795 & 0.3116 & 0.944 & 0.0848 & 0.0847 & 0.3322 & 0.950 \\
\bottomrule
\end{tabular}
\caption{Performance of $\hat\theta_0^{\mathrm{cf}}$ under the baseline design, over $500$ replications.}
\label{tab:simu_summary}
\end{table}

We evaluate the estimator across four simulation designs:
\begin{itemize}
    \item \textbf{Baseline design.} This applies the data-generating process described above over the full grid of $n$, $p = q \in \{500, 1000, 2500, 5000\}$ and $\sigma \in \{0.1, 1\}$.
    
    \item \textbf{Signal strength.} The baseline design fixes a single non-zero value for every parameter. To let the mediated pathway range from absent to strong, we proceed as follows. Fix a direction vector $g \in \mathbb{R}^{\bar q}$ with entries drawn once from $\mathcal U(0.5, 1.5)$ and set
    \[
    \gamma_1 = \gamma_0 = c\, g, \qquad c \in \{0, 0.1, 0.25, 0.5, 1\},
    \]
    leaving all other components of the data-generating process unchanged, so that $\theta_0 = 1 + c\,\delta_0^\top g$ and the mediated contribution $\delta_0^\top \gamma_1$ vanishes at $c = 0$ and grows linearly in $c$. We report results for $p = q \in \{1000, 2500, 5000\}$ at $\sigma = 0.1$, so that the accuracy of $\hat\theta_0$ can be traced as the mediator signal weakens relative to the noise.
    
    \item \textbf{Heavy-tailed errors.} Separately, we replace all three error terms by variance-matched Student-$t$ variates: writing $t \sim t_\nu$ with $\nu = 6$, each coordinate of $\mathbf u_X$ and $\mathbf u_M$ and the outcome error $\epsilon$ is set to $\sigma\, t/\sqrt{\nu/(\nu-2)} = \sigma\, t/\sqrt{3/2}$, which has variance exactly $\sigma^2$ and so isolates tail thickness from any change of scale. All other components of the design and the full grid of $n$, $p = q$ and $\sigma$ are unchanged.

    \begin{figure}[t]
    \centering
    \begin{subfigure}[t]{0.48\linewidth}
        \centering
        \includegraphics[width=\linewidth]{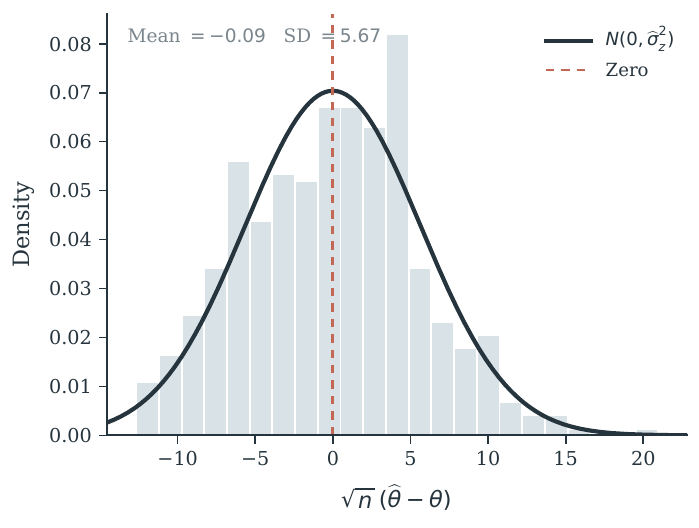}
        \caption{Histogram}
        \label{fig:histqqplot}
    \end{subfigure}
    \hfill
    \begin{subfigure}[t]{0.48\linewidth}
        \centering
        \includegraphics[width=\linewidth]{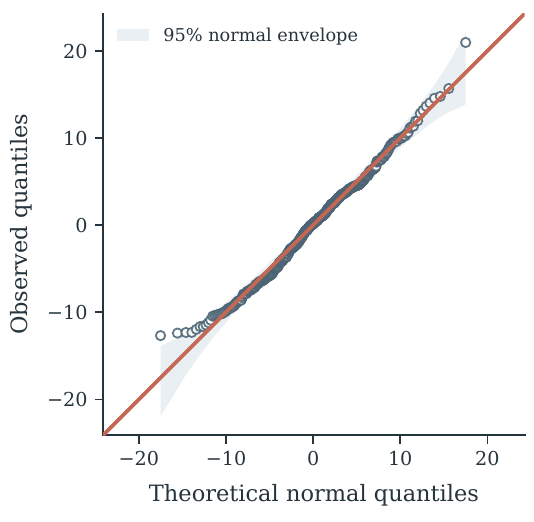}
        \caption{Normal Q--Q plot}
        \label{fig:normality_qq}
    \end{subfigure}
    \caption{Histogram and Q--Q plot of $\sqrt{n}(\widehat{\theta}_0^{\mathrm{cf}}-\theta_0)$ under the base design with $p=q=2500$, $n=1500$, $\sigma=1$, and $\widetilde p=\widetilde q=7$, based on 500 Monte Carlo replications.}
    \label{fig:normality_base}
\end{figure}
    
    \item \textbf{Sensitivity to the working factor dimensions.} Algorithm~\ref{algo:FARMA} requires the numbers $\tilde p$ and $\tilde q$ of principal components extracted from $X$ and from $M$, and $\bar p$, $\bar q$ are unknown in any application. Setting $\kappa = \tilde p = \tilde q$, we vary
    \[ \kappa \in \{1, 2, 3, 5, 7, 10, 15, 20, 30\} \]
    across all five sample sizes and $p = q \in \{1000, 2500, 5000\}$ at $\sigma = 0.1$, retaining the baseline design in all other respects. This brackets the truth $\bar p = \bar q = 3$ on both sides and reaches an order of magnitude above it. We also vary $\tilde p$ and $\tilde q$ separately over $\{3, 7, 20\}$ at $n = 2500$.
\end{itemize}

\begin{figure}[t]
    \centering
    \includegraphics[width=1\linewidth]{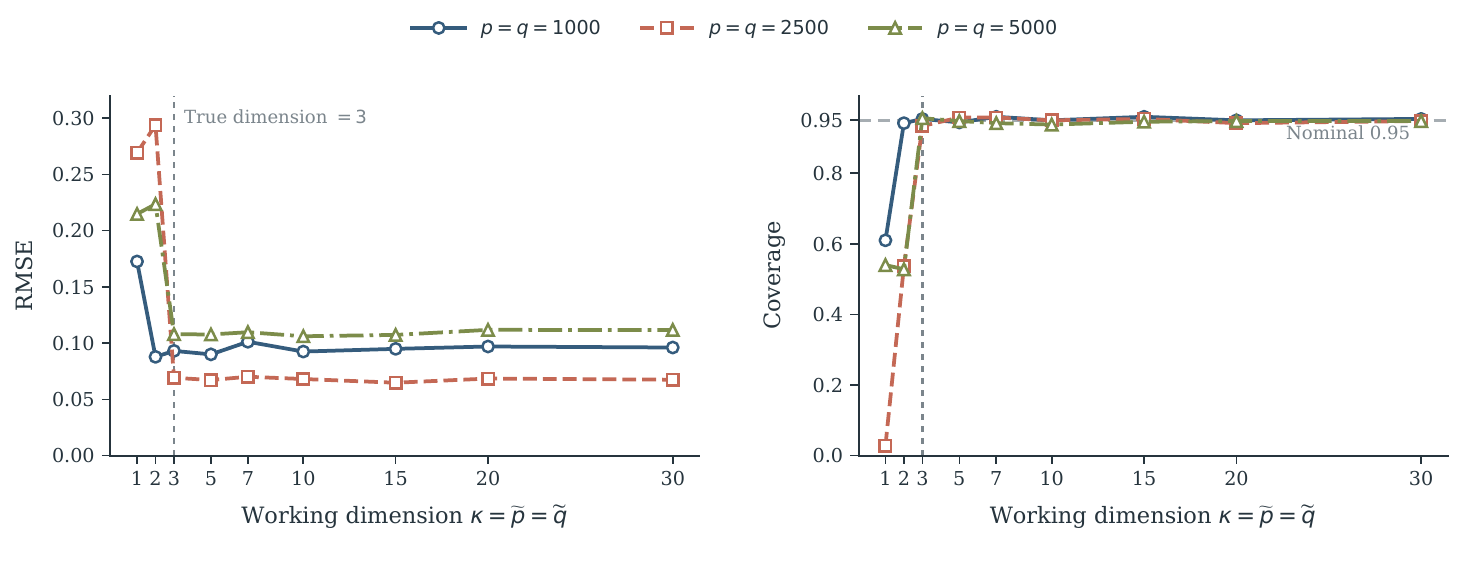}
    \caption{Sensitivity Analysis of the working factor dimensions.}
    \label{fig:sen1}
\end{figure}

Overall, the simulation results provide consistent support for the accuracy of $\widehat{\theta}_0^{\mathrm{cf}}$ and the validity of the associated inference. Under the baseline design, the RMSE closely tracks the empirical standard deviation, both decrease as $n$ increases, and the confidence intervals become shorter while maintaining coverage near the nominal $95\%$ level, even when $p=q$ substantially exceeds $n$. The same pattern persists under variance-matched $t_6$ errors; see Table~\ref{tab:simu_heavy}. The histogram and Q--Q plot are also broadly consistent with the Gaussian distribution; see Figure~\ref{fig:histqqplot}. As the mediated pathway varies from absent to strong, the estimator continues to exhibit decreasing estimation error with increasing $n$ and coverage close to the nominal level; see Table~\ref{tab:simu_signal}. The sensitivity analysis further shows that the estimator is relatively insensitive to moderate or substantial overestimation of the common working factor dimension; see Table~\ref{tab:simu_kappa} and Figure~\ref{fig:sen1}. Similar stability is observed when $\widetilde p$ and $\widetilde q$ are varied separately; see Table~\ref{tab:simu_pq_cross}. In contrast, severe underestimation of the working dimensions can produce persistent bias and substantial undercoverage in the more high-dimensional settings, as shown in Table~\ref{tab:simu_kappa}. These findings are consistent with the asymptotic theory and indicate that reliable estimation and inference require the working representations to capture the relevant latent factor spaces.

\section{Real Data Experiments}
\label{section:7realdata}

Depression is among the most consistently reported risk factors for cognitive decline in Alzheimer's disease, yet the biological route through which it acts remains poorly understood \citep{livingston2024dementia}. A hypothesis is that part of the effect is biological, transmitted through epigenetic changes such as DNA methylation, yet direct evidence for such mediation remains scarce \citep{zhang2026individualized, fransquet2018blood}. We apply our proposed method to investigate how depressive symptoms influence cognitive decline in Alzheimer's disease, and whether part of this relationship may operate through an underlying biological process represented by high-dimensional DNA methylation measurements.

As mentioned in Section~\ref{section:introduction}, we view oxidative stress as the underlying biological process of interest and DNA methylation as a high-dimensional representation of this process. Since reactive oxygen species are short lived and highly reactive, oxidative stress is difficult to measure directly \citep{katerji2019approaches}. DNA methylation provides more stable molecular signals that may reflect changes associated with oxidative activity \citep{miranda2000role, pandya2013antioxidants, black2015depression}. These changes are not expected to be represented by a single CpG site, but rather by patterns distributed across many sites. We therefore use the joint methylation profile as a collection of high-dimensional surrogates for the latent mediator, allowing the underlying biological signal to be summarized across the methylation measurements.

We conduct our analysis using data from the Alzheimer's Disease Neuroimaging Initiative (ADNI), a longitudinal, multi-center study designed to validate biomarkers for Alzheimer's disease clinical trials \citep{mueller2005alzheimer}. Our treatment is the Geriatric Depression Scale (GDS), a self-reported measure of depressive symptoms ranging from $0$ to $15$, with higher scores indicating more severe symptoms. Our outcome is the Alzheimer's Disease Assessment Scale-Cognitive Subscale (ADAS-Cog), which ranges from $0$ to $85$, where higher scores reflect greater cognitive impairment. The mediator surrogates are DNA methylation measurements recorded at hundreds of thousands of CpG sites across the genome. Further details of the study are available at \url{https://adni.loni.usc.edu/}.

\paragraph{Data preprocessing.}
We study whether part of the relationship between depressive symptoms and cognition may operate through the latent biological process represented by DNA methylation, using measurements from the earliest available visit date during the ADNI-GO and ADNI-2 phases, for which methylation data are available. The dataset\footnote{The data are accessible at \url{https://ida.loni.usc.edu/login.jsp}.} includes DNA methylation data for $653$ individuals, ADAS-Cog scores for $1{,}212$ individuals, GDS scores for $1{,}800$ individuals, demographic information for $1{,}895$ individuals, and medical history data for $1{,}673$ individuals. Of these, $649$ subjects had complete data across all components and form our analysis sample. We preprocess the data as follows:
\begin{itemize}
    \item The treatment $A$ is a binarized Geriatric Depression Scale, where a GDS score below $5$ indicates a normal mood state \citep{greenberg2012geriatric}, so subjects with GDS $> 5$ are classified as depressed and the rest as non-depressed. We include ten clinically relevant covariates $X$, such as gender, weight, and education level, of which seven are numerical and three are categorical (see Appendix~\ref{covariates} for the full list).
    \item We impute missing covariate values with \texttt{mice} \citep{van2011mice}. To increase model expressiveness, we add second- through sixth-order polynomial terms and interactions between the categorical and numerical covariates, giving a covariate vector of dimension $171$.
    \item The raw DNA methylation data give marker intensities in red and green channels for each sample, from the Illumina HumanMethylationEPIC BeadChip array. We compute the methylation ratio, ranging continuously from $0$ to $1$, using the \texttt{minfi} package \citep{aryee2014minfi}.
    \item The full methylation vector has dimension $866{,}836$. We retain only sites with no missing values and CpG sites reported as significantly associated with Alzheimer's disease in large-scale EWAS \citep{zhang2020epigenome, lunnon2014methylomic, battram2022ewas}, which reduces the mediators to dimension $3{,}206$.
\end{itemize}
After preprocessing, we retain $n=649$ subjects, each with a $171$-dimensional covariate-surrogate vector $X$ and $3{,}206$ DNA methylation measurements forming the high-dimensional mediator-surrogate vector $M$.

\paragraph{Estimation.}
To estimate the natural indirect effect, natural direct effect, and total effect of GDS on ADAS-Cog, we apply Algorithms~\ref{algo:FARMA} and~\ref{algo:FARMA-effects} to the counterfactual means $\mathbb{E}[Y^{(1,1)}](=\theta_{11})$, $\mathbb{E}[Y^{(0,0)}](=\theta_{00})$, and $\mathbb{E}[Y^{(1,0)}](=\theta_{10})$ (see Appendix~\ref{effects} for the decomposition). We set the projection dimensions $\tilde p$ and $\tilde q$ to the number of leading principal components that explain $50\%$ of the cumulative variance, yielding $(\tilde p ,\tilde q) = (4,11)$. We quantify uncertainty with a nonparametric bootstrap. We resample the $n=649$ subjects with replacement. For each resample, we apply FARMA using ten independent random two-fold splits and average the resulting direct, indirect, and total effect estimates to obtain one bootstrap replicate. Repeating this procedure across $B$ resamples yields a bootstrap distribution for each effect, from which we report the bootstrap mean and the $95\%$ percentile confidence interval defined by the empirical $2.5\%$ and $97.5\%$ quantiles.

\paragraph{Results.}
The full-sample point estimates and the bootstrap summaries are given in Table~\ref{tab:combined_effects}.

\begin{table}[H]
\centering
\small
\begin{tabular}{lccc}
\hline
\textbf{Effect} & \textbf{Estimate} & \textbf{Bootstrap Mean} & \textbf{Quantile CI (95\%)}\\
\hline
Direct   & 3.5891 & 3.7306 & $[1.5699, 6.1491]$ \\
Indirect & 0.0354 & 0.0043 & $[-0.5131, 0.4899]$ \\
Total    & 3.6244 & 3.7349 & $[1.6816, 6.1864]$ \\
\hline
\end{tabular}
\caption{Estimated natural direct, natural indirect, and total effects with $B = 300$.}
\label{tab:combined_effects}
\end{table}

The estimated total effect of GDS-defined depressive status on ADAS-Cog was $3.6244$, with a $95\%$ percentile bootstrap confidence interval of $[1.6816, 6.1864]$. The positive estimate indicates that participants with elevated depressive symptoms had an estimated ADAS-Cog score approximately $3.62$ points higher than those without elevated symptoms, corresponding to worse cognitive function. The natural direct effect was estimated at $3.5891$ with a $95\%$ confidence interval of $[1.5699, 6.1491]$, whereas the natural indirect effect through the latent pathway represented by the DNA methylation measurements was $0.0354$ with an interval of $[-0.5131, 0.4899]$. Thus, the intervals for the total and direct effects excluded zero, while the interval for the indirect effect included zero. Based on the point estimates, the indirect effect accounted for approximately $1.0\%$ of the total effect.

Overall, the estimated total effect was primarily attributable to the direct effect. The near-zero indirect-effect estimate and its confidence interval provide little evidence of mediation through the latent biological pathway represented by the DNA methylation measurements considered in this analysis. However, these results do not establish the absence of a mediated effect and do not rule out mediation through biological processes not adequately captured by the selected methylation features or through other biological and behavioral pathways.
\section{Conclusion and Future Work}
\label{section:8conclusion}

In this paper, we introduced Factor-Adjusted Regression for Mediation Analysis (FARMA), a factor-based framework for estimating mediation functionals when the covariates and mediators are latent and observed only through high-dimensional noisy surrogates. The main idea is that when the factor structure is pervasive, low-dimensional projections of the observed surrogates keep the information needed for causal mediation analysis. Within this framework, we showed that the mediation functional is approximately identified at finite proxy dimensions, and that the approximation error goes to zero as the numbers of surrogate covariates and mediators grow. We then built a cross-fitted estimator that separates the construction of the projection directions from the regression steps and does not need the true numbers of latent factors to be known. Under suitable regularity and growth conditions, this estimator is $\sqrt{n}$-consistent and asymptotically normal, so we can form valid Wald confidence intervals for the mediation functional, and hence for the natural direct and indirect effects. The simulations support the finite-sample accuracy of the estimator and its confidence intervals. In the ADNI application, we find that depressive symptoms are associated with greater cognitive impairment, with little evidence of mediation through the selected DNA methylation measurements.

Several directions remain open for future research. The present theory requires the dimensions of the observed variables to grow faster than the sample size, and extending it to the proportional regime, in which $\frac{p}{n} \to c_X$ and $\frac{q}{n} \to c_M$ for finite positive constants $c_X$ and $c_M$ that would require accounting for projection errors that persist asymptotically rather than vanish. A second direction is to relax sequential exchangeability to accommodate confounding beyond the latent structure captured by the available surrogates, together with the sensitivity analysis needed to assess robustness to such departures. Finally, we focused on linear structural models, and extending FARMA to generalized linear and other nonlinear outcome and mediator models would widen the range of problems it can handle.

\bibliographystyle{plainnat}
\bibliography{references} 

\newpage
\appendix 
\section{Appendix \label{sec:appendix-identification}}

This appendix contains the proofs of the two main results, along with notation and some auxiliary lemmas they rely on. Section~\ref{proof:id} proves the identification theorem, Theorem~\ref{thm:id} from Section~\ref{section:3identification}. The argument is carried out at the population level and is built around the projected functional $\theta_{0,p,q}=T_1+T_3T_2$, whose approximation error we control through three remainder terms $R_\alpha$, $R_\gamma$, and $R_\delta$. Section~\ref{Prop proof} proves the asymptotic normality theorem, Theorem~\ref{thm:clt} from Section~\ref{section:5theory}, which gives the sampling theory for our cross-fitted estimator. The supporting results for identification and asymptotic analysis are gathered in Sections~\ref{sec:id-prelim} and~\ref{sec:prelim}, respectively. The two matrix identities used in the identification proof are collected in Section~\ref{sec:technical-lemmas}.

The notation of Section~\ref{sec:prelim-notation} is used throughout both proofs. Any change is specified at the point where it is introduced. The identification argument involves no sampling. The asymptotic analysis uses the two-fold cross-fitting scheme of Algorithm~\ref{algo:FARMA}, where the projection matrices and the regression coefficients are estimated on separate folds and then the two fold-specific estimators are averaged. We describe full details in the Section~\ref{Prop proof}.

\subsection{Notation}\label{sec:prelim-notation}

For notational convenience, let $\mathbf{f}_{X_i}\in \mathbb{R}^{\bar{p}}$ and $\mathbf{f}_{M_i}\in \mathbb{R}^{\bar{q}}$ denote the latent factors for observation $i$, and define $\mathbf f_i=(\mathbf f_{X_i}^\top,\mathbf f_{M_i}^\top)^\top\in\mathbb{R}^{\bar p+\bar q}$.  Let $\mathbf F_X\in\mathbb{R}^{n\times\bar p}$ and $\mathbf F_M\in\mathbb{R}^{n\times\bar q}$ be the matrices for the corresponding latent factors across observations. Define $\mathbf F=(\mathbf F_X,\mathbf F_M)\in\mathbb{R}^{n\times(\bar p+\bar q)}$ and the augmented factor matrix $\mathbf F_1=(\mathbf 1,\mathbf F)\in\mathbb{R}^{n\times(1+\bar p+\bar q)}$, where $\mathbf 1\in\mathbb{R}^n$ is the vector of ones. 

Let $\mathbf H_X\in\mathbb{R}^{\tilde p\times\bar p}$ and $\mathbf H_M\in\mathbb{R}^{\tilde q\times\bar q}$ be the transformation matrices. Define $\mathbf H=\operatorname{diag}(\mathbf H_X,\mathbf H_M)\in\mathbb{R}^{(\tilde p+\tilde q)\times(\bar p+\bar q)}$ and $\mathbf H_1=\operatorname{diag}(1,\mathbf H_X,\mathbf H_M)\in\mathbb{R}^{(1+\tilde p+\tilde q)\times(1+\bar p+\bar q)}$. 
The true design matrix on the projected scale is $\mathbf Z=\mathbf F_1\mathbf H_1^\top\in\mathbb{R}^{n\times(1+\tilde p+\tilde q)}$, or equivalently $\mathbf Z^\top=\mathbf H_1\mathbf F_1^\top\in\mathbb{R}^{(1+\tilde p+\tilde q)\times n}$.

Next, let $\mathbf{u}_{X_i}\in \mathbb{R}^p$ and $\mathbf{u}_{M_i}\in \mathbb{R}^q$ denote the idiosyncratic errors, and define $\mathbf u_i=(\mathbf u_{X_i}^\top,\mathbf u_{M_i}^\top)^\top\in\mathbb{R}^{p+q}$. Let $\mathbf{U}_X \in \mathbb{R}^{n\times p}$ and $\mathbf{U}_M\in \mathbb{R}^{n\times q}$ be the matrices of these errors across observations, and write $\mathbf U=(\mathbf U_X,\mathbf U_M)\in\mathbb{R}^{n\times(p+q)}$. Let $\mathbf V_X\in\mathbb{R}^{p\times\tilde p}$ and $\mathbf V_M\in\mathbb{R}^{q\times\tilde q}$ be as defined in Sections \ref{section:3identification} and \ref{section:5theory}, respectively. Define $\mathbf W_X=\sqrt p\,\mathbf V_X$, $\mathbf W_M=\sqrt q\,\mathbf V_M$, and $\mathbf W=\operatorname{diag}(p^{-1}\mathbf W_X,q^{-1}\mathbf W_M)\in\mathbb{R}^{(p+q)\times(\tilde p+\tilde q)}.$ For each observation, define the projected errors $\mathbf{e}_{X_i}=p^{-1}\mathbf{W}_X^\top \mathbf{u}_{X_i}\in \mathbb{R}^{\tilde p}$, $\mathbf{e}_{M_i}=q^{-1}\mathbf{W}_M^\top \mathbf{u}_{M_i}\in \mathbb{R}^{\tilde q}$ and $\mathbf e_i=\mathbf W^\top\mathbf u_i\in\mathbb{R}^{\tilde p+\tilde q}$. Their augmented versions are $\mathbf e_{0i}=(0,\mathbf e_i^\top)^\top\in\mathbb{R}^{1+\tilde p+\tilde q}$ and $\mathbf e_{X0,i}=(0,\mathbf e_{X_i}^\top)^\top\in\mathbb{R}^{1+\tilde p}$. Define the projected estimation error matrix $\mathbf E=\mathbf U\mathbf W\in\mathbb{R}^{n\times(\tilde p+\tilde q)}$, and its augmented version $\mathbf E_0=(\mathbf 0_n,\mathbf E)\in\mathbb{R}^{n\times(1+\tilde p+\tilde q)}$, where $\mathbf 0_n\in\mathbb{R}^n$ is the zero column vector.

The estimated factors are collected in $\hat{\mathbf F}=(\hat{\mathbf F}_X,\hat{\mathbf F}_M)\in\mathbb{R}^{n\times(\tilde p+\tilde q)}$ where $\hat{\mathbf F}_X\in\mathbb{R}^{n\times\tilde p}$ and $\hat{\mathbf F}_M\in\mathbb{R}^{n\times\tilde q}$. The estimated design matrix is $\hat{\mathbf Z}=(\mathbf 1,\hat{\mathbf F})\in\mathbb{R}^{n\times(1+\tilde p+\tilde q)}$. Throughout, we use the decomposition
\begin{align}
\hat{\mathbf Z}&=\mathbf Z+\mathbf E_0,
\label{equation:zhat}
\end{align}
where $\hat{\mathbf Z}$, $\mathbf Z$, and $\mathbf E_0$ all lie in $\mathbb R^{n\times(1+\tilde p+\tilde q)}$.

\medskip\noindent For the outcome model, let $\Delta=(\alpha_1,\beta_1^\top,\gamma_1^\top)^\top\in\mathbb{R}^{1+\bar p+\bar q}$, where $\alpha_1\in\mathbb{R}$, $\beta_1\in\mathbb{R}^{\bar p}$, and $\gamma_1\in\mathbb{R}^{\bar q}$ denote the true coefficients on the latent factor scale. The corresponding coefficient vector in the projected factor scale is
\begin{align*}
\delta&=(\mathbf H_1^\dagger)^\top\Delta=\begin{pmatrix}\alpha_1\\ (\mathbf H_X^\dagger)^\top\beta_1 \\
(\mathbf H_M^\dagger)^\top\gamma_1\end{pmatrix}\in\mathbb{R}^{1+\tilde p+\tilde q}.
\end{align*}
The estimated coefficient vector is $\hat\delta=(\hat\alpha_1,\hat\beta_1^\top,\hat\gamma_1^\top)^\top\in\mathbb{R}^{1+\tilde p+\tilde q}$, where $\hat\alpha_1\in\mathbb{R}$, $\hat\beta_1\in\mathbb{R}^{\tilde p}$, and $\hat\gamma_1\in\mathbb{R}^{\tilde q}$. Let $\mathbf Y=(Y_1,\ldots,Y_n)^\top\in\mathbb{R}^n$ and $\boldsymbol \epsilon_1=(\epsilon_{1,1},\ldots,\epsilon_{1,n})^\top\in\mathbb{R}^n$. The treatment-specific outcome equation is
\begin{align}
\mathbf Y&=\mathbf Z\delta+\boldsymbol \epsilon_1,
\label{equation:yhat}
\end{align}
where $\mathbf Z\in\mathbb{R}^{n\times(1+\tilde p+\tilde q)}$ and $\delta\in\mathbb{R}^{1+\tilde p+\tilde q}$.

\medskip\noindent For the mediator block, define the selector matrix $\mathbf S_{\tilde q}=\begin{bmatrix}0_{\tilde q\times(1+\tilde p)} & I_{\tilde q}\end{bmatrix} \in\mathbb R^{\tilde q\times(1+\tilde p+\tilde q)},$ so that $\mathbf S_{\tilde q}$ extracts the last $\tilde q$ coordinates of any vector in the projected design scale. Also, define $\mathbf P_{\bar q}=\begin{bmatrix}0_{\bar q\times(1+\bar p)} & I_{\bar q}\end{bmatrix} \in\mathbb R^{\bar q\times(1+\bar p+\bar q)},$ so that $\mathbf H_M^\top\mathbf S_{\tilde q}=\mathbf P_{\bar q}\mathbf H_1^\top$. Throughout, $e_{\mathbf 1}$ denotes the first standard basis vector of whatever dimension the surrounding matrix product requires.

\subsection{Supporting Results}\label{sec:id-prelim}

This subsection collects the invariance lemma and the well-posedness lemma for the population regressions used in the proof of Theorem~\ref{thm:id}. The two matrix identities, the intercept-augmented block inversion formula and the Woodbury identity, are given in Section~\ref{sec:technical-lemmas}. We first fix notation and write the moment identities that the lemmas and the main proof in Section~\ref{proof:id} draw on repeatedly. Define
\begin{align}\label{eq:LQN}
L &=\begin{pmatrix}1 \\
\frac{1}{p}  \mathbf W_X^\top X \\
\frac{1}{q}  \mathbf W_M^\top M\end{pmatrix}; \quad Q=\begin{pmatrix}1 \\
\frac{1}{p}  \mathbf W_X^\top X\end{pmatrix}; \quad N = \frac{1}{q}  \mathbf W_M^\top M; \quad Z = \begin{pmatrix}1 \\
\mathbf{H}_X \mathbf{f}_X \\
\mathbf{H}_M \mathbf{f}_M\end{pmatrix}\,.
\end{align}
Substituting the factor model $X=\mathbf\Gamma_X\mathbf f_X+\mathbf u_X$ and $M=\mathbf\Gamma_M\mathbf f_M+\mathbf u_M$ from Equation~\eqref{eq:factors} into $L$, and using $\mathbf H_X=\frac1p\mathbf W_X^\top\mathbf\Gamma_X$, $\mathbf H_M=\frac1q\mathbf W_M^\top\mathbf\Gamma_M$ from Definition~\ref{definition:H}, gives the signal-plus-noise decomposition as follows:
\begin{align}
L&=Z+\mathbf e_0=\mathbf H_1\mathbf f_1+\mathbf e_0,
\label{eq:LL}
\end{align}
where $\mathbf f_1=(1,\mathbf f_X^\top,\mathbf f_M^\top)^\top$, $\mathbf H_1=\operatorname{diag}(1,\mathbf H_X,\mathbf H_M)$, $e_0=(0,\mathbf e_X^\top,\mathbf e_M^\top)^\top$, $\mathbf e_X=(\mathbf W_X^\top\mathbf u_X)/p\in\mathbb R^{\tilde p}$, and $\mathbf e_M=(\mathbf W_M^\top\mathbf u_M)/q \in\mathbb R^{\tilde q}$. Since $Q$ consists of the leading $1+\tilde p$ coordinates of $L$, the same computation restricted to the $X$-block gives
\begin{align}
Q&=\mathbf H_{X1}\mathbf f_{X1}+\mathbf e_{X0},\quad \mathbf f_{X1}=(1,\mathbf f_X^\top)^\top,\quad \mathbf H_{X1}=\operatorname{diag}(1,\mathbf H_X),\quad \mathbf e_{X0}=(0,\mathbf e_X^\top)^\top.
\label{eq:QQ}
\end{align}

We record the following moment identities, which hold for both treatment groups and are used throughout the Section \ref{proof:id}. By Assumption~\ref{Assumption:covariances}(ii), $\mathbb E(\mathbf u_X)=\mathbb E(\mathbf u_M)=0$, so $\mathbb E(\mathbf e_0)=0$, and the independence property gives $\mathbb E[\mathbf f_1\mathbf e_0^\top\mid A]=0.$ Also by Assumption~\ref{Assumption:covariances}(ii), $\mathbb E(\mathbf u_X\mathbf u_M^\top)=0$, hence $\mathbb E(\mathbf e_X\mathbf e_M^\top)=0$, so
\begin{align*}
\Sigma_{\mathbf e_0}&:=\mathbb E(\mathbf e_0\mathbf e_0^\top)=\operatorname{diag}(0,\Sigma_{\mathbf e_X},\Sigma_{\mathbf e_M}),\quad \Sigma_{\mathbf e_X}=\tfrac1p\mathbf V_X^\top\Sigma_{\mathbf u_X}\mathbf V_X,\quad \Sigma_{\mathbf e_M}=\tfrac1q\mathbf V_M^\top\Sigma_{\mathbf u_M}\mathbf V_M,
\end{align*}
using $\mathbf W_X=\sqrt p\,\mathbf V_X$ and $\mathbf W_M=\sqrt q\,\mathbf V_M$ from Definition~\ref{definition:H}. The conditional second moments of $\mathbf u_X,\mathbf u_M$ in Assumption~\ref{Assumption:covariances}(i) and (ii) do not depend on $(\mathbf f_X,\mathbf f_M,A)$, so these identities hold unconditionally. Now, restricting to $A=0$ gives the $X$-block analogues
\begin{align*}
&\mathbb E[\mathbf f_{X1}\mathbf e_{X0}^\top\mid A=0]=0, \quad \mathbb E[\mathbf e_{X0}\mathbf e_{X0}^\top]=\operatorname{diag}(0,\Sigma_{\mathbf e_X}).
\end{align*}

\medskip

The first lemma establishes that the target functional depends on the latent factors only through their diversified projections.

\begin{lemma}[Invariance under linear transformation of the factors]
\label{lemma:A1}
For any $\bH_X, \bH_M$ with full column rank, we have: 
    \begin{align*}
&\mathbb{E}\Big[ \mathbb{E}\big[ \mathbb{E}[Y \mid \mathbf f_M, \mathbf f_X, A = 1] \,\big|\, \mathbf f_X, A = 0 \big] \Big]=\mathbb{E}\Big[ \mathbb{E}\big[ \mathbb{E}[Y \mid \mathbf H_M \mathbf f_M, \mathbf H_X \mathbf f_X, A = 1] \,\big|\, \mathbf H_X \mathbf f_X, A = 0 \big] \Big].
\end{align*}
\end{lemma}

\begin{proof}
Since $\mathbf H_X\mathbf f_X$ and $\mathbf H_M\mathbf f_M$ are measurable functions of $\mathbf f_X$ and $\mathbf f_M$ respectively, we have $\sigma(\mathbf H_X\mathbf f_X,\mathbf H_M\mathbf f_M) \subseteq\sigma(\mathbf f_X,\mathbf f_M)$.  On the other hand, as $\mathbf H_X$ and $\mathbf H_M$ have full column rank they admit left inverses $\mathbf H_X^\dagger$ and $\mathbf H_M^\dagger$ respectively, which satisfy  $\mathbf H_X^\dagger\mathbf H_X=I_{\bar p}, \mathbf H_M^\dagger\mathbf H_M=I_{\bar q}$. Consequently we have $\mathbf f_X=\mathbf H_X^\dagger(\mathbf H_X\mathbf f_X)$ and $\mathbf f_M=\mathbf H_M^\dagger(\mathbf H_M\mathbf f_M)$ and hence $\sigma(\mathbf f_X,\mathbf f_M)\subseteq\sigma(\mathbf H_X\mathbf f_X,\mathbf H_M\mathbf f_M)$. Therefore, $\sigma(\mathbf f_X,\mathbf f_M) = \sigma(\mathbf H_X\mathbf f_X,\mathbf H_M\mathbf f_M)$.  
This argument also yields $\sigma(\mathbf f_X) = \sigma(\mathbf H_X\mathbf f_X)$.  As the conditional expectation of a random variable given another random variable is the projection of the former onto the sigma-field generated by the latter, we have: 
$$
\sigma(\mathbf f_X,\mathbf f_M) = \sigma(\mathbf H_X\mathbf f_X,\mathbf H_M\mathbf f_M) \implies  \mathbb E[Y\mid \mathbf H_M\mathbf f_M,\mathbf H_X\mathbf f_X,A=1]=\mathbb E[Y\mid \mathbf f_M,\mathbf f_X,A=1]\quad\text{a.s.} \,.
$$
Furthermore, from  $\sigma(\mathbf f_X) = \sigma(\mathbf H_X\mathbf f_X)$, we have: 
$$
\bbE\big[\mathbb E[Y\mid \mathbf H_M\mathbf f_M,\mathbf H_X\mathbf f_X,A=1]\mid \mathbf H_X\mathbf f_X,A=0\big] =\mathbb E\big[\mathbb E[Y\mid \mathbf f_M,\mathbf f_X,A=1]\mid \mathbf f_X,A=0\big], \quad \text{a.s.} \,.
$$
Finally, taking another expectation with respect to $\mathbf H \mathbf f_X$ and $\mathbf f_X$ on the left and right sides of the above equality, respectively, we conclude the proof. 
\end{proof}
\noindent Recall the definition of $\theta_{0,p,q}$ in Section~\ref{section:3identification}:
\begin{align*}
T_1&= e_{\mathbf 1}^\top \mathbb{E}[L L^\top \mid A = 1]^{-1}\mathbb{E}[L Y \mid A = 1],\quad T_2=\mathbf S_{\tilde q}\mathbb{E}[L L^\top\mid A = 1]^{-1}\mathbb{E}[L Y\mid A = 1]\in\mathbb R^{\tilde q},
\end{align*}
\begin{align*}
T_3&=e_{\mathbf 1}^\top\mathbb{E}[Q Q^\top\mid A = 0]^{-1}\mathbb{E}[Q N^\top\mid A = 0]\in\mathbb R^{1\times\tilde q}.
\end{align*}
The next lemma guarantees that these population regression coefficients in the projected functional are well defined.

\begin{lemma}[Well-posedness of the population regressions]\label{lemma:wellposed}
Let $\mathbf{W}_X$ and $\mathbf{W}_M$ be full-projection matrices as defined in Definition~\ref{definition:H}. Under Assumptions~\ref{Assumption:covariances} and \ref{Assumption:pervasiveness}, the matrices $\mathbb E[LL^\top\mid A=1]$ and $\mathbb E[QQ^\top\mid A=0]$ are positive definite, and hence invertible. Thus, $T_1,T_2$ and $T_3$ are well defined.
\end{lemma}

\begin{proof}
For any nonzero $w\in\mathbb R^{\tilde p}$, we have that
\begin{align*}
w^\top\Sigma_{\mathbf e_X}w&=\frac1p(\mathbf V_X w)^\top\Sigma_{\mathbf u_X}(\mathbf V_X w)\ge\frac{\lambda_{\min}(\Sigma_{\mathbf u_X})}{p}\|\mathbf V_X w\|^2\ge\frac{\lambda_{\min}(\mathbf V_X^\top\mathbf V_X)\,\lambda_{\min}(\Sigma_{\mathbf u_X})}{p}\|w\|^2_2>0,
\end{align*}
using $\lambda_{\min}(\mathbf V_X^\top\mathbf V_X)\ge c$ from Definition~\ref{definition:H} and $\lambda_{\min}(\Sigma_{\mathbf u_X})\ge c>0$ from Assumption~\ref{Assumption:covariances}(v). Hence, $\Sigma_{\mathbf e_X}\succ 0$ and the same argument gives $\Sigma_{\mathbf e_M}\succ 0$.
\\

\noindent Next, we show that the augmented factor second-moment matrices are positive definite. Recall $\Omega_1=\mathbb E[\mathbf f_1\mathbf f_1^\top\mid A=1]$. Now, take any nonzero vector $v=(v_0,\,b^\top)^\top\in\mathbb R^{1+\bar p+\bar q}$, where $b=(b_X^\top,b_M^\top)^\top$. We have that $v^\top\Omega_1 v=\mathbb E\big[(v_0+b^\top\mathbf f)^2\mid A=1\big].$ If $b\neq 0$, let $\mu_{\mathbf f\mid 1}=\mathbb E[\mathbf f\mid A=1]$ be the conditional mean under treatment group of the factors. Now, by the definition of variance, we have that
\begin{align*}
&\operatorname{Var}(b^\top\mathbf f\mid A=1)=\mathbb E\Big[\big(b^\top\mathbf f-\mathbb E[b^\top\mathbf f\mid A=1]\big)^2\;\Big|\;A=1\Big] =\mathbb E\Big[\big(b^\top(\mathbf f-\mu_{\mathbf f\mid 1})\big)^2\;\Big|\;A=1\Big]
\end{align*}
Since $b^\top(\mathbf f-\mu_{\mathbf f\mid 1})$ is a scalar, so we can write its square as $\big(b^\top(\mathbf f-\mu_{\mathbf f\mid 1})\big)^2=b^\top(\mathbf f-\mu_{\mathbf f\mid 1})(\mathbf f-\mu_{\mathbf f\mid 1})^\top b.$ When we take the conditional expectation, we get that
\begin{align*}
&b^\top\,\mathbb E\Big[(\mathbf f-\mu_{\mathbf f\mid 1})(\mathbf f-\mu_{\mathbf f\mid 1})^\top\;\Big|\;A=1\Big]\,b=b^\top \operatorname{Cov}(\mathbf f\mid A=1) b=b^\top\Sigma_{\mathbf f\mid A=1}\,b,
\end{align*}
By Assumption~\ref{Assumption:covariances}(iv), we have that $\lambda_{\min}(\Sigma_{\mathbf f\mid A=1})\ge c>0$, and so
\begin{align*}
&\operatorname{Var}(b^\top\mathbf f\mid A=1)=b^\top\Sigma_{\mathbf f\mid A=1}\,b\ge\lambda_{\min}(\Sigma_{\mathbf f\mid A=1})\,\|b\|^2\ge c\,\|b\|^2>0.
\end{align*}
Since $\mathbb E[Z^2]\ge\operatorname{Var}(Z)$ for any $Z$ and adding the constant $v_0$ does not change the variance, we get that
\begin{align*}
v^\top\Omega_1 v&=\mathbb E\big[(v_0+b^\top\mathbf f)^2\mid A=1\big]\ge\operatorname{Var}(b^\top\mathbf f\mid A=1)>0.
\end{align*}
If $b=0$, then $v\ne0$ implies $v_0\ne0$, so $v^\top\Omega_1v=v_0^2>0$. Hence, $\Omega_1\succ0$. The same argument applied to $\mathbf f_{X1}$ shows that $\Omega_{X0}:=\mathbb E[\mathbf f_{X1}\mathbf f_{X1}^\top\mid A=0]\succ0$. 
\\
By Equation~\ref{eq:LL} and the moment identities in Section~\ref{sec:id-prelim}, we have
\begin{align*}\mathbb E[LL^\top\mid A=1]=\mathbf H_1\Omega_1\mathbf H_1^\top+\Sigma_{\mathbf e_0}.\end{align*}
Let $\Omega_1^{1/2}$ denote the symmetric positive definite square root of $\Omega_1$. For any $v=(v_0,v_X^\top,v_M^\top)^\top\in\mathbb R^{1+\tilde p+\tilde q}$, we obtain
\begin{align*}
v^\top\mathbb E[LL^\top\mid A=1]v
&=\big\|\Omega_1^{1/2}\mathbf H_1^\top v\big\|^2+v_X^\top\Sigma_{\mathbf e_X}v_X+v_M^\top\Sigma_{\mathbf e_M}v_M\ge0.
\end{align*}
Suppose this sum vanishes. Since each term is nonnegative, all three terms must be zero. The positive definiteness of $\Sigma_{\mathbf e_X}$ and $\Sigma_{\mathbf e_M}$ implies that the equalities $v_X^\top\Sigma_{\mathbf e_X}v_X=0$ and $v_M^\top\Sigma_{\mathbf e_M}v_M=0$ yield $v_X=0$ and $v_M=0$, respectively. The first term gives $\|\Omega_1^{1/2}\mathbf H_1^\top v\|^2=0$, which implies $\Omega_1^{1/2}\mathbf H_1^\top v=0$. Since $\Omega_1^{1/2}$ is invertible, we obtain $\mathbf H_1^\top v=0$.
\\

\noindent Since $\mathbf H_1=\operatorname{diag}(1,\mathbf H_X,\mathbf H_M)$ and $v_X=v_M=0$, 
$$\mathbf H_1^\top v=(v_0,\,\mathbf H_X^\top v_X,\,\mathbf H_M^\top v_M)^\top=(v_0,0,0)^\top,$$
so $\mathbf H_1^\top v=0$ forces $v_0=0$ as well, and hence $v=0$. Therefore, we get that $\mathbb E[LL^\top\mid A=1]\succ 0$.
\\

\noindent For the control part, Equation \ref{eq:QQ} gives
$\mathbb E[QQ^\top\mid A=0]=\mathbf H_{X1}\,\Omega_{X0}\,\mathbf H_{X1}^\top+\operatorname{diag}(0,\Sigma_{\mathbf e_X}).$
The same argument with $v=(v_0,v_X^\top)^\top$ gives
\begin{align*}
&v^\top\mathbb E[QQ^\top\mid A=0]v=\big\|\Omega_{X0}^{1/2}\mathbf H_{X1}^\top v\big\|^2+v_X^\top\Sigma_{\mathbf e_X}v_X\ge 0,
\end{align*}
which vanishes only if $v_X=0$ and $\mathbf H_{X1}^\top v=(v_0,0)^\top=0$, forcing $v_0=0$ and hence $v=0$. Therefore,  $\mathbb E[QQ^\top\mid A=0]\succ 0$. Therefore, both matrices are invertible, and thus, $T_1,T_2$ and $T_3$ are well defined.
\end{proof}
The block-inversion formula for the intercept and noise signal shared by $\mathbb E[LL^\top\mid A=1]$ and $\mathbb E[QQ^\top\mid A=0]$ is given in Lemma~\ref{lemma:augmented-inverse} in Section~\ref{sec:technical-lemmas}.

\section{Proof of Theorem \ref{thm:id}}\label{proof:id}

\begin{proof}
Recall the definition of $(L, Q, N)$ from \eqref{eq:LQN}. 
We write $\theta_{0,p,q}$ in terms of the distributions of $L$, $Q$, and $N$ as
\begin{align*}
\theta_{0,p,q} &=e_1^\top \mathbb{E}[L L^\top \mid A = 1]^{-1}\mathbb{E}[L Y \mid A = 1] \\
& \hspace{10em} + \left(e_1^\top \mathbb{E}[Q Q^\top \mid A = 0]^{-1}\mathbb{E}[Q N^\top \mid A = 0]\right) \ \left(\mathbf S_{\tilde q} \mathbb{E}[L L^\top \mid A = 1]^{-1}\mathbb{E}[L Y \mid A = 1]\right) \\
& := T_1 + T_3 T_2 \ ,
\end{align*}
By Lemma \ref{lemma:wellposed}, $\mathbb E[LL^\top\mid A=1]$ and $\mathbb E[QQ^\top\mid A=0]$ are invertible, so $T_1,T_2$ and $T_3$ are well defined. In the rest of the proof, we will establish a bound on the right-hand side of the above inequality. 
\begin{align*}
T_1&=\alpha_1+R_{\alpha},~\quad |R_{\alpha}|\lesssim C_1\frac{1}{\min\{p\nu_{\min}^2(\mathbf H_X),q\nu_{\min}^2(\mathbf H_M)\}}, \\
\mathbf H_M^\top T_2&=\gamma_1+R_{\gamma},\quad
\quad \|R_{\gamma}\|\lesssim C_{2,X}\frac{1}{p\nu_{\min}^2(\mathbf H_X)}+C_{2,M}\frac{1}{q\nu_{\min}^2(\mathbf H_M)}, \\
T_3&=\delta_0^\top\mathbf H_M^\top+R_{\delta},
\end{align*}
and that $R_\delta$ factors as $R_\delta=\widetilde R_\delta\mathbf H_M^\top$ for a matrix $\widetilde R_\delta$ satisfying $\|\widetilde R_\delta\|\lesssim C_3/(p\nu_{\min}^2(\mathbf H_X))$. These are proved in Steps \hyperref[step:id-T1]{1}, \hyperref[step:id-T2]{2}, and \hyperref[step:id-T3]{3} respectively. The constants are
\begin{align*}
C_1 &:= \sqrt{\kappa(\Sigma_{\mathbf f\mid1})}\,
   \frac{\|\mu_{\mathbf f\mid 1}\|\,\|\Delta'\|\,\bar\lambda}{\lambda_{\min}(\Sigma_{\mathbf f\mid 1})}, 
\quad C_{2,X}:= C_{H_M}\,\sqrt{\kappa(\Sigma_{\mathbf f_X\mid1})}\,    \frac{\|\Sigma_{\mathbf f_M\mathbf f_X^\top\mid 1}\|\,\|\beta_1\|\,\bar\lambda_{V_X}\bar\lambda_{\mathbf u_X}}        {\lambda_{\min}(\Sigma_{\mathbf u_{XM,1}})\,\lambda_{\min}(\Sigma_{\mathbf f_X\mid 1})}, \\\quad
C_{2,M}&:= \sqrt{\kappa(D)}\, \frac{\|\gamma_1\|\,\bar\lambda_{V_M}\bar\lambda_{\mathbf u_M}}{\lambda_{\min}(\Sigma_{\mathbf u_{XM,1}})}, 
~~\quad C_3:= \sqrt{\kappa(\Sigma_{\mathbf f_X\mid0})}\,   \frac{\|\mu_{\mathbf f_{X}\mid 0}\|\,\|\Phi\|\,\bar\lambda_{\mathbf u_X}}{\lambda_{\min}(\Sigma_{\mathbf f_{X}\mid 0})}.
\end{align*}
Here, we have $\bar\lambda_{\mathbf u_X}:=\lambda_{\max}(\Sigma_{\mathbf u_X})$, $\bar\lambda_{\mathbf u_M}:=\lambda_{\max}(\Sigma_{\mathbf u_M})$, $\bar\lambda_{V_X}:=\allowbreak \lambda_{\max}(\mathbf V_X^\top\mathbf V_X)$, $\bar\lambda_{V_M}:=\allowbreak \lambda_{\max}(\mathbf V_M^\top\mathbf V_M)$, and $\bar\lambda:=\max\{\bar\lambda_{V_X}\bar\lambda_{\mathbf u_X},\,\bar\lambda_{V_M}\bar\lambda_{\mathbf u_M}\}$. The factor $C_{H_M}$ is the $O(1)$ bound on $\|\mathbf H_M^\top(\mathbf H_M D\mathbf H_M^\top+\Sigma_{\mathbf e_M})^{-1}\mathbf H_M\|$ established in Step~\hyperref[step:id-T2]{2}, the matrix $D$ is defined in Equation~\eqref{equation:M2} of Step~\hyperref[step:id-T2]{2}, and $\Phi$ is defined in Equation~\eqref{eq:Phi-prime} of Step~\hyperref[step:id-T3]{3}. Each constant is a fixed population quantity bounded using Assumptions~\ref{Assumption:covariances}--\ref{Assumption:pervasiveness} and Definition~\ref{definition:H}, and does not depend on $p$ or $q$. Since $\mathbf H_M^\top T_2=\gamma_1+R_{\gamma}$, the error term expands as
\begin{align*}
\theta_{0,p,q}-\theta_0
&=T_1+T_3T_2-(\alpha_1+\gamma_1^\top\delta_0) \\&=\alpha_1+R_\alpha+(\delta_0^\top\mathbf H_M^\top+R_\delta)T_2-(\alpha_1+\delta_0^\top\gamma_1) \\
&=\alpha_1+R_\alpha+\delta_0^\top(\gamma_1+R_\gamma)+R_\delta T_2-(\alpha_1+\delta_0^\top\gamma_1)\\ &=R_\alpha+\delta_0^\top R_\gamma+R_\delta T_2.
\end{align*}
The matrix $\mathbf H_M^\top$ acts on $T_2\in\mathbb R^{\tilde q}$, and since $\mathbf H_M$ has full column rank with $\bar q$, the kernel of $\mathbf H_M^\top$ is nontrivial. The relation $\mathbf H_M^\top T_2=\gamma_1+R_\gamma$ determines only the component of $T_2$ lying in $\operatorname{range}(\mathbf H_M)$, and leaves the component of $T_2$ in $\ker(\mathbf H_M^\top)$ unconstrained. Thus, the final term $R_\delta T_2$ must be controlled without separately bounding $\|T_2\|$. This can be shown using the factorization $R_\delta=\widetilde R_\delta\mathbf H_M^\top$ established in Step \hyperref[step:id-T3]{3}. Substituting this factorization gives
\begin{align*}
R_\delta T_2&=\widetilde R_\delta\,\mathbf H_M^\top T_2=\widetilde R_\delta\,(\gamma_1+R_\gamma)
\end{align*}
and since $\|\gamma_1\|\le C$ by Assumption \ref{Assumption:pervasiveness}(i), we get that
\begin{align*}
|R_\delta T_2|&\le\|\widetilde R_\delta\|\,(\|\gamma_1\|+\|R_\gamma\|) \lesssim\frac{1}{p\,\nu_{\min}^2(\mathbf H_X)}+\frac{1}{p\,\nu_{\min}^2(\mathbf H_X)}\left(\frac{1}{p\nu_{\min}^2(\mathbf H_X)}+\frac{1}{q\nu_{\min}^2(\mathbf H_M)}\right)
\end{align*}
Therefore,
\begin{align*}
|\theta_0-\theta_{0,p,q}|
&\lesssim|R_\alpha|+\|\delta_0\|\,\|R_\gamma\|+|R_\delta T_2| \\
&\lesssim C_1\frac{1}{\min\{p\nu_{\min}^2\left(\mathbf H_X),\,q\nu_{\min}^2(\mathbf H_M\right)\}}
+C_2\left(\frac{1}{p\nu_{\min}^2(\mathbf H_X)}+\frac{1}{q\nu_{\min}^2(\mathbf H_M)}\right) \\
&\quad+C_3 \left(\frac{1}{p\,\nu_{\min}^2(\mathbf H_X)}+\frac{1}{p\nu_{\min}^2(\mathbf H_X)}\left(\frac{1}{p\nu_{\min}^2(\mathbf H_X)}+\frac{1}{q\nu_{\min}^2(\mathbf H_M)}\right)\right) \\*
&\le C_*\left(\frac{1}{p\nu_{\min}^2(\mathbf H_X)}+\frac{1}{q\nu_{\min}^2(\mathbf H_M)}\right)\left(1+\frac{1}{p\nu_{\min}^2(\mathbf H_X)}\right).
\end{align*}
The stated rate in Theorem \ref{thm:id} follows.
\\

\noindent\textbf{Step 1: $T_1$.}
\label{step:id-T1}

\noindent Recall the outcome regression \eqref{eq:factor} under the treatment group. The intercept $\alpha_1$ admits the representation
\begin{align*}
&e_{\mathbf 1}^\top \mathbb E[LL^\top\mid A=1]^{-1}\mathbb E[LY\mid A=1]={\alpha_1}+\underbrace{(-K_1+K_2)}_{R_{\alpha}}.
\end{align*}
We derive this representation as follows
\begin{align}
&e_{\mathbf 1}^\top \mathbb{E}[L L^\top\mid A=1]^{-1} \mathbb{E}[L Y\mid A=1] \notag \\*
&\overset{(1)}{=}e_{\mathbf 1}^\top \mathbb{E}[L L^\top\mid A=1]^{-1} \mathbb{E}[ L L^\top\mid A=1] \delta+ e_{\mathbf 1}^\top \mathbb{E}[L L^\top\mid A=1]^{-1} \mathbb{E}[ L Y-L L^\top\delta\mid A=1] \notag \\*
&=\alpha_1+ e_{\mathbf 1}^\top \mathbb{E}[L L^\top\mid A=1]^{-1} \mathbb{E}[ L Y-L L^\top \delta\mid A=1] \notag \\
&\overset{(2)}{=}\alpha_1+ e_{\mathbf 1}^\top \mathbb{E}[L L^\top\mid A=1]^{-1} \mathbb{E}[ L (Z - L)^\top \delta\mid A=1] + e_{\mathbf 1}^\top \mathbb{E}[L L^\top\mid A=1]^{-1} \mathbb{E}[L \epsilon_1\mid A=1] \notag \\
&=\alpha_1- \underbrace{ e_{\mathbf 1}^\top \mathbb{E}[L L^\top\mid A=1]^{-1} \mathbb{E}[ L \mathbf{e}_0^\top \delta\mid A=1] }_{K_1}+ \underbrace{e_{\mathbf 1}^\top \mathbb{E}[L L^\top\mid A=1]^{-1} \mathbb{E}[L \epsilon_1\mid A=1]}_{K_2}\,.
\label{eq:alpha1}
\end{align}
Here, $(1)$ uses that $\delta$ is a fixed vector, so $\mathbb E[LL^\top\delta\mid A=1]=\mathbb E[LL^\top\mid A=1]\delta$. Equality $(2)$ uses the outcome model $Y=Z^\top\delta+\epsilon_1$ from~\eqref{eq:factor} under $A=1$. The remaining steps use $Z-L=-\mathbf e_0$ from~\eqref{eq:LL}. We show that $K_1$ and $K_2$ are each asymptotically
negligible.
\\

\noindent\textbf{Negligibility of $K_{1}$.} From $L=\mathbf H_1\mathbf f_1+\mathbf e_0$, we have
\begin{align*}
K_1
&= e_{\mathbf 1}^\top \mathbb{E}[L L^\top\mid A=1]^{-1} \mathbb{E}[ L \mathbf{e}_0^\top \delta\mid A=1] \\
&=e_{\mathbf 1}^\top \mathbb{E}[L L^\top\mid A=1]^{-1}\mathbb{E}[ \mathbf{H}_1\mathbf{f}_1 \mathbf{e}_0^\top\mid A=1]\delta+e_{\mathbf 1}^\top \mathbb{E}[L L^\top\mid A=1]^{-1}\Sigma_{\mathbf e_0}\delta \\
&=e_{\mathbf 1}^\top \mathbb{E}[L L^\top\mid A=1]^{-1}\Sigma_{\mathbf e_0}\delta,
\end{align*}
where the first term vanishes by $\mathbb E[\mathbf f_1\mathbf e_0^\top\mid A=1]=0$ from Section~\ref{sec:id-prelim}. With $\Lambda_{\mathbf f_1\mid1}:=\mathbb E[\mathbf f_1\mathbf f_1^\top\mid A=1]$ for the uncentered second moment and $\Delta'=(\beta_1^\top,\gamma_1^\top)^\top$, we get
\begin{align*}
\|K_1\|
&=\big\|e_{\mathbf 1}^\top \big( \mathbf H_1 \Lambda_{\mathbf f_1\mid1} \mathbf H_1^\top +\Sigma_{\mathbf e_0} \big)^{-1} \Sigma_{\mathbf e_0} \mathbf H_1(\mathbf H_1^\top \mathbf H_1)^{-1}\Delta\big\|.
\end{align*}
Using Lemma~\ref{lemma:augmented-inverse}, whose formula is stated in the
mean $\mu_{\mathbf f\mid1}$ and covariance $\Sigma_{\mathbf f\mid1}$, and noting
that $\Sigma_{\mathbf e_0}=\operatorname{diag}(0,\Sigma_{\mathbf e})$ so that
the intercept component of $\Delta$ drops out and that only $\Delta'$ survives, we get
\begin{align*}
\|K_1\|
&= \big\| -\mu_{\mathbf f\mid1}^\top \mathbf H^\top(\mathbf H\Sigma_{\mathbf f\mid1}\mathbf H^\top + \Sigma_{\mathbf e})^{-1} \Sigma_{\mathbf e} \mathbf H(\mathbf H^\top \mathbf H)^{-1} \Delta' \big\| \\
&\overset{(1)}{=} \big\| -\mu_{\mathbf f\mid1}^\top \mathbf H^\top(\Sigma_{\mathbf e}^{-1}-\Sigma_{\mathbf e}^{-1} \mathbf H(\Sigma_{\mathbf f\mid1}^{-1}+\mathbf H^\top\Sigma_{\mathbf e}^{-1}\mathbf H)^{-1} \mathbf H^\top\Sigma_{\mathbf e}^{-1}) \Sigma_{\mathbf e} \mathbf H(\mathbf H^\top \mathbf H)^{-1} \Delta' \big\| \\*
&= \big\| \mu_{\mathbf f\mid1}^\top \Delta' - \mu_{\mathbf f\mid1}^\top \mathbf H^\top\Sigma_{\mathbf e}^{-1} \mathbf H(\Sigma_{\mathbf f\mid1}^{-1}+\mathbf H^\top\Sigma_{\mathbf e}^{-1}\mathbf H)^{-1} \Delta' \big\|\\
&= \big\| \mu_{\mathbf f\mid1}^\top \Sigma_{\mathbf f\mid1}^{-1}(\Sigma_{\mathbf f\mid1}^{-1}+\mathbf H^\top\Sigma_{\mathbf e}^{-1}\mathbf H)^{-1} \Delta' \big\| \\
&= \big\| \mu_{\mathbf f\mid1}^\top (\mathbf{I}_{\bar p+\bar q}+\mathbf H^\top\Sigma_{\mathbf e}^{-1}\mathbf H\Sigma_{\mathbf f\mid1})^{-1} \Delta' \big\|
\end{align*}
where $(1)$ uses the Woodbury identity of Lemma~\ref{lemma:woodbury}. The matrix $A:=\mathbf H^\top\Sigma_{\mathbf e}^{-1}\mathbf H$ is symmetric positive semidefinite, but $A\Sigma_{\mathbf f\mid1}$ inside $(\mathbf I+A\Sigma_{\mathbf f\mid1})^{-1}$ is not symmetric. So, we set $S:=\Sigma_{\mathbf f\mid1}^{1/2}A\Sigma_{\mathbf f\mid1}^{1/2}$, which is symmetric positive semidefinite,
\begin{align*}
(\mathbf I+A\Sigma_{\mathbf f\mid1})^{-1}&=\Sigma_{\mathbf f\mid1}^{-1/2}(\mathbf I+S)^{-1}\Sigma_{\mathbf f\mid1}^{1/2}, \quad \|(\mathbf I+A\Sigma_{\mathbf f\mid1})^{-1}\|\le \sqrt{\kappa(\Sigma_{\mathbf f\mid1})}\,\frac{1}{1+\lambda_{\min}(S)},
\end{align*}
with $\lambda_{\min}(S)\ge \lambda_{\min}(\Sigma_{\mathbf f|1})\,\lambda_{\min}(A)$ and $\kappa(\Sigma_{\mathbf f|1})=\lambda_{\max}(\Sigma_{\mathbf f|1})/\lambda_{\min}(\Sigma_{\mathbf f|1})=O(1)$ by Assumption \ref{Assumption:covariances}(iv). This gives us that
\begin{align*}
\|K_1\|&\le \sqrt{\kappa(\Sigma_{\mathbf f\mid1})}\,\frac{\|\mu_{\mathbf f\mid1}\|\,\|\Delta'\|}{1+\lambda_{\min}(\Sigma_{\mathbf f\mid1})\,\lambda_{\min}(\mathbf H^\top \Sigma_{\mathbf e}^{-1}\mathbf H)}.
\end{align*}
It remains to bound $\lambda_{\min}(\mathbf H^\top\Sigma_{\mathbf e}^{-1}\mathbf H)$ from below. Since $\mathbf H=\operatorname{diag}(\mathbf H_X,\mathbf H_M)$ and $\Sigma_{\mathbf e}=\allowbreak \operatorname{diag}(\Sigma_{\mathbf e_X},\Sigma_{\mathbf e_M})$
are block diagonal,
\begin{align*}
\mathbf H^\top\Sigma_{\mathbf e}^{-1}\mathbf H&=\operatorname{diag}\Big(\mathbf H_X^\top\big(\tfrac1p\mathbf V_X^\top\Sigma_{\mathbf u_X}\mathbf V_X\big)^{-1}\mathbf H_X,\ \mathbf H_M^\top\big(\tfrac1q\mathbf V_M^\top\Sigma_{\mathbf u_M}\mathbf V_M\big)^{-1}\mathbf H_M\Big),
\end{align*}
so $\lambda_{\min}(\mathbf H^\top\Sigma_{\mathbf e}^{-1}\mathbf H)$ is the smaller of the two blocks minimum eigenvalues. For the first block,
\begin{align*}
\lambda_{\min}( \mathbf{H}_X^\top \big(\tfrac{1}{p}\mathbf{V}_X^\top \Sigma_{\mathbf{u}_X}\mathbf{V}_X\big)^{-1}\mathbf{H}_X )
&\ge \lambda_{\min}(\mathbf{H}_X^\top \mathbf{H}_X)\,\lambda_{\min}\big(\tfrac{1}{p}\mathbf{V}_X^\top\Sigma_{\mathbf{u}_X}\mathbf{V}_X\big)^{-1} \\
&\ge \frac{p\,\lambda_{\min}(\mathbf{H}_X^\top \mathbf{H}_X)}{\lambda_{\max}(\mathbf{V}_X^\top\mathbf{V}_X)\,\lambda_{\max}(\Sigma_{\mathbf{u}_X})} \\
&=\frac{p\,\nu_{\min}^2(\mathbf{H}_X)}{\lambda_{\max}(\mathbf{V}_X^\top\mathbf{V}_X)\,\lambda_{\max}(\Sigma_{\mathbf{u}_X})},
\end{align*}
using $\lambda_{\max}(\mathbf V_X^\top\mathbf V_X)\le C$ from Definition~\ref{definition:H}, and the identical computation gives 
\begin{align*}
\lambda_{\min}(\mathbf H_M^\top(\tfrac1q\mathbf V_M^\top\Sigma_{\mathbf u_M}\mathbf V_M)^{-1}\mathbf H_M)\ge \frac{{(q\,\nu_{\min}^2(\mathbf H_M))}}{{\lambda_{\max}(\mathbf V_M^\top\mathbf V_M)\,\lambda_{\max}(\Sigma_{\mathbf u_M})}}.
\end{align*}
Then
\begin{align*}
\lambda_{\min}(\mathbf H^\top\Sigma_{\mathbf e}^{-1}\mathbf H)
&\ge\min \left\{ {\frac{p\nu_{\min}^2(\mathbf H_X)}{\bar\lambda_{\mathbf u_X}\bar\lambda_{V_X}},\frac{q\nu_{\min}^2(\mathbf H_M)}{\bar\lambda_{\mathbf u_M}\bar\lambda_{V_M}}}\right\}\ge\frac{\min\{ p\nu_{\min}^2(\mathbf H_X),\, q\nu_{\min}^2(\mathbf H_M)\}}{\max\{\bar\lambda_{\mathbf u_X}\bar\lambda_{V_X},\,\bar\lambda_{\mathbf u_M}\bar\lambda_{V_M}\}}.
\end{align*}
We write $\underline\lambda_f:=\lambda_{\min}(\Sigma_{\mathbf f\mid1})$, and then substituting the lower bound gives
\begin{align*}
\|K_1\|
&\le\sqrt{\kappa(\Sigma_{\mathbf f\mid1})}\,\|\mu_{\mathbf f\mid 1}\|\,\|\Delta'\|\cdot\frac{1}{1+\underline\lambda_f \min\{ p\nu_{\min}^2(\mathbf H_X),\, q\nu_{\min}^2(\mathbf H_M)\}/\bar\lambda} \\
&=\sqrt{\kappa(\Sigma_{\mathbf f\mid1})}\,\|\mu_{\mathbf f\mid 1}\|\,\|\Delta'\|\cdot\frac{\bar\lambda}{\bar\lambda+\underline\lambda_f \min\{ p\nu_{\min}^2(\mathbf H_X),\, q\nu_{\min}^2(\mathbf H_M)\}} \\
&\le\sqrt{\kappa(\Sigma_{\mathbf f\mid1})}\,\frac{\|\mu_{\mathbf f\mid 1}\|\,\|\Delta'\|\,\bar\lambda}{\underline\lambda_f}\cdot\frac{1}{\min\{ p\nu_{\min}^2(\mathbf H_X),\, q\nu_{\min}^2(\mathbf H_M)\}} \\
&= C_1\,\frac{1}{\min\{p\nu_{\min}^2(\mathbf H_X),\,q\nu_{\min}^2(\mathbf H_M)\}},
\end{align*}
where the third inequality drops the positive term $\bar\lambda$ from the denominator.
\\

\noindent\textbf{Negligibility of $K_2$.}  From $L=Z+\mathbf e_0$,
$\mathbb E[L\epsilon_1\mid A=1]=\mathbb E[Z\epsilon_1\mid A=1]+\mathbb E[\mathbf e_0\epsilon_1\mid A=1].$
The first term vanishes because $\mathbb E[\epsilon_1]=0$ by Assumption~\ref{Assumption:covariances}(ii), so
\begin{align*}
\mathbb E[Z\epsilon_1\mid A=1]&=\mathbb E\big[Z\,\mathbb E[\epsilon_1]\mid A=1\big] =0.
\end{align*}
The second term vanishes by Assumption~\ref{Assumption:covariances}(ii), since $\mathbf e_0$ is a linear function of $(\mathbf u_X,\mathbf u_M)$, which are orthogonal to $\epsilon_1$ under $A=1$. Hence $K_2=e_{\mathbf 1}^\top\mathbb E[LL^\top\mid A=1]^{-1}\cdot 0=0$ exactly. Since $R_{\alpha}=-K_1+K_2$, we have that
\begin{align*}
|R_{\alpha}| &\le |K_1| + |K_2| = \mathcal{O}\left( \frac{1}{\min \{ p \nu_{\min}^2(\mathbf{H}_X),\, q \nu_{\min}^2(\mathbf{H}_M)\}} \right).
\end{align*}

\noindent\textbf{Step 2: $T_2$.}\
\label{step:id-T2}

\noindent Using the outcome model exactly as in Step \hyperref[step:id-T1]{1}, we get
\begin{align*}
\mathbf{H}_M^\top& \mathbf S_{\tilde q}\mathbb{E}[L L^\top\mid A=1]^{-1} \mathbb{E}[L Y\mid A=1]\\
&= \mathbf{H}_M^\top \mathbf S_{\tilde q}\mathbb{E}[L L^\top\mid A=1]^{-1} \mathbb{E}[L L^\top\delta\mid A=1]
+ \mathbf{H}_M^\top \mathbf S_{\tilde q}\mathbb{E}[L L^\top\mid A=1]^{-1}\mathbb{E}[L (Z-L)^\top\delta\mid A=1] \\*
&\quad\quad + \mathbf{H}_M^\top \mathbf S_{\tilde q}\mathbb{E}[L L^\top\mid A=1]^{-1} \mathbb{E}[L \epsilon_1\mid A=1] \\*
&= \gamma_1 - \underbrace{\mathbf{H}_M^\top \mathbf S_{\tilde q}\mathbb{E}[L L^\top\mid A=1]^{-1} \mathbb{E}[L \mathbf{e}_0^\top\delta\mid A=1]}_{K_3} +\underbrace{\mathbf{H}_M^\top \mathbf S_{\tilde q}\mathbb{E}[L L^\top\mid A=1]^{-1} \mathbb{E}[L \epsilon_1\mid A=1]}_{K_4},
\end{align*}
so that $\mathbf H_M^\top T_2=\gamma_1+\underbrace{(-K_3+K_4)}_{R_\gamma}$. It suffices to prove the negligibility of $K_3$ and $K_4$.
\\

\noindent\textbf{Negligibility of $K_{3}$.} From $L=\mathbf H_1\mathbf f_1+\mathbf e_0$
and the cancellation $\mathbb E[\mathbf f_1\mathbf e_0^\top\mid A=1]=0$ of
Section~\ref{sec:id-prelim},
\begin{align*}
K_3&=\mathbf{H}_M^\top \mathbf S_{\tilde q} \mathbb{E}[L L^\top\mid A=1]^{-1}\Sigma_{\mathbf{e}_0} \delta.
\end{align*}
To bound the norm of $K_3$, write the block inverse of $\mathbb E[LL^\top\mid A=1]$, partitioned according to the intercept, $X$-block, and $M$-block. Since $\mathbf H_M^\top\mathbf S_{\tilde q}=\mathbf {P}_{\bar q}\mathbf H_1^\top$, where $\mathbf S_{\tilde q}$ selects the last $\tilde q$ projected coordinates and $\mathbf {P}_{\bar q}$ selects the last $\bar q$ true-factor coordinates, and using $\Sigma_{\mathbf e_0}=\operatorname{diag}(0,\Sigma_{\mathbf e_X},\Sigma_{\mathbf e_M})$ together with $\delta=(\alpha_1,(\mathbf H_X^\dagger)^\top\beta_1,(\mathbf H_M^\dagger)^\top\gamma_1)$, we get that
\begin{align*}
\|K_3\| &= \left\| \mathbf P_{\bar q} \mathbf{H}_1^\top  \mathbb{E}[L L^\top| A=1]^{-1}\Sigma_{\mathbf{e}_0} \delta \right\| \\
&= \left\| \mathbf P_{\bar q} \begin{bmatrix}1 & 0 & 0 \\
0 & \mathbf H_{X}^\top & 0 \\
0 & 0 & \mathbf H_{M}^\top\end{bmatrix} \begin{bmatrix}B_{11} & B_{12} & B_{13} \\
B_{21} & B_{22} & B_{23} \\
B_{31} & B_{32} & B_{33}\end{bmatrix}\begin{bmatrix}0 & 0 & 0 \\
0 & \Sigma_{\mathbf{e}_{X}} & 0 \\
0 & 0 & \Sigma_{\mathbf{e}_{M}}\end{bmatrix}\begin{bmatrix}0 \\
(\mathbf{H}_{X}^\dagger)^\top\beta_1 \\
(\mathbf{H}_M^\dagger)^\top \gamma_1\end{bmatrix} \right\| \\
&= \left\| \mathbf H_{M}^\top B_{32} \Sigma_{\mathbf{e}_{X}}(\mathbf{H}_{X}^\dagger)^\top\beta_1 + \mathbf H_{M}^\top B_{33} \Sigma_{\mathbf{e}_M} (\mathbf{H}_M^\dagger)^\top \gamma_1 \right\| \\
&\le \underbrace{\| \mathbf H_{M}^\top B_{32} \Sigma_{\mathbf{e}_{X}}(\mathbf{H}_{X}^\dagger)^\top\beta_1 \|}_{M_1} + \underbrace{\| \mathbf H_{M}^\top B_{33} \Sigma_{\mathbf{e}_M} (\mathbf{H}_M^\dagger)^\top \gamma_1 \|}_{M_2}
\end{align*}
The forms of $B_{32}$ and $B_{33}$ follow from
Lemma~\ref{lemma:augmented-inverse} and a further application of
Block Inverse formula to the $(X,M)$ partition of
$\mathbf H\Sigma_{\mathbf f\mid1}\mathbf H^\top+\Sigma_{\mathbf e}$. The Block Inverse formula gives
\begin{align*}
B_{32}&=-B_{33}\,\mathbf H_M\Sigma_{\mathbf f_M\mathbf f_X^\top\mid 1}\mathbf H_X^\top(\mathbf H_X\Sigma_{\mathbf f_X\mid 1}\mathbf H_X^\top+\Sigma_{\mathbf e_X})^{-1}, \\
B_{33}&= \big(\mathbf{H}_M\Sigma_{\mathbf{f}_M\mid 1}\mathbf{H}_M^\top + \Sigma_{\mathbf{e}_M}-\mathbf{H}_M\Sigma_{\mathbf{f}_M\mathbf{f}_X^\top\mid 1}\mathbf{H}_X^\top(\mathbf{H}_X\Sigma_{\mathbf{f}_X\mid 1}\mathbf{H}_X^\top + \Sigma_{\mathbf{e}_X})^{-1}\mathbf{H}_X\Sigma_{\mathbf{f}_X\mathbf{f}_M^\top\mid 1}\mathbf{H}_M^\top\big)^{-1}.
\end{align*}
Since $M_1$ and $M_2$ are norms, the leading minus sign in $B_{32}$ does not
affect the bounds.
\\\\
\noindent\textbf{Bound on $M_2$.} We have
\begin{align}
& \|M_2\| \notag \\
&= \big\| \mathbf H_{M}^\top B_{33} \Sigma_{\mathbf e_{M}} (\mathbf{H}_{M}^\dagger)^\top \gamma_1 \big\| \notag \\
&= \big\| \mathbf H_{M}^\top \Big(\mathbf{H}_M \underbrace{\big(\Sigma_{\mathbf{f}_M\mid1}-\Sigma_{\mathbf{f}_M\mathbf{f}_X^\top\mid1}\mathbf{H}_X^\top (\mathbf{H}_X\Sigma_{\mathbf{f}_X\mid1}\mathbf{H}_X^\top + \Sigma_{\mathbf{e}_X})^{-1} \mathbf{H}_X\Sigma_{\mathbf{f}_X\mathbf{f}_M^\top\mid1}\big)}_{D}\mathbf{H}_M^\top \notag + \Sigma_{\mathbf{e}_M} \Big)^{-1} \Sigma_{\mathbf{e}_{M}} (\mathbf{H}_{M}^\dagger)^\top \gamma_1 \big\| \notag \\*
&= \big\| \mathbf H_{M}^\top \big(\mathbf{H}_M {D}\mathbf{H}_M^\top + \Sigma_{\mathbf{e}_M} \big)^{-1} \Sigma_{\mathbf{e}_{M}} (\mathbf{H}_{M}^\dagger)^\top \gamma_1 \big\| \notag \\
&\overset{(1)}{=} \big\| \mathbf H_{M}^\top \big(\Sigma_{\mathbf{e}_M}^{-1} - \Sigma_{\mathbf{e}_M}^{-1}\mathbf{H}_M(D^{-1} + \mathbf{H}_M^\top\Sigma_{\mathbf{e}_M}^{-1}\mathbf{H}_M)^{-1}\mathbf{H}_M^\top\Sigma_{\mathbf{e}_M}^{-1}\big) \Sigma_{\mathbf{e}_{M}} \mathbf{H}_{M}(\mathbf{H}_{M}^\top \mathbf{H}_{M})^{-1}\gamma_1 \big\| \notag \\*
&= \big\| (\mathbf{I}_{\bar q} + \mathbf{H}_M^\top\Sigma_{\mathbf{e}_M}^{-1}\mathbf{H}_M D)^{-1} \gamma_1 \big\|
\label{equation:M2}
\end{align}
where step $(1)$ applies the Woodbury identity of Lemma~\ref{lemma:woodbury} to the term $(\mathbf H_M D\mathbf H_M^\top+\Sigma_{\mathbf e_M})^{-1}$, and the remaining equalities collect terms as in the bound for $\|K_1\|$ in Step~\hyperref[step:id-T1]{1}, using $(\mathbf H_M^\dagger)^\top=\mathbf H_M(\mathbf H_M^\top\mathbf H_M)^{-1}$, and then applying the symmetric-similarity reduction used for $K_1$ with $A_M:=\mathbf H_M^\top\Sigma_{\mathbf e_M}^{-1}\mathbf H_M$, we get 
\begin{align*}
\|M_2\|&\le \sqrt{\kappa(D)}\,\frac{\|\gamma_1\|}{1+\lambda_{\min}(D)\lambda_{\min}(A_M)}.
\end{align*}
The two-sided bound on $D$ established below gives $\lambda_{\min}(D)\ge\lambda_{\min}(\Sigma_{\mathbf u_{XM,1}})$ and $\kappa(D)=O(1)$. For $\lambda_{\min}(A_M)$, the identical computation as in Step~\hyperref[step:id-T1]{3} gives
\begin{align}
\lambda_{\min}(A_M)&\ge\frac{q\,\nu_{\min}^2(\mathbf H_M)}{\bar\lambda_{V_M}\bar\lambda_{\mathbf u_M}}.
\label{eq:AM-bound}
\end{align}
Now, combining these,
\begin{align*}
&\lambda_{\min}(D)\,\lambda_{\min}(A_M)\ge\lambda_{\min}(\Sigma_{\mathbf u_{XM,1}})\cdot\frac{q\,\nu_{\min}^2(\mathbf H_M)}{\bar\lambda_{V_M}\bar\lambda_{\mathbf u_M}},
\end{align*}
so, since $t\mapsto1/(1+t)$ is decreasing and dropping the positive term $\bar\lambda_{V_M}\bar\lambda_{\mathbf u_M}$ from the denominator only enlarges the bound,
\begin{align*}
\|M_2\| &\le C_{2,M}\,\frac{1}{q\,\nu_{\min}^2(\mathbf H_M)}.
\end{align*}

\medskip
\noindent\textbf{Two-sided bound on $D$.}
We now bound $D$ above and below, uniformly in $p,q$. We can simplify this expression using that $\mathbf f_M \mid A = 1$ has the same distribution as $\delta_1 + \bB_1 \mathbf f_X + \bu_{XM, 1} \mid A = 1$ (see Equation \eqref{eq:factor})
\begin{align*}
\Sigma_{\mathbf{f}_M|1}
&= \var(\mathbf f_M \mid A = 1) = \var(\delta_1 + \bB_1 \mathbf f_X + \bu_{XM, 1} \mid A = 1) = \bB_1 \Sigma_{\mathbf{f}_X |1}\mathbf{B}_1^\top+\Sigma_{\mathbf{u}_{XM,1}},
\end{align*}
where in the last equality we use the fact that $\bu_{XM, 1} \independent \mathbf f_X \mid A = 1$.  Similarly, $\Sigma_{\mathbf f_M\mathbf f_X^\top\mid1}=\mathbf B_1\Sigma_{\mathbf f_X\mid1}$. Also, since $\Sigma_{\mathbf f_X\mid1}$ is symmetric, we have $\Sigma_{\mathbf f_X\mathbf f_M^\top\mid1}=\Sigma_{\mathbf f_M\mathbf f_X^\top\mid1}^\top=\Sigma_{\mathbf f_X\mid1}\mathbf B_1^\top.$ Now, substituting these identities into
\begin{align*}D=\Sigma_{\mathbf f_M\mid1}-\Sigma_{\mathbf f_M\mathbf f_X^\top\mid1}\mathbf H_X^\top(\mathbf H_X\Sigma_{\mathbf f_X\mid1}\mathbf H_X^\top+\Sigma_{\mathbf e_X})^{-1}\mathbf H_X\Sigma_{\mathbf f_X\mathbf f_M^\top\mid1},\end{align*}
we get that
\begin{align}
D&=\mathbf B_1\Sigma_{\mathbf f_X\mid1}\mathbf B_1^\top+\Sigma_{\mathbf u_{XM,1}}-\mathbf B_1\Sigma_{\mathbf f_X\mid1}\mathbf H_X^\top(\mathbf H_X\Sigma_{\mathbf f_X\mid1}\mathbf H_X^\top+\Sigma_{\mathbf e_X})^{-1}\mathbf H_X\Sigma_{\mathbf f_X\mid1}\mathbf B_1^\top\notag\\
&=\Sigma_{\mathbf u_{XM,1}}+\mathbf B_1\left(\underbrace{\Sigma_{\mathbf f_X\mid1}-\Sigma_{\mathbf f_X\mid1}\mathbf H_X^\top(\mathbf H_X\Sigma_{\mathbf f_X\mid1}\mathbf H_X^\top+\Sigma_{\mathbf e_X})^{-1}\mathbf H_X\Sigma_{\mathbf f_X\mid1}}_{E}\right)\mathbf B_1^\top\notag\\
&=\Sigma_{\mathbf u_{XM,1}}+\mathbf B_1E\mathbf B_1^\top.
\label{eq:D-decomp}
\end{align}

\noindent Now, we first show that $E\succeq0$. We know that $\mathbf H_X\mathbf f_X+\mathbf e_X$ is the $X$-block of $L$ from \eqref{eq:LL}. By Assumption~\ref{Assumption:covariances}(ii) and the tower property, we have
\begin{align*}\mathrm{Cov}(\mathbf H_X\mathbf f_X+\mathbf e_X\mid A=1)=\mathbf H_X\Sigma_{\mathbf f_X\mid1}\mathbf H_X^\top+\Sigma_{\mathbf e_X},\qquad \mathrm{Cov}(\mathbf f_X,\mathbf H_X\mathbf f_X+\mathbf e_X\mid A=1)=\Sigma_{\mathbf f_X\mid1}\mathbf H_X^\top.\end{align*}
Therefore,
\begin{align*}\Omega:=\mathrm{Cov}\left(\begin{pmatrix}\mathbf f_X\\ \mathbf H_X\mathbf f_X+\mathbf e_X\end{pmatrix}\Bigm|A=1\right)=\begin{pmatrix}\Sigma_{\mathbf f_X\mid1}&\Sigma_{\mathbf f_X\mid1}\mathbf H_X^\top\\ \mathbf H_X\Sigma_{\mathbf f_X\mid1}&\mathbf H_X\Sigma_{\mathbf f_X\mid1}\mathbf H_X^\top+\Sigma_{\mathbf e_X}\end{pmatrix}\succeq0.\end{align*}
The bottom-right block is invertible since $\mathbf H_X\Sigma_{\mathbf f_X\mid1}\mathbf H_X^\top\succeq0$ and $\Sigma_{\mathbf e_X}\succ0$. Now, fix $a$ and let
\begin{align*}b:=-(\mathbf H_X\Sigma_{\mathbf f_X\mid1}\mathbf H_X^\top+\Sigma_{\mathbf e_X})^{-1}\mathbf H_X\Sigma_{\mathbf f_X\mid1}a.\end{align*}
Since $\Omega\succeq0$, we have
\begin{align*}
0&\le\binom{a}{b}^\top\Omega\binom{a}{b}=a^\top\Sigma_{\mathbf f_X\mid1}a-a^\top\Sigma_{\mathbf f_X\mid1}\mathbf H_X^\top(\mathbf H_X\Sigma_{\mathbf f_X\mid1}\mathbf H_X^\top+\Sigma_{\mathbf e_X})^{-1}\mathbf H_X\Sigma_{\mathbf f_X\mid1}a=a^\top Ea.
\end{align*}
Since $a$ was arbitrary, $E\succeq0$. Also, $\mathbf B_1E\mathbf B_1^\top\succeq0$. Thus, from \eqref{eq:D-decomp},
\begin{align*}D-\Sigma_{\mathbf u_{XM,1}}=\mathbf B_1E\mathbf B_1^\top\succeq0.\end{align*}
For the lower bound, let $v$ be a unit eigenvector of $D$ corresponding to $\lambda_{\min}(D)$. Then,
\begin{align*}
\lambda_{\min}(D)&=v^\top Dv=v^\top\Sigma_{\mathbf u_{XM,1}}v+v^\top(D-\Sigma_{\mathbf u_{XM,1}})v\ge v^\top\Sigma_{\mathbf u_{XM,1}}v\ge\lambda_{\min}(\Sigma_{\mathbf u_{XM,1}}).
\end{align*}
Hence, using Assumption~\ref{Assumption:covariances}(ii),
\begin{align*}\lambda_{\min}(D)\ge\lambda_{\min}(\Sigma_{\mathbf u_{XM,1}})\ge c>0.\end{align*}

\noindent
For the upper bound, for any matrix $\mathbf C$ and symmetric positive definite $K$, $\mathbf C^\top K^{-1}\mathbf C\succeq0$. Now, take $\mathbf C=\mathbf H_X\Sigma_{\mathbf f_X\mid1}$ and $K=\mathbf H_X\Sigma_{\mathbf f_X\mid1}\mathbf H_X^\top+\Sigma_{\mathbf e_X}$. Then the subtracted term in $E$ is positive semidefinite, and hence $E\preceq\Sigma_{\mathbf f_X\mid1}.$ Therefore, $\mathbf B_1E\mathbf B_1^\top\preceq\mathbf B_1\Sigma_{\mathbf f_X\mid1}\mathbf B_1^\top.$ Using \eqref{eq:D-decomp} and $\Sigma_{\mathbf f_M\mid1}=\mathbf B_1\Sigma_{\mathbf f_X\mid1}\mathbf B_1^\top+\Sigma_{\mathbf u_{XM,1}}$, we have
\begin{align*}
D=\Sigma_{\mathbf u_{XM,1}}+\mathbf B_1E\mathbf B_1^\top\preceq\Sigma_{\mathbf u_{XM,1}}+\mathbf B_1\Sigma_{\mathbf f_X\mid1}\mathbf B_1^\top=\Sigma_{\mathbf f_M\mid1}.
\end{align*}
Let $u$ be a unit eigenvector of $D$ corresponding to $\lambda_{\max}(D)$. Then, using Assumption~\ref{Assumption:covariances}(iv),
\begin{align*}
\lambda_{\max}(D)=u^\top Du
=u^\top\Sigma_{\mathbf f_M\mid1}u-u^\top(\Sigma_{\mathbf f_M\mid1}-D)u\le u^\top\Sigma_{\mathbf f_M\mid1}u\le\lambda_{\max}(\Sigma_{\mathbf f_M\mid1})\le C,
\end{align*}
\\
\noindent
Lastly, combining the two bounds, we get
\begin{align}
0<c&\le\lambda_{\min}(D)\le\lambda_{\max}(D)\le C<\infty,\quad \kappa(D)=\frac{\lambda_{\max}(D)}{\lambda_{\min}(D)}\le\frac{C}{c}=O(1).
\label{eq:D-two-sided}
\end{align}
Since $\lambda_{\min}(D)>0$, $D\succ0$, which justifies the use of $D^{1/2}$ and $D^{-1/2}$ above.

\medskip
\noindent\textbf{Bound on $M_1$.} We proceed in a similar way as for $M_2$. First, we substitute the
expression for $B_{32}$ and apply the Woodbury identity of
Lemma~\ref{lemma:woodbury} to
$(\mathbf H_X\Sigma_{\mathbf f_X\mid1}\mathbf H_X^\top+\Sigma_{\mathbf e_X})^{-1}$. This gives
\begin{align*}
&\|M_1\| \\
&= \big\| \mathbf H_{M}^\top B_{32} \Sigma_{\mathbf{e}_X} (\mathbf{H}_X^\dagger)^\top \beta_1 \big\| \\
&= \big\| \mathbf H_{M}^\top\big(\mathbf{H}_M{D}\mathbf{H}_M^\top+ \Sigma_{\mathbf{e}_M}\big)^{-1} \mathbf{H}_M \Sigma_{\mathbf f_M \mathbf f_X^\top\mid1} \mathbf{H}_X^\top (\mathbf{H}_X \Sigma_{\mathbf{f}_X\mid1}\mathbf{H}_X^\top + \Sigma_{\mathbf e_X})^{-1} \Sigma_{\mathbf{e}_X} (\mathbf{H}_X^\dagger)^\top \beta_1\big\| \\
&\overset{(1)}{=} \big\| \mathbf H_{M}^\top\big(\mathbf{H}_M{D}\mathbf{H}_M^\top+ \Sigma_{\mathbf{e}_M}\big)^{-1} \mathbf{H}_M \Sigma_{\mathbf f_M \mathbf f_X^\top\mid1} \mathbf{H}_X^\top \big(\Sigma_{\mathbf{e}_X}^{-1} - \Sigma_{\mathbf{e}_X}^{-1}\mathbf{H}_X(\Sigma_{\mathbf{f}_X\mid1}^{-1} + \mathbf{H}_X^\top\Sigma_{\mathbf{e}_X}^{-1}\mathbf{H}_X)^{-1}\mathbf{H}_X^\top\Sigma_{\mathbf{e}_X}^{-1}\big) \Sigma_{\mathbf{e}_X} (\mathbf{H}_X^\dagger)^\top \beta_1\big\|.
\end{align*}
Now, expanding the inner factor and collecting terms exactly as in the derivation of
$\|K_1\|$ in Step~\hyperref[step:id-T1]{1}, we get that
\begin{align}
&\| M_1\| \notag \\*
&= \big\| \mathbf H_{M}^\top\Big(\mathbf{H}_M{D}\mathbf{H}_M^\top+ \Sigma_{\mathbf{e}_M}\Big)^{-1} \mathbf{H}_M \Sigma_{\mathbf f_M \mathbf f_X^\top |1} \beta_1 \notag \\
&\quad - \mathbf H_{M}^\top\Big(\mathbf{H}_M{D}\mathbf{H}_M^\top+ \Sigma_{\mathbf{e}_M}\Big)^{-1}\mathbf{H}_M \Sigma_{\mathbf f_M \mathbf f_X^\top |1} \mathbf{H}_X^\top\Sigma_{\mathbf{e}_X}^{-1}\mathbf{H}_X(\Sigma_{\mathbf{f}_X|1}^{-1} + \mathbf{H}_X^\top\Sigma_{\mathbf{e}_X}^{-1}\mathbf{H}_X)^{-1} \beta_1\big\| \notag \\*
&= \big\|  \mathbf H_{M}^\top\Big(\mathbf{H}_M{D}\mathbf{H}_M^\top+ \Sigma_{\mathbf{e}_M}\Big)^{-1}\mathbf{H}_M \Sigma_{\mathbf f_M \mathbf f_X^\top |1} \Sigma_{
     \mathbf f_X|1}^{-1} (\Sigma_{\mathbf{f}_X|1}^{-1} + \mathbf{H}_X^\top\Sigma_{\mathbf{e}_X}^{-1}\mathbf{H}_X)^{-1}\beta_1 \big\| \notag \\
&= \big\|  \mathbf H_{M}^\top\Big(\mathbf{H}_M{D}\mathbf{H}_M^\top+ \Sigma_{\mathbf{e}_M}\Big)^{-1}\mathbf{H}_M \Sigma_{\mathbf f_M \mathbf f_X^\top |1}  (\mathbf{I}_{\bar p} + \mathbf{H}_X^\top\Sigma_{\mathbf{e}_X}^{-1}\mathbf{H}_X \Sigma_{\mathbf{f}_X|1})^{-1}\beta_1 \big\| \notag \\
&\le \big\|  \mathbf H_{M}^\top\Big(\mathbf{H}_M{D}\mathbf{H}_M^\top+ \Sigma_{\mathbf{e}_M}\Big)^{-1}\mathbf{H}_M \big\| \cdot \big\|\Sigma_{\mathbf f_M \mathbf f_X^\top |1}  (\mathbf{I}_{\bar p} + \mathbf{H}_X^\top\Sigma_{\mathbf{e}_X}^{-1}\mathbf{H}_X \Sigma_{\mathbf{f}_X|1})^{-1}\beta_1 \big\|
\label{equation:M1}
\end{align}
We apply the same symmetric similarity argument as for $K_1$ above, and we get that
\begin{align*}
\|M_1\|
&\le \big\|\mathbf H_M^\top(\mathbf H_M D \mathbf H_M^\top + \Sigma_{\mathbf e_M})^{-1}\mathbf H_M\big\|\cdot\sqrt{\kappa(\Sigma_{\mathbf f_X\mid1})} \cdot\|\Sigma_{\mathbf f_M \mathbf f_X^\top \mid 1}\|\,\|\beta_1\|\,\frac{\bar\lambda_{V_X}\bar\lambda_{\mathbf u_X}}{\lambda_{\min}(\Sigma_{\mathbf f_X\mid1})}\cdot\frac{1}{p\nu_{\min}^2(\mathbf H_X)}.
\end{align*}
The last equality comes from $\lambda_{\min}( \mathbf{H}_X^\top\Sigma_{\mathbf{e}_X}^{-1}\mathbf{H}_X ) \ge\frac{p\nu_{\min}^2(\mathbf{H}_X)}{\bar{\lambda}_{V_X}\bar{\lambda}_{\mathbf u_X}}$ as derived earlier. It remains to show the leading factor is ${\Bigg\|  \mathbf H_{M}^\top\Big(\mathbf{H}_M{D}\mathbf{H}_M^\top+ \Sigma_{\mathbf{e}_M}\Big)^{-1}\mathbf{H}_M \Bigg\|}=O(1)$. Applying the Woodbury identity
to $(\mathbf H_M D\mathbf H_M^\top+\Sigma_{\mathbf e_M})^{-1}$ and collecting terms, we get that
\begin{align*}
&\|\mathbf H_{M}^\top\big(\mathbf{H}_M{D}\mathbf{H}_M^\top+ \Sigma_{\mathbf{e}_M}\big)^{-1}\mathbf{H}_M\| \\*
&= \|\mathbf H_{M}^\top( \Sigma_{\mathbf{e}_M}^{-1}-\Sigma_{\mathbf{e}_M}^{-1}\mathbf{H}_M ({D}^{-1}+\mathbf{H}_M^\top \Sigma_{\mathbf{e}_M}^{-1}\mathbf{H}_M)^{-1}\mathbf{H}_M^\top \Sigma_{\mathbf e_M}^{-1}) \mathbf{H}_M \| \\
&= \|\mathbf H_{M}^\top \Sigma_{\mathbf{e}_M}^{-1} \mathbf{H}_M-  \mathbf{H}_M^\top\Sigma_{\mathbf{e}_M}^{-1}\mathbf{H}_M ({D}^{-1}+\mathbf{H}_M^\top \Sigma_{\mathbf{e}_M}^{-1}\mathbf{H}_M)^{-1}\mathbf{H}_M^\top \Sigma_{\mathbf e_M}^{-1} \mathbf{H}_M \| \\
&=\|\mathbf{H}_M^\top\Sigma_{\mathbf{e}_M}^{-1}\mathbf{H}_M (\mathbf{I}_{\bar q}+D\mathbf{H}_M^\top \Sigma_{\mathbf{e}_M}^{-1}\mathbf{H}_M)^{-1}   \| \\
&=\|D^{-1}(\mathbf{I}_{\bar q}+D\mathbf{H}_M^\top\Sigma_{\mathbf{e}_M}^{-1}\mathbf{H}_M -\mathbf{I}_{\bar q} )(\mathbf{I}_{\bar q}+D\mathbf{H}_M^\top \Sigma_{\mathbf{e}_M}^{-1}\mathbf{H}_M)^{-1}   \| \\
&\le \|D^{-1}\|  + \|  D^{-1} (\mathbf{I}_{\bar q}+D\mathbf{H}_M^\top \Sigma_{\mathbf{e}_M}^{-1}\mathbf{H}_M)^{-1} \|\\
&\le \frac{1}{\lambda_{\min}(D)} + \|  D^{-1} (\mathbf{I}_{\bar q}+D\mathbf{H}_M^\top \Sigma_{\mathbf{e}_M}^{-1}\mathbf{H}_M)^{-1} \| \\
&\le \frac{1}{\lambda_{\min}(\Sigma_{\mathbf{u}_{XM,1}})} + \frac{1}{\lambda_{\min}(\Sigma_{\mathbf{u}_{XM,1}}) }\cdot \| (\mathbf{I}_{\bar q}+D\mathbf{H}_M^\top \Sigma_{\mathbf{e}_M}^{-1}\mathbf{H}_M)^{-1} \|
\end{align*}
Now, using the similar steps as for $K_1$, we get that
\begin{align*}
&\|\mathbf H_{M}^\top\big(\mathbf{H}_M{D}\mathbf{H}_M^\top+ \Sigma_{\mathbf{e}_M}\big)^{-1}\mathbf{H}_M\| \le   \frac{1}{\lambda_{\min}(\Sigma_{\mathbf{u}_{XM,1}})} + \frac{1}{\lambda_{\min}(\Sigma_{\mathbf{u}_{XM,1}})}\cdot\sqrt{\kappa(D)}\cdot \frac{1}{1+\lambda_{\min}(D)\frac{(q \nu_{\min}^2(\mathbf{H}_M))}{\bar{\lambda}_{\mathbf u_M}\bar{\lambda}_{V_M}}}
\end{align*}
We name this upper bound as $C_{H_M}$. Thus, we have that  ${\Big\|  \mathbf H_{M}^\top\Big(\mathbf{H}_M{D}\mathbf{H}_M^\top+ \Sigma_{\mathbf{e}_M}\Big)^{-1}  \mathbf{H}_M \Big\|}= O(1)$. Substituting this bound into the inequality \eqref{equation:M1}, we get
\begin{align*}
\| M_1\| &\le C_{H_M} \sqrt{\kappa(\Sigma_{\mathbf f_X|1})}
\frac{\|\Sigma_{\mathbf f_M \mathbf f_X^\top \mid 1}\|
\|\beta_1\|\bar{\lambda}_{V_X}\bar{\lambda}_{\mathbf u_X}}{\lambda_{\min}(\Sigma_{\mathbf f_X \mid 1})} \cdot \frac{1}{p \nu_{\min}^2(\mathbf{H}_X)}= C_{2,X}\cdot \frac{1}{p \nu_{\min}^2(\mathbf{H}_X)}.
\end{align*}
Using the bounds established above for $M_1$ and $M_2$ in $\|K_3\| \le M_1 + M_2$, it follows that
\begin{align*}
\|K_3\| &\le M_1+M_2 = \mathcal{O} \left(\frac{1}{p \nu^2_{\min}(\mathbf{H}_X)}+ \frac{1}{q \nu^2_{\min}(\mathbf{H}_M)} \right).
\end{align*}

\noindent 
\textbf{For $K_4$.} We have that
\begin{align*}
K_4 &= \mathbf{H}_M^\top \mathbf S_{\tilde q} \mathbb{E}[L L^\top| A=1]^{-1} \mathbb{E}[ L \epsilon_1| A=1] = 0.
\end{align*}
By the same decomposition $L=Z+\mathbf e_0$ and the same two conditional-mean-zero arguments used for $K_2$, $\mathbb E[L\epsilon_1\mid A=1]=0$ exactly. Since $R_\gamma=-K_3+K_4$, we have that
\begin{align*}
\|R_{\gamma}\| &\le \|K_3\|+ \|K_4\| = \mathcal{O}\left( \frac{1}{p \nu^2_{\text{min}}(\mathbf{H}_X)}+ \frac{1}{q \nu^2_{\text{min}}(\mathbf{H}_M)}\right)
\end{align*}

\noindent\textbf{Step 3: $T_3$.}\label{step:id-T3}\\
To identify $\delta_0$ up to the linear transformation $\mathbf H_M$, we decompose $N$ under the control group. Using $M=\mathbf\Gamma_M\mathbf f_M+\mathbf u_M$ and the mediator regression $\mathbf f_M=\delta_0+\mathbf B_0\mathbf f_X+\mathbf u_{XM,0}$ under $A=0$,
\begin{align}
N
&= \frac{1}{q}\mathbf{W}_M^\top M \\
&= \frac{1}{q}\mathbf{W}_M^\top \Gamma_M \mathbf{f}_M + \frac{1}{q}\mathbf{W}_M^\top \mathbf{u}_M \notag \\
&= \frac{1}{q}\mathbf{W}_M^\top \Gamma_M (\delta_0 + \mathbf{B}_0 \mathbf{f}_X+ \mathbf{u}_{XM,0})+ \frac{1}{q}\mathbf{W}_M^\top \mathbf{u}_M \notag \\
&= \mathbf{H}_M (\delta_0 + \mathbf{B}_0 \mathbf{f}_X+ \mathbf{u}_{XM,0})+ \mathbf{e}_M \notag \\
&= \mathbf{H}_M \delta_0 + \mathbf{H}_M\mathbf{B}_0 \mathbf{H}_X^\dagger \mathbf{H}_X\mathbf{f}_X+ \mathbf{H}_M\mathbf{u}_{XM,0}+ \mathbf{e}_M \notag \\
&= \phi_{XM,0}^\top \mathbf{H}_{X1}\mathbf{f}_{X1} + \mathbf{e}_{XM,0},
\label{equation : R hat 2}
\end{align}
so that $N^\top=\mathbf f_{X1}^\top\mathbf H_{X1}^\top\phi_{XM,0}+\mathbf e_{XM,0}^\top$, where
\begin{align*}
\phi_{XM,0}&=\begin{pmatrix}\delta_0^\top\mathbf H_M^\top \\(\mathbf H_M\mathbf B_0\mathbf H_X^\dagger)^\top\end{pmatrix}=(\mathbf H_{X1}^\dagger)^\top\Phi\mathbf H_M^\top,\quad \mathbf e_{XM,0}^\top=\mathbf u_{XM,0}^\top\mathbf H_M^\top+\mathbf e_M^\top,
\end{align*}
and where, since $(\mathbf H_M\mathbf B_0\mathbf H_X^\dagger)^\top=(\mathbf H_X^\dagger)^\top\mathbf B_0^\top\mathbf H_M^\top$, we have
\begin{align}
\Phi&:=\begin{pmatrix}\delta_0^\top \\\mathbf B_0^\top\end{pmatrix}\in\mathbb R^{(1+\bar p)\times\bar q},\quad\Phi' := \mathbf{B}_0^\top, \quad \|\Phi'\| =\|\mathbf B_0\|_{\mathrm{op}}\le C
\label{eq:Phi-prime}
\end{align}
by Assumption \ref{Assumption:pervasiveness}(i). The matrix $\Phi'$ denotes the lower $\bar p$-row block of $\Phi$, the analogue of the truncation $\Delta'=(\beta_1^\top,\gamma_1^\top)^\top$ of $\Delta=(\alpha_1,\beta_1^\top,\gamma_1^\top)^\top$ used in Step \hyperref[step:id-T1]{1}. The quantity $\Phi'$ is what appears below wherever $\Sigma_{\mathbf e_X}$ acts on $\Phi$.
\\
\noindent
We have
\begin{align*}
&e_1^\top  \mathbb{E}[Q Q^\top|A=0]^{-1} \mathbb{E}[Q {N}^\top|A=0] \\*
&=e_1^\top  \mathbb{E}[Q Q^\top|A=0]^{-1} \mathbb{E}[ Q Q^\top \phi_{XM,0}|A=0 ]+e_1^\top  \mathbb{E}[Q Q^\top|A=0]^{-1}\mathbb{E}[Q {N}^\top - Q Q^\top \phi_{XM,0}|A=0] \\*
&=\delta_0^\top\mathbf{H}_M^\top +e_1^\top  \mathbb{E}[Q Q^\top|A=0]^{-1}  \mathbb{E}[Q (\mathbf{f}_{X1}^\top \mathbf{H}_{X1}^\top\phi_{XM,0}+\mathbf{e}_{XM,0}^\top ) - Q Q^\top \phi_{XM,0}|A=0] \\
&=\delta_0^\top \mathbf{H}_M^\top +e_1^\top  \mathbb{E}[Q Q^\top|A=0]^{-1}  \mathbb{E}[Q (\mathbf{f}_{X1}^\top \mathbf{H}_{X1}^\top - Q^\top )\phi_{XM,0}|A=0] +e_1^\top  \mathbb{E}[Q Q^\top|A=0]^{-1}\mathbb{E}[Q \mathbf{e}_{XM,0}^\top|A=0] \\*
&=\delta_0^\top\mathbf{H}_M^\top +e_1^\top  \mathbb{E}[Q Q^\top|A=0]^{-1}  \mathbb{E}[Q (-\mathbf{e}_{X0}^\top )\phi_{XM,0}|A=0] +e_1^\top \mathbb{E}[Q Q^\top|A=0]^{-1} \mathbb{E}[Q (\mathbf{e}_{XM,0}^\top)|A=0] \\*
&\overset{(1)}{=}\delta_0^\top\mathbf{H}_M^\top -\underbrace{e_1^\top  \mathbb{E}[Q Q^\top|A=0]^{-1}  \mathbb{E}[\mathbf{H}_{X1}\mathbf{f}_{X1} \mathbf{e}_{X0}^\top \phi_{XM,0}|A=0]}_{=0} -e_1^\top  \mathbb{E}[Q Q^\top|A=0]^{-1}  \mathbb{E}[\mathbf{e}_{X0} \mathbf{e}_{X0}^\top \phi_{XM,0}|A=0] \\*
&~~~~+\underbrace{e_1^\top \mathbb{E}[Q Q^\top|A=0]^{-1} \mathbb{E}[\mathbf{H}_{X1}\mathbf{f}_{X1} (\mathbf{e}_{XM,0}^\top)|A=0]}_{=0 }+e_1^\top \mathbb{E}[Q Q^\top|A=0]^{-1} \mathbb{E}[\mathbf{e}_{X0}\mathbf{e}_{XM,0}^\top|A=0] \\*
&= \delta_0^\top\mathbf{H}_M^\top -\underbrace{e_1^\top  \mathbb{E}[Q Q^\top|A=0]^{-1}  \mathbb{E}[ \mathbf{e}_{X0}\mathbf{e}_{X0}^\top]\phi_{XM,0}}_{K_5}+\underbrace{e_1^\top  \mathbb{E}[Q Q^\top|A=0]^{-1}\mathbb{E}[ \mathbf{e}_{X0} \mathbf{e}_{XM,0}^\top]}_{K_6}
\end{align*}
Here, $(1)$ uses $\mathbb E[\mathbf f_{X1}\mathbf e_{X0}^\top\mid A=0]=0$ and $\mathbb E[\mathbf f_{X1}\mathbf e_{XM,0}^\top\mid A=0]=0$. These follow from $\mathbb E[\mathbf e_{X0}\mid \mathbf f_X,A=0]=0$ and $\mathbb E[\mathbf e_{XM,0}\mid\mathbf f_X,A=0]=0$, respectively, via the tower property. Since $\mathbf e_{XM,0}=\mathbf H_M\mathbf u_{XM,0}+\mathbf e_M$, we have
\begin{align*}
&\mathbb E[\mathbf e_{XM,0}^\top\mid \mathbf f_X,A=0]=\mathbb E[\mathbf u_{XM,0}^\top\mid \mathbf f_X,A=0]\mathbf H_M^\top+\mathbb E[\mathbf e_M^\top\mid\mathbf f_X,A=0]=0,
\end{align*}
using Assumptions \ref{Assumption:covariances}(ii). Consider that $R_\delta = -K_5 + K_6$, and we need to bound both the terms.
\\

\noindent
\textbf{Negligibility of $R_{\delta}$.}\\
Recall that we need to consider the negligibility of $\tilde R_\delta$ when $R_\delta = \tilde R_\delta \mathbf{H}_M^\top$ as mentioned at the start of the proof. From the factor representation $Q = \mathbf{H}_{X1}\mathbf{f}_{X1} + \mathbf{e}_{X0}$, we have 
\begin{align*}
K_5
&= e_{\mathbf 1}^\top \mathbb{E}[Q Q^\top\mid A=0]^{-1}\mathbb{E}[\mathbf{e}_{X0} \mathbf{e}_{X0}^\top]\phi_{XM,0} \\
&\overset{(1)}{=} e_{\mathbf 1}^\top \big(\mathbb{E}[\mathbf{H}_{X1}\mathbf{f}_{X1}\mathbf{f}_{X1}^\top \mathbf{H}_{X1}^\top\mid A=0]  +\mathbb{E}[ \mathbf{e}_{X0}\mathbf{e}_{X0}^\top\mid A=0]\big)^{-1} \mathbb{E}[\mathbf{e}_{X0}\mathbf{e}_{X0}^\top]\phi_{XM,0} \\*
&= e_{\mathbf 1}^\top \big(\mathbf{H}_{X1} \Lambda_{\mathbf{f}_{X1}\mid0} \mathbf{H}_{X1}^\top  +\Sigma_{\mathbf{e}_{X0}} \big)^{-1} \Sigma_{\mathbf{e}_{X0}} \phi_{XM,0} \\
&= e_{\mathbf 1}^\top \big(\mathbf{H}_{X1} \Lambda_{\mathbf{f}_{X1}\mid0} \mathbf{H}_{X1}^\top  +\Sigma_{\mathbf{e}_{X0}} \big)^{-1} \Sigma_{\mathbf{e}_{X0}} (\mathbf{H}_{X1}^\dagger)^\top \Phi \mathbf{H}_M^\top.
\end{align*}
Here, $(1)$ follows from $\mathbb E[\mathbf f_{X1}\mathbf e_{X0}^\top\mid A=0]=0$, so the two cross terms vanish, and where $\Lambda_{\mathbf f_{X1}\mid0}:=\mathbb E[\mathbf f_{X1}\mathbf f_{X1}^\top\mid A=0]$ is the uncentered second moment. We now rewrite the expression for $K_5$.
\begin{align*}
K_5 &= \underbrace{e_1^\top (\mathbf{H}_{X1} \Lambda_{\mathbf{f}_{X1}\mid0} \mathbf{H}_{X1}^\top  +\Sigma_{\mathbf{e}_{X0}} )^{-1} \Sigma_{\mathbf{e}_{X0}} \mathbf{H}_{X1}(\mathbf{H}_{X1}^\top \mathbf{H}_{X1})^{-1}\Phi}_{\tilde{K}_5} \mathbf{H}_M^\top =\tilde{K}_5 \mathbf{H}_M^\top
\end{align*}
Now, we bound the $\tilde K_5$,
\begin{align*}
&\|\tilde K_5\| \\
&= \left\| e_1^\top  \begin{bmatrix}\begin{gathered}
1+\mu^\top_{\mathbf{f}_{X}|0}\mathbf{H}_X^\top(\mathbf{H}_X\Sigma_{\mathbf{f}_{X}|0}\mathbf{H}_X^\top \\ +\Sigma_{\mathbf{e}_{X}})^{-1}\mathbf{H}_X\mu_{\mathbf{f}_{X}|0}
\end{gathered} & -\mu_{\mathbf{f}_{X}|0}^\top \mathbf{H}_X^\top(\mathbf{H}_X\Sigma_{\mathbf{f}_{X}|0}\mathbf{H}_X^\top + \Sigma_{\mathbf{e}_{X}})^{-1} \\
(-\mu_{\mathbf{f}_{X}|0}^\top\mathbf{H}_X^\top(\mathbf{H}_X\Sigma_{\mathbf{f}_{X}|0}\mathbf{H}_X^\top + \Sigma_{\mathbf{e}_{X}})^{-1})^\top & (\mathbf{H}_X\Sigma_{\mathbf{f}_{X}|0}\mathbf{H}_X^\top+ \Sigma_{\mathbf{e}_{X}})^{-1}\end{bmatrix}  \begin{bmatrix}0 \\
\Sigma_{\mathbf{e}_{X}} \mathbf{H}_X(\mathbf{H}_X^\top \mathbf{H}_X)^{-1}{\Phi'}\end{bmatrix} \right\| \\*
&= \| - \mu_{\mathbf{f}_{X}|0}^\top \mathbf{H}_X^\top(\mathbf{H}_X\Sigma_{\mathbf{f}_{X}|0}\mathbf{H}_X^\top + \Sigma_{\mathbf{e}_{X}})^{-1} \Sigma_{\mathbf{e}_{X}} \mathbf{H}_X(\mathbf{H}_X^\top \mathbf{H}_X)^{-1}{\Phi'} \| \\
&= \| - \mu_{\mathbf{f}_{X}|0}^\top \mathbf{H}_X^\top(\Sigma_{\mathbf{e}_{X}}^{-1}-\Sigma_{\mathbf{e}_{X}}^{-1} \mathbf{H}_X(\Sigma_{\mathbf{f}_{X}|0}^{-1}+\mathbf{H}_X^\top\Sigma_{\mathbf{e}_{X}}^{-1}\mathbf{H}_X)^{-1} \mathbf{H}_X^\top\Sigma_{\mathbf{e}_{X}}^{-1}) \Sigma_{\mathbf{e}_{X}} \mathbf{H}_X(\mathbf{H}_X^\top \mathbf{H}_X)^{-1}{\Phi'}  \| \\
&= \| \mu_{\mathbf{f}_{X}|0}^\top {\Phi'} - \mu_{\mathbf{f}_{X}|0}^\top ((\Sigma_{\mathbf{f}_{X}|0}^{-1}+\mathbf{H}_X^\top \Sigma_{\mathbf{e}_{X}}^{-1} \mathbf{H}_X)-\Sigma_{\mathbf{f}_{X}|0}^{-1})(\Sigma_{\mathbf{f}_{X}|0}^{-1}+\mathbf{H}_X^\top\Sigma_{\mathbf{e}_{X}}^{-1}\mathbf{H}_X)^{-1} {\Phi'} \| \\
&= \| \mu_{\mathbf{f}_{X}|0}^\top \Sigma_{\mathbf{f}_{X}|0}^{-1}(\Sigma_{\mathbf{f}_{X}|0}^{-1}+\mathbf{H}_X^\top\Sigma_{\mathbf{e}_{X}}^{-1}\mathbf{H}_X)^{-1} {\Phi'}  \| \\
&= \| \mu_{\mathbf{f}_{X}|0}^\top (\mathbf{I}_{\bar p}+\mathbf{H}_X^\top\Sigma_{\mathbf{e}_{X}}^{-1}\mathbf{H}_X\Sigma_{\mathbf{f}_{X}|0})^{-1}{\Phi'}  \| \\
&\le \sqrt{\kappa(\Sigma_{\mathbf f_X|0})}\frac{\| \mu_{\mathbf{f}_{X}|0}^\top \|  \| {\Phi'} \|}{1 + \lambda_{\min}(\Sigma_{\mathbf{f}_{X}|0}){\lambda_{\min}(\mathbf{H}_X^\top \Sigma_{\mathbf{e}_{X}}^{-1}\mathbf{H}_X)}}
\end{align*}
Using the similar steps as above, we have
\begin{align*}
\| \tilde K_5\| &\le \sqrt{\kappa(\Sigma_{\mathbf f_X|0})}\frac{\| \mu_{\mathbf{f}_{X}|0}^\top \|  \| {\Phi'} \| \bar{\lambda}_{V_X}\bar{\lambda}_{\mathbf{u}_X}}{ \lambda_{\min}(\Sigma_{\mathbf{f}_{X}|0})} \cdot \frac{1}{p \nu_{\min}^2 (\mathbf{H}_X)} = C_3 \cdot \frac{1}{p \nu_{\min}^2 (\mathbf{H}_X)}.
\end{align*}

\noindent \textbf{For $K_6$.} Since $\mathbf e_{X0}$ is a linear function of $\mathbf u_X$,
\begin{align*}
K_6 &= e_1^\top  \mathbb{E}[Q Q^\top|A=0]^{-1}\mathbb{E}[ \mathbf{e}_{X0} \mathbf{e}_{XM,0}^\top] \\
&= e_1^\top  \mathbb{E}[Q Q^\top|A=0]^{-1}\mathbb{E}[ \mathbf{e}_{X0} (\mathbf{u}_{XM,0}^\top\mathbf{H}_M^\top+ \mathbf{e}_{M}^\top)] \\
&= { e_1^\top  \mathbb{E}[Q Q^\top|A=0]^{-1}\mathbb{E}[ \mathbf{e}_{X0} \mathbf{u}_{XM,0}^\top\mathbf{H}_M^\top]}+{e_1^\top  \mathbb{E}[Q Q^\top|A=0]^{-1}\mathbb{E}[ \mathbf{e}_{X0}\mathbf{e}_M^\top]} \\
&=0
\end{align*}
the first and second term vanishing by Assumptions \ref{Assumption:covariances}(ii).
\\

\noindent Recall that $R_{\delta}=-K_5+K_6$, so $R_{\delta} = -K_5 = -\tilde K_5 \mathbf{H}_M^\top$. For notational convenience, take $\tilde R_\delta = -\tilde K_5$, and we have $\|\tilde R_{\delta}\| = \mathcal{O} \left(p^{-1}\nu^{-2}_{\min}(\bH_X)\right)$ as $p\nu^2_{\min}(\mathbf{H}_X)\to \infty$.
\end{proof}

\subsection{Proof of Corollary \ref{cor:ate}}
\begin{proof}
For $a\in\{0,1\}$, define $\mu_a := \mathbb E[Y^{(a)}] = \alpha_a + \gamma_a^\top \delta_a$. Here each $\mu_a=\alpha_a+\gamma_a^\top\delta_a$ takes both its outcome coefficients $(\alpha_a,\gamma_a)$ and its mediator coefficient $\delta_a$ within the same arm $a$, so that $\mu_1$ and $\mu_0$ are the within-arm cases $(a_Y,a_M)=(1,1)$ and $(0,0)$, whereas $\theta_0=\alpha_1+\gamma_1^\top\delta_0$ of Theorem~\ref{thm:id} is the cross-world case $(1,0)$. Since Assumptions~\ref{Assumption:covariances}(iv)--(v) and~\ref{Assumption:pervasiveness} hold for each arm separately, the argument of Theorem~\ref{thm:id} applies to each with the arm index relabeled. We therefore write $\mu_a$ for this quantity in Theorem \ref{thm:id}. Then, we have
$\theta_{\mathrm{ATE}} = \mu_1 - \mu_0$ and $ \theta_{\mathrm{ATE},p,q} = \mu_{1,p,q} - \mu_{0,p,q}.$ Using the same notation as in Theorem \ref{thm:id}, let
\begin{align*}
L&=\begin{pmatrix}1 \\
\frac{1}{p}\mathbf W_X^\top X \\\frac{1}{q}\mathbf W_M^\top M\end{pmatrix}, \quad Q=\begin{pmatrix}1 \\\frac{1}{p}\mathbf W_X^\top X\end{pmatrix}, \quad N=\frac{1}{q}\mathbf W_M^\top M.
\end{align*}
By the same population-level argument as in Theorem \ref{thm:id}, applied separately under $A=1$ and $A=0$, $\mu_{a,p,q}$ admits the representation given in the corollary statement for each $a\in\{0,1\}$.  It follows that
\begin{align*}
\theta_{\mathrm{ATE}} - \theta_{\mathrm{ATE},p,q}&=(\mu_1 - \mu_{1,p,q}) - (\mu_0 - \mu_{0,p,q}),
\end{align*}
so, by the triangle inequality,
\begin{align*}
\bigl|\theta_{\mathrm{ATE}} - \theta_{\mathrm{ATE},p,q}\bigr| &\le |\mu_1 - \mu_{1,p,q}| + |\mu_0 - \mu_{0,p,q}|.
\end{align*}
Applying Theorem \ref{thm:id} to $a=1$ and $a=0$, there exist constants $C_1', C_0' > 0$, independent of $p$ and $q$, such that
\begin{align*}
|\mu_a - \mu_{a,p,q}|&\le C_a'\left(
\frac{1}{p\,\nu_{\min}^2(\mathbf H_X)}+ \frac{1}{q\,\nu_{\min}^2(\mathbf H_M)} \right)\left(1+\frac{1}{p\,\nu_{\min}^2(\mathbf H_X)}\right), \quad a\in\{0,1\}.
\end{align*}
Combining the two bounds yields
\begin{align*}
\bigl|\theta_{\mathrm{ATE}} - \theta_{\mathrm{ATE},p,q}\bigr| &\le (C_1' + C_0')\left(\frac{1}{p\,\nu_{\min}^2(\mathbf H_X)}+\frac{1}{q\,\nu_{\min}^2(\mathbf H_M)}\right)\left(1+\frac{1}{p\,\nu_{\min}^2(\mathbf H_X)}\right).
\end{align*}
As $p$ and $q$ grow large, the right-hand side converges to zero, so $\theta_{\mathrm{ATE},p,q} \to \theta_{\mathrm{ATE}}$ as $p\nu^2_{\min}(\mathbf{H}_X)\to \infty$ and $q\nu^2_{\min}(\mathbf{H}_M)\to \infty$.
\end{proof}

\section{Asymptotic Analysis}\label{Prop proof}

To prove Theorem~\ref{thm:clt}, we first establish a few results that will be used in the main proof. Recall that FARMA uses two-fold cross-fitting. We consider one regression fold at a time, with the projection matrices estimated using the other fold. Within the regression fold, the treated units are used to estimate $\hat\alpha_1$ and $\hat\gamma_1$, while the control units are used for the second-stage bridge regression. We derive the expansion for one fold and combine the two fold-specific estimators at the end.

For fold $k$, let $n_k^t$ and $n_k^c$ denote the numbers of treated and control units respectively. Since the two folds have approximately equal size and the treatment probabilities are bounded away from zero and one, $n_k^t\asymp n, n_k^c\asymp n$ and $n_k\asymp n.$ Thus, rates written in terms of $n_k^t$ or $n_k^c$ are of the same order as rates in $n$.

The supporting results below are used in the main proof. Lemmas~\ref{lem:Wbound}-\ref{lem:masterbias} give the bounds needed to control the error terms, while Lemma~\ref{lemma:approxgamma1} gives the expansion of $\hat\gamma_1$. We use these results in the proof of Theorem~\ref{thm:clt}.

\subsection{Supporting Results}
\label{sec:prelim}

\subsubsection*{Proof of Proposition~\ref{prop:loading-cond}}\label{app:loading-cond}
\begin{proof}[Proof of Proposition~\ref{prop:loading-cond}]

We first prove the result for $\mathbf H_X$. 
We write $S_X=\Gamma_X\Sigma_{\mathbf f_X}\Gamma_X^\top$ and $\Sigma_X=S_X+\Sigma_{\mathbf u_X}$. Since $\mathbf W_X=\sqrt p\,\mathbf V_X$,
\begin{align*}
\mathbf H_X=\frac1p\mathbf W_X^\top\Gamma_X=\frac1{\sqrt p}\mathbf V_X^\top\Gamma_X,\quad \mathbf H_X^\top\mathbf H_X = \frac1p\Gamma_X^\top \bV_X \bV_X^\top \Gamma_X =  \frac1p\Gamma_X^\top P_{\mathbf V_X}\Gamma_X,
\end{align*}
Therefore, it is enough to bound the minimum and maximum eigenvalues of $\mathbf H_X^\top\mathbf H_X$. For notational simplicity, throughout the proof we write $n=n_k$ and relabel the observations in $\mathcal D_k$ as $i= 1, 2, \cdots, n$. It is immediate from the property of the projection matrix: 
\begin{align*}
\nu_{\max}^2(\mathbf H_X)=\lambda_{\max}\left(\frac1p\Gamma_X^\top P_{\mathbf V_X}\Gamma_X\right)\le\lambda_{\max}\left(\frac1p\Gamma_X^\top\Gamma_X\right)\le C_\Gamma.
\end{align*}
Hence, $\nu_{\max}(\mathbf H_X)\le\sqrt C_\Gamma.$ Next, we establish the lower bound. Towards that end, first, it is immediate from Assumption \ref{Assumption:covariances} (iii) and (iv):
$$
\Sigma_{\mathbf f} = \pi \Sigma_{\mathbf f \mid 1} + (1-\pi) \Sigma_{\mathbf f \mid 0} + \pi(1-\pi)(\mu_0 - \mu_1)(\mu_0 - \mu_1)^\top \implies c_f \le \lambda_{\min}(\Sigma_{\mathbf f}) \le \lambda_{\max}(\Sigma_{\mathbf f}) \le C_f \,. 
$$
Hence, using the fact that $\mathbf f_X$ is the first $\bar p$ elements of $\mathbf f$, we have: 
\begin{equation}
\label{eq:f_x_eigen_bound}
c_f \le \lambda_{\min}(\Sigma_{\mathbf f_X}) \le \lambda_{\max}(\Sigma_{\mathbf f_X}) \le C_f \,.
\end{equation}
Next, consider $S_X=\Gamma_X\Sigma_{\mathbf f_X}\Gamma_X^\top = \var(\Gamma_X \mathbf f_X)$. 
Clearly $\rank(S_X) \le \rank(\Sigma_{\mathbf f_X}) \le \bar p$. Furthermore, 
$S_X$ and $\Sigma_{\mathbf f_X}^{1/2}\Gamma_X^\top\Gamma_X\Sigma_{\mathbf f_X}^{1/2}$ has same set of non-zero eigenvalues. Furthemore, from Assumption \ref{Assumption:pervasiveness} and Equation \eqref{eq:f_x_eigen_bound}, we have:  
$$
p c_\Gamma c_f \bI_{\bar p} \preceq p c_\Gamma \Sigma_{\mathbf f_X} \preceq p \ \Sigma_{\mathbf f_X}^{1/2}\frac{\Gamma_X^\top\Gamma_X}{p}\Sigma_{\mathbf f_X}^{1/2} = \Sigma_{\mathbf f_X}^{1/2}\Gamma_X^\top\Gamma_X\Sigma_{\mathbf f_X}^{1/2} \preceq p C_\Gamma \Sigma_{\mathbf f_X} \preceq  p C_\Gamma C_f \bI_{\bar p}\,.
$$
Therefore, the eigengap at $\bar p$ satisfies
\begin{equation}
\label{eq:lb_SX}
\lambda_{\bar p}(S_X)-\lambda_{\bar p+1}(S_X)=\lambda_{\bar p}(S_X)\ge pc.
\end{equation}
Finally, since $S_X=\Gamma_X(\Sigma_{\mathbf f_X}\Gamma_X^\top)$, we have $\operatorname{Col}(S_X)\subseteq\operatorname{Col}(\Gamma_X)$. Since both spaces have dimension $\bar p$, it follows that $\operatorname{Col}(S_X)=\operatorname{Col}(\Gamma_X)$. 
Hence, the leading $\bar p$-dimensional eigenspace of $S_X$ is $\operatorname{Col}(\Gamma_X)$.

\medskip 

Next, we show $\|\hat \Sigma_X - S_X\|_{\rm op} = o_p(p)$ where $\Sigma_X$ is the sample covariance matrix of $\{X_i\}_{1 \le i \le n}$. Towards that end, recall that $X_i = \Gamma_X \mathbf f_{X, i} + \bu_{X, i}$ where both $\mathbf f_{X, i}$ and $\bu_{X, i}$ are centered random variable (Assumption \ref{Assumption:covariances}). An expansion of $\hat \Sigma_X - S_X$ yields:  
\begin{align}
\hat\Sigma_X-S_X = \frac1n\sum_{i=1}^n\mathbf x_i\mathbf x_i^\top-\bar{\mathbf x}\bar{\mathbf x}^\top-S_X &=\frac1n\sum_{i=1}^n\bigl(\Gamma_X\mathbf f_{X,i}+\mathbf u_{X,i}\bigr)\bigl(\Gamma_X\mathbf f_{X,i}+\mathbf u_{X,i}\bigr)^\top-\bar{\mathbf x}\bar{\mathbf x}^\top-S_X\notag\\
&=\underbrace{\Gamma_X\Bigl(\frac1n\sum_{i=1}^n\mathbf f_{X,i}\mathbf f_{X,i}^\top-\Sigma_{\mathbf f_X}\Bigr)\Gamma_X^\top}_{S^{(1)}} +\underbrace{\Bigl(\frac1n\sum_{i=1}^n\mathbf u_{X,i}\mathbf u_{X,i}^\top-\Sigma_{\mathbf u_X}\Bigr)}_{S^{(2)}}\notag\\
& \quad +\underbrace{\Gamma_X\frac1n\sum_{i=1}^n\mathbf f_{X,i}\mathbf u_{X,i}^\top}_{S^{(3)}} +\Bigl(\Gamma_X\frac1n\sum_{i=1}^n\mathbf f_{X,i}\mathbf u_{X,i}^\top\Bigr)^\top +\Sigma_{\mathbf u_X} -\bar{\mathbf x}\bar{\mathbf x}^\top \notag \\
& := S^{(1)} + S^{(2)} + S^{(3)} + (S^{(3)})^\top + \Sigma_{\mathbf u_X} -\bar{\mathbf x}\bar{\mathbf x}^\top \,.
\label{eq:EX-decomp}
\end{align}
Let $\mathbf F_X=[\mathbf f_{X,1},\ldots,\mathbf f_{X,n}]^\top$ and $\mathbf U_X=[\mathbf u_{X,1},\ldots,\mathbf u_{X,n}]^\top$. We bound each term in \eqref{eq:EX-decomp} separately.
\\\\
\noindent
\textbf{\underline{Bound for $S^{(1)}$.}} Define $\bk_i=\Sigma_{\mathbf f_X}^{-1/2}\mathbf f_{X,i}$.
By Assumption~\ref{Assumption:covariances} (i) and (iii), $\mathbf k_i$ are independent, with $\bbE(\mathbf k_i) = 0$ and $\var(\bk_i) = \bI_{\bar p}$. 
To check sub-Gaussianity, for any unit vector $\mathbf v\in\mathbb R^{\bar p}$,
\begin{align*}
\|\mathbf v^\top\mathbf k_i\|_{\psi_2}=\left\|\mathbf v^\top\Sigma_{\mathbf f_X}^{-1/2}\mathbf f_{X,i}\right\|_{\psi_2}=\left\|(\Sigma_{\mathbf f_X}^{-1/2}\mathbf v)^\top\mathbf f_{X,i}\right\|_{\psi_2} \le \frac{1}{\sqrt{c_{f}}}\|\mathbf f_X\|_{\psi_2} \le C.
\end{align*}
Here, in the second-to-last inequality, we use the bound from Equation \eqref{eq:f_x_eigen_bound}, and in the last inequality we use the fact that $\mathbf f_X$ is sub-Gaussian (Assumption \ref{assump:subgaussian}). Hence $\bk_i$'s are i.i.d. centered isotropic sub-Gaussian random vectors. 
Let $\mathbf K_X=[\mathbf k_1,\ldots,\mathbf k_n]^\top.$ Since $\mathbf f_{X,i}=\Sigma_{\mathbf f_X}^{1/2}\mathbf k_i$, we have $\mathbf F_X=\mathbf K_X\Sigma_{\mathbf f_X}^{1/2}$. 
Hence, we have
\begin{align}
\label{eq:KF}
\Bigl\|\frac1n\mathbf F_X^\top\mathbf F_X-\Sigma_{\mathbf f_X}\Bigr\|_{\mathrm{op}}
 =\Bigl\|\Sigma_{\mathbf f_X}^{1/2}\left(\frac1n\mathbf K_X^\top\mathbf K_X-I_{\bar p}\right)\Sigma_{\mathbf f_X}^{1/2}\Bigr\|_{\mathrm{op}} 
&\le C_f \Bigl\|\frac1n\mathbf K_X^\top\mathbf K_X-I_{\bar p}\Bigr\|_{\mathrm{op}} \,.
\end{align}
Using the Equation~(4.27) in the proof of Theorem~4.6.1 of \citet{Vershynin_2026} applied to $\mathbf K_X$, we have
\begin{align*} 
\bbP\left(\left\| \frac1n\mathbf K_X^\top\mathbf K_X-I_{\bar p} \right\|_{\mathrm{op}} > C_K \delta_{t, 1}\right) \le 2e^{-t}, \qquad \delta_{t, 1}=C\left(\sqrt{\frac{\bar p +t}{n}}+\frac{\bar p + t}{n}\right).
\end{align*}
Call the complementary event $A_{1, t}$. One $A_{1, t}$, we have: 
\begin{align*}
\|S^{(1)}\|_{\mathrm{op}}
=\left\|\Gamma_X\left(\frac1n\mathbf F_X^\top\mathbf F_X-\Sigma_{\mathbf f_X}\right)\Gamma_X^\top\right\|_{\mathrm{op}} &\le\|\Gamma_X\|_{\mathrm{op}}^2\left\|\frac1n\mathbf F_X^\top\mathbf F_X-\Sigma_{\mathbf f_X}\right\|_{\mathrm{op}}\\
&\le p \ C_KC C_f \delta_{t, 1} := C_1 p \delta_{t, 1} \,.
\end{align*}

\medskip
\noindent
\textbf{\underline{Bound for $S^{(2)}$}.}
Define $\mathbf l_i=\Sigma_{\mathbf u_X}^{-1/2}\mathbf u_{X,i}$. 
By Assumptions~\ref{Assumption:covariances}(i)-(ii), \ref{Assumption:covariances}(v), and~\ref{assump:subgaussian}(i), the vectors $\mathbf l_i$ are independent, mean-zero, isotropic sub-Gaussian vectors. Let $\mathbf L_X=[\mathbf l_1,\ldots,\mathbf l_n]^\top$. Using this, we have: 
\begin{align*}
\|S^{(2)}\|_{\mathrm{op}} =\Bigl\|\frac1n\mathbf U_X^\top\mathbf U_X-\Sigma_{\mathbf u_X}\Bigr\|_{\mathrm{op}} &=\Bigl\|\Sigma_{\mathbf u_X}^{1/2}\Bigl(\frac1n\mathbf L_X^\top\mathbf L_X-I_p\Bigr)\Sigma_{\mathbf u_X}^{1/2}\Bigr\|_{\mathrm{op}}\\
&\le\|\Sigma_{\mathbf u_X}\|_{\mathrm{op}}\,\Bigl\|\frac1n\mathbf L_X^\top\mathbf L_X-I_p\Bigr\|_{\mathrm{op}} \\
&\le C_U\Bigl\|\frac1n\mathbf L_X^\top\mathbf L_X-I_p\Bigr\|_{\mathrm{op}}.
\end{align*}
Another application of Equation~(4.27) in the proof of Theorem~4.6.1 of
\citet{Vershynin_2026} to the $n\times p$ matrix $\mathbf L_X$ yields, on an event $A_{2, t}$ where $\bbP(A_{2, t}) \ge 1-2e^{-t}$: 
\begin{align*}
\left\| \frac1n\mathbf L_X^\top\mathbf L_X-I_p \right\|_{\mathrm{op}} \le C_L\left(\sqrt{\frac{p+t}{n}}+\frac{p+t}{n}\right) := C_L \delta_{t, 2}.
\end{align*}
Consequently, on this event, we have $\|S^{(2)}\|_{\mathrm{op}} \le C_U C_L \delta_{t, 2} := C_2 \delta_{t, 2}$. 
\\\\
\noindent
\textbf{\underline{Bound for $S^{(3)}$}.} Again, by applying Theorem~4.6.1 of \citet{Vershynin_2026} on $\mathbf K_X$ and $\mathbf L_X$ (as they have isotropic, centered sub-Gaussian rows), we have, on an event $A_{3, t}$, with $\bbP(A_{3, t}) \ge 1 - Ce^{-t}$:  
\begin{align*}
\|\mathbf K_X\|_{\mathrm{op}}\le C(\sqrt n+\sqrt{\bar p}+\sqrt t),\qquad \|\mathbf L_X\|_{\mathrm{op}}\le C(\sqrt n+\sqrt p+\sqrt t).
\end{align*}
Therefore, on this event: 
\begin{align*}
\|S^{(3)}\|_{\mathrm{op}} =\left\|\Gamma_X\frac1n\mathbf F_X^\top\mathbf U_X\right\|_{\mathrm{op}} \le\|\Gamma_X\|_{\mathrm{op}}\frac1n\|\mathbf F_X\|_{\mathrm{op}}\|\mathbf U_X\|_{\mathrm{op}}&\le C_3\frac{\sqrt p}{n}(\sqrt n+\sqrt t)(\sqrt n+\sqrt p+\sqrt t) \\
& = C_3\sqrt{p}\left(1 + \sqrt{\frac{t}{n}}\right)\left(1 + \sqrt{\frac{p}{n}} + \sqrt{\frac{t}{n}}\right) := C_3 \delta_{t, 3} \,.
\end{align*}
Therefore, going back to Equation \eqref{eq:EX-decomp}, to bound $\|\hat \Sigma_X - S_X\|_{\rm op}$, we need to bound $\|\Sigma_{\bu_X}\|_{\rm op}$ and $\|\bar \bx \bar \bx^\top\|_{\rm op} = \|\bar \bx\|_2^2$. For the former, we use $\|\Sigma_{\mathbf u_X}\|_{\mathrm{op}}\le C_u$ (Assumption \ref{Assumption:covariances}(v)). 

For the later, note that, $\bar{\mathbf x}=\Gamma_X\bar{\mathbf f}_X+\bar{\mathbf u}_X$. 
We therefore bound $\bar{\mathbf f}_X$ and $\bar{\mathbf u}_X$ separately. As $\mathbf f_{X,i} \in \reals^{\bar p}$ are i.i.d., centered sub-Gaussian random variables, we have, on an event $A_{41, t}$, with $\bbP(A_{41, t}) \ge 1-Ce^{-t}$: 
\begin{align*}
\|\bar{\mathbf f}_X\|_2\le C\sqrt{\frac{\bar p+t}{n}}.
\end{align*}
Now consider $\bar{\mathbf u}_X$. Now given any $v \in \bbS^{p-1}$, we know $v^\top {\bu}_X$ is a centered sub-Gaussian random variable. Therefore, we have, 
$$
\bbP\left(|v^\top \bar \bu_\bX| \ge t\right) \le 2e^{-cnt^2} $$
Next, consider $\cA_{1/2}$, a $1/2$-net of $\bbS^{p-1}$, i.e., for any $v \in \bbS^{p-1}$, there exists $v_i \in \cA_{1/2}$ such that $\|v - v_i\|_2 \le 1/2$. Then $|\cA_{1/2}| \le C_1^p$ for some constant $C_1 > 0$. Therefore,
$$
\bbP\left(\max_{v \in \cA_{1/2}} |v^\top \bar \bu_X| \ge t\right) \le 2C_1^pe^{-cnt^2} = 2e^{p\log{C_1} - cnt^2}.
$$
Finally, using that given $v \in \bbS^{p-1}$, there exists $\omega(v) \in \cA_{1/2}$ such that $\|v - \omega(v)\|_2 \le 1/2$, we have:  
$$
\|\bar \bu_X\|_2 = \sup_{v \in \bbS^{p-1}} |v^\top \bar \bu_X| = \sup_{v \in \bbS^{p-1}} |(v - \omega(v) + \omega(v))^\top \bar \bu_X| \le \frac{1}{2} \|\bar u_X\|_2 + \sup_{v \in \cA_{1/2}}|v^\top \bar \bu_X| \,.
$$
Hence, $\|\bar \bu_X\|_2 \le 2 \sup\{|v^\top \bar \bu_X|: v \in \cA_{1/2}\}$. Therefore,  
$$
\bbP(\|\bar u_X\|_2 > t) \le 2e^{p\log{C_1} - \frac{c}{4}nt^2} \implies \bbP\left(\|\bar u_X\|_2 > 2\sqrt{\frac{t + p\log{C_1}}{cn}}\right) \le  2e^{-t} \,.
$$
Call the complementary event $A_{42, t}$.  Now, using $\bar{\mathbf x}=\Gamma_X\bar{\mathbf f}_X+\bar{\mathbf u}_X$, we have on $A_{4, t} = A_{41, t} \cap A_{42, t}$ (with $\bbP(A_{4, t}) \ge 1 - C'e^{-t}$ for some constant $C' > 0$): 
\begin{align*}
\|\bar{\mathbf x}\bar{\mathbf x}^\top\|_{\mathrm{op}} = \|\bar{\mathbf x}\|^2_2 & \le 2\|\Gamma_X\bar{\mathbf f}_X\|^2_2+ 2\|\bar{\mathbf u}_X\|^2_2 \le 2\|\Gamma_X\|^2_{\mathrm{op}}\|\bar{\mathbf f}_X\|^2_2+2\|\bar{\mathbf u}_X\|^2_2 
\le C_4\left(\frac{p(\bar p + t)}{n} + \frac{p+t}{n}\right) \le C_4 \frac{p(1 + t)}{n}. 
\end{align*}
Define $A_n = \cap_{j = 1}^4 A_{j, t}$ with $t = K \log{n}$ for some large constant $K > 0$. It is immediate that $\bbP(A_n) \ge 1 - n^{-K_1}$ for some constant $K_1 > 0$.  Then, on the event $A_n$, we have, for $C \ge \max\{C_j: 1 \le j \le 4\}$: 
\begin{align}
\|\hat \Sigma_X - S_X\|_{\rm op} & \le C \left\{p\left(\sqrt{\frac{\bar p +  K \log{n}}{n}}+\frac{\bar p +  K \log{n}}{n}\right) + \left(\sqrt{\frac{p+ K \log{n}}{n}}+\frac{p+ K \log{n}}{n}\right) \right. \notag \\
& \qquad \left.  + \sqrt{p}\left(1 + \sqrt{\frac{ K \log{n}}{n}}\right)\left(1 + \sqrt{\frac{p}{n}} + \sqrt{\frac{ K \log{n}}{n}}\right) + \frac{p(1+ K \log{n})}{n}\right\} \le C p \sqrt{\frac{\log{n}}{n}} \,,
\label{eq:EX-op}
\end{align}
where we absorb all constants in $C$. This completes the proof of the upper bound on $\|\hat \Sigma_X - S_X\|_{\rm op}$. 
For the rest of the proof, we use the notation $P_A$ to denote the projection matrix on the column space of $A$. 
$P_{{\mathbf V}_{X,\bar p}}$ denotes the projection onto the leading $\bar p$-dimensional eigenspace of $\hat\Sigma_X$. 
Now using the fact that $\lambda_{\bar p}(S_X)-\lambda_{\bar p+1}(S_X)\ge pc_\Gamma c_f$ (Equation \eqref{eq:lb_SX}), an application of Davis-Kahan sin $\Theta$ theorem (Theorem 1 of \cite{daviskaha}) yields, on $A_n$: 
\begin{align*}
     \bigl\|P_{{\mathbf V}_{X,\bar p}} -P_{S_X}\bigr\|_{\mathrm{op}} = \bigl\|P_{{\mathbf V}_{X,\bar p}} -P_{\Gamma_X}\bigr\|_{\mathrm{op}} \le \frac{2\|\hat\Sigma_X-S_X\|_{\mathrm{op}}}{\lambda_{\bar p}(S_X)-\lambda_{\bar p+1}(S_X)} & \le\frac{2Cp\sqrt{\frac{\log{n}}{n}}}{pc_\Gamma c_f} = \frac{2C}{c_\Gamma c_f}\sqrt{\frac{\log{n}}{n}} \,.
\end{align*}
In the first equality, we have used the fact that columns space of $S_X$ is same as that of $\Gamma_X$.  
As a consequence, we have, on $A_n$: 
$$\left\|\frac1p\Gamma_X^\top\left(P_{{\mathbf V}_{X,\bar p}}-P_{\Gamma_X}\right)\Gamma_X\right\|_{\rm op} \le \lambda_{\max}\left(\frac{\Gamma_X^\top \Gamma_X}{p}\right) \bigl\|P_{{\mathbf V}_{X,\bar p}} -P_{\Gamma_X}\bigr\|_{\mathrm{op}} \le\frac{2C C_\Gamma }{c_\Gamma c_f}\sqrt{\frac{\log{n}}{n}} \,.$$
Now, we present the lower bound on $\nu_{\min}(\bH_X)$. 
\begin{align*}
\nu_{\min}^2(\mathbf H_X) =\lambda_{\min}\left(\frac1p\Gamma_X^\top P_{{\mathbf V}_X}\Gamma_X\right) & \ge\lambda_{\min}\left(\frac1p\Gamma_X^\top P_{{\mathbf V}_{X,\bar p}}\Gamma_X\right)\\
&\ge\lambda_{\min}\left(\frac1p\Gamma_X^\top\Gamma_X\right)-\left\|\frac1p\Gamma_X^\top\left(P_{{\mathbf V}_{X,\bar p}}-P_{\Gamma_X}\right)\Gamma_X\right\|_{\rm op} \\
& \ge c_\Gamma -\frac{2C C_{\Gamma,X} }{c_\Gamma c_f}\sqrt{\frac{\log{n}}{n}} \ge \frac{c_\Gamma}{2}.
\end{align*}
The last inequality holds for all large $n$. This completes the proof.

\medskip\noindent
For the mediator block, $\mathbb E[\mathbf f_M]$ need not be zero. Since $\hat\Sigma_M$ is computed after demeaning, we apply the same argument to $\mathbf f_M-\mathbb E[\mathbf f_M]$. By Assumption~\ref{Assumption:covariances}(iii)--(iv), the eigenvalues of $\operatorname{Cov}(\mathbf f_M)$ are bounded away from zero and infinity. Let us assume that $A_{M,n}$ denote the event obtained by applying the preceding argument to the mediator block with $(q,\mathbf \Gamma_M,\operatorname{Cov}(\mathbf f_M), \Sigma_{\mathbf u_M},\bar q)$ in place of $(p,\mathbf \Gamma_X,\Sigma_{\mathbf f_X},\Sigma_{\mathbf u_X},\bar p)$ using $t=K\log n$. Then, for sufficiently large $n$, we have \(P(A_{M,n})\ge 1-Cn^{-K_1},\) and on $A_{M,n}$, \(c_M\le \nu_{\min}(\mathbf H_M)\le \nu_{\max}(\mathbf H_M)\le C_M\) for some positive constants $c_M$ and $C_M$.

Recall that the corresponding bounds for $\mathbf H_X$ hold on the event $A_n$ as above. By the union bound, \(P(A_n\cap A_{M,n}) \ge 1-(1+C)n^{-K_1}.\) Thus, we have that
\begin{align*}
c\le \nu_{\min}(\mathbf H_X) \le \nu_{\max}(\mathbf H_X)\le C,\qquad c\le \nu_{\min}(\mathbf H_M)\le \nu_{\max}(\mathbf H_M)\le C
\end{align*}
hold with probability at least $1-n^{-K_1}$ for some $K_1>0$ and all sufficiently large $n$. This completes the proof.
\end{proof}

\begin{lemma}[Operator norm of the weighting matrix]\label{lem:Wbound}
Let $\mathbf W=\operatorname{diag}(p^{-1}\mathbf W_X,q^{-1}\mathbf W_M)$, where $\mathbf W_X=\sqrt p\,\mathbf V_X$ and $\mathbf W_M=\sqrt q\,\mathbf V_M$, with $\mathbf V_X^\top\mathbf V_X=I_{\tilde p}$ and $\mathbf V_M^\top\mathbf V_M=I_{\tilde q}$. Then, $\|\mathbf W\|_{\mathrm{op}}^2 = (p \wedge q)^{-1}$. 
\end{lemma}

\begin{proof}
Since $\mathbf W_X=\sqrt p\,\mathbf V_X$ and $\mathbf W_M=\sqrt q\,\mathbf V_M$, we have $\mathbf W=\operatorname{diag}(p^{-1/2}\mathbf V_X,q^{-1/2}\mathbf V_M).$ Because $\mathbf V_X$ and $\mathbf V_M$ have orthonormal columns, we have that
\begin{align*}\|p^{-1/2}\mathbf V_X\|_{\mathrm{op}}^2=\frac1p,\quad\|q^{-1/2}\mathbf V_M\|_{\mathrm{op}}^2=\frac1q.\end{align*}
As the operator norm of a block-diagonal matrix is the maximum of the operator norms of its diagonal blocks, so
\begin{align*}
\|\mathbf W\|_{\mathrm{op}}^2=\max\left\{p^{-1},q^{-1}\right\}=(p \wedge q)^{-1}.
\end{align*}
\end{proof}

\begin{lemma}\label{lem:E0bounds}
Suppose Assumptions~\ref{Assumption:covariances}--\ref{Assumption:pervasiveness} and~\ref{assump:subgaussian}--\ref{assump:growth} hold. Then, conditional on the training fold, as $n,p,q\to\infty$,
\begin{enumerate}
\item[(i)] $\left\|\frac1n\mathbb E(\mathbf E_0^\top\mathbf E_0)\right\|_{\text{op}}=O\left(\frac1{p\wedge q}\right), \|\mathbf E_0\|_{\text{op}}=O_p\left(\sqrt{\frac{n}{p\wedge q}}\right).$
\item[(ii)] $\left\|\frac1{\sqrt n}\mathbf F_1^\top\mathbf E_0\right\|_{\text{op}}=O_p\left(\frac1{\sqrt{p\wedge q}}\right),\left\|\frac1n\mathbf 1^\top\mathbf E_0\right\|_2=O_p\left(\frac1{\sqrt{n(p\wedge q)}}\right).$
\item[(iii)] $\left\|\frac1{\sqrt n}\left\{\mathbf E_0^\top\mathbf E_0-\mathbb E(\mathbf E_0^\top\mathbf E_0)\right\}\right\|_{\text{op}}=O_p\left(\frac1{p\wedge q}\right), \left\|\frac1{\sqrt n}\mathbf E_0^\top\mathbf 1\mathbf 1^\top\mathbf E_0\right\|_{\text{op}}=O_p\left(\frac{\sqrt n}{p\wedge q}\right).$
\item[(iv)] $\left\|\frac1{\sqrt n}\mathbf E_0^\top\boldsymbol\epsilon_1\right\|_2=O_p\left(\frac1{\sqrt{p\wedge q}}\right).$
\end{enumerate}
Here $n$ denotes the number of rows of the relevant sub-sample. When the lemma is applied to the treated outcome regression, $n$ is replaced by $n_k^t$, and when it is applied to the control bridge regression, $n$ is replaced by $n_k^c$. Since $n_k^t\asymp n_k^c\asymp n_k\asymp n$, every rate above is order-equivalent under these substitutions.
\end{lemma}

\begin{proof}
Recall from Section~\ref{sec:prelim-notation} that $\mathbf E_0=(\mathbf 0_n,\mathbf E)$ with $\mathbf E=\mathbf U\mathbf W$. Since the first column of $\mathbf E_0$ is zero, it is enough to prove the corresponding bounds for $\mathbf E$. We use the inequalities $\|\mathbf A\|_{\text{op}}\le \|\mathbf A\|_F$, $\|\mathbf A\mathbf B\|_{\text{op}}\le \|\mathbf A\|_{\text{op}}\|\mathbf B\|_{\text{op}}$, and $\operatorname{tr}(\mathbf A^\top\mathbf B\mathbf A)\le \|\mathbf B\|_{\text{op}}\|\mathbf A\|_F^2$ for positive semidefinite $\mathbf B$. By Lemma~\ref{lem:Wbound}, $\|\mathbf W\|_{\text{op}}^2={1}/({p\wedge q})$. Since $\mathbf W$ has at most $\tilde p+\tilde q$ columns and $\tilde p,\tilde q$ are fixed,
\begin{align*}
\|\mathbf W\|_F^2&\le(\tilde p+\tilde q)\|\mathbf W\|_{\text{op}}^2 =O\left(\frac{1}{p\wedge q}\right).
\end{align*}
Because $\mathbf W$ is fixed conditional on the training fold, the rows of $\mathbf E=\mathbf U\mathbf W$ are independent across $i$ within the regression fold.

\medskip\noindent\textbf{Proof of part (i).}
Let $\Sigma_{\mathbf u}=\mathbb E(\mathbf u_i\mathbf u_i^\top)$. By Assumption~\ref{Assumption:covariances}(ii) and (v), $\|\Sigma_{\mathbf u}\|_{\mathrm{op}}\le C$. Since $\mathbf E=\mathbf U\mathbf W$, we have $\mathbf E^\top\mathbf E=\mathbf W^\top\mathbf U^\top\mathbf U\mathbf W.$ The rows of $\mathbf U$ are i.i.d., so $\mathbb E(\mathbf U^\top\mathbf U)=n\Sigma_u$, and therefore $\mathbb E(\mathbf E^\top\mathbf E)=n\mathbf W^\top\Sigma_u\mathbf W.$  It follows that
\begin{align*}
\left\|\frac{1}{n}\mathbb E(\mathbf E^\top\mathbf E)\right\|_{\text{op}}
&=\left\|\mathbf W^\top\Sigma_u\mathbf W\right\|_{\text{op}} \le \|\mathbf W\|_{\text{op}}^2\|\Sigma_u\|_{\text{op}}=O\left(\frac{1}{p\wedge q}\right),
\end{align*}
Since $\mathbf E_0$ only adds a zero column to $\mathbf E$, the same bound holds for $\mathbf E_0$.

\medskip\noindent Next, using $\|\mathbf E\|_{\mathrm{op}}\le\|\mathbf E\|_F$, we get
\begin{align*}
\mathbb E\|\mathbf E\|_{\mathrm{op}}^2\le\mathbb E\|\mathbf E\|_F^2=\mathbb E\operatorname{tr}(\mathbf E^\top\mathbf E)=n\operatorname{tr}(\mathbf W^\top\Sigma_{\mathbf u}\mathbf W)\le n\|\Sigma_{\mathbf u}\|_{\mathrm{op}}\|\mathbf W\|_F^2=O\left(\frac{n}{p\wedge q}\right).
\end{align*}
Therefore, by Markov's inequality, we have that $\|\mathbf E\|_{\mathrm{op}}=O_p\left(\sqrt{{n}/{(p\wedge q)}}\right).$ Since adding the zero column does not change the operator norm, we have
\begin{align*}\|\mathbf E_0\|_{\mathrm{op}}=\|\mathbf E\|_{\mathrm{op}}=O_p\left(\sqrt{\frac{n}{p\wedge q}}\right).\end{align*}
This proves part (i).

\medskip
\noindent\textbf{Proof of part (ii).}
Since $\mathbf F_1=(\mathbf 1 \quad \mathbf F)$, we first bound $\frac{1}{\sqrt{n}}\mathbf 1^\top\mathbf E$ and $\frac{1}{\sqrt{n}}\mathbf F^\top\mathbf E$ in order to find the bound of $\frac{1}{\sqrt{n}}\mathbf{F}_1^\top \mathbf{E}_0$. For the first term,
\begin{align*}
\mathbb E\left\|\frac1{\sqrt n}\mathbf 1^\top\mathbf E\right\|_2^2=\mathbb E\left\|\frac1{\sqrt n}\sum_{i=1}^n\mathbf W^\top\mathbf u_i\right\|_2^2=\frac1n\sum_{i=1}^n\mathbb E\|\mathbf W^\top\mathbf u_i\|_2^2=\operatorname{tr}(\mathbf W^\top\Sigma_{\mathbf u}\mathbf W)\le\|\Sigma_{\mathbf u}\|_{\mathrm{op}}\|\mathbf W\|_F^2=O\left(\frac1{p\wedge q}\right),
\end{align*}
where the cross terms vanish because the observations are independent and $\mathbb E(\mathbf u_i)=0$. Thus, we have
\begin{align*}\left\|\frac1{\sqrt n}\mathbf 1^\top\mathbf E\right\|_2=O_p\left(\frac1{\sqrt{p\wedge q}}\right).\end{align*}
For the second term, note that $\mathbf F^\top\mathbf E=\sum_{i=1}^n\mathbf f_i\mathbf u_i^\top\mathbf W.$ By Assumption~\ref{Assumption:covariances}, $\mathbb E(\mathbf f_i\mathbf u_i^\top)=0$, so the cross terms vanish. Hence, we have
\begin{align*}
\mathbb E\left\|\frac{1}{\sqrt n}\mathbf F^\top\mathbf E\right\|_F^2
= \frac1n\sum_{i=1}^n\mathbb E\|\mathbf f_i\mathbf u_i^\top\mathbf W\|_F^2 & = \mathbb E\left[\|\mathbf f_i\|_2^2\|\mathbf W^\top\mathbf u_i\|_2^2\right]\overset{(1)}\le \{\mathbb E\|\mathbf f_i\|_2^4\}^{1/2}\{\mathbb E\|\mathbf W^\top\mathbf u_i\|_2^4\}^{1/2} \overset{(2)}=O\left(\frac{1}{p\wedge q}\right),
\end{align*}
where $(1)$ uses Cauchy-Schwarz and $(2)$ uses Assumption~\ref{assump:subgaussian}
and Lemma~\ref{lem:Wbound}. Since sub-Gaussian random variables have finite moments of all orders, the fourth moments above are finite. Now, combining the bounds for $\mathbf 1^\top\mathbf E$ and $\mathbf F^\top\mathbf E$, and using $\|\cdot\|_{\text{op}}\le\|\cdot\|_F$, Markov's inequality gives us
\begin{align*}\left\|\frac1{\sqrt n}\mathbf F_1^\top\mathbf E_0\right\|_{\mathrm{op}}=O_p\left(\frac1{\sqrt{p\wedge q}}\right).\end{align*}
Moreover, we have
\begin{align*}
\mathbb E\left\|\frac1n\mathbf 1^\top\mathbf E\right\|_2^2&=\frac1n\mathbb E\left\|\frac1{\sqrt n}\mathbf 1^\top\mathbf E\right\|_2^2 =O\left(\frac{1}{n(p\wedge q)}\right),
\end{align*}
Thus, we simply have $\left\|\mathbf 1^\top\mathbf E_0/n\right\|_2=O_p\left(1/{\sqrt{n(p\wedge q)}}\right).$ This proves part (ii).

\medskip\noindent\textbf{Proof of part (iii).}
Let $\mathbf w_a$ denote the $a$th column of $\mathbf W$, for $a=1,\ldots,\tilde p+\tilde q$. Since $\mathbf E=\mathbf U\mathbf W$, the $(a,b)$ entry of $\mathbf E^\top\mathbf E$ is
\begin{align*}(\mathbf E^\top\mathbf E)_{ab}=\sum_{i=1}^n(\mathbf w_a^\top\mathbf u_i)(\mathbf w_b^\top\mathbf u_i).\end{align*}
Hence, we have
\begin{align*}\frac1{\sqrt n}\left\{(\mathbf E^\top\mathbf E)_{ab}-\mathbb E[(\mathbf E^\top\mathbf E)_{ab}]\right\}=\frac1{\sqrt n}\sum_{i=1}^n\left\{(\mathbf w_a^\top\mathbf u_i)(\mathbf w_b^\top\mathbf u_i)-\mathbb E[(\mathbf w_a^\top\mathbf u_i)(\mathbf w_b^\top\mathbf u_i)]\right\}.\end{align*}
The summands are independent and mean zero. Therefore,
\begin{align*}
\operatorname{Var}\left[\frac1{\sqrt n}\left\{(\mathbf E^\top\mathbf E)_{ab}-\mathbb E[(\mathbf E^\top\mathbf E)_{ab}]\right\}\right]
&=\operatorname{Var}\left[(\mathbf w_a^\top\mathbf u_1)(\mathbf w_b^\top\mathbf u_1)\right]\\
&\le\mathbb E\left[(\mathbf w_a^\top\mathbf u_1)^2(\mathbf w_b^\top\mathbf u_1)^2\right]\\
&\le\{\mathbb E(\mathbf w_a^\top\mathbf u_1)^4\}^{1/2}
\{\mathbb E(\mathbf w_b^\top\mathbf u_1)^4\}^{1/2}\\
&\le C\|\mathbf w_a\|_2^2\|\mathbf w_b\|_2^2\\
&\le\frac{C}{(p\wedge q)^2},
\end{align*}
where the fourth-moment bound follows from Assumption~\ref{assump:subgaussian}(i), and the last inequality uses $\|\mathbf w_a\|_2,\|\mathbf w_b\|_2\le\|\mathbf W\|_{\mathrm{op}}$ together with Lemma~\ref{lem:Wbound}. By Chebyshev's inequality, for every $a,b$,
\begin{align*}\frac1{\sqrt n}\left\{(\mathbf E^\top\mathbf E)_{ab}-\mathbb E[(\mathbf E^\top\mathbf E)_{ab}]\right\}=O_p\left(\frac1{p\wedge q}\right).\end{align*}
Since $\tilde p+\tilde q$ is fixed, the matrix has a fixed number of entries. Also, for a fixed-dimensional matrix $\mathbf A$,
\begin{align*}\|\mathbf A\|_{\mathrm{op}}\le\|\mathbf A\|_F\le(\tilde p+\tilde q)\max_{a,b}|A_{ab}|.\end{align*}
Therefore, we have that
\begin{align*}\left\|\frac1{\sqrt n}\left\{\mathbf E^\top\mathbf E-\mathbb E(\mathbf E^\top\mathbf E)\right\}\right\|_{\mathrm{op}}=O_p\left(\frac1{p\wedge q}\right).\end{align*}
Since $\mathbf E_0$ only adds a zero column to $\mathbf E$, the same bound holds for $\mathbf E_0$. This proves the first half of part (iii).

\medskip\noindent For the second part, note that
$\mathbf E^\top\mathbf 1\mathbf 1^\top\mathbf E=(\mathbf E^\top\mathbf 1)(\mathbf E^\top\mathbf 1)^\top$ is a rank-one matrix, so we have
\begin{align*}\left\|\frac{1}{\sqrt n}\mathbf E^\top\mathbf 1\mathbf 1^\top\mathbf E\right\|_{\text{op}}=\frac{1}{\sqrt n}\|\mathbf E^\top\mathbf 1\|_2^2.\end{align*}
Now we connect $\|\mathbf E^\top\mathbf 1\|_2$ to the quantity already bounded in part (ii). Since $\mathbf E^\top\mathbf 1=n\cdot(\mathbf 1^\top\mathbf E/n)^\top$, we get $\|\mathbf E^\top\mathbf 1\|_2^2=n^2\left\|\mathbf 1^\top\mathbf E/n\right\|_2^2.$ Substituting this, we get that
\begin{align*}
\left\|\frac{1}{\sqrt n}\mathbf E^\top\mathbf 1\mathbf 1^\top\mathbf E\right\|_{\text{op}} &=\frac{n^2}{\sqrt n}\left\|\frac1n\mathbf 1^\top\mathbf E\right\|_2^2 =n^{3/2}\left\|\frac1n\mathbf 1^\top\mathbf E\right\|_2^2.
\end{align*}
By part (ii), $\left\|\mathbf 1^\top\mathbf E/n\right\|_2=O_p\left(1/{\sqrt{n(p\wedge q)}}\right),$ and hence $\left\|\mathbf 1^\top\mathbf E/n\right\|_2^2=O_p\left(1/{n(p\wedge q)}\right).$ Therefore, we have
\begin{align*}
\left\|\frac1{\sqrt n}\mathbf E^\top\mathbf 1\mathbf 1^\top\mathbf E\right\|_{\mathrm{op}}=n^{3/2}O_p\left(\frac1{n(p\wedge q)}\right)=O_p\left(\frac{\sqrt n}{p\wedge q}\right).
\end{align*}
Since $\mathbf E_0$ only adds a zero column to $\mathbf E$, the same bound holds for $\mathbf E_0$. This proves part (iii).

\medskip\noindent\textbf{Proof of part (iv).}
We have  $\mathbf E^\top\boldsymbol\epsilon_1=\sum_{i=1}^n\mathbf W^\top\mathbf u_i\epsilon_{1,i}.$ Using independence across observations and the tower-property condition $\mathbb E(\mathbf u_i\epsilon_{1,i})=0$, the cross terms vanish. Hence, we have
\begin{align*}
\mathbb E\left\|\frac{1}{\sqrt n}\mathbf E^\top\boldsymbol\epsilon_1\right\|_2^2=\frac1n\sum_{i=1}^n\mathbb E\|\mathbf W^\top\mathbf u_i\epsilon_{1,i}\|_2^2=\mathbb E\left[\epsilon_{1,i}^2\|\mathbf W^\top\mathbf u_i\|_2^2\right]\le \{\mathbb E|\epsilon_{1,i}|^4\}^{1/2}\{\mathbb E\|\mathbf W^\top\mathbf u_i\|_2^4\}^{1/2} =O\left(\frac{1}{p\wedge q}\right),
\end{align*}
by Cauchy--Schwarz, Assumption~\ref{assump:subgaussian}(i), and Lemma~\ref{lem:Wbound}. Therefore, by Markov's inequality,
\begin{align*}\left\|\frac{1}{\sqrt n}\mathbf E_0^\top\boldsymbol\epsilon_1\right\|_2=O_p\left(\frac{1}{\sqrt{p\wedge q}}\right).\end{align*}
This proves part (iv).
\end{proof}

\begin{lemma}\label{1PE1}
Suppose Assumptions~\ref{Assumption:covariances}--\ref{Assumption:pervasiveness} and~\ref{assump:subgaussian}--\ref{assump:growth} hold. Then, conditional on the training fold, $\mathbf 1^\top P_S\mathbf 1=O_p(1)$, where $P_S=\mathbf E_0S(S^\top\mathbf E_0^\top\mathbf E_0S)^{-1}S^\top\mathbf E_0^\top.$
\end{lemma}

\begin{proof}
The idea is to use the special form of $\mathbf E_0$. Since the first block of $\mathbf E_0$ is zero, $\mathbf E_0S=\mathbf E\bar S$, where $\bar S$ is the lower block of $S$. We then normalize $\mathbf E\bar S$ by its population covariance and work with the resulting whitened matrix. Let $n$ be the sample size of the regression sub-sample. The quantities $p=p_n$ and $q=q_n$ may grow with $n$, while $\tilde p,\tilde q,\bar p,\bar q$ are fixed. Define $d=(\tilde p-\bar p)+(\tilde q-\bar q)$, which is fixed. For the $X$ part, let $\mathbf U_X\in\mathbb R^{n\times p}$ and $\mathbf W_X\in\mathbb R^{p\times\tilde p}$, and recall $\mathbf W_X=\sqrt p\,\mathbf V_X$ with $\mathbf V_X^\top\mathbf V_X=I_{\tilde p}$, so that
\begin{align*}
 \mathbf E_X&=\frac{1}{p}\mathbf U_XW_X =\frac{1}{p}\mathbf U_X(\sqrt p\,V_X) =\frac{1}{\sqrt p}\mathbf U_X \mathbf V_X\in\mathbb R^{n\times \tilde p}.
\end{align*}
Similarly, for the $M$ part,
\begin{align*} \mathbf E_M&=\frac{1}{q}\mathbf U_M \mathbf W_M =\frac{1}{\sqrt q}\mathbf U_M \mathbf V_M\in\mathbb R^{n\times \tilde q},
\end{align*}
where $W_M=\sqrt q\,V_M$ and $\mathbf V_M^\top \mathbf V_M=I_{\tilde q}$. Also, $\mathbf E =  \begin{pmatrix}\mathbf E_X & \mathbf E_M   \end{pmatrix}   \in\mathbb R^{n\times(\tilde p+\tilde q)}$ and $\mathbf E_0  =   \begin{pmatrix}
0 & \mathbf E
\end{pmatrix} \in\mathbb R^{n\times(1+\tilde p+\tilde q)}.$ Let $S_X\in\mathbb R^{\tilde p\times(\tilde p-\bar p)}$ be an orthonormal basis for the orthogonal complement of $\operatorname{Col}(\mathbf H_X)$, and let $S_M\in\mathbb R^{\tilde q\times(\tilde q-\bar q)}$ be an orthonormal basis for the orthogonal complement of $\operatorname{Col}(\mathbf H_M)$. Hence $S_X^\top S_X=I_{\tilde p-\bar p}$ and $S_M^\top S_M=I_{\tilde q-\bar q}.$ We take $\bar S=\text{diag}(S_X  \quad S_M)\in\mathbb R^{(\tilde p+\tilde q)\times d}.$ If a different orthonormal basis is used for the same space, the projection is unchanged because only the column space matters. By Assumption~\ref{Assumption:covariances}, the second moments $\Sigma_{\mathbf u_X}=\mathbb E[\frac{1}{n}\mathbf U_X^\top\mathbf U_X]$ and $\Sigma_{\mathbf u_M}=\mathbb E[\frac{1}{n}\mathbf U_M^\top\mathbf U_M]$ satisfy $\mathbb E[\mathbf U_X^\top\mathbf U_M]=0$ and
\begin{align*}\lambda_{\min}(\Sigma_{\mathbf u_X})\ge c_X>0,\quad \lambda_{\min}(\Sigma_{\mathbf u_M})\ge c_M>0.\end{align*} Assumption~\ref{assump:subgaussian} provides the fourth-moment bounds used below.

\bigskip\noindent\textbf{\underline{Step 1. Reduce $\mathbf E_0S$ to $\mathbf E\bar S$.}}
\noindent Write $S$ by separating its first row from the lower block
$S= \begin{pmatrix}s_0 & \bar S^\top\end{pmatrix}^\top s_0\in\mathbb R^d$ and $\bar S\in\mathbb R^{(\tilde p+\tilde q)\times d}.$ Since $\mathbf E_0=(0\ \mathbf E)$, we have
\begin{align*}
\mathbf E_0S&=(0\ \mathbf E)
\begin{pmatrix}s_0^\top\\\bar S
\end{pmatrix} =0\cdot s_0^\top+\mathbf E\bar S =\mathbf E\bar S.
\end{align*}
So $\mathbf E_0S$ and $\mathbf E\bar S$ are the same matrix, giving $P_S=P_{\mathbf E\bar S}$ and $\mathbf 1^\top P_S\mathbf 1=\mathbf 1^\top P_{\mathbf E\bar S}\mathbf 1$. It is therefore enough to prove $\mathbf 1^\top P_{\mathbf E\bar S}\mathbf 1=\mathcal O_P(1)$.

\medskip
\noindent
\textbf{\underline{Step 2. Population covariance matrix using block matrices.}}
\noindent
Define $\Sigma =\mathbb E\left[\bar S^\top\mathbf E^\top\mathbf E\bar S/n
\right].$ We compute this directly from the block form of $\mathbf E$. Since $\mathbf E =\begin{pmatrix}\mathbf E_X & \mathbf E_M\end{pmatrix},$ we have
\begin{align*}\mathbf E^\top\mathbf E=\begin{pmatrix}\mathbf E_X^\top\mathbf E_X&\mathbf E_X^\top\mathbf E_M
\\\mathbf E_M^\top\mathbf E_X&\mathbf E_M^\top\mathbf E_M\end{pmatrix} ~ \text{ and }~\bar S=\begin{pmatrix}S_X & 0\\0 & S_M\end{pmatrix},\end{align*}
we get
\begin{align*}
\frac{1}{n}\bar S^\top\mathbf E^\top\mathbf E\bar S
&=\frac{1}{n}
\begin{pmatrix}S_X^\top & 0\\0 & S_M^\top
\end{pmatrix}\begin{pmatrix}\mathbf E_X^\top\mathbf E_X&\mathbf E_X^\top\mathbf E_M\\
\mathbf E_M^\top\mathbf E_X&\mathbf E_M^\top\mathbf E_M
\end{pmatrix}
\begin{pmatrix}S_X & 0\\0 & S_M\end{pmatrix} \\&=
\begin{pmatrix}\frac{1}{n}S_X^\top\mathbf E_X^\top\mathbf E_XS_X&\frac{1}{n}S_X^\top\mathbf E_X^\top\mathbf E_MS_M\\\frac{1}{n}S_M^\top\mathbf E_M^\top\mathbf E_XS_X&\frac{1}{n}S_M^\top\mathbf E_M^\top\mathbf E_MS_M
\end{pmatrix}.
\end{align*} Now we compute each block. Since
$\mathbf E_X=\frac{1}{\sqrt p}\mathbf U_X \mathbf  V_X ~\text{and}~\mathbf E_M=\frac{1}{\sqrt q}\mathbf U_M\mathbf V_M,$ and taking expectations gives us
\begin{align*}
\mathbb E\left[\frac1n\mathbf E_X^\top\mathbf E_X\right]&=\frac1p\mathbf V_X^\top\Sigma_{\mathbf u_X}\mathbf V_X,\quad \mathbb E\left[\frac1n\mathbf E_M^\top\mathbf E_M\right]=\frac1q\mathbf V_M^\top\Sigma_{\mathbf u_M}\mathbf V_M,\quad \mathbb E\left[\frac1n\mathbf E_X^\top\mathbf E_M\right]=0.
\end{align*}
Therefore, $\Sigma=\text{diag}( p^{-1}\Omega_X \quad q^{-1}\Omega_M),$ where $\Omega_X=S_X^\top\mathbf V_X^\top\Sigma_{\mathbf u_X}\mathbf V_XS_X,$ and $\Omega_M=S_M^\top\mathbf V_M^\top\Sigma_{\mathbf u_M}\mathbf V_MS_M.$  We now show that $\Omega_X$ and $\Omega_M$ are uniformly positive definite. For any $a\in\mathbb R^{\tilde p-\bar p}$,
\begin{align*}
a^\top\Omega_Xa=(\mathbf V_XS_Xa)^\top\Sigma_{\mathbf u_X}(\mathbf V_XS_Xa)\ge c\|\mathbf V_XS_Xa\|_2^2,
\end{align*}
where the last inequality follows from Assumption~\ref{Assumption:covariances}. Since $\mathbf V_X^\top\mathbf V_X=I_{\tilde p}$ and $S_X^\top S_X=I_{\tilde p-\bar p}$,
\begin{align*}
\|\mathbf V_XS_Xa\|_2^2=a^\top S_X^\top\mathbf V_X^\top\mathbf V_XS_Xa=a^\top S_X^\top S_Xa=\|a\|_2^2.
\end{align*}
Hence, $a^\top\Omega_Xa\ge c\|a\|_2^2,$ so $\lambda_{\min}(\Omega_X)\ge c$. The same argument gives $\lambda_{\min}(\Omega_M)\ge c$. Therefore, $\lambda_{\min}(p^{-1}\Omega_X)\ge c/p,$ and  $\lambda_{\min}(q^{-1}\Omega_M)\ge c/q,$ and hence $\Sigma$ is positive definite and invertible.

\medskip\noindent\textbf{\underline{Step 3: Whiten the matrix and rewrite the projection.}}
Define $\widetilde{\mathbf E}=\mathbf E\bar S\Sigma^{-1/2}.$ Since $\Sigma$ is positive definite, $\Sigma^{-1/2}$ is invertible. Therefore, multiplying $\mathbf E\bar S$ on the right by $\Sigma^{-1/2}$ does not change its column space, so $\operatorname{Col}(\widetilde{\mathbf E})=\operatorname{Col}(\mathbf E\bar S)$ and $P_{\widetilde{\mathbf E}}=P_{\mathbf E\bar S}.$ Hence, we have
\begin{align*}
\mathbf 1^\top P_{\mathbf E\bar S}\mathbf 1&=\mathbf 1^\top P_{\widetilde{\mathbf E}}\mathbf 1=\mathbf 1^\top\widetilde{\mathbf E}(\widetilde{\mathbf E}^\top\widetilde{\mathbf E})^{-1}\widetilde{\mathbf E}^\top\mathbf 1.
\end{align*}
Now, using the normalization by $n$, we get
\begin{align*}
\mathbf 1^\top P_{\mathbf E\bar S}\mathbf 1
&=\left(\frac1{\sqrt n}\mathbf 1^\top\widetilde{\mathbf E}\right)\left(\frac1n\widetilde{\mathbf E}^\top\widetilde{\mathbf E}\right)^{-1}\left(\frac1{\sqrt n}\widetilde{\mathbf E}^\top\mathbf 1\right).
\end{align*}
Define $A_n=\mathbf 1^\top\widetilde{\mathbf E}/{\sqrt n},$ and $B_n=\widetilde{\mathbf E}^\top\widetilde{\mathbf E}/n.$ Then, we get $\mathbf 1^\top P_{\mathbf E\bar S}\mathbf 1=A_nB_n^{-1}A_n^\top.$ Thus, it is enough to show that $\|A_n\|_2=O_p(1)$ and $ \|B_n^{-1}\|_{\mathrm{op}}=O_p(1).$

\medskip\noindent\textbf{\underline{Step 4: Whitened rows have covariance $I_d$.}}
Let $\widetilde e_i^\top$ denote the $i$th row of $\widetilde{\mathbf E}$ and let $e_i^\top$ denote the $i$th row of $\mathbf E$. From the definition of $\widetilde{\mathbf E}$, $\widetilde e_i^\top=e_i^\top\bar S\Sigma^{-1/2}.$ Since $\mathbb E(e_i)=0$, we also have $\mathbb E(\widetilde e_i)=0$. Recall that $\Sigma=\mathbb E\left[(1/n)\bar S^\top\mathbf E^\top\mathbf E\bar S\right].$ Since $\mathbf E^\top\mathbf E=\sum_{i=1}^n e_ie_i^\top,$ and the rows are identically distributed within the regression sub-sample,
\begin{align*}
\Sigma&=\frac1n\sum_{i=1}^n\mathbb E\left[\bar S^\top e_ie_i^\top\bar S\right]=\mathbb E\left[\bar S^\top e_ie_i^\top\bar S\right].
\end{align*}
Therefore, we have
\begin{align*}
\operatorname{Cov}(\widetilde e_i)=\mathbb E(\widetilde e_i\widetilde e_i^\top)=\Sigma^{-1/2}\mathbb E\left[\bar S^\top e_ie_i^\top\bar S\right]\Sigma^{-1/2}=\Sigma^{-1/2}\Sigma\Sigma^{-1/2}=I_d.
\end{align*}
Thus, the rows of $\widetilde{\mathbf E}$ are mean zero with covariance $I_d$.

\bigskip \noindent
\textbf{\underline{Step 5. Fourth moments remain bounded after whitening.}}
\noindent
Since $\Sigma=\text{diag} \begin{pmatrix} p^{-1}\Omega_X \quad  q^{-1}\Omega_M\end{pmatrix},$ and thus, $\Sigma^{-1/2} =(\sqrt p\,\Omega_X^{-1/2} \quad \sqrt q\,\Omega_M^{-1/2}).$ Also, we have
\begin{align*}\begin{aligned}
\mathbf E\bar S = \begin{pmatrix} \mathbf E_XS_X & \mathbf E_MS_M
\end{pmatrix}= \begin{pmatrix} \frac{1}{\sqrt p}\mathbf U_X \mathbf V_XS_X & \frac{1}{\sqrt q}\mathbf U_M \mathbf  V_MS_M
\end{pmatrix}.
\end{aligned}\end{align*}
Therefore, we have
\begin{align*}
\widetilde E =\mathbf E\bar S\Sigma^{-1/2} &=
\begin{pmatrix}\frac{1}{\sqrt p}\mathbf U_X \mathbf V_XS_X&\frac{1}{\sqrt q}\mathbf U_M \mathbf V_MS_M
\end{pmatrix}\begin{pmatrix}
\sqrt p\,\Omega_X^{-1/2} & 0\\
0 & \sqrt q\,\Omega_M^{-1/2}
\end{pmatrix} \\&=
\begin{pmatrix}\mathbf U_X \mathbf V_XS_X\Omega_X^{-1/2}&\mathbf U_M \mathbf V_MS_M\Omega_M^{-1/2}
\end{pmatrix}.
\end{align*}
We write the $i$th whitened row as $\widetilde e_i=\begin{pmatrix}\widetilde g_{X,i}^\top& \widetilde g_{M,i}^\top\end{pmatrix}^\top,$ where $\widetilde g_{X,i}$ and $\widetilde g_{M,i}$ denote the $X$ and $M$ blocks, respectively. Consider one coordinate of $\widetilde g_{X,i}$. It can be written as $a^\top\mathbf u_{X,i},$ where $a$ is a column of $\mathbf V_XS_X\Omega_X^{-1/2}$. Since $\mathbf V_X$ and $S_X$ have orthonormal columns,
\begin{align*}
\|a\|_2
&\le\|\mathbf V_X\|_{\mathrm{op}}\|S_X\|_{\mathrm{op}}\|\Omega_X^{-1/2}\|_{\mathrm{op}}=\|\Omega_X^{-1/2}\|_{\mathrm{op}}=\lambda_{\min}(\Omega_X)^{-1/2}
\le C,
\end{align*}
where the last inequality follows from Step~2.

\medskip\noindent By Assumption~\ref{assump:subgaussian}(i), linear forms of $\mathbf u_{X,i}$ have uniformly bounded fourth moments. Hence, $\mathbb E(a^\top\mathbf u_{X,i})^4\le C.$ Since $\tilde p-\bar p$ is fixed, it follows that $\mathbb E\|\widetilde g_{X,i}\|_2^4\le C.$ The same argument applied to the $M$ block gives $\mathbb E\|\widetilde g_{M,i}\|_2^4\le C.$ Finally, since $\|\widetilde e_i\|_2^2=\|\widetilde g_{X,i}\|_2^2+\|\widetilde g_{M,i}\|_2^2,$ we have
\begin{align*}\|\widetilde e_i\|_2^4\le2\|\widetilde g_{X,i}\|_2^4+2\|\widetilde g_{M,i}\|_2^4.\end{align*}
Therefore, $\mathbb E\|\widetilde e_i\|_2^4\le C$ uniformly in $n,p,q$.

\bigskip\noindent\textbf{\underline{Step 6. $B_n^{-1}=\mathcal O_P(1)$.}}
\noindent
Recall that
\begin{align*} B_n = \frac{1}{n}\widetilde E^\top\widetilde E =  \frac{1}{n}\sum_{i=1}^n\widetilde e_i\widetilde e_i^\top.\end{align*} 
From Step 4, $\mathbb E[\widetilde e_i\widetilde e_i^\top]=I_d.$ For the $(j,k)$ entry, $(B_n)_{jk}= \frac{1}{n}\sum_{i=1}^n\widetilde e_{ij}\widetilde e_{ik}.$ Hence, we have $\mathbb E[(B_n)_{jk}]=(I_d)_{jk}.$  Also, using independence across $i$ and the uniform fourth moment bound,
\begin{align*}
\operatorname{Var}\{(B_n)_{jk}\}&=\operatorname{Var}\left(\frac{1}{n}\sum_{i=1}^n\widetilde e_{ij}\widetilde e_{ik}\right) =\frac{1}{n^2}\sum_{i=1}^n\operatorname{Var}(\widetilde e_{ij}\widetilde e_{ik}) \le\frac{1}{n^2}\sum_{i=1}^n\mathbb E[\widetilde e_{ij}^2\widetilde e_{ik}^2]\le\frac{1}{n^2}\sum_{i=1}^n\mathbb E\|\widetilde e_i\|^4 \le\frac{C_4}{n}\to 0.
\end{align*}
Therefore, we have  $(B_n)_{jk}\xrightarrow{p}(I_d)_{jk}$ for each fixed pair $(j,k)$. Since $d$ is fixed, this gives $B_n\xrightarrow{p}I_d. $ By continuity of matrix inversion at $I_d$,
$B_n^{-1}\xrightarrow{p}I_d.$ Therefore $B_n^{-1}=\mathcal O_P(1).$

\medskip\noindent \textbf{\underline{Step 7. $A_n=\mathcal O_P(1)$.}}
\noindent
Recall that
\begin{align*}A_n^\top=\frac{1}{\sqrt n}\widetilde E^\top\mathbf{1}=\frac{1}{\sqrt n}\sum_{i=1}^n\widetilde e_i.\end{align*}
Since the rows are mean zero and $\operatorname{Cov}(\widetilde e_i)=I_d,$ we have
\begin{align*}
\mathbb E\|A_n^\top\|^2 &=\mathbb E\left\|\frac{1}{\sqrt n}\sum_{i=1}^n\widetilde e_i\right\|^2 =\operatorname{tr}\left[\operatorname{Cov}\left(\frac{1}{\sqrt n}\sum_{i=1}^n\widetilde e_i\right)\right]=\operatorname{tr}(I_d) =d.
\end{align*}
Therefore, by Markov's inequality applied to $\|A_n\|^2$, we get
\begin{align*}
\mathbb P(\|A_n\|>M) &=  \mathbb P(\|A_n\|^2>M^2) \le\frac{\mathbb E\|A_n\|^2}{M^2} =\frac{d}{M^2}.
\end{align*}
Now, letting $M\to\infty$, we get
$A_n=\mathcal O_P(1).$ Using Step 3, we get
\begin{align*}\mathbf{1}^\top P_S\mathbf{1}&=\mathbf{1}^\top P_{\mathbf E\bar S}\mathbf{1} = A_nB_n^{-1}A_n^\top \le
\|A_n\|^2\|B_n^{-1}\|_{\mathrm{op}}.
\end{align*}
From Step 6, $B_n^{-1}=\mathcal O_P(1),$ and from Step 7, $A_n=\mathcal O_P(1).$ Thus, we have that
\begin{align*}
    \|A_n\|^2\|B_n^{-1}\|_{\mathrm{op}} &= \mathcal O_P(1)\mathcal O_P(1)= \mathcal O_P(1).
\end{align*}
Therefore, $\mathbf{1}^\top P_S\mathbf{1}= \mathcal O_P(1).$ This proves the result.
\end{proof}

\begin{lemma}[Trace bound for a fixed-rank projection]\label{lem:tracebound}
Suppose the conditions of Lemma~\ref{1PE1} hold. Let \begin{align*}P_{S^*}=\mathbf E^*S^*(S^{*\top}\mathbf E^{*\top}\mathbf E^*S^*)^{-1}S^{*\top}\mathbf E^{*\top},\end{align*}
be the projection onto $\operatorname{Col}(\mathbf E^*S^*)$ for an error matrix $\mathbf E^*$ and $S^*$ has a fixed number of columns, and let $\widetilde{\mathbf F}^*$ be centered with bounded second moments and independent of $\mathbf E^*$. Then,
\begin{align*}\left\|\widetilde{\mathbf F}^{*\top}P_{S^*}\widetilde{\mathbf F}^*\right\|_{\mathrm{op}}=O_p(1).\end{align*}
\end{lemma}

\begin{proof}
From the same whitening argument as in Lemma~\ref{1PE1}, write $P_{S^*}=\widetilde{\mathbf E}(\widetilde{\mathbf E}^\top\widetilde{\mathbf E})^{-1}\widetilde{\mathbf E}^\top,$ where the rows $\widetilde e_i$ of $\widetilde{\mathbf E}$ are mean zero with covariance $I_d$. Define $C_n=(\widetilde{\mathbf F}^{*\top}\widetilde{\mathbf E})/{\sqrt n}$ and $B_n=(\widetilde{\mathbf E}^\top\widetilde{\mathbf E})/n.$ Then, we have $\widetilde{\mathbf F}^{*\top}P_{S^*}\widetilde{\mathbf F}^*=C_nB_n^{-1}C_n^\top.$ We first show that $\|C_n\|_F=O_p(1)$. 

For each entry, we have
\begin{align*}(C_n)_{jk}=\frac1{\sqrt n}\sum_{i=1}^n\widetilde f^*_{ij}\widetilde e_{ik}.\end{align*}
Since $\widetilde{\mathbf F}^*$ is centered and independent of $\widetilde{\mathbf E}$, $\mathbb E(\widetilde f^*_{ij}\widetilde e_{ik})=0.$ Also, using independence across $i$,
\begin{align*}
\operatorname{Var}\{(C_n)_{jk}\}&=\frac1n\sum_{i=1}^n\mathbb E\left[(\widetilde f^*_{ij})^2\widetilde e_{ik}^2\right]=\mathbb E(\widetilde f^*_{ij})^2\,\mathbb E(\widetilde e_{ik}^2)\le C,
\end{align*}
where $\mathbb E(\widetilde e_{ik}^2)=1$ by Lemma~\ref{1PE1}. Hence $(C_n)_{jk}=O_p(1)$. Since both dimensions are fixed, $\|C_n\|_F=O_p(1).$ By Lemma~\ref{1PE1}, $\|B_n^{-1}\|_{\mathrm{op}}=O_p(1).$ Therefore,
\begin{align*}
\operatorname{tr}\left(\widetilde{\mathbf F}^{*\top}P_{S^*}\widetilde{\mathbf F}^*\right)=\operatorname{tr}\left(C_nB_n^{-1}C_n^\top\right)\le\|C_n\|_F^2\|B_n^{-1}\|_{\mathrm{op}}=O_p(1).
\end{align*}
Since $\widetilde{\mathbf F}^{*\top}P_{S^*}\widetilde{\mathbf F}^*$ is positive semidefinite, we get
\begin{align*}\left\|\widetilde{\mathbf F}^{*\top}P_{S^*}\widetilde{\mathbf F}^*\right\|_{\mathrm{op}}\le\operatorname{tr}\left(\widetilde{\mathbf F}^{*\top}P_{S^*}\widetilde{\mathbf F}^*\right)=O_p(1).\end{align*}
\end{proof}

\begin{lemma}[Change of basis for the bias terms]\label{lem:masterbias}
Let $\hat{\mathbf Z}$ be defined as in Equation~\ref{equation:zhat}. Suppose Assumptions~\ref{Assumption:covariances}-\ref{Assumption:pervasiveness} and~\ref{assump:subgaussian}-\ref{assump:growth} hold. Then, conditional on the training fold,
\begin{align*}\left\|\mathbf H_1^\top\left(\frac1{n_t}\hat{\mathbf Z}^\top\hat{\mathbf Z}\right)^{-1}\mathbf H_1\right\|_{\mathrm{op}}=O_p(1).\end{align*}
\end{lemma}

\begin{proof}
Recall $M = \frac{1}{n_t}\hat{\mathbf{Z}}^\top\hat{\mathbf{Z}}$, with all matrices restricted to treated rows.
\\

\noindent\textbf{\underline{Step 1: Change of basis.}}
Since $\mathbf H_1$ is full column rank, write $\mathbf H_1=UR$, where $U\in\mathbb R^{(1+\tilde p+\tilde q)\times(1+\bar p+\bar q)}$ has orthonormal columns and $R$ is invertible. By Proposition~\ref{prop:loading-cond}, $\nu_{\min}(R)\ge c,$ and $\|R\|_{\mathrm{op}}\le C.$ Now, we extend $U$ to a full orthonormal basis of $\mathbb{R}^{1+\tilde{p}+\tilde{q}}$ using a matrix $S\in\mathbb{R}^{(1+\tilde{p}+\tilde{q})\times(\tilde{p}-\bar{p}+\tilde{q}-\bar{q})}$ whose columns are orthonormal to $U$, so that $J = [U\ \ S]$ is square and orthogonal. Thus, $S^\top\mathbf{H}_1 = S^\top U R = 0$.

\medskip\noindent Since $J$ is orthogonal, $J^{-1}=J^\top$ and
$(J^\top MJ)^{-1}=J^\top M^{-1}J$, so $M^{-1}=J(J^\top MJ)^{-1}J^\top$. Also,
\begin{align*}
J^\top\mathbf{H}_1 &= J^\top UR=\begin{pmatrix}U^\top U\\ S^\top U\end{pmatrix}R=\begin{pmatrix}I\\0\end{pmatrix}R=\begin{pmatrix}R\\0\end{pmatrix},
\end{align*}
Similarly, we have that $\mathbf H_1^\top J = \begin{pmatrix}R^\top & 0\end{pmatrix}.$
\\

\noindent\textbf{\underline{Step 2: Matrix $J^\top MJ$.}} We write $M = \mathbf{H}_1\Sigma_{\mathbf{F}}\mathbf{H}_1^\top + \mathbf{H}_1 B + B^\top\mathbf{H}_1^\top + C$, where $\Sigma_{\mathbf{F}}=n_t^{-1}\mathbf{F}_1^\top \mathbf{F}_1 ,$ $B ={n_t}^{-1} \mathbf{F}_1^\top\mathbf{E}_0,$ and $C = \mathbf{E}_0^\top\mathbf{E}_0/{n_t}.$ Now, using $\mathbf{H}_1=UR$ and the orthogonality of $S^\top U=0$, we get that
\begin{align*}
J^\top MJ=\begin{bmatrix} \begin{gathered}\underbrace{R\Sigma_{\mathbf{F}}R^\top}_{=\,D}+\underbrace{RBU+U^\top B^\top R^\top+U^\top CU}_{=\,K_{11}}
\end{gathered}
& \underbrace{RBS+U^\top CS}_{=\,K_{12}}\\ \underbrace{S^\top B^\top R^\top+S^\top CU}_{=\,K_{21}}
& \underbrace{S^\top CS}_{=\,K_{22}}
\end{bmatrix}.
\end{align*}
\\

\noindent\textbf{\underline{Step 3: Schur complement and $E_n$.}} By the block matrix inversion formula, the $(1,1)$ block of $(J^\top MJ)^{-1}$ is the inverse of the Schur complement
\begin{align*} D + \underbrace{K_{11}-K_{12}K_{22}^{-1}K_{21}}_{=\,E_n} = D + E_n.\end{align*}
Now, substituting the explicit expressions for $B$ and $C$, the cross term $K_{12}K_{22}^{-1}K_{21}$ simplifies as follows.  First, factor as
\begin{align*}
K_{12} &= \frac{1}{n_t}(R\mathbf{F}_1^\top+U^\top\mathbf{E}_0^\top)\mathbf{E}_0 S,\qquad K_{21} = \frac{1}{n_t}S^\top\mathbf{E}_0^\top(\mathbf{F}_1 R^\top+\mathbf{E}_0 U),\qquad  K_{22} = \frac{1}{n_t}S^\top\mathbf{E}_0^\top\mathbf{E}_0 S.
\end{align*}
Therefore, we will get
\begin{align*}
K_{12}K_{22}^{-1}K_{21} &= \frac{1}{n_t}(R\mathbf{F}_1^\top+U^\top\mathbf{E}_0^\top)\mathbf{E}_0 S (S^\top\mathbf{E}_0^\top\mathbf{E}_0 S)^{-1} S^\top\mathbf{E}_0^\top (\mathbf{F}_1 R^\top+\mathbf{E}_0 U)\\
&= \frac{1}{n_t}(R\mathbf{F}_1^\top+U^\top\mathbf{E}_0^\top) P_S (\mathbf{F}_1 R^\top+\mathbf{E}_0 U),
\end{align*}
where
$P_S = \mathbf{E}_0 S(S^\top\mathbf{E}_0^\top\mathbf{E}_0 S)^{-1}
 S^\top\mathbf{E}_0^\top$
is the orthogonal projection onto the column space of $\mathbf{E}_0 S$. Thus,
\begin{align*}
E_n &= K_{11}-K_{12}K_{22}^{-1}K_{21} = \frac{1}{n_t}R\mathbf{F}_1^\top\mathbf{E}_0 U + \frac{1}{n_t}U^\top\mathbf{E}_0^\top\mathbf{F}_1 R^\top + \frac{1}{n_t}U^\top\mathbf{E}_0^\top\mathbf{E}_0 U\\
&\quad - \frac{1}{n_t}(R\mathbf{F}_1^\top+U^\top\mathbf{E}_0^\top)   P_S(\mathbf{F}_1 R^\top+\mathbf{E}_0 U).
\end{align*}
To simplify, we observe that the first three terms can be written as
\begin{align*}
\frac{1}{n_t}(R\mathbf{F}_1^\top+U^\top\mathbf{E}_0^\top)
(\mathbf{F}_1 R^\top+\mathbf{E}_0 U)
-\frac{1}{n_t}R\mathbf{F}_1^\top\mathbf{F}_1 R^\top,
\end{align*}
so that
\begin{align*}
E_n &= \frac{1}{n_t}(R\mathbf{F}_1^\top+U^\top\mathbf{E}_0^\top) (I-P_S)(\mathbf{F}_1 R^\top+\mathbf{E}_0 U)- \frac{1}{n_t}R\mathbf{F}_1^\top\mathbf{F}_1 R^\top.
\end{align*}
Now, expanding $(I-P_S)$ and collecting terms by whether they involve $P_S$ or not, this becomes
\begin{align}
\label{eqEn_expanded}
 E_n &= -\frac{1}{n_t}R\mathbf{F}_1^\top P_S\mathbf{F}_1 R^\top + \frac{1}{n_t}R\mathbf{F}_1^\top(I-P_S)\mathbf{E}_0 U\notag\\&\quad + \frac{1}{n_t}U^\top\mathbf{E}_0^\top(I-P_S)\mathbf{F}_1 R^\top + \frac{1}{n_t}U^\top\mathbf{E}_0^\top(I-P_S)\mathbf{E}_0 U.
\end{align}

\noindent\textbf{\underline{Step 4: Invertibility of $D+E_n$.}} We show that $\lambda_{\min}(D+E_n)$ is bounded away from zero with probability tending to one.  Thus, we apply Weyl's inequality
\begin{align*}\lambda_{\min}(D+E_n)\ge\lambda_{\min}(D)-\|E_n\|.\end{align*}
The argument shows that $D$ has smallest eigenvalue bounded away from zero, while $\|E_n\|=o_p(1)$.

\noindent\textbf{Lower bound on $\lambda_{\min}(D)$.}
Since $D=\frac{1}{n_t}R\mathbf{F}_1^\top\mathbf{F}_1 R^\top$,
\begin{align*}\lambda_{\min}(D)\ge\nu_{\min}^2(R)\,\lambda_{\min}\left(\frac{1}{n_t}\mathbf{F}_1^\top\mathbf{F}_1\right).\end{align*}
The treated-unit factor Gram matrix concentrates around the conditional second-moment matrix $\Omega_1=\mathbb E[\mathbf f_{1,i}\mathbf f_{1,i}^\top\mid A_i=1]$, which is positive definite by Assumption~\ref{Assumption:covariances}, so $\lambda_{\min}(\frac1{n_t}\mathbf F_1^\top\mathbf F_1)\ge c>0$ with probability tending to one. With $\nu_{\min}(R)\ge c_R>0$ by construction, so
\begin{align*}
\lambda_{\min}(D)\ge c \cdot c_R^2 > 0
\end{align*}
with probability tending to one.

\medskip
\noindent\textbf{Upper bound on $\|E_n\|$.}
From the expansion \eqref{eqEn_expanded}, the triangle inequality gives
\begin{align*}\begin{aligned}
 \|E_n\| &\le
\underbrace{\left\|\frac{1}{n_t}R\mathbf{F}_1^\top P_S\mathbf{F}_1 R^\top\right\|}_{T_1} +\underbrace{\left\|\frac{1}{n_t}R\mathbf{F}_1^\top(I-P_S)\mathbf{E}_0 U\right\|}_{T_2}+\underbrace{\left\|\frac{1}{n_t}U^\top\mathbf{E}_0^\top(I-P_S)\mathbf{F}_1 R^\top\right\|}_{T_3} +\underbrace{\left\|\frac{1}{n_t}U^\top\mathbf{E}_0^\top(I-P_S)\mathbf{E}_0 U\right\|}_{T_4}.
\end{aligned}\end{align*}
We bound each term in turn.

\medskip\noindent\textit{Bound on $T_1$.}
Since $R$ has bounded operator norm,
\begin{align*}T_1\le \|R\|^2\left\|\frac{1}{n_t}\mathbf{F}_1^\top P_S\mathbf{F}_1\right\|.\end{align*}
Write $\mathbf{F}_1 = \mathbf{1}\mu_{\mathbf{F}}^\top+\widetilde{\mathbf{F}}_1$, where $\widetilde{\mathbf{F}}_1$ is the centred part.  Then
\begin{align*}
\mathbf{F}_1^\top P_S\mathbf{F}_1 &=\mu_{\mathbf{F}}\mathbf{1}^\top P_S\mathbf{1}\mu_{\mathbf{F}}^\top+\mu_{\mathbf{F}}\mathbf{1}^\top P_S\widetilde{\mathbf{F}}_1+\widetilde{\mathbf{F}}_1^\top P_S\mathbf{1}\mu_{\mathbf{F}}^\top+\widetilde{\mathbf{F}}_1^\top P_S\widetilde{\mathbf{F}}_1,
\end{align*}
By the triangle inequality,
\begin{align*}
 \left\|\frac{1}{n_t}\mathbf{F}_1^\top P_S\mathbf{F}_1\right\|_{\mathrm{op}} &\le\underbrace{\frac{1}{n_t}\left\|\mu_{\mathbf{F}}\mathbf{1}^\top P_S\mathbf{1}\mu_{\mathbf{F}}^\top\right\|_{\mathrm{op}}}_{A_1} +\underbrace{\frac{1}{n_t}\left\|\mu_{\mathbf{F}}\mathbf{1}^\top P_S\widetilde{\mathbf{F}}_1\right\|_{\mathrm{op}}}_{A_2} +\underbrace{\frac{1}{n_t}\left\|\widetilde{\mathbf{F}}_1^\top P_S\mathbf{1}\mu_{\mathbf{F}}^\top\right\|_{\mathrm{op}}}_{A_3} +\underbrace{\frac{1}{n_t}\left\|\widetilde{\mathbf{F}}_1^\top P_S\widetilde{\mathbf{F}}_1\right\|_{\mathrm{op}}}_{A_4}.
\end{align*}
For $A_1$, by Lemma~\ref{1PE1}, $\mathbf 1^\top P_S\mathbf 1=O_p(1).$ Thus,
\begin{align*}
A_1=\frac1{n_t}\left\|\mu_{\mathbf F}\mathbf 1^\top P_S\mathbf 1\mu_{\mathbf F}^\top\right\|_{\mathrm{op}}=\frac1{n_t}\|\mu_{\mathbf F}\|_2^2\,\mathbf 1^\top P_S\mathbf 1=O_p\left(\frac{\|\mu_{\mathbf F}\|_2^2}{n_t}\right).
\end{align*}
For $A_2$, since $P_S$ is an orthogonal projection, $\mathbf 1^\top P_S\widetilde{\mathbf F}_1=(P_S\mathbf 1)^\top(P_S\widetilde{\mathbf F}_1)$, so
\begin{align*}
A_2\le\frac1{n_t}\|\mu_{\mathbf F}\|_2\,\|P_S\mathbf 1\|_2\,\|P_S\widetilde{\mathbf F}_1\|_{\mathrm{op}}=\frac1{n_t}\|\mu_{\mathbf F}\|_2\big(\mathbf 1^\top P_S\mathbf 1\big)^{1/2}\big\|\widetilde{\mathbf F}_1^\top P_S\widetilde{\mathbf F}_1\big\|_{\mathrm{op}}^{1/2}=O_p\left(\frac{\|\mu_{\mathbf F}\|_2}{n_t}\right),
\end{align*}
where the last equality uses $A_1$ and Lemma~\ref{lem:tracebound}. Since the term $A_3$ is the transpose of the term in $A_2$, $A_3=O_p\left({\|\mu_{\mathbf F}\|_2}/n_t\right).$ Lastly, for $A_4$, Lemma~\ref{lem:tracebound} gives $\|\widetilde{\mathbf F}_1^\top P_S\widetilde{\mathbf F}_1\|_{\mathrm{op}}=O_p(1)$, so $A_4=O_p\left(1/{n_t}\right).$ Finally, combining the four terms, we get that
\begin{align}
\left\|\frac1{n_t}\mathbf F_1^\top P_S\mathbf F_1\right\|_{\mathrm{op}}&\le A_1+A_2+A_3+A_4=O_p\left(\frac{1+\|\mu_{\mathbf F}\|_2+\|\mu_{\mathbf F}\|_2^2}{n_t}\right).\label{eqT1bound}
\end{align}
Since $\|R\|_{\mathrm{op}}\le C$, it follows that $T_1=O_p({{n_t}^{-1}(1+\|\mu_{\mathbf F}\|_2+\|\mu_{\mathbf F}\|_2^2)}).$

\medskip\noindent\textit{Bound on $T_2$ and $T_3$.}
Since $\|U\|=1$, $\|R\|=\mathcal{O}(1)$, we get
\begin{align*}
T_2&\le\frac{1}{n_t}\|R\mathbf{F}_1^\top (I - P_S) \mathbf{E}_0 U\|\le\frac{1}{n_t}\|R\|\left(\|\mathbf{F}_1^\top\mathbf{E}_0 \| + \|\mathbf{F}_1^\top P_S\|~\| \mathbf{E}_0 \|\right)\|U\|\le\frac{C}{n_t}\left(\|\mathbf{F}_1^\top\mathbf{E}_0 \| + \|\mathbf{F}_1^\top P_S\|~\| \mathbf{E}_0 \|\right).
\end{align*}
By Lemma~\ref{lem:E0bounds}(ii), $\left\|\frac{1}{\sqrt {n_t}}\mathbf{F}_1^\top\mathbf{E}_0\right\|_{\text{op}}=O_p\{(p\wedge q)^{-1/2}\}$, by Lemmas~\ref{1PE1}--\ref{lem:tracebound}, $\|\mathbf{F}_1^\top P_S \mathbf{F}_1\|_{\text{op}}=\mathcal{O}_P(1)$ and by Lemma~\ref{lem:E0bounds}(i), we have $\|\mathbf{E}_0\|_{\text{op}}=\mathcal{O}_P(\sqrt{n_t/(p \wedge q)})$. So, we have
\begin{align*}
T_2 = \mathcal{O}_P\left(\frac{1}{n_t}\sqrt{\frac{n_t}{p\wedge q}}\right)= \mathcal{O}_P\left(\frac{1}{\sqrt{n_t(p\wedge q)}}\right)
\end{align*}
so we have $T_2, T_3= \mathcal{O}_p(1/{\sqrt{n_t(p\wedge q)}}).$

\medskip\noindent\textit{Bound on $T_4$.}
Using $\|U\|=1$ and $\|I-P_S\|\le 1$,
\begin{align}T_4\le\left\|\frac{1}{n_t}\mathbf{E}_0^\top\mathbf{E}_0\right\|= \mathcal{O}_p\left(\frac{1}{p\wedge q}\right)\label{eqT4bound}\end{align}
where the last equality follows from Lemma \ref{lem:E0bounds}. Hence, $T_4 = \mathcal{O}_p\left(1/({p\wedge q})\right).$

\medskip\noindent\textit{Combining the bounds.}
Now, collecting Equations \eqref{eqT1bound} and \eqref{eqT4bound}, we get
\begin{equation}\label{eqEn_rate}
\|E_n\|=\mathcal O_p\left(\frac{1+\|\mu_{\mathbf F}\|_2+\|\mu_{\mathbf F}\|_2^2}{n_t}+\frac1{\sqrt{n_t(p\wedge q)}}+\frac1{p\wedge q}\right).
\end{equation}
Each term vanishes once $n_t\to\infty$ and $p\wedge q\to\infty$, so $\|E_n\|=o_p(1)$ under the growth conditions of the theorem, with no relative rate between $n_t$ and $p\wedge q$ required. Weyl's inequality then gives
\begin{align*}
\lambda_{\min}(D+E_n)&\ge c\,c_R^2-\|E_n\|\ge c\,c_R^2-o_p(1) \ge\frac{c\,c_R^2}{2}>0
\end{align*}
with probability tending to one, and therefore
\begin{equation}\label{eqDplusEn_inv}
\|(D+E_n)^{-1}\|_{\text{op}}=O_p(1).
\end{equation}
In conclusion, the change of basis gives $M^{-1}=J(J^\top MJ)^{-1}J^\top$ and $\mathbf H_1^\top J=\begin{pmatrix}R^\top & 0\end{pmatrix}$, and the $(1,1)$ block of $(J^\top MJ)^{-1}$ is the inverse Schur complement $(D+E_n)^{-1}$. Putting these together,
\begin{align*}
\mathbf H_1^\top M^{-1}\mathbf H_1
=\mathbf H_1^\top J(J^\top MJ)^{-1}J^\top\mathbf H_1
=\begin{pmatrix}R^\top & 0\end{pmatrix}(J^\top MJ)^{-1}\begin{pmatrix}R\\ 0\end{pmatrix}=R^\top(D+E_n)^{-1}R.
\end{align*}
Since $\|R\|_{\text{op}}=\mathcal O(1)$ and $\|(D+E_n)^{-1}\|_{\text{op}}=O_p(1)$ by Equation~\eqref{eqDplusEn_inv},
\begin{align*}
\left\|\mathbf H_1^\top M^{-1}\mathbf H_1\right\|_{\text{op}}
&\le\|R\|_{\text{op}}^2\,\|(D+E_n)^{-1}\|_{\text{op}} =O_p(1),
\end{align*}
which is what the lemma claims. The proof also gives two facts we use later, the bound $\|(D+E_n)^{-1}\|_{\text{op}}=O_p(1)$ and the orthogonality $e_{\mathbf 1}^\top S=0$. The second holds because the first column of $\mathbf H_1=\operatorname{diag}(1,\mathbf H_X,\mathbf H_M)$ is $e_{\mathbf 1}$, so $e_{\mathbf 1}$ lies in $\operatorname{Col}(\mathbf H_1)=\operatorname{Col}(U)$, and the columns of $S$ are orthogonal to that space by construction.
\end{proof}

\subsubsection*{Consistency and linear expansion of $\hat\gamma_1$}
\label{sec:gammalemma}

We next establish the linear expansion of the first-stage mediator coefficient $\hat\gamma_1$. The quantity $\mathbf H_M^\top\hat\gamma_1$ enters the second-stage bridge regression, so we need both its convergence rate and its leading term.

For this lemma and its proof, every sum over $i$ runs over the treated units $\{i:A_i=1\}$, the matrices $\hat{\mathbf Z},\mathbf F_1,\mathbf E_0,\boldsymbol\epsilon_1$ collect treated rows only, and $M=\frac1{n_t}\hat{\mathbf Z}^\top\hat{\mathbf Z}$. The treated regression involves the conditional second-moment matrix $\Omega_1=\mathbb E[\mathbf f_{1,i}\mathbf f_{1,i}^\top\mid A_i=1]$, positive definite by Assumption~\ref{Assumption:covariances}(iii)--(iv).

\begin{lemma}\label{lemma:approxgamma1}
Suppose Assumptions~\ref{Assumption:covariances}--\ref{Assumption:pervasiveness} and~\ref{assump:subgaussian}--\ref{assump:growth} hold. Then,
\begin{align*}\mathbf H_M^\top\hat\gamma_1-\gamma_1=\mathbf P_{\bar q}\Omega_1^{-1}\frac1{n_t}\sum_{i:A_i=1}\mathbf f_{1,i}\epsilon_{1,i}+r_\gamma,\qquad\|r_\gamma\|_2=o_p(n_t^{-1/2}).\end{align*}
\end{lemma}
\begin{proof}
We want to show that
\begin{align*}\sqrt {n_t}\left({\bf H}_M^\top \hat{\gamma}_1-\gamma_1\right)=\mathcal O_p(1),\end{align*}
Let $\mathbf S_{\tilde q}$ be the projected mediator selector and $\mathbf P_{\bar q}$ be the true mediator selector defined in the notation section, so that $\mathbf H_M^\top\mathbf S_{\tilde q}=\mathbf P_{\bar q}\mathbf H_1^\top$. From $\hat\gamma_1$ on the treated units, we have
\begin{align*}
{\bf H}_M^\top\hat{\gamma}_1
&={\bf H}_M^\top \mathbf S_{\tilde q}\left(\frac1{n_t}\hat{\mathbf Z}^\top\hat{\mathbf Z}\right)^{-1}\frac1{n_t}\hat{\mathbf Z}^\top Y .
\end{align*}
Now, using the outcome model $Y=\mathbf Z\delta+\epsilon_1$ on the treated units, we get that
\begin{align*}
{\bf H}_M^\top\hat{\gamma}_1 &={\bf H}_M^\top \mathbf S_{\tilde q}\left(\frac1{n_t}\hat{\mathbf Z}^\top \hat{\mathbf Z}\right)^{-1}\frac1{n_t}\hat{\mathbf Z}^\top(\mathbf Z\delta+\epsilon_1)\\
&={\bf H}_M^\top \mathbf S_{\tilde q}\left(\frac1{n_t}\hat{\mathbf Z}^\top \hat{\mathbf Z}\right)^{-1}\frac1{n_t}\hat{\mathbf Z}^\top\{\hat{\mathbf Z}\delta+(\mathbf Z-\hat{\mathbf Z})\delta+\epsilon_1\}\\
&= \gamma_1 +{\bf H}_M^\top \mathbf S_{\tilde q}\left(\frac1{n_t}\hat{\mathbf Z}^\top \hat{\mathbf Z}\right)^{-1}\frac1{n_t}\hat{\mathbf Z}^\top(\mathbf Z-\hat{\mathbf Z})\delta+{\bf H}_M^\top \mathbf S_{\tilde q}
\left(\frac1{n_t}\hat{\mathbf Z}^\top \hat{\mathbf Z}\right)^{-1}\frac1{n_t}\hat{\mathbf Z}^\top\epsilon_1 .
\end{align*}
Since $\mathbf Z-\hat{\mathbf Z}=-\mathbf E_0$, we have
\begin{align}
\label{eq:gammahat_decomp}
\sqrt{n_t}(&{\bf H}_M^\top\hat{\gamma}_1-\gamma_1) \\
&= -{\bf H}_M^\top \mathbf S_{\tilde q} \left(\frac1{n_t}\hat{\mathbf Z}^\top \hat{\mathbf Z}\right)^{-1} \frac1{\sqrt{n_t}}\hat{\mathbf Z}^\top{\bf E}_0\delta +{\bf H}_M^\top \mathbf S_{\tilde q} \left( \frac1{n_t}\hat{\mathbf Z}^\top \hat{\mathbf Z} \right)^{-1} \notag\frac1{\sqrt{n_t}}\hat{\mathbf Z}^\top\epsilon_1\\
&= - \mathbf P_{\bar q} {\bf { H}}_1^\top \left(\frac1{n_t}\hat{\mathbf{Z}}^\top \hat{\mathbf{Z}}\right)^{-1} \frac1 {\sqrt {n_t}} ({\bf{H}}_1\mathbf{F}_{1}^\top + \mathbf{E}_0^\top) {\bf{E}}_0 \delta + \mathbf P_{\bar q} {\bf {H}}_1^\top \left(\frac1{n_t}\hat{\mathbf{Z}}^\top \hat{\mathbf{Z}}\right)^{-1} \frac1 {\sqrt {n_t}} ({\bf{H}}_1\mathbf{F}_{1}^\top + \mathbf{E}_0^\top) \epsilon_1\notag\displaybreak[1]\\
&= - \underbrace{\mathbf P_{\bar q} {\bf {H}}_1^\top  \left(\frac1{n_t}\hat{\mathbf{Z}}^\top \hat{\mathbf{Z}}\right)^{-1} {\bf{H}}_1 \frac1 {\sqrt{n_t}} \mathbf{F}_{1}^\top  {\bf{E}}_0 \delta}_{=J_1} - \underbrace{\mathbf P_{\bar q} {\bf {H}}_1^\top \left(\frac1{n_t}\hat{\mathbf{Z}}^\top \hat{\mathbf{Z}}\right)^{-1} \frac1 {\sqrt{n_t}} \mathbf{E}_0^\top {\bf{E}}_0 \delta}_{=J_2}\notag\displaybreak[1]\\
& \quad + \underbrace{\mathbf P_{\bar q} {\bf {H}}_1^\top   \left(\frac1{n_t}\hat{\mathbf{Z}}^\top \hat{\mathbf{Z}}\right)^{-1} {\bf{H}}_1 \frac1 {\sqrt{n_t}} \mathbf{F}_{1}^\top \epsilon_1}_{=J_3} + \underbrace{\mathbf P_{\bar q} {\bf { H}}_1^\top  \left(\frac1{n_t}\hat{\mathbf{Z}}^\top \hat{\mathbf{Z}}\right)^{-1} \frac1 {\sqrt{n_t}} \mathbf{E}_0^\top \epsilon_1}_{= J_4}\,.\notag
\end{align}
write $M=\frac1{n_t}\hat{\mathbf Z}^\top\hat{\mathbf Z}.$

\medskip\noindent\textbf{\underline{Term $J_1$.}}
We have
\begin{align*}J_1=\mathbf P_{\bar q}\mathbf H_1^\top M^{-1}\mathbf H_1\frac1{\sqrt{n_t}}\mathbf F_1^\top\mathbf E_0\delta.\end{align*}
By Lemma~\ref{lem:masterbias}, $\|\mathbf H_1^\top M^{-1}\mathbf H_1\|_{\mathrm{op}}=O_p(1)$, so we have
\begin{align*}
\|J_1\|&\le\|\mathbf P_{\bar q}\|_{\mathrm{op}}\,\|\mathbf H_1^\top M^{-1}\mathbf H_1\|_{\mathrm{op}}\left\|\frac1{\sqrt{n_t}}\mathbf F_1^\top\mathbf E_0\delta\right\| =O_p(1)\left\|\frac1{\sqrt{n_t}}\mathbf F_1^\top\mathbf E_0\delta\right\|.
\end{align*}
By Lemma~\ref{lem:E0bounds}(ii) and $\|\delta\|=O(1)$, we get
\begin{align*}
\left\|\frac1{\sqrt{n_t}}\mathbf F_1^\top\mathbf E_0\delta\right\|&\le\left\|\frac1{\sqrt{n_t}}\mathbf F_1^\top\mathbf E_0\right\|_{\mathrm{op}}\|\delta\| =O_p\left\{\left(\frac1{p\wedge q}\right)^{1/2}\right\},
\end{align*}
Thus $J_1=o_p(1)$ as $p\wedge q\to\infty$.

\medskip
\noindent\textbf{\underline{Term $J_2$.}}
\begin{align*}J_2=\mathbf P_{\bar q}\mathbf H_1^\top M^{-1}\frac1{\sqrt {n_t}}\mathbf E_0^\top\mathbf E_0\delta .\end{align*}
This term differs from $J_1$ because there is no $\mathbf H_1$ on the right of $M^{-1}$. After the change of basis the left $S$-block disappears, but the right side keeps an $S$-part, so we use the block-inverse identity directly. As in Lemma~\ref{lem:masterbias}, write $\mathbf H_1=UR$ with $U^\top U=I$ and complete $U$ to an orthogonal $J=[U\ \ S]$, so that $\mathbf H_1^\top J=(R^\top\ 0)$ and $M^{-1}=J(J^\top MJ)^{-1}J^\top$. From the block inverse formula, the first block row of $(J^\top MJ)^{-1}$ is
\begin{align*}\begin{pmatrix}
(D+E_n)^{-1}&-(D+E_n)^{-1}K_{12}K_{22}^{-1}
\end{pmatrix}.\end{align*}
Therefore, using $\mathbf H_1^\top J=\begin{pmatrix}R^\top&0\end{pmatrix}$ and $J^\top=\begin{pmatrix}U^\top\\S^\top\end{pmatrix}$, we have
\begin{align*}
\mathbf H_1^\top M^{-1}
&=R^\top(D+E_n)^{-1}
\left(U^\top-K_{12}K_{22}^{-1}S^\top\right)\\
&=R^\top(D+E_n)^{-1}\Bigg[U^\top
-\left\{\frac1{n_t}R\mathbf F_1^\top\mathbf E_0S+\frac1{n_t}U^\top\mathbf E_0^\top\mathbf E_0S\right\}\left\{\frac1{n_t}S^\top\mathbf E_0^\top\mathbf E_0S\right\}^{-1}S^\top\Bigg],
\end{align*}
with $\|(D+E_n)^{-1}\|_{\mathrm{op}}=O_p(1)$ by Lemma~\ref{lem:masterbias}. We apply this identity to the $(1/\sqrt {n_t})\mathbf E_0^\top\mathbf E_0\delta$ and use $P_S=\mathbf E_0S(S^\top\mathbf E_0^\top\mathbf E_0S)^{-1}S^\top\mathbf E_0^\top$.
Then,
\begin{align*}
J_2&=\mathbf P_{\bar q}R^\top(D+E_n)^{-1}
\Bigg[\frac1{\sqrt{n_t}}U^\top\mathbf E_0^\top\mathbf E_0\delta-\left(\frac1{n_t}R\mathbf F_1^\top\mathbf E_0S+\frac1{n_t}U^\top\mathbf E_0^\top\mathbf E_0S\right)\left(\frac1{n_t}S^\top\mathbf E_0^\top\mathbf E_0S\right)^{-1}\frac1{\sqrt{n_t}}S^\top\mathbf E_0^\top\mathbf E_0\delta\Bigg]\\
&=\mathbf P_{\bar q}R^\top(D+E_n)^{-1}\Bigg[\frac1{\sqrt{n_t}}U^\top\mathbf E_0^\top\mathbf E_0\delta-\frac1{\sqrt{n_t}}\left(R\mathbf F_1^\top+U^\top\mathbf E_0^\top\right)P_S\mathbf E_0\delta\Bigg]\\
&=\mathbf P_{\bar q}R^\top(D+E_n)^{-1}\Bigg[\frac1{\sqrt{n_t}}U^\top\mathbf E_0^\top(I-P_S)\mathbf E_0\delta-\frac1{\sqrt{n_t}}R\mathbf F_1^\top P_S\mathbf E_0\delta\Bigg].
\end{align*}
The outside factor is bounded because $\|\mathbf P_{\bar q}\|=1$, $\|R\|_{\mathrm{op}}=O(1)$, and $\|(D+E_n)^{-1}\|_{\mathrm{op}}=O_p(1)$. Thus it is enough to bound the two terms inside the square bracket.
\\

\noindent For the first term, using $\|U\|_{\mathrm{op}}=1$ and $\|I-P_S\|_{\mathrm{op}}\le1$, we get that
\begin{align*}
\left\|\frac1{\sqrt {n_t}}U^\top\mathbf E_0^\top(I-P_S)\mathbf E_0\delta\right\|&\le \frac1{\sqrt {n_t}}\|\mathbf E_0^\top(I-P_S)\mathbf E_0\|_{\mathrm{op}}\|\delta\| \le\sqrt {n_t}\left\|\frac1{n_t}\mathbf E_0^\top\mathbf E_0\right\|_{\mathrm{op}}\|\delta\|.
\end{align*}
Since $\left\|\frac1{n_t}\mathbf E_0^\top\mathbf E_0\right\|_{\mathrm{op}}=O_p\left(\frac1{p\wedge q}\right), $ from Lemma \ref{lem:E0bounds}(i) the first term is $O_p\left( \frac{\sqrt {n_t}}{p\wedge q} \right).$ For the second term, we have
\begin{align*}
\left\|\frac1{\sqrt {n_t}}R\mathbf F_1^\top P_S\mathbf E_0\delta\right\|
&\le\|R\|_{\mathrm{op}}\,\frac1{\sqrt {n_t}}\|\mathbf F_1^\top P_S\|_{\mathrm{op}}\|\mathbf E_0\delta\|.
\end{align*}
Here $\|R\|_{\mathrm{op}}=O(1)$. We write $\mathbf F_1=\mathbf 1\mu_{\mathbf F}^\top+\widetilde{\mathbf F}_1$. By Lemma~\ref{1PE1}, $\|P_S\mathbf 1\|_2^2=\mathbf 1^\top P_S\mathbf 1=O_p(1)$, and by Lemma~\ref{lem:tracebound}, $\|\widetilde{\mathbf F}_1^\top P_S\|_{\mathrm{op}}^2=\|\widetilde{\mathbf F}_1^\top P_S\widetilde{\mathbf F}_1\|_{\mathrm{op}}=O_p(1)$. Since $\|\mu_{\mathbf F}\|=O(1)$ by Assumption~\ref{Assumption:covariances}(iii), we have $\|\mathbf F_1^\top P_S\|_{\mathrm{op}}\le\|\mu_{\mathbf F}\|\,\|P_S\mathbf 1\|_2+\|\widetilde{\mathbf F}_1^\top P_S\|_{\mathrm{op}}=O_p(1)$. Also,
\begin{align*}
\|\mathbf E_0\delta\|&\le\|\mathbf E_0^\top\mathbf E_0\|_{\mathrm{op}}^{1/2}\|\delta\| =\sqrt {n_t}\left\|\frac1{n_t}\mathbf E_0^\top\mathbf E_0\right\|_{\mathrm{op}}^{1/2}\|\delta\| =O_p\left(\sqrt{\frac {n_t}{p\wedge q}}\right).
\end{align*}
Therefore, the second term is $O_p\left(\frac1{\sqrt{p\wedge q}}\right).$  Now, combining the two pieces inside the bracket and using the bounded outside factor, we get
\begin{align*}\|J_2\|=O_p\left(\frac{\sqrt {n_t}}{p\wedge q}+\frac1{\sqrt{p\wedge q}}\right).\end{align*}
Thus $J_2=o_p(1)$ under Assumption~\ref{assump:growth}(ii), since $(p\wedge q)/n_t\to\infty$ gives $\sqrt{n_t}/(p\wedge q)\to0$.

\medskip
\noindent\textbf{\underline{Term $J_3$.}}
We have \begin{align*}J_3=\mathbf P_{\bar q}\mathbf H_1^\top M^{-1}\mathbf H_1\frac1{\sqrt {n_t}}\mathbf F_1^\top\epsilon_1 .\end{align*}
By the treated-arm exogeneity and mean zero condition $\mathbb{E}(\epsilon_{1,i})=\mathbb E(\epsilon_{1,i}|\mathbf{f}_{X,i},\mathbf{f}_{M,i},A_i=1)=0$ of Assumption~\ref{Assumption:covariances}(ii), we have
\begin{align*}\mathbb E\left(\frac1{\sqrt {n_t}}\mathbf F_1^\top\epsilon_1\,\middle|\,\mathbf F_1\right)=0.\end{align*}
Also, using the bounded conditional variance of Assumption~\ref{assump:subgaussian}(i), we get
\begin{align*}
\mathbb E\left[\left\|\frac1{\sqrt {n_t}}\mathbf F_1^\top\epsilon_1\right\|^2\,\middle|\,\mathbf F_1\right]
&=\operatorname{tr}\left\{\frac1{n_t}\mathbf F_1^\top\mathbb E(\epsilon_1\epsilon_1^\top\mid\mathbf F_1)\mathbf F_1\right\} \le C\operatorname{tr}\left(\frac1{n_t}\mathbf F_1^\top\mathbf F_1
\right).
\end{align*}
The last term is $O_p(1)$ by the bounded factor second moments of Assumption~\ref{Assumption:covariances}(iii)--(iv). Chebyshev's inequality then gives $\|\frac1{\sqrt{n_t}}\mathbf F_1^\top\epsilon_1\|=O_p(1)$. Note that $\mathbf H_1^\top M^{-1}\mathbf H_1=R^\top(D+E_n)^{-1}R$ from Lemma~\ref{lem:masterbias}, where $D=R\Sigma_{\mathbf F}R^\top$ with $\Sigma_{\mathbf F}=\frac1{n_t}\mathbf F_1^\top\mathbf F_1\to_p\Omega_1$ and $\|E_n\|=o_p(1)$. Hence, we have
\begin{align*}
\mathbf H_1^\top M^{-1}\mathbf H_1
&=R^\top(D+E_n)^{-1}R\xrightarrow{p}R^\top(R\Omega_1R^\top)^{-1}R=\Omega_1^{-1}.
\end{align*}
Therefore, we get
\begin{align*}
J_3 &=\mathbf P_{\bar q}\Omega_1^{-1}\frac1{\sqrt{n_t}}\mathbf F_1^\top\epsilon_1+\mathbf P_{\bar q}\big(\mathbf H_1^\top M^{-1}\mathbf H_1-\Omega_1^{-1}\big)\frac1{\sqrt{n_t}}\mathbf F_1^\top\epsilon_1\\&=\mathbf P_{\bar q}\Omega_1^{-1}\frac1{\sqrt{n_t}}\sum_{i:A_i=1}\mathbf f_{1,i}\epsilon_{1,i}+o_p(1),
\end{align*}
since the first factor of the second term is $o_p(1)$ in operator norm and the
second factor is $O_p(1)$.

\medskip\noindent \textbf{\underline{Term $J_4$.}}
\begin{align*}J_4=\mathbf P_{\bar q}\mathbf H_1^\top M^{-1}\frac1{\sqrt {n_t}}\mathbf E_0^\top\epsilon_1.\end{align*}
Similar to $J_2$, this is a one-sided term because there is no $\mathbf H_1$ on the
right. So, the block-inverse identity gives
\begin{align*}\begin{aligned}
\mathbf H_1^\top M^{-1}\frac1{\sqrt {n_t}}\mathbf E_0^\top\epsilon_1 =R^\top(D+E_n)^{-1}\Biggl[\frac1{\sqrt {n_t}}U^\top\mathbf E_0^\top(I-P_S)\epsilon_1 -\frac1{\sqrt {n_t}}R\mathbf F_1^\top P_S\epsilon_1\Biggr].
\end{aligned}\end{align*}
For the first term, conditional on $\mathbf E_0$, the matrix $P_S$ is fixed. Also, by the independence and moment conditions on $\epsilon_1$ in Assumptions~\ref{Assumption:covariances}(i)--(ii) and~\ref{assump:subgaussian}(i), we have $\mathbb E(\epsilon_1\mid\mathbf E_0)=0$ and $\mathbb E(\epsilon_1\epsilon_1^\top\mid\mathbf E_0)\preceq CI$. Therefore, we have
\begin{align*}
\mathbb E\left[\left\|\frac1{\sqrt{n_t}}U^\top\mathbf E_0^\top(I -P_S)\epsilon_1\right\|_2^2\,\middle|\,\mathbf E_0\right]&=\frac1{n_t}\operatorname{tr}\left\{U^\top\mathbf E_0^\top(I-P_S)\mathbb E(\epsilon_1\epsilon_1^\top\mid\mathbf E_0)(I-P_S)\mathbf E_0U\right\}\\
&\le\frac{C}{n_t}\operatorname{tr}\left\{U^\top\mathbf E_0^\top(I-P_S)\mathbf E_0U\right\}\\&\le\frac{C}{n_t}\operatorname{tr}\left(U^\top\mathbf E_0^\top\mathbf E_0U\right).
\end{align*}
Since $U$ has a fixed number of orthonormal columns,
\begin{align*}
\frac1{n_t}\operatorname{tr}\left( U^\top\mathbf E_0^\top\mathbf E_0U \right)&\le C\left\|\frac1{n_t}\mathbf E_0^\top\mathbf E_0
\right\|_{\mathrm{op}}=O_p\left(\frac1{p\wedge q}\right)
\end{align*}
by Lemma~\ref{lem:E0bounds}(i). Hence, we have
\begin{align*}\left\|\frac1{\sqrt{n_t}}U^\top\mathbf E_0^\top(I-P_S)\epsilon_1\right\|_2=O_p\left(\frac1{\sqrt{p\wedge q}}\right).\end{align*}
For the second term $\frac1{\sqrt {n_t}}R\mathbf F_1^\top P_S\epsilon_1$, we again write $\mathbf F_1=\mathbf 1\mu_{\mathbf F}^\top+\widetilde{\mathbf F}_1.$ The mean part is $\frac1{\sqrt {n_t}}R\mu_{\mathbf F}\mathbf 1^\top P_S\epsilon_1.$ By Assumptions~\ref{Assumption:covariances}(i)--(ii), $\epsilon_{1,i}$ is independent of the latent factors and proxy errors generating $\mathbf f_{1,i}$ and $\mathbf E_0$, and $\mathbb E(\epsilon_{1,i})=0$. Hence, $\mathbb E(\epsilon_{1,i}\mid\mathbf f_{1,i},\mathbf E_0,A_i=1)=\mathbb E(\epsilon_{1,i})=0.$ Since $P_S$ is a function of $\mathbf E_0$, we have $\mathbb E(\epsilon_1\mid P_S)=0$, hence $\mathbb E(\mathbf 1^\top P_S\epsilon_1\mid P_S)=0,$ and
\begin{align*}
\mathbb E\left[(\mathbf 1^\top P_S\epsilon_1)^2 \mid P_S \right]
&= \mathbf 1^\top P_S \mathbb E(\epsilon_1\epsilon_1^\top\mid P_S)P_S\mathbf 1 \le C\mathbf 1^\top P_S^2\mathbf 1=C\mathbf 1^\top P_S\mathbf 1.
\end{align*}
By Lemma~\ref{1PE1}, $\mathbf 1^\top P_S\mathbf 1=O_p(1)$. Therefore, we have $\mathbf 1^\top P_S\epsilon_1=O_p(1),$ and so
\begin{align*}\left\|\frac1{\sqrt {n_t}}R\mu_{\mathbf F}\mathbf 1^\top P_S\epsilon_1\right\|=
O_p\left(\frac1{\sqrt {n_t}}\right).\end{align*}
For the centered part, conditional on $\widetilde{\mathbf F}_1$ and $\mathbf E_0$, we get that
\begin{align*}
&\mathbb E\left[\left\|\frac1{\sqrt{n_t}}R\widetilde{\mathbf F}_1^\top P_S\epsilon_1\right\|_2^2\,\middle|\,\widetilde{\mathbf F}_1,\mathbf E_0\right]\le\frac{C}{n_t}\operatorname{tr}\left(\widetilde{\mathbf F}_1^\top P_S\widetilde{\mathbf F}_1\right).
\end{align*}
By Lemma~\ref{lem:tracebound}, we have $\left\|\widetilde{\mathbf F}_1^\top P_S\widetilde{\mathbf F}_1\right\|_{\mathrm{op}}=O_p(1).$ Since the factor dimension is fixed, its trace is also $O_p(1)$. Hence,
\begin{align*}\left\|\frac1{\sqrt{n_t}}R\widetilde{\mathbf F}_1^\top P_S\epsilon_1\right\|_2=O_p\left(\frac1{\sqrt{n_t}}\right).\end{align*}
Therefore, we have
\begin{align*}\|J_4\| = O_p\left\{\left(\frac1{\sqrt{p\wedge q}}\right) + \frac1{\sqrt {n_t}} \right\}.\end{align*}
Thus, $J_1$, $J_2$ and $J_4$ are $o_p(1)$. Now, combining these with the expansion of $J_3$, we get that
\begin{align*}\begin{aligned}
\sqrt {n_t}\left({\bf H}_M^\top\hat{\gamma}_1-\gamma_1\right) &=\mathbf P_{\bar q}\,\Omega_1^{-1}\frac1{\sqrt {n_t}}\sum_{i:A_i=1}\mathbf f_{1,i}\epsilon_{1,i}+o_p(1) =\mathcal O_p(1).
\end{aligned}\end{align*}
Now, dividing by $\sqrt{n_t}$ gives the stated linear expansion with $r_\gamma=o_p(n_t^{-1/2})$, and thus, ${\bf H}_M^\top\hat{\gamma}_1-\gamma_1=\mathcal O_p(n_t^{-1/2}).$
\end{proof}

\begin{lemma}\label{lemma:nullspace}
Suppose Assumptions~\ref{Assumption:covariances}--\ref{Assumption:pervasiveness} and~\ref{assump:subgaussian}--\ref{assump:growth} hold. Let $\mathbf {P}_M=\mathbf H_M(\mathbf H_M^\top\mathbf H_M)^{-1}\mathbf H_M^\top$ and $\hat\gamma_{1,\perp}=(I-\mathbf {P}_M)\hat\gamma_1$. Then
\[\|\hat\gamma_{1,\perp}\|_2=O_p\left(\sqrt{\frac q{n_t}}\right).\]
\end{lemma}

\begin{proof}
We write the treated regression in coordinates that separate $\operatorname{Col}(\mathbf H_1)$ from its orthogonal complement. Step~1 sets up these coordinates and shows that the complement directions of the design contain only proxy errors. Step~2 shows that the Gram matrix is well conditioned. Step~3 expands the rescaled null-space coefficient into four terms, Step~4 and 5 bounds them for $\hat\gamma_{1,\perp}$.

\medskip\noindent\textbf{\underline{Step 1: Coordinates.}}
We begin with the least-squares representation of the treated regression,
\begin{align*}
\hat\delta=\left(\frac1{n_t}\hat{\mathbf Z}^\top\hat{\mathbf Z}\right)^{-1}\frac1{n_t}\hat{\mathbf Z}^\top\mathbf Y.
\end{align*}
As in Lemma~\ref{lem:masterbias}, write $\mathbf H_1=UR$ with $U^\top U=I$, and complete $U$ to the orthogonal matrix $J=[\,U\ \ S\,]$. Since $\mathbf H_1=\operatorname{diag}(1,\mathbf H_X,\mathbf H_M)$, we may choose
\begin{align*}
S=\begin{pmatrix}0&0\\S_X&0\\0&S_M\end{pmatrix},
\end{align*}
where the columns of $S_X$ and $S_M$ form orthonormal bases for $\ker(\mathbf H_X^\top)$ and $\ker(\mathbf H_M^\top)$, respectively. Let $d_X=\tilde p-\bar p$, $d_M=\tilde q-\bar q$, and $d=d_X+d_M$. Since $\delta=(\mathbf H_1^\dagger)^\top\Delta$, we have $\delta\in\operatorname{Col}(\mathbf H_1)=\operatorname{Col}(U)$. We write
\begin{align*}
J^\top\hat\delta=\begin{pmatrix}\hat a\\\hat\kappa\end{pmatrix}=\begin{pmatrix}U^\top\hat\delta\\S^\top\hat\delta\end{pmatrix},\qquad\hat\kappa=\begin{pmatrix}\hat\kappa_X\\\hat\kappa_M\end{pmatrix},
\end{align*}
so that $\hat\delta=U\hat a+S\hat\kappa$. Moreover, since $S_MS_M^\top=I-\mathbf {P}_M$ and $\hat\kappa_M=S_M^\top\hat\gamma_1$, we have
\begin{align*}
S_M\hat\kappa_M=(I-\mathbf {P}_M)\hat\gamma_1=\hat\gamma_{1,\perp}.
\end{align*}
Next, $\hat{\mathbf Z}J=[\,\hat{\mathbf Z}U\ \ \hat{\mathbf Z}S\,]$. Since $\mathbf ZS=\mathbf F_1\mathbf H_1^\top S=0$, Equation \eqref{equation:zhat} gives $\hat{\mathbf Z}S=\mathbf E_0S$. Thus, the directions orthogonal to $\operatorname{Col}(\mathbf H_1)$ contain only projected proxy errors. Since $\mathbf E_X=p^{-1/2}\mathbf U_X\mathbf V_X$ and $\mathbf E_M=q^{-1/2}\mathbf U_M\mathbf V_M$, their corresponding covariance scales are $p^{-1}$ and $q^{-1}$. We therefore rescale these directions in the next step.

\medskip\noindent\textbf{\underline{Step 2: Normalization and Gram matrix.}}
Define $\Sigma_S=\mathbb E[n_t^{-1} S^\top\mathbf E_0^\top\mathbf E_0S ]$ conditional on the training fold. Using the block form of $S$ and $\mathbf E_0$, and the independence of $\mathbf u_X$ and $\mathbf u_M$ in Assumption~\ref{Assumption:covariances}(ii), we have
\begin{align*}
\Sigma_S&=\operatorname{diag}(p^{-1}\Omega_X,q^{-1}\Omega_M),\qquad
\Omega_X=S_X^\top \mathbf V_X^\top\Sigma_{\mathbf u_X}\mathbf V_XS_X,\qquad
\Omega_M=S_M^\top \mathbf V_M^\top\Sigma_{\mathbf u_M}\mathbf V_MS_M.
\end{align*}
By Assumption~\ref{Assumption:covariances}(v), the eigenvalues of $\Omega_X$ and $\Omega_M$ are bounded above and away from zero, so $\Sigma_S$ is invertible. We then define the normalized error matrix $\widetilde{\mathbf E}_S=\mathbf E_0S\Sigma_S^{-1/2}$. Lemma~\ref{1PE1} gives $n_t^{-1}\widetilde{\mathbf E}_S^\top\widetilde{\mathbf E}_S\overset{p}{\longrightarrow}I_d.$ We use this normalization only as an invertible reparameterization of the same least-squares problem. Since $\hat{\mathbf Z}U=\mathbf F_1R^\top+\mathbf E_0U$ and $\hat{\mathbf Z}S=\widetilde{\mathbf E}_S\Sigma_S^{1/2}$, we have
\begin{align*}
\hat{\mathbf Z}J=\left[\hat{\mathbf Z}U,\,\widetilde{\mathbf E}_S\right]
\begin{pmatrix}I&0\\0&\Sigma_S^{1/2}\end{pmatrix}.
\end{align*}
Thus, define $\bar{\mathbf Z}=[\hat{\mathbf Z}U,\,\widetilde{\mathbf E}_S]$, which gives the treated regression of Algorithm~\ref{algo:FARMA} in rescaled $J$-coordinates. Before expanding the least-squares estimator, let $\bar M=n_t^{-1}\bar{\mathbf Z}^\top\bar{\mathbf Z}$ and first show that the normalized Gram matrix is well conditioned. Its upper-left block is
\begin{align*}
\frac1{n_t}(\hat{\mathbf Z}U)^\top(\hat{\mathbf Z}U)
&=R\left(\frac1{n_t}\mathbf F_1^\top\mathbf F_1\right)R^\top
+R\frac1{n_t}\mathbf F_1^\top\mathbf E_0U
+U^\top\frac1{n_t}\mathbf E_0^\top\mathbf F_1R^\top
+U^\top\frac1{n_t}\mathbf E_0^\top\mathbf E_0U.
\end{align*}
By Lemma~\ref{lem:E0bounds}(i)--(ii), the last three terms are $o_p(1)$. Therefore, we have
\begin{align*}
\frac1{n_t}(\hat{\mathbf Z}U)^\top(\hat{\mathbf Z}U)
=R\left(\frac1{n_t}\mathbf F_1^\top\mathbf F_1\right)R^\top+o_p(1)
\overset{p}{\longrightarrow}R\Omega_1R^\top.
\end{align*}
For the lower-right block, Lemma~\ref{1PE1} gives $n_t^{-1}\widetilde{\mathbf E}_S^\top\widetilde{\mathbf E}_S=I_d+o_p(1)$. For the off-diagonal block,
\begin{align*}
\frac1{n_t}(\hat{\mathbf Z}U)^\top\widetilde{\mathbf E}_S=R\frac1{n_t}\mathbf F_1^\top\widetilde{\mathbf E}_S+U^\top\frac1{n_t}\mathbf E_0^\top\widetilde{\mathbf E}_S.
\end{align*}
The first term is $O_p(n_t^{-1/2})$ because the normalized error rows are mean zero, have bounded second moments, and are independent of the true factors. For the second term, if $\tilde e_{S,i}$ denotes the $i$th row of $\widetilde{\mathbf E}_S$ and $e_{0,i}$ the $i$th row of $\mathbf E_0$, then $\mathbb E\|e_{0,i}\|^2=O(1/(p\wedge q))$ and $\mathbb E\|\tilde e_{S,i}\|^2=O(1)$. Hence, by Cauchy--Schwarz, $\|\mathbb E(e_{0,i}\tilde e_{S,i}^\top)\|_{\mathrm{op}}=O((p\wedge q)^{-1/2})=o(1),$ and the centered fluctuation is also $o_p(1)$ since the dimension is fixed. Thus, $(1/{n_t})(\hat{\mathbf Z}U)^\top\widetilde{\mathbf E}_S=o_p(1)$. Now, combining the three blocks,
\begin{align*}
\bar M\overset{p}{\longrightarrow}
\begin{pmatrix}R\Omega_1R^\top&0\\0&I_d\end{pmatrix}.
\end{align*}
The limit is positive definite since $\lambda_{\min}(\Omega_1)$ is bounded away from zero by Assumption~\ref{Assumption:covariances}(iii)--(iv) and $\nu_{\min}(R)\ge c_R>0$ by Lemma~\ref{lem:masterbias}. Therefore, $\|\bar M^{-1}\|_{\mathrm{op}}=O_p(1)$.

\medskip\noindent\textbf{\underline{Step 3: Expansion of the null-space coefficient.}}
We expand the null-space coefficient in the same way as the terms $J_1,\ldots,J_4$ in Lemma~\ref{lemma:approxgamma1}. From Step~2, the least-squares estimator in the normalized coordinates satisfies
\begin{align*}\begin{pmatrix}\hat a\\ \Sigma_S^{1/2}\hat\kappa\end{pmatrix}=\bar M^{-1}\frac1{n_t}\bar{\mathbf Z}^\top \mathbf Y.\end{align*}
Since $\delta\in\operatorname{Col}(U)$, we have $\delta=UU^\top\delta$, so $\bar{\mathbf Z}(\delta^\top U,\ 0^\top)^\top=\hat{\mathbf Z}\delta=\mathbf Z\delta+\mathbf E_0\delta$. Thus, by Equation \eqref{equation:yhat}, we have
\begin{align*}\mathbf Y=\bar{\mathbf Z}\begin{pmatrix}U^\top\delta\\0\end{pmatrix}-\mathbf E_0\delta+\boldsymbol\epsilon_1.\end{align*}
Now, substituting this into the least-squares expression gives
\begin{align*}
\begin{pmatrix}\hat a-U^\top\delta\\ \Sigma_S^{1/2}\hat\kappa\end{pmatrix}&=-\bar M^{-1}\frac1{n_t}\bar{\mathbf Z}^\top\mathbf E_0\delta+\bar M^{-1}\frac1{n_t}\bar{\mathbf Z}^\top\boldsymbol\epsilon_1.
\end{align*}
Since $\bar{\mathbf Z}=\left[\mathbf F_1R^\top+\mathbf E_0U,\,\widetilde{\mathbf E}_S\right]$, we have
\begin{align*}\bar{\mathbf Z}^\top\mathbf E_0\delta=\begin{pmatrix}R\mathbf F_1^\top\mathbf E_0\delta+U^\top\mathbf E_0^\top\mathbf E_0\delta\\\widetilde{\mathbf E}_S^\top\mathbf E_0\delta\end{pmatrix},\quad \bar{\mathbf Z}^\top\boldsymbol\epsilon_1=\begin{pmatrix}R\mathbf F_1^\top\boldsymbol\epsilon_1+U^\top\mathbf E_0^\top\boldsymbol\epsilon_1\\\widetilde{\mathbf E}_S^\top\boldsymbol\epsilon_1\end{pmatrix}.\end{align*}
Let $P_d=\begin{pmatrix}0_{d\times(1+\bar p+\bar q)}&I_d\end{pmatrix}$ select the last $d$ coordinates. Now, taking the null-space coordinates and
multiplying by $\sqrt{n_t}$ gives
\begin{align}
\sqrt{n_t}\Sigma_S^{1/2}\hat\kappa
&=-\underbrace{P_d\bar M^{-1}\begin{pmatrix}R\frac1{\sqrt{n_t}}\mathbf F_1^\top\mathbf E_0\delta\\0\end{pmatrix}}_{K_1}-\underbrace{P_d\bar M^{-1}\begin{pmatrix}\frac1{\sqrt{n_t}}U^\top\mathbf E_0^\top\mathbf E_0\delta\\\frac1{\sqrt{n_t}}\widetilde{\mathbf E}_S^\top\mathbf E_0\delta\end{pmatrix}}_{K_2}\notag\\
&\quad+\underbrace{P_d\bar M^{-1}
\begin{pmatrix}R\frac1{\sqrt{n_t}}\mathbf F_1^\top\boldsymbol\epsilon_1\\0
\end{pmatrix}}_{K_3}+\underbrace{P_d\bar M^{-1}\begin{pmatrix}\frac1{\sqrt{n_t}}U^\top\mathbf E_0^\top\boldsymbol\epsilon_1\\\frac1{\sqrt{n_t}}\widetilde{\mathbf E}_S^\top\boldsymbol\epsilon_1\end{pmatrix}}_{K_4}.
\label{eq:null-K-decomp}
\end{align}

\medskip\noindent\textbf{\underline{Step 4: Bounds on $K_1,\ldots,K_4$.}}

\noindent\textit{Term $K_1$.}
Using $\|\bar M^{-1}\|_{\mathrm{op}}=O_p(1)$, the bounded operator norm of $R$, $\|\delta\|=O_p(1)$, and Lemma~\ref{lem:E0bounds}(ii),
\begin{align*}
\|K_1\|
&\le O_p(1)\left\|\frac1{\sqrt{n_t}}\mathbf F_1^\top\mathbf E_0\delta\right\|\le O_p(1)\left\|\frac1{\sqrt{n_t}}\mathbf F_1^\top\mathbf E_0\right\|_{\mathrm{op}}\|\delta\|=O_p\left(\frac1{\sqrt{p\wedge q}}\right) =o_p(1).\end{align*}

\noindent\textit{Term $K_2$.}
By Lemma~\ref{lem:E0bounds}(i), the first block of $K_2$ satisfies
\begin{align*}
\left\|\frac1{\sqrt{n_t}}U^\top\mathbf E_0^\top\mathbf E_0\delta\right\|
&\le \sqrt{n_t}\left\|\frac1{n_t}\mathbf E_0^\top\mathbf E_0\right\|_{\mathrm{op}}\|\delta\|=O_p\left(\frac{\sqrt{n_t}}{p\wedge q}\right) =o_p(1).\end{align*}
For the second block of $K_2$, we expand the normalized null-space error directly. By definition, $(1/{\sqrt{n_t}})\widetilde{\mathbf E}_S^\top\mathbf E_0\delta=(1/{\sqrt{n_t}})\Sigma_S^{-1/2}S^\top\mathbf E_0^\top\mathbf E_0\delta.$ Since $\mathbf E_0=\left(\mathbf 0_n\quad\mathbf E_X\quad\mathbf E_M \right)$, we have $\mathbf E_0\delta=\mathbf E_X(\mathbf H_X^\dagger)^\top\beta_1+\mathbf E_M(\mathbf H_M^\dagger)^\top\gamma_1.$ Hence, using the block forms of $S$ and $\Sigma_S^{-1/2}$, we have
\begin{align*}\frac1{\sqrt{n_t}}\widetilde{\mathbf E}_S^\top\mathbf E_0\delta
&=
\begin{pmatrix}
\displaystyle \frac{\sqrt p}{\sqrt{n_t}}\Omega_X^{-1/2}S_X^\top\mathbf E_X^\top\mathbf E_X(\mathbf H_X^\dagger)^\top\beta_1+\frac{\sqrt p}{\sqrt{n_t}}\Omega_X^{-1/2}S_X^\top\mathbf E_X^\top\mathbf E_M(\mathbf H_M^\dagger)^\top\gamma_1
\\
\displaystyle
\frac{\sqrt q}{\sqrt{n_t}}\Omega_M^{-1/2}S_M^\top\mathbf E_M^\top\mathbf E_X(\mathbf H_X^\dagger)^\top\beta_1+\frac{\sqrt q}{\sqrt{n_t}}\Omega_M^{-1/2}S_M^\top\mathbf E_M^\top\mathbf E_M(\mathbf H_M^\dagger)^\top\gamma_1
\end{pmatrix}.
\end{align*}
For these four terms, we have
\begin{align*}\mathbb E(\mathbf E_X^\top\mathbf E_X)=\frac{n_t}{p}\mathbf V_X^\top\Sigma_{\mathbf u_X}\mathbf V_X,\quad \mathbb E(\mathbf E_M^\top\mathbf E_M)=\frac{n_t}{q}\mathbf V_M^\top\Sigma_{\mathbf u_M}\mathbf V_M.\end{align*}
Moreover, from Assumption~\ref{Assumption:covariances}(ii), $\mathbb E(\mathbf E_X^\top\mathbf E_M)=0$. Since the projected dimensions are fixed and the proxy errors have uniformly bounded fourth moments by Assumption~\ref{assump:subgaussian}(i), we get
\begin{align*}\left\|\mathbf E_X^\top\mathbf E_X-\mathbb E(\mathbf E_X^\top\mathbf E_X)\right\|_{\mathrm{op}}=O_p\left(\frac{\sqrt{n_t}}{p}\right),\quad \left\|\mathbf E_M^\top\mathbf E_M-\mathbb E(\mathbf E_M^\top\mathbf E_M)\right\|_{\mathrm{op}}=O_p\left(\frac{\sqrt{n_t}}{q}\right),\end{align*}
and the cross-product satisfies $\|\mathbf E_X^\top\mathbf E_M\|_{\mathrm{op}}=O_p\big(\sqrt{n_t/(pq)}\big)$. Consider first the $\mathbf E_X^\top\mathbf E_X$ term. Splitting it into its mean and centered parts,
\begin{align*}
&\frac{\sqrt p}{\sqrt{n_t}}\Omega_X^{-1/2}S_X^\top\mathbf E_X^\top\mathbf E_X(\mathbf H_X^\dagger)^\top\beta_1\\
&\quad=\frac{\sqrt p}{\sqrt{n_t}}\Omega_X^{-1/2}S_X^\top\left\{\mathbf E_X^\top\mathbf E_X-\mathbb E(\mathbf E_X^\top\mathbf E_X)\right\}(\mathbf H_X^\dagger)^\top\beta_1+\frac{\sqrt p}{\sqrt{n_t}}\Omega_X^{-1/2}S_X^\top\mathbb E(\mathbf E_X^\top\mathbf E_X)(\mathbf H_X^\dagger)^\top\beta_1.
\end{align*}
Since $\|\Omega_X^{-1/2}\|_{\mathrm{op}}=O(1)$, $\|S_X\|_{\mathrm{op}}=1$, and $\|\mathbf H_X^\dagger\|_{\mathrm{op}}=O_p(1)$ by Proposition~\ref{prop:loading-cond}, the centered part is $O_p(p^{-1/2})$ and the mean part is $O_p(\sqrt{n_t/p})$. The same argument applies to the $\mathbf E_M^\top\mathbf E_M$ term, so
\begin{align*}\frac{\sqrt p}{\sqrt{n_t}}\Omega_X^{-1/2}S_X^\top\mathbf E_X^\top\mathbf E_X(\mathbf H_X^\dagger)^\top\beta_1=O_p\left(\sqrt{\frac{n_t}{p}}+\frac1{\sqrt p}\right), \quad \frac{\sqrt q}{\sqrt{n_t}}\Omega_M^{-1/2}S_M^\top\mathbf E_M^\top\mathbf E_M(\mathbf H_M^\dagger)^\top\gamma_1=O_p\left(\sqrt{\frac{n_t}{q}}+\frac1{\sqrt q}\right).\end{align*}
For the first cross-product term, we have that
\begin{align*}
\left\|\frac{\sqrt p}{\sqrt{n_t}}\Omega_X^{-1/2}S_X^\top\mathbf E_X^\top\mathbf E_M(\mathbf H_M^\dagger)^\top\gamma_1
\right\|&\le O_p(1)\frac{\sqrt p}{\sqrt{n_t}}
\left\|\mathbf E_X^\top\mathbf E_M\right\|_{\mathrm{op}}=O_p(1)\frac{\sqrt p}{\sqrt{n_t}}
O_p\left(\sqrt{\frac{n_t}{pq}}\right)=O_p(q^{-1/2}),
\end{align*}
and similarly the second cross-product term is $O_p(p^{-1/2})$. Now, combining the four terms gives us
\begin{align*}\left\|\frac1{\sqrt{n_t}}\widetilde{\mathbf E}_S^\top\mathbf E_0\delta\right\|=O_p\left(\sqrt{\frac{n_t}{p}}+\sqrt{\frac{n_t}{q}}+\frac1{\sqrt p}+\frac1{\sqrt q}\right).\end{align*}
Under Assumption~\ref{assump:growth}(ii), this is $o_p(1)$. Therefore, together with the bound for the first block, $K_2=o_p(1)$.

\noindent\textit{Term $K_3$.}
By the same conditional second-moment argument used for $J_3$ in Lemma~\ref{lemma:approxgamma1}, $\|n_t^{-1/2}\mathbf F_1^\top\boldsymbol\epsilon_1\|=O_p(1)$. Since $ \|P_d\|_{\mathrm{op}}\| =1$, $\|\bar M^{-1}\|_{\mathrm{op}}\|=\mathcal{O}_P(1)$ and $\|R\|_{\text{op}}=O(1)$, it follows that $\|K_3\|= \mathcal{O}_P(1)$.

\noindent\textit{Term $K_4$.}
For the first block, Lemma~\ref{lem:E0bounds}(iv) gives $\left\|(1/{\sqrt{n_t}})~U^\top\mathbf E_0^\top\boldsymbol\epsilon_1\right\|=O_p\left({(p\wedge q)}^{-1/2}\right).$
For the second block, using $\widetilde{\mathbf E}_S=\mathbf E_0S\Sigma_S^{-1/2}$, the block forms of $S$ and $\Sigma_S^{-1/2}$, and the forms of $\mathbf E_X$ and $\mathbf E_M$ in Step~1, we get that
\begin{align*}\frac1{\sqrt{n_t}}\widetilde{\mathbf E}_S^\top\boldsymbol\epsilon_1=\begin{pmatrix}\frac1{\sqrt{n_t}}\Omega_X^{-1/2}S_X^\top \mathbf V_X^\top\mathbf U_X^\top\boldsymbol\epsilon_1\\\frac1{\sqrt{n_t}}\Omega_M^{-1/2}S_M^\top \mathbf V_M^\top\mathbf U_M^\top\boldsymbol\epsilon_1\end{pmatrix}.\end{align*}
Using Assumption~\ref{Assumption:covariances}(ii), together with the bounded second moments and the bounded eigenvalues of $\Omega_X^{-1}$ and $\Omega_M^{-1}$, both blocks are $O_p(1)$, so $\|n_t^{-1/2}\widetilde{\mathbf E}_S^\top\boldsymbol\epsilon_1\|=O_p(1)$. Since $\|P_d\|_{\mathrm{op}}=1$ and $\|\bar M^{-1}\|_{\mathrm{op}}=O_p(1)$, it follows that $K_4=O_p(1)$. Now, combining the four terms in Equation\eqref{eq:null-K-decomp}, we have $\sqrt{n_t}\|\Sigma_S^{1/2}\hat\kappa\|=O_p(1)$, and hence, we have $\|\Sigma_S^{1/2}\hat\kappa\|=O_p(n_t^{-1/2})$.

\medskip\noindent\textbf{\underline{Step 5: Rate for $\hat\gamma_{1,\perp}$.}} Since we have $\Sigma_S^{1/2}=\operatorname{diag}(p^{-1/2}\Omega_X^{1/2},q^{-1/2}\Omega_M^{1/2})$, the mediator null-space block satisfies that $\|q^{-1/2}\Omega_M^{1/2}\hat\kappa_M\|=O_p(n_t^{-1/2})$. Since the eigenvalues of $\Omega_M$ are bounded above and away from zero, $\|\hat\kappa_M\|=O_p\left(\sqrt{ q/{n_t}}\right)$. Since $S_M$ has orthonormal columns, we  have
\begin{equation}\label{eq:null-gamma-rate}
\|\hat\gamma_{1,\perp}\|=\|S_M\hat\kappa_M\|= O_p\left(\sqrt{\frac q{n_t}}\right).
\end{equation}
\end{proof}

\subsection{Proof of Theorem \ref{thm:clt}} \label{sec:main}

\begin{proof}
Recall that $\hat\theta_0=\hat\alpha_1+\widehat{\gamma_1^\top\delta_0}$ and $\theta_0=\alpha_1+\gamma_1^\top\delta_0$. Since the second-stage response depends on the first-stage estimator $\hat\gamma_1$, we separate the bridge estimation error from the error due to estimating $\gamma_1$. For one regression fold, $\hat\alpha_1$ and $\hat\gamma_1$ are estimated using the treated units, while $\widehat{\gamma_1^\top\delta_0}$ is estimated using the control units. The whitening matrices are estimated on the other fold and are therefore fixed conditional on the training fold. We write $n_t$ and $n_c$ for the treated and control sample sizes in the regression fold. Since $n_t,n_c\asymp n$, the ratios $n/n_t$ and $n/n_c$ are bounded in probability.

Since $\hat\gamma_1$ is the coefficient on the projected mediator factors, it is a $\tilde q$-dimensional vector, while $\gamma_1$ is recovered through $\mathbf H_M^\top\hat\gamma_1$. We write $\Delta_{\gamma,n}:=\mathbf H_M^\top\hat\gamma_1-\gamma_1$. The population bridge intercept conditional on $\hat\gamma_1$ is $\hat\gamma_1^\top\mathbf H_M\delta_0$. Also, $\hat\gamma_1^\top\mathbf H_M\delta_0-\gamma_1^\top\delta_0=\delta_0^\top(\mathbf H_M^\top\hat\gamma_1-\gamma_1).$ Therefore, we have
\begin{align}
\sqrt n(\hat\theta_0-\theta_0)&=\sqrt n(\hat\alpha_1-\alpha_1)+\sqrt n\big(\widehat{\gamma_1^\top\delta_0}-\gamma_1^\top\delta_0\big)\notag\\
&=\sqrt n(\hat\alpha_1-\alpha_1)+\sqrt n\big(\widehat{\gamma_1^\top\delta_0}-\hat\gamma_1^\top\mathbf H_M\delta_0\big)+\sqrt n(\hat\gamma_1^\top\mathbf H_M\delta_0-\gamma_1^\top\delta_0)\notag\\
&=\sqrt n(\hat\alpha_1-\alpha_1)+\sqrt n\big(\widehat{\gamma_1^\top\delta_0}-\hat\gamma_1^\top\mathbf H_M\delta_0\big)+\sqrt n\,\delta_0^\top(\mathbf H_M^\top\hat\gamma_1-\gamma_1).
\label{eqtheta_decomp_start}
\end{align}
We consider the three terms in Equation \eqref{eqtheta_decomp_start} separately. For the first term,
\begin{align*}\sqrt n(\hat\alpha_1-\alpha_1)=\sqrt{\frac n{n_t}}\left(-G_1-G_2+G_3+G_4\right).\end{align*}
The second term is expanded over the control units on the $\sqrt{n_c}$ scale into the terms $L_1,\ldots,L_6$. For the third term, Lemma~\ref{lemma:approxgamma1} gives $\Delta_{\gamma,n}=O_p(n_t^{-1/2})$, so $\sqrt n\,\Delta_{\gamma,n}^\top\delta_0=O_p(1)$. Write this term as $\sqrt{n/n_c}\,L_7$ where $L_7=\sqrt{n_c}\,\Delta_{\gamma,n}^\top\delta_0$.

\medskip \noindent\textbf{Part 1: Linear Expansions.} We first expand the two estimation errors into their bias and first-order terms.

\medskip\noindent\textit{\underline{Expansion of $\hat\alpha_1$.}}  Under $A=1$, the outcome model in \eqref{eq:factor} can be written as $Y=\mathbf Z\delta+\epsilon_1$. Recall that $\hat\alpha_1$ is the first coordinate of the least-squares coefficient from regressing $Y$ on $\hat{\mathbf Z}$ using the treated units. Let $e_{\mathbf 1}=(1,0,\ldots,0)^\top$. Then, we have that
\begin{align*}
\hat\alpha_1&=e_\mathbf{1}^\top\left(\hat{\mathbf Z}^\top\hat{\mathbf Z}\right)^{-1}\hat{\mathbf Z}^\top Y\\
&=e_\mathbf{1}^\top\left(\frac1{n_t}\hat{\mathbf Z}^\top\hat{\mathbf Z}\right)^{-1}\frac1{n_t}\hat{\mathbf Z}^\top Y\\
&=e_\mathbf{1}^\top\left(\frac1{n_t}\hat{\mathbf Z}^\top\hat{\mathbf Z}\right)^{-1}\frac1{n_t}\hat{\mathbf Z}^\top(\mathbf Z\delta+\epsilon_1)\\
&=e_\mathbf{1}^\top\left(\frac1{n_t}\hat{\mathbf Z}^\top\hat{\mathbf Z}\right)^{-1}\frac1{n_t}\hat{\mathbf Z}^\top(\hat{\mathbf Z}\delta+\mathbf Z\delta-\hat{\mathbf Z}\delta+\epsilon_1)\\
&=e_\mathbf{1}^\top\left(\frac1{n_t}\hat{\mathbf Z}^\top\hat{\mathbf Z}\right)^{-1}\frac1{n_t}\hat{\mathbf Z}^\top\hat{\mathbf Z}\delta+e_\mathbf{1}^\top\left(\frac1{n_t}\hat{\mathbf Z}^\top\hat{\mathbf Z}\right)^{-1}\frac1{n_t}\hat{\mathbf Z}^\top(\mathbf Z-\hat{\mathbf Z})\delta+e_\mathbf{1}^\top\left(\frac1{n_t}\hat{\mathbf Z}^\top\hat{\mathbf Z}\right)^{-1}\frac1{n_t}\hat{\mathbf Z}^\top\epsilon_1.
\end{align*}
The first term is $e_{\mathbf 1}^\top\delta=\alpha_1$. Recall that $\hat{\mathbf Z}=\mathbf Z+\mathbf E_0$, so $\mathbf Z-\hat{\mathbf Z}=-\mathbf E_0$. Now, taking $\alpha_1$ on the left side and multiplying the whole equation by $\sqrt{n_t}$ gives
\begin{equation}\label{eqG3}
\begin{aligned}
\sqrt{n_t}(\hat\alpha_1-\alpha_1)
&=-e_\mathbf{1}^\top\left(\frac1{n_t}\hat{\mathbf Z}^\top\hat{\mathbf Z}\right)^{-1}\frac1{\sqrt{n_t}}\hat{\mathbf Z}^\top\mathbf E_0\delta+e_\mathbf{1}^\top\left(\frac1{n_t}\hat{\mathbf Z}^\top\hat{\mathbf Z}\right)^{-1}\frac1{\sqrt{n_t}}\hat{\mathbf Z}^\top\epsilon_1.
\end{aligned}
\end{equation}
Also, since $\mathbf Z=\mathbf F_1\mathbf H_1^\top$, we have $\hat{\mathbf Z}^\top=\mathbf H_1\mathbf F_1^\top+\mathbf E_0^\top$. Therefore, we have
\begin{align*}
\sqrt{n_t}(\hat\alpha_1-\alpha_1)
&=-\underbrace{e_\mathbf{1}^\top\left(\frac1{n_t}\hat{\mathbf Z}^\top\hat{\mathbf Z}\right)^{-1}\mathbf H_1\frac1{\sqrt{n_t}}\mathbf F_1^\top\mathbf E_0\delta}_{=G_1}
-\underbrace{e_\mathbf{1}^\top\left(\frac1{n_t}\hat{\mathbf Z}^\top\hat{\mathbf Z}\right)^{-1}\frac1{\sqrt{n_t}}\mathbf E_0^\top\mathbf E_0\delta}_{=G_2}\\
&\quad+\underbrace{e_\mathbf{1}^\top\left(\frac1{n_t}\hat{\mathbf Z}^\top\hat{\mathbf Z}\right)^{-1}\mathbf H_1\frac1{\sqrt{n_t}}\mathbf F_1^\top\epsilon_1}_{=G_3}
+\underbrace{e_\mathbf{1}^\top\left(\frac1{n_t}\hat{\mathbf Z}^\top\hat{\mathbf Z}\right)^{-1}\frac1{\sqrt{n_t}}\mathbf E_0^\top\epsilon_1}_{=G_4}.
\end{align*}
Here, $G_3$ is the signal term, while $G_1,G_2$, and $G_4$ involve the estimation error $\mathbf E_0$. In Part~2, we show that $G_1,G_2,G_4=o_p(1)$, while $G_3$ remains.
\\

\noindent{\textit{\underline{Bridge equation for the second stage.}}} By Equation \eqref{eq:estimation}, $\hat{\mathbf f}_M=\mathbf H_M\mathbf f_M+\frac1q\mathbf W_M^\top\mathbf u_M$, and since $\hat\gamma_1^\top\mathbf H_M=\gamma_1^\top+\Delta_{\gamma,n}^\top$, we have
\begin{align}
b=\left(\gamma_1^\top+\Delta_{\gamma,n}^\top\right)\mathbf f_M+\hat\gamma_1^\top\frac1q\mathbf W_M^\top\mathbf u_M.
\label{eq:b-start}
\end{align}
To control the mediator-whitening term, Lemma~\ref{lemma:approxgamma1} controls $\mathbf H_M^\top\hat\gamma_1-\gamma_1$ but not the full vector $\hat\gamma_1$, so we split $\hat\gamma_1$ into its row-space and null-space parts. Let $\mathbf {P}_M=\mathbf H_M(\mathbf H_M^\top\mathbf H_M)^{-1}\mathbf H_M^\top$ be the projection onto $\operatorname{Col}(\mathbf H_M)$, and write
\begin{align*}\hat\gamma_1=\mathbf {P}_M\hat\gamma_1+(I-\mathbf {P}_M)\hat\gamma_1=\hat\gamma_{1,\parallel}+\hat\gamma_{1,\perp}.\end{align*}
Since $\mathbf H_M^\top\hat\gamma_1=\gamma_1+\Delta_{\gamma,n}$, the row-space part is $\hat\gamma_{1,\parallel}=(\mathbf H_M^\dagger)^\top(\gamma_1+\Delta_{\gamma,n})$. Hence, the whitening term splits as
\begin{equation}
\label{eq:whitening-split}
\hat\gamma_1^\top\frac1q\mathbf W_M^\top\mathbf u_M
=\gamma_1^\top\mathbf H_M^\dagger\frac1q\mathbf W_M^\top\mathbf u_M+\Delta_{\gamma,n}^\top\mathbf H_M^\dagger\frac1q\mathbf W_M^\top\mathbf u_M+\hat\gamma_{1,\perp}^\top\frac1q\mathbf W_M^\top\mathbf u_M .
\end{equation}
By Proposition~\ref{prop:loading-cond}, $\|\mathbf H_M^\dagger\|_{\mathrm{op}}=O_p(1)$, while $\frac1q\mathbf W_M^\top\mathbf u_M=O_p(q^{-1/2})$. Therefore, the two row-space terms in \eqref{eq:whitening-split} have orders $\mathcal{O}_P(q^{-1/2})$ and $\mathcal{O}_P((n_t q)^{-1/2})$, respectively.

For the null-space term, Lemma~\ref{lemma:nullspace} gives $\|\hat\gamma_{1,\perp}\|=O_p(\sqrt{q/n_t})$. Let $\mathcal G_t$ denote the sigma-field generated by the training fold and the treated observations used in the first-stage regression. When we condition on $\mathcal G_t$, $\hat\gamma_{1,\perp}$ and $\mathbf W_M$ are fixed, while the control proxy errors $\{\mathbf u_{M,i}:A_i=0\}$ are independent, mean zero, and independent of the treated sample. For a control observation, define $h_i=\hat\gamma_{1,\perp}^\top\frac1q\mathbf W_M^\top\mathbf u_{M,i}$. Using $\|\Sigma_{\mathbf u_M}\|_{\mathrm{op}}\le C$ from Assumption~\ref{Assumption:covariances}(v) and $\mathbf W_M^\top\mathbf W_M=qI_{\tilde q}$, we obtain
\begin{align}\label{eq:null-whitening-var}
\operatorname{Var}\left(h_i\mid\mathcal G_t\right)=\frac1{q^2}
\hat\gamma_{1,\perp}^\top\mathbf W_M^\top\Sigma_{\mathbf u_M}\mathbf W_M\hat\gamma_{1,\perp}\le\frac{C}{q^2}\|\mathbf W_M\hat\gamma_{1,\perp}\|^2=\frac Cq\|\hat\gamma_{1,\perp}\|^2=O_p\left(
\frac1{n_t}\right).
\end{align}
Therefore, using the bounded second moments of the fixed-dimensional control factors and the independence of $\mathbf f_{X1,i}$ from $\mathbf u_{M,i}$,
\begin{align*}
\mathbb E\left[\left\|\frac1{\sqrt{n_c}}\sum_{i:A_i=0}\mathbf f_{X1,i}h_i\right\|^2\;\middle|\;\mathcal G_t\right]
&=\mathbb E\|\mathbf f_{X1,i}\|^2\,\operatorname{Var}\left(h_i\mid\mathcal G_t\right)\le C\,\operatorname{Var}\left(h_i\mid\mathcal G_t\right)=O_p\left(\frac1{n_t}\right),
\end{align*}
Conditional on $\mathcal G_t$, the summands are independent with mean zero, and $h_i$ is independent of $\mathbf f_{X1,i}$. Assumption~\ref{Assumption:covariances}(iii)-(iv), then gives us
\begin{equation}
\label{eq:null-whitening-F}\frac1{\sqrt{n_c}}\sum_{i:A_i=0}\mathbf f_{X1,i}
\hat\gamma_{1,\perp}^\top\frac1q\mathbf W_M^\top\mathbf u_{M,i}=O_p\left(
\frac1{\sqrt{n_t}}\right)=o_p(1).
\end{equation}
Since $\mathbb E\|\mathbf e_{X,i}\|^2=O(1/p)$ and $\mathbf u_{X,i}$ is independent of $\mathbf u_{M,i}$ by Assumption~\ref{Assumption:covariances}(ii), the $\mathbf e_{X,i}h_i$ are mean zero across the control units. Moreover, conditional on $\mathcal G_t$, we have that
\begin{align*}
\mathbb E\left[\|\mathbf e_{X,i}h_i\|^2\;\middle|\;\mathcal G_t\right]&=\mathbb E\|\mathbf e_{X,i}\|^2\,\mathbb E\left[h_i^2\mid\mathcal G_t\right]=O\left(\frac1p\right)O_p\left(\frac1{n_t}\right)=O_p\left(\frac1{pn_t}\right).
\end{align*}
Therefore, we have
\begin{equation}
\label{eq:null-whitening-E}
\frac1{\sqrt{n_c}}\mathbf E_{X0}^\top\left(\frac1q\mathbf U_M\mathbf W_M\hat\gamma_{1,\perp}\right)=O_p\left(\frac1{\sqrt{pn_t}}\right)=o_p(1).
\end{equation}
Now, combining Equations \eqref{eq:null-whitening-F} and \eqref{eq:null-whitening-E} with the two row-space terms in Equation \eqref{eq:whitening-split}, the complete mediator-whitening contribution has order $o_P(1)$ under the normalized contractions used in the second-stage regression. Lastly, substituting the mediator equation $\mathbf f_M=\delta_0+\mathbf B_0\mathbf f_X+\mathbf u_{XM,0}$ into Equation \eqref{eq:b-start}, we obtain
\begin{align}
b&=\gamma_1^\top\delta_0+\Delta_{\gamma,n}^\top\delta_0+\gamma_1^\top\mathbf B_0\mathbf f_X+\gamma_1^\top\mathbf u_{XM,0}+\Delta_{\gamma,n}^\top\mathbf B_0\mathbf f_X+\Delta_{\gamma,n}^\top\mathbf u_{XM,0}+\hat\gamma_1^\top\frac1q\mathbf W_M^\top\mathbf u_M,
\label{eqbridge}
\end{align}
where the last term is the whitening term shown above to be negligible.

\medskip\noindent{\textit{\underline{Expansion of $\widehat{\gamma_1^\top\delta_0}$.}}} With the bridge equation in hand, the expansion is as it was for $\hat\alpha_1$ except that the term $\Delta_{\gamma,n}^\top\mathbf B_0\mathbf f_X$ needs care. Since $\hat{\mathbf f}_{X_i}=\mathbf H_X\mathbf f_{X_i}+\mathbf e_{X_i}$, we have
\begin{align*}
\Delta_{\gamma,n}^\top\mathbf B_0\mathbf f_{X_i}=\tilde c^\top\hat{\mathbf R}_i-\Delta_{\gamma,n}^\top\mathbf B_0\mathbf H_X^\dagger\mathbf e_{X_i},\quad \tilde c=\big(0,\ \Delta_{\gamma,n}^\top\mathbf B_0\mathbf H_X^\dagger\big)^\top .
\end{align*}
Since $\tilde c$ is constant across the control units with zero intercept, the least squares fits $\hat{\mathbf R}\tilde c$ exactly through the slope, and its contribution to the intercept is $e_{\mathbf 1}^\top\tilde c=0$. We therefore omit it from $\mathbf b$, and collect the remaining noise in $\mathcal R_n$, whose $i$th entry is
\begin{align*}
(\mathcal R_n)_i=\Delta_{\gamma,n}^\top\mathbf u_{XM,0,i}-\Delta_{\gamma,n}^\top\mathbf B_0\mathbf H_X^\dagger\mathbf e_{X_i}+\hat\gamma_1^\top\mathbf e_{M_i}.
\end{align*}

We estimate $\gamma_1^\top\delta_0$ by regressing $\mathbf b$ on $\hat{\mathbf R}$ over the control units, using the decomposition of Equation~\eqref{eqbridge},
\begin{align*}
\widehat{\gamma_1^\top\delta_0}
&=e_{\mathbf 1}^\top(\hat{\mathbf R}^\top\hat{\mathbf R})^{-1}\hat{\mathbf R}^\top b\\
&=e_{\mathbf 1}^\top\left(\frac1{n_c}\hat{\mathbf R}^\top\hat{\mathbf R}\right)^{-1}\frac1{n_c}\hat{\mathbf R}^\top b\\
&=e_{\mathbf 1}^\top\left(\frac1{n_c}\hat{\mathbf R}^\top\hat{\mathbf R}\right)^{-1}\frac1{n_c}\hat{\mathbf R}^\top\left(\mathbf R\phi+\Delta_{\gamma,n}^\top\delta_0\cdot\mathbf 1_{n_c}+\boldsymbol\eta+\mathcal R_n\right)\\
&=e_{\mathbf 1}^\top\left(\frac1{n_c}\hat{\mathbf R}^\top\hat{\mathbf R}\right)^{-1}\frac1{n_c}\hat{\mathbf R}^\top\left(\hat{\mathbf R}\phi+\mathbf R\phi-\hat{\mathbf R}\phi+\Delta_{\gamma,n}^\top\delta_0\cdot\mathbf 1_{n_c}+\boldsymbol\eta+\mathcal R_n\right)\\
&=e_{\mathbf 1}^\top\left(\frac1{n_c}\hat{\mathbf R}^\top\hat{\mathbf R}\right)^{-1}\frac1{n_c}\hat{\mathbf R}^\top\hat{\mathbf R}\phi +e_{\mathbf 1}^\top\left(\frac1{n_c}\hat{\mathbf R}^\top\hat{\mathbf R}\right)^{-1}\frac1{n_c}\hat{\mathbf R}^\top(\mathbf R-\hat{\mathbf R})\phi+e_{\mathbf 1}^\top\left(\frac1{n_c}\hat{\mathbf R}^\top\hat{\mathbf R}\right)^{-1}\frac1{n_c}\hat{\mathbf R}^\top\left(\Delta_{\gamma,n}^\top\delta_0\cdot\mathbf 1_{n_c}\right)\\
&\quad +e_{\mathbf 1}^\top\left(\frac1{n_c}\hat{\mathbf R}^\top\hat{\mathbf R}\right)^{-1}\frac1{n_c}\hat{\mathbf R}^\top\boldsymbol\eta+e_{\mathbf 1}^\top\left(\frac1{n_c}\hat{\mathbf R}^\top\hat{\mathbf R}\right)^{-1}\frac1{n_c}\hat{\mathbf R}^\top\mathcal R_n.
\end{align*}
The first term is $\phi_1=\gamma_1^\top\delta_0$, since
$(\frac1{n_c}\hat{\mathbf R}^\top\hat{\mathbf R})^{-1}\frac1{n_c}\hat{\mathbf R}^\top\hat{\mathbf R}=I$.
Since $\mathbf 1_{n_c}=\hat{\mathbf R}e_{\mathbf 1}$, the third term simplifies exactly as $\Delta_{\gamma,n}^\top\delta_0,$
so a constant added to every response of a regression with an intercept column is
absorbed entirely and exactly into the fitted intercept. Hence
\begin{align*}
\widehat{\gamma_1^\top\delta_0}
&=\gamma_1^\top\delta_0+\Delta_{\gamma,n}^\top\delta_0
+e_{\mathbf 1}^\top\left(\frac1{n_c}\hat{\mathbf R}^\top\hat{\mathbf R}\right)^{-1}\frac1{n_c}\hat{\mathbf R}^\top(\mathbf R-\hat{\mathbf R})\phi\\
&\quad +e_{\mathbf 1}^\top\left(\frac1{n_c}\hat{\mathbf R}^\top\hat{\mathbf R}\right)^{-1}\frac1{n_c}\hat{\mathbf R}^\top\boldsymbol\eta+e_{\mathbf 1}^\top\left(\frac1{n_c}\hat{\mathbf R}^\top\hat{\mathbf R}\right)^{-1}\frac1{n_c}\hat{\mathbf R}^\top\mathcal R_n\\
&=\gamma_1^\top\delta_0+\Delta_{\gamma,n}^\top\delta_0
-e_{\mathbf 1}^\top\left(\frac1{n_c}\hat{\mathbf R}^\top\hat{\mathbf R}\right)^{-1}\frac1{n_c}\hat{\mathbf R}^\top\mathbf E_{X0}\phi\\
&\quad +e_{\mathbf 1}^\top\left(\frac1{n_c}\hat{\mathbf R}^\top\hat{\mathbf R}\right)^{-1}\frac1{n_c}\hat{\mathbf R}^\top\boldsymbol\eta+e_{\mathbf 1}^\top\left(\frac1{n_c}\hat{\mathbf R}^\top\hat{\mathbf R}\right)^{-1}\frac1{n_c}\hat{\mathbf R}^\top\mathcal R_n,
\end{align*}
where the last equality uses $\mathbf R-\hat{\mathbf R}=-\mathbf E_{X0}$. Now, substituting $\hat{\mathbf R}^\top=\mathbf R^\top+\mathbf E_{X0}^\top=\mathbf H_{X1}\mathbf F_{X1}^\top+\mathbf E_{X0}^\top$
exactly as for $\hat\alpha_1$, and scaling by $\sqrt{n_c}$,
\begin{align}\label{eqL3}
&\sqrt{n_c}\left(\widehat{\gamma_1^\top\delta_0}-\gamma_1^\top\delta_0\right)\\
&=\sqrt{n_c}\,\Delta_{\gamma,n}^\top\delta_0 -\underbrace{e_{\mathbf 1}^\top\left(\frac1{n_c}\hat{\mathbf R}^\top\hat{\mathbf R}\right)^{-1}\mathbf H_{X1}\frac1{\sqrt{n_c}}\mathbf F_{X1}^\top\mathbf E_{X0}\phi}_{=L_1} -\underbrace{e_{\mathbf 1}^\top\left(\frac1{n_c}\hat{\mathbf R}^\top\hat{\mathbf R}\right)^{-1}\frac1{\sqrt{n_c}}\mathbf E_{X0}^\top\mathbf E_{X0}\phi}_{=L_2}\notag\\
&\quad+\underbrace{e_{\mathbf 1}^\top\left(\frac1{n_c}\hat{\mathbf R}^\top\hat{\mathbf R}\right)^{-1}\mathbf H_{X1}\frac1{\sqrt{n_c}}\mathbf F_{X1}^\top\boldsymbol\eta}_{=L_3} +\underbrace{e_{\mathbf 1}^\top\left(\frac1{n_c}\hat{\mathbf R}^\top\hat{\mathbf R}\right)^{-1}\frac1{\sqrt{n_c}}\mathbf E_{X0}^\top\boldsymbol\eta}_{=L_4}\notag\displaybreak[1]\\
&\quad+\underbrace{e_{\mathbf 1}^\top\left(\frac1{n_c}\hat{\mathbf R}^\top\hat{\mathbf R}\right)^{-1}\mathbf H_{X1}\frac1{\sqrt{n_c}}\mathbf F_{X1}^\top\mathcal R_n}_{=L_5} +\underbrace{e_{\mathbf 1}^\top\left(\frac1{n_c}\hat{\mathbf R}^\top\hat{\mathbf R}\right)^{-1}\frac1{\sqrt{n_c}}\mathbf E_{X0}^\top\mathcal R_n}_{=L_6}.\notag
\end{align}
We write $L_7=\sqrt{n_c}\,\Delta_{\gamma,n}^\top\delta_0=\sqrt{n_c}\,\delta_0^\top(\mathbf H_M^\top\hat\gamma_1-\gamma_1)$,
\begin{align*}\sqrt{n_c}\left(\widehat{\gamma_1^\top\delta_0}-\gamma_1^\top\delta_0\right)=L_7-L_1-L_2+L_3+L_4+L_5+L_6.
\end{align*}
The terms $L_1,\ldots,L_4$ are the analogues of $G_1,\ldots,G_4$ under 
\begin{align*}
\begin{aligned}
&(\mathbf Z,\mathbf H_1,\delta,\epsilon_1,\mathbf F_1,\mathbf E_0,p\wedge q)\mapsto(\mathbf R,\mathbf H_{X1},\phi,\boldsymbol\eta,\mathbf F_{X1},\mathbf E_{X0},p),
\end{aligned}
\end{align*}
 the terms $L_5,L_6$ come from the remainder $\mathcal R_n$, and $L_7$ is the first-order cost of estimating $\gamma_1$.
\\

\noindent\textbf{Part 2: Asymptotic Negligibility of the Bias Terms.}\label{part2}

\medskip\noindent We prove that $G_1,G_2,G_4,L_1,L_2,L_4,L_5,L_6$ are each $o_p(1)$, and that $L_7=\sqrt{n_c/n_t}\delta_0^\top J_3+o_p(1)$. Throughout, $M=\frac1{n_t}\hat{\mathbf Z}^\top\hat{\mathbf Z}$ denotes the treated Gram matrix and $\widetilde M=\frac1{n_c}\hat{\mathbf R}^\top\hat{\mathbf R}$ the control Gram matrix, and every operator-norm bound is conditional on the training fold on which the whitening matrices are fixed, so that the rows of $\mathbf E_0$ and $\mathbf E_{X0}$ are independent within each stratum. Since $n_k^t\asymp n_k^c\asymp n_k\asymp n$, every rate written in $n_k^t$ or $n_k^c$ is order equivalent to the corresponding rate in $n$.
Here, the bias terms fall into two structural patterns. The two-sided terms $G_1$ and $L_1$ carry a loading matrix on both sides of the Gram inverse, while the one-sided terms $G_2,G_4,L_2$, and $L_4$ carry a loading matrix only on the left. We record the two block inverse identities used below.

\medskip\noindent\textbf{(BI-1) Two-sided identity.} Since $J=[U\ S]$ is orthogonal, $M^{-1}=J(J^\top MJ)^{-1}J^\top$, and from $J^\top\mathbf H_1=\binom{R}{0}$ and $\mathbf H_1^\top J=(R^\top\ 0)$,
\begin{equation}
\begin{aligned}\label{eq:BI1}
\mathbf H_1^\top M^{-1}\mathbf H_1
&=\big(R^\top\ \ 0\big)(J^\top MJ)^{-1}\binom{R}{0}\\&=R^\top\big[(J^\top MJ)^{-1}\big]_{11}R\\
&=R^\top(D+E_n)^{-1}R,
\end{aligned}
\end{equation}
where $(D+E_n)^{-1}$ is the upper-left block of the inverse from the Schur complement of $D+E_n$. With $\|R\|_{\text{op}}=O(1)$ and $\|(D+E_n)^{-1}\|_{\text{op}}=O_p(1)$ from
Lemma~\ref{lem:masterbias},
\begin{equation}\label{eq:BI1-bound}
\big\|\mathbf H_1^\top M^{-1}\mathbf H_1\big\|_{\text{op}} \le\|R\|_{\text{op}}^2\,\|(D+E_n)^{-1}\|_{\text{op}}=O_p(1).
\end{equation}

\medskip
\noindent\textbf{(BI-2) One-sided identity.}
For the one-sided pattern we need the full first block-row of $(J^\top MJ)^{-1}$. We write the partitioned matrix as follows:
\begin{align*}J^\top MJ=\begin{pmatrix}K_{11}&K_{12}\\ K_{21}&K_{22}\end{pmatrix},
\end{align*}
the standard block-inverse formula gives, for the first block-row,
\begin{align*}
\begin{aligned}
\big[(J^\top MJ)^{-1}\big]_{1*} &=\Big((D+E_n)^{-1},\ -(D+E_n)^{-1}K_{12}K_{22}^{-1}\Big),
\end{aligned}
\end{align*}
where $D+E_n=K_{11}-K_{12}K_{22}^{-1}K_{21}$. Therefore, using $\mathbf H_1^\top J=(R^\top\ 0)$,
\begin{equation}\label{eq:BI2}
\begin{aligned}
\mathbf H_1^\top M^{-1}
&=\big(R^\top\ \ 0\big)(J^\top MJ)^{-1}J^\top\\&=R^\top(D+E_n)^{-1}\big(I,\ -K_{12}K_{22}^{-1}\big)J^\top\\&=R^\top(D+E_n)^{-1}\big[U^\top-K_{12}K_{22}^{-1}S^\top\big].
\end{aligned}
\end{equation}
Now, substituting the blocks $K_{12}=\frac1{n_t}(R\mathbf F_1^\top+U^\top\mathbf E_0^\top)\mathbf E_0S$ and $K_{22}=\frac1{n_t}S^\top\mathbf E_0^\top\mathbf E_0S$, we get that
\begin{equation}
\begin{aligned}
\label{eq:BI2-explicit}
\mathbf H_1^\top M^{-1} &=R^\top(D+E_n)^{-1}\Big[U^\top-\Big(\tfrac1{n_t}R\mathbf F_1^\top\mathbf E_0S +\tfrac1{n_t}U^\top\mathbf E_0^\top\mathbf E_0S\Big)\Big(\tfrac1{n_t}S^\top\mathbf E_0^\top\mathbf E_0S\Big)^{-1}S^\top\Big].
\end{aligned}
\end{equation}
The two $S$-dependent factors combine into the projection $P_S$, so for any conformable vector $w$,
\begin{equation}
\begin{aligned}
&\label{eq:PS-fold}
\Big(\tfrac1{n_t}R\mathbf F_1^\top\mathbf E_0S+\tfrac1{n_t}U^\top\mathbf E_0^\top\mathbf E_0S\Big)\Big(\tfrac1{n_t}S^\top\mathbf E_0^\top\mathbf E_0S\Big)^{-1}\tfrac1{\sqrt{n_t}}S^\top\mathbf E_0^\top w=\tfrac1{\sqrt{n_t}}\big(R\mathbf F_1^\top+U^\top\mathbf E_0^\top\big)P_Sw,
\end{aligned}
\end{equation}
because $\mathbf E_0S(S^\top\mathbf E_0^\top\mathbf E_0S)^{-1}S^\top\mathbf E_0^\top=P_S$ and the factors of $n_t$ cancel. Both identities have exact analogues on the control side, with $(\mathbf H_{X1},\widetilde U,\widetilde R,\widetilde S,\widetilde M,\mathbf F_{X1},\mathbf E_{X0},\widetilde P_S,\widetilde D+\widetilde E_n,n_c)$ in place of $(\mathbf H_1,U,R,S,M,\mathbf F_1,\mathbf E_0,P_S,D+E_n,n_t)$. 
\\

\noindent\textit{Terms from $\hat\alpha_1$ (treated units).}\quad
\\

\noindent\textbf{\underline{$G_1=o_p(1)$.}}\quad
Let $\mathbf v=\tfrac1{\sqrt{n_t}}\mathbf F_1^\top\mathbf E_0\delta$. Since the first column of $\mathbf H_1=\operatorname{diag}(1,\mathbf H_X,\mathbf H_M)$ is $e_{\mathbf 1}$, we have $e_{\mathbf 1}\in\operatorname{Col}(\mathbf H_1)=\operatorname{Col}(U)$. Thus, $e_{\mathbf 1}^\top S=0$. Writing $a^\top=e_{\mathbf 1}^\top U$ and using $M^{-1}=J(J^\top MJ)^{-1}J^\top$, we obtain
\begin{align*}
G_1&=e_{\mathbf 1}^\top M^{-1}\mathbf H_1\mathbf v=(a^\top,0)(J^\top MJ)^{-1}\binom{R}{0}\mathbf v=a^\top(D+E_n)^{-1}R\mathbf v.
\end{align*}
It remains to bound $\|\mathbf v\|$. By Lemma~\ref{lem:E0bounds}(ii) and $\|\delta\|=O_p(1)$, we have
\begin{align*}
\|\mathbf v\|&=\Big\|\tfrac1{\sqrt{n_t}}\mathbf F_1^\top\mathbf E_0\delta\Big\|
\le\Big\|\tfrac1{\sqrt{n_t}}\mathbf F_1^\top\mathbf E_0\Big\|_{\mathrm{op}}\|\delta\|=O_p\big((p\wedge q)^{-1/2}\big).
\end{align*}
Now, combining this with $\|a\|\le1$, $\|(D+E_n)^{-1}\|_{\mathrm{op}}=O_p(1)$, and $\|R\|_{\mathrm{op}}=O(1)$, we have
\begin{align*}
|G_1|&\le\|a\|\,\|(D+E_n)^{-1}\|_{\mathrm{op}}\|R\|_{\mathrm{op}}\|\mathbf v\|=O_p\big((p\wedge q)^{-1/2}\big)=o_p(1),
\end{align*}
as $p\wedge q\to\infty$.

\medskip\noindent\textbf{\underline{$G_2=o_p(1)$.}}\quad
By definition, $G_2=e_{\mathbf 1}^\top M^{-1}\tfrac1{\sqrt{n_t}}\mathbf E_0^\top\mathbf E_0\delta$. Using the block-inverse identity and \eqref{eq:PS-fold} with $w=\mathbf E_0\delta$, together with $e_{\mathbf 1}^\top S=0$, we obtain
\begin{align}
G_2&=a^\top(D+E_n)^{-1}\Big[\underbrace{\tfrac1{\sqrt{n_t}}U^\top\mathbf E_0^\top(I-P_S)\mathbf E_0\delta}_{=:G_{2,1}}-\underbrace{\tfrac1{\sqrt{n_t}}R\mathbf F_1^\top P_S\mathbf E_0\delta}_{=:G_{2,2}}\Big].
\label{eq:G2-split}
\end{align}
Since $\|a\|\le1$ and $\|(D+E_n)^{-1}\|_{\mathrm{op}}=O_p(1)$, it suffices to bound $\|G_{2,1}\|$ and $\|G_{2,2}\|$. For $G_{2,1}$, using $\|U\|_{\mathrm{op}}=1$, $\|I-P_S\|_{\mathrm{op}}\le1$, and Lemma~\ref{lem:E0bounds}(i), (iii), we have
\begin{align}
\|G_{2,1}\|&\le\tfrac1{\sqrt{n_t}}\big\|\mathbf E_0^\top(I-P_S)\mathbf E_0\big\|_{\mathrm{op}}\|\delta\|\le\sqrt{n_t}\Big\|\tfrac1{n_t}\mathbf E_0^\top\mathbf E_0\Big\|_{\mathrm{op}}\|\delta\|=O_p\Big(\tfrac{\sqrt{n_t}}{p\wedge q}\Big).
\label{eq:G21}
\end{align}
For $G_{2,2}$, we use $\|R\|_{\mathrm{op}}=O(1)$, $\|P_S\|_{\mathrm{op}}\le1$, and $\|\tfrac1{\sqrt{n_t}}\mathbf F_1^\top\|_{\mathrm{op}}=O_p(1)$ by Assumption~\ref{Assumption:covariances}(iii)--(iv). Moreover,
\begin{align*}
\|\mathbf E_0\delta\|&\le\|\mathbf E_0^\top\mathbf E_0\|_{\mathrm{op}}^{1/2}\|\delta\|=\sqrt{n_t}\Big\|\tfrac1{n_t}\mathbf E_0^\top\mathbf E_0\Big\|_{\mathrm{op}}^{1/2}\|\delta\|=O_p\Big(\sqrt{\tfrac{n_t}{p\wedge q}}\Big).
\end{align*}
Therefore, we have
\begin{align}
\|G_{2,2}\|&\le\|R\|_{\mathrm{op}}\Big\|\tfrac1{\sqrt{n_t}}\mathbf F_1^\top\Big\|_{\mathrm{op}}\|P_S\|_{\mathrm{op}}\|\mathbf E_0\delta\|=O_p\Big(\sqrt{\tfrac{n_t}{p\wedge q}}\Big).
\label{eq:G22}
\end{align}
Now, combining Equations \eqref{eq:G21} and \eqref{eq:G22}, the second rate dominates as $p\wedge q\to\infty$. Hence, we get
\begin{align*}
|G_2|&=O_p\Big(\tfrac{\sqrt{n_t}}{p\wedge q}+\sqrt{\tfrac{n_t}{p\wedge q}}\Big)=O_p\Big(\sqrt{\tfrac{n_t}{p\wedge q}}\Big)=o_p(1),
\end{align*}
since $n_t/(p\wedge q)\to0$.

\medskip\noindent\textbf{\underline{$G_4=o_p(1)$.}}\quad
By definition, we have $G_4=e_{\mathbf 1}^\top\mathbf H_1^\top M^{-1}\tfrac1{\sqrt{n_t}}\mathbf E_0^\top\epsilon_1$. Applying the one-sided identity \eqref{eq:BI2-explicit} and using Equation \eqref{eq:PS-fold} with $w=\epsilon_1$, we obtain
\begin{align}
G_4&=e_{\mathbf 1}^\top R^\top(D+E_n)^{-1}\Big[\underbrace{\tfrac1{\sqrt{n_t}}U^\top\mathbf E_0^\top(I-P_S)\epsilon_1}_{=:G_{4,1}}-\underbrace{\tfrac1{\sqrt{n_t}}R\mathbf F_1^\top P_S\epsilon_1}_{=:G_{4,2}}\Big].
\label{eq:G4-split}
\end{align}
The outside factor is $O_p(1)$ as shown before. By Assumption~\ref{Assumption:covariances}(i)--(ii), conditional on the training fold, $\epsilon_1$ is independent of $(\mathbf F_1,\mathbf E_0)$ and has mean zero. Moreover, Assumption~\ref{assump:subgaussian}(i) gives $\mathbb E(\epsilon_1\epsilon_1^\top\mid\mathbf F_1,\mathbf E_0)\preceq CI_{n_t}$, where the training fold is held fixed throughout.

For $G_{4,1}$, since $I-P_S$ is an orthogonal projection, we have
\begin{align}
\mathbb E\left[\|G_{4,1}\|^2\mid\mathbf F_1,\mathbf E_0\right]
\le\tfrac C{n_t}\operatorname{tr}\left(U^\top\mathbf E_0^\top(I-P_S)\mathbf E_0U\right)\le\tfrac C{n_t}\operatorname{tr}\left(U^\top\mathbf E_0^\top\mathbf E_0U\right)=O_p\left(\frac1{p\wedge q}\right),
\label{eq:G41}
\end{align}
where the last bound follows from Lemma~\ref{lem:E0bounds}(i) and the fixed projected dimension. Therefore, $\|G_{4,1}\|=O_p((p\wedge q)^{-1/2})$. For $G_{4,2}$, we first decompose $\mathbf F_1=\mathbf 1\mu_{\mathbf F}^\top+\widetilde{\mathbf F}_1$ into mean and centered parts, so
\begin{align*}
\tfrac1{\sqrt{n_t}}R\mathbf F_1^\top P_S\epsilon_1=\tfrac1{\sqrt{n_t}}R\mu_{\mathbf F}\mathbf 1^\top P_S\epsilon_1+\tfrac1{\sqrt{n_t}}R\widetilde{\mathbf F}_1^\top P_S\epsilon_1.
\end{align*}
For the mean part, using the conditional mean-zero property and bounded variance of $\epsilon_1$, we have
\begin{align}
\mathbb E\left[(\mathbf 1^\top P_S\epsilon_1)^2\mid\mathbf F_1,\mathbf E_0\right]
&\le C\mathbf 1^\top P_S^2\mathbf 1=C\mathbf 1^\top P_S\mathbf 1=O_p(1),
\label{eq:meanpart-var}
\end{align}
where the last equality follows from Lemma~\ref{1PE1} and $P_S^2=P_S$. Hence, $\mathbf 1^\top P_S\epsilon_1=O_p(1)$. Since $\|R\|_{\mathrm{op}}=O(1)$ and $\|\mu_{\mathbf F}\|_2=O(1)$ by Assumption~\ref{Assumption:covariances}(iii),
\begin{align}
\Big\|\tfrac1{\sqrt{n_t}}R\mu_{\mathbf F}\mathbf 1^\top P_S\epsilon_1\Big\|
&\le\tfrac1{\sqrt{n_t}}\|R\|_{\mathrm{op}}\|\mu_{\mathbf F}\|_2|\mathbf 1^\top P_S\epsilon_1|=O_p(n_t^{-1/2}).
\label{eq:G42-mean}
\end{align}
For the centered part, conditional on $(\widetilde{\mathbf F}_1,\mathbf E_0)$, we have
\begin{align}
&\mathbb E\left[\Big\|\tfrac1{\sqrt{n_t}}R\widetilde{\mathbf F}_1^\top P_S\epsilon_1\Big\|^2\;\middle|\;\widetilde{\mathbf F}_1,\mathbf E_0\right]=\tfrac1{n_t}\operatorname{tr}\left(R\widetilde{\mathbf F}_1^\top P_S\mathbb E(\epsilon_1\epsilon_1^\top\mid\widetilde{\mathbf F}_1,\mathbf E_0)P_S\widetilde{\mathbf F}_1R^\top\right)\le\tfrac C{n_t}\|R\|_{\mathrm{op}}^2\operatorname{tr}\left(\widetilde{\mathbf F}_1^\top P_S\widetilde{\mathbf F}_1\right)=O_p(n_t^{-1}),
\label{eq:G42-centered}
\end{align}
where $\operatorname{tr}(\widetilde{\mathbf F}_1^\top P_S\widetilde{\mathbf F}_1)=O_p(1)$ by Lemma~\ref{lem:tracebound}. Thus, $\|\tfrac1{\sqrt{n_t}}R\widetilde{\mathbf F}_1^\top P_S\epsilon_1\|=O_p(n_t^{-1/2})$. Now, combining the mean and centered parts, $\|G_{4,2}\|=O_p(n_t^{-1/2})$. Together with Equation \eqref{eq:G41} and the $O_p(1)$ outside factor, we obtain
\begin{align*}
|G_4|=O_p\left(\frac1{\sqrt{p\wedge q}}+\frac1{\sqrt{n_t}}\right)=o_p(1),
\end{align*}
as $p\wedge q\to\infty$ and $n_t\to\infty$.

\medskip\noindent\textit{Terms from $\widehat{\gamma_1^\top\delta_0}$ (control units).}

\medskip\noindent Before bounding $L_1,L_2,L_4$, we first obtain the control-side versions of the results used for $G_1,G_2,G_4$ under the substitution of $(\mathbf F_1,\mathbf E_0,S,M,P_S,p\wedge q)$ to $(\mathbf F_{X1},\mathbf E_{X0},\widetilde S,\widetilde M,\widetilde P_S,p),$ where $p$ replaces $p\wedge q$ because the second-stage regression involves only the $X$-side error block $\mathbf E_X$. Repeating the proof of Lemma~\ref{lem:masterbias} gives the control Schur complement $\widetilde D+\widetilde E_n$ of $\widetilde J^\top\widetilde M\widetilde J$, where $\widetilde D=n_c^{-1}\widetilde R\mathbf F_{X1}^\top\mathbf F_{X1}\widetilde R^\top$. As before, $\lambda_{\min}(\widetilde D)\ge c\,c_{\widetilde R}^2>0$ by Assumption~\ref{Assumption:covariances}(iii)--(iv). The four-piece bound in Equation \eqref{eqEn_expanded} then gives
\begin{align}
\|\widetilde E_n\|
&=O_p\Big(\tfrac{1+\|\mu_{\mathbf F_X}\|_2+\|\mu_{\mathbf F_X}\|_2^2}{n_c}
+\tfrac1{\sqrt{n_cp}}+\tfrac1p\Big)
\implies \|(\widetilde D+\widetilde E_n)^{-1}\|_{\mathrm{op}}=O_p(1).
\label{eq:Etil-rate}
\end{align}
The block-inverse identities \eqref{eq:BI1}--\eqref{eq:PS-fold} also hold under this substitution. Similarly, Lemma~\ref{1PE1} applies to $\widetilde P_S$, and the rates in Lemma~\ref{lem:E0bounds} hold with $p$ in place of $p\wedge q$. Finally, $\|\phi\|=O_p(1)$ since $\phi=(\gamma_1^\top\delta_0,\ \gamma_1^\top\mathbf B_0\mathbf H_X^\dagger)^\top$, where the structural parameters are bounded by Assumption~\ref{Assumption:pervasiveness}(i) and $\|\mathbf H_X^\dagger\|_{\mathrm{op}}=O_p(1)$ by Proposition~\ref{prop:loading-cond}.

\medskip\noindent\textbf{\underline{$L_1=o_p(1)$.}}\quad
Let $\widetilde{\mathbf v}=\tfrac1{\sqrt{n_c}}\mathbf F_{X1}^\top\mathbf E_{X0}\phi$, so that $L_1=e_{\mathbf 1}^\top\widetilde M^{-1}\mathbf H_{X1}\widetilde{\mathbf v}$. Since the first column of $\mathbf H_{X1}=\operatorname{diag}(1,\mathbf H_X)$ is $e_{\mathbf 1}$, we have $e_{\mathbf 1}\in\operatorname{Col}(\mathbf H_{X1})=\operatorname{Col}(\widetilde U)$. Thus, $e_{\mathbf 1}^\top\widetilde S=0$. Writing $\widetilde a^\top=e_{\mathbf 1}^\top\widetilde U$ and using the control-side block-inverse identity, we obtain
\begin{align*}
L_1&=(\widetilde a^\top,0)(\widetilde J^\top\widetilde M\widetilde J)^{-1}\binom{\widetilde R}{0}\widetilde{\mathbf v}=\widetilde a^\top(\widetilde D+\widetilde E_n)^{-1}\widetilde R\,\widetilde{\mathbf v}.
\end{align*}
By the $X$-only version of Lemma~\ref{lem:E0bounds}(ii) and $\|\phi\|=O_p(1)$, we have
\begin{align*}
\|\widetilde{\mathbf v}\|
&\le\Big\|\tfrac1{\sqrt{n_c}}\mathbf F_{X1}^\top\mathbf E_{X0}\Big\|_{\mathrm{op}}\|\phi\|=O_p\Big(\frac1{\sqrt p}\Big).
\end{align*}
Now, combining this with $\|\widetilde a\|\le1$, $\|(\widetilde D+\widetilde E_n)^{-1}\|_{\mathrm{op}}=O_p(1)$, and $\|\widetilde R\|_{\mathrm{op}}=O(1)$, we have as $p\to\infty$,
\begin{align*}
|L_1|&\le\|\widetilde a\|\,\|(\widetilde D+\widetilde E_n)^{-1}\|_{\mathrm{op}}\,\|\widetilde R\|_{\mathrm{op}}\,\|\widetilde{\mathbf v}\|=O_p\Big(\frac1{\sqrt p}\Big)=o_p(1),
\end{align*}

\noindent\textbf{\underline{$L_2=o_p(1)$.}}\quad
By definition, $L_2=e_{\mathbf 1}^\top\mathbf H_{X1}^\top\widetilde M^{-1}\tfrac1{\sqrt{n_c}}\mathbf E_{X0}^\top\mathbf E_{X0}\phi$. This has the same structure as $G_2$, with $\mathbf H_{X1}$ on the left only. Applying the control analogue of Equation \eqref{eq:PS-fold} with $w=\mathbf E_{X0}\phi$, we obtain
\begin{align*}
L_2&=e_{\mathbf 1}^\top\widetilde R^\top(\widetilde D+\widetilde E_n)^{-1}\Big[\underbrace{\tfrac1{\sqrt{n_c}}\widetilde U^\top\mathbf E_{X0}^\top(I-\widetilde P_S)\mathbf E_{X0}\phi}_{=:L_{2,1}}-\underbrace{\tfrac1{\sqrt{n_c}}\widetilde R\mathbf F_{X1}^\top\widetilde P_S\mathbf E_{X0}\phi}_{=:L_{2,2}}\Big].
\end{align*}
The outside factor is $O_p(1)$, so it suffices to bound $\|L_{2,1}\|$ and $\|L_{2,2}\|$. By the $X$-only version of Lemma~\ref{lem:E0bounds}(iii), $\|n_c^{-1}\mathbf E_{X0}^\top\mathbf E_{X0}\|_{\mathrm{op}}=O_p(p^{-1})$. For $L_{2,1}$, using $\|\widetilde U\|_{\mathrm{op}}=1$, $\|I-\widetilde P_S\|_{\mathrm{op}}\le1$, and $\|\phi\|=O_p(1)$, we have
\begin{align*}
\|L_{2,1}\|&\le\tfrac1{\sqrt{n_c}}\|\mathbf E_{X0}^\top(I-\widetilde P_S)\mathbf E_{X0}\|_{\mathrm{op}}\|\phi\|
\le\sqrt{n_c}\Big\|\tfrac1{n_c}\mathbf E_{X0}^\top\mathbf E_{X0}\Big\|_{\mathrm{op}}\|\phi\|
=O_p\Big(\tfrac{\sqrt{n_c}}p\Big),
\end{align*}
exactly as in Equation \eqref{eq:G21}. For $L_{2,2}$, we use $\|\widetilde R\|_{\mathrm{op}}=O(1)$, $\|\tfrac1{\sqrt{n_c}}\mathbf F_{X1}^\top\|_{\mathrm{op}}=O_p(1)$, and $\|\widetilde P_S\|_{\mathrm{op}}\le1$. Moreover, we have
\begin{align*}
\|\mathbf E_{X0}\phi\|&\le\|\mathbf E_{X0}^\top\mathbf E_{X0}\|_{\mathrm{op}}^{1/2}\|\phi\|=\sqrt{n_c}\Big\|\tfrac1{n_c}\mathbf E_{X0}^\top\mathbf E_{X0}\Big\|_{\mathrm{op}}^{1/2}\|\phi\|=O_p\Big(\sqrt{\tfrac{n_c}{p}}\Big).
\end{align*}
Therefore,
\begin{align*}
\|L_{2,2}\|&\le\|\widetilde R\|_{\mathrm{op}}\Big\|\tfrac1{\sqrt{n_c}}\mathbf F_{X1}^\top\Big\|_{\mathrm{op}}\|\widetilde P_S\|_{\mathrm{op}}\|\mathbf E_{X0}\phi\|=O_p\Big(\sqrt{\tfrac{n_c}{p}}\Big).
\end{align*}
Now, combining the two bounds, the second rate dominates as $p\to\infty$. Hence, we have
\begin{align*}
|L_2|&=O_p\Big(\tfrac{\sqrt{n_c}}p+\sqrt{\tfrac{n_c}{p}}\Big)=O_p\Big(\sqrt{\tfrac{n_c}{p}}\Big)=o_p(1),
\end{align*}
since $n_c/p\to0$.

\medskip\noindent\textbf{\underline{$L_4=o_p(1)$.}}\quad
By definition, $L_4=e_{\mathbf 1}^\top\mathbf H_{X1}^\top\widetilde M^{-1}\tfrac1{\sqrt{n_c}}\mathbf E_{X0}^\top\boldsymbol\eta$, where $\eta_i=\gamma_1^\top\mathbf u_{XM,0,i}$ for the control units. This has the same structure as $G_4$, with $\boldsymbol\eta$ in place of $\epsilon_1$. Applying the control one-sided identity and Equation \eqref{eq:PS-fold} with $w=\boldsymbol\eta$, we obtain
\begin{align*}
L_4&=e_{\mathbf 1}^\top\widetilde R^\top(\widetilde D+\widetilde E_n)^{-1}\Big[\underbrace{\tfrac1{\sqrt{n_c}}\widetilde U^\top\mathbf E_{X0}^\top(I-\widetilde P_S)\boldsymbol\eta}_{=:L_{4,1}}-\underbrace{\tfrac1{\sqrt{n_c}}\widetilde R\mathbf F_{X1}^\top\widetilde P_S\boldsymbol\eta}_{=:L_{4,2}}\Big].
\end{align*}
The outside factor is $O_p(1)$, so it suffices to bound $\|L_{4,1}\|$ and $\|L_{4,2}\|$. By Assumption~\ref{Assumption:covariances}(i)--(ii), $\boldsymbol\eta$ is independent of $(\mathbf F_{X1},\mathbf E_{X0})$ and has mean zero, conditional on the training fold. Moreover, Assumptions~\ref{Assumption:covariances}(v) and~\ref{Assumption:pervasiveness}(i) give $\operatorname{Var}(\eta_i)=\gamma_1^\top\Sigma_{\mathbf u_{XM,0}}\gamma_1\le C$.

For $L_{4,1}$, conditional on $\mathbf E_{X0}$, and using that $I-\widetilde P_S$ is an orthogonal projection, we have
\begin{align*}
\mathbb E\left[\|L_{4,1}\|^2\mid\mathbf E_{X0}\right]&\le\tfrac C{n_c}\operatorname{tr}\left(\widetilde U^\top\mathbf E_{X0}^\top(I-\widetilde P_S)\mathbf E_{X0}\widetilde U\right)\le\tfrac C{n_c}\|\mathbf E_{X0}\|_{\mathrm{op}}^2=O_p(p^{-1}),
\end{align*}
where the last bound follows from the $X$-only version of Lemma~\ref{lem:E0bounds}(iii), since the projected dimension is fixed. Thus, $\|L_{4,1}\|=O_p(p^{-1/2})$.

For $L_{4,2}$, we split $\mathbf F_{X1}=\mathbf 1\mu_{\mathbf F_X}^\top+\widetilde{\mathbf F}_{X1}$ into mean and centered parts:
\begin{align*}
\tfrac1{\sqrt{n_c}}\widetilde R\mathbf F_{X1}^\top\widetilde P_S\boldsymbol\eta=\tfrac1{\sqrt{n_c}}\widetilde R\mu_{\mathbf F_X}\mathbf 1^\top\widetilde P_S\boldsymbol\eta+\tfrac1{\sqrt{n_c}}\widetilde R\widetilde{\mathbf F}_{X1}^\top\widetilde P_S\boldsymbol\eta.
\end{align*}
For the mean part, the conditional mean-zero property and bounded variance of $\boldsymbol\eta$ give, as in Equation \eqref{eq:meanpart-var},
\begin{align*}
\mathbb E\left[(\mathbf 1^\top\widetilde P_S\boldsymbol\eta)^2\mid\widetilde P_S\right]&\le C\mathbf 1^\top\widetilde P_S^2\mathbf 1=C\mathbf 1^\top\widetilde P_S\mathbf 1=O_p(1),
\end{align*}
where the last equality follows from Lemma~\ref{1PE1}. Hence, $\mathbf 1^\top\widetilde P_S\boldsymbol\eta=O_p(1)$. Since $\|\widetilde R\|_{\mathrm{op}}=O(1)$ and $\|\mu_{\mathbf F_X}\|_2=O(1)$ by Assumption~\ref{Assumption:covariances}(iii), we have
\begin{align*}
\Big\|\tfrac1{\sqrt{n_c}}\widetilde R\mu_{\mathbf F_X}\mathbf 1^\top\widetilde P_S\boldsymbol\eta\Big\|&\le\tfrac1{\sqrt{n_c}}\|\widetilde R\|_{\mathrm{op}}\|\mu_{\mathbf F_X}\|_2|\mathbf 1^\top\widetilde P_S\boldsymbol\eta|
=O_p\left(\frac1{\sqrt{n_c}}\right).
\end{align*}
For the centered part, conditional on $(\widetilde{\mathbf F}_{X1},\mathbf E_{X0})$, the same argument as in Equation \eqref{eq:G42-centered} gives us that
\begin{align*}
\mathbb E\left[\Big\|\tfrac1{\sqrt{n_c}}\widetilde R\widetilde{\mathbf F}_{X1}^\top\widetilde P_S\boldsymbol\eta\Big\|^2\;\middle|\;\widetilde{\mathbf F}_{X1},\mathbf E_{X0}\right]\le\tfrac C{n_c}\|\widetilde R\|_{\mathrm{op}}^2\operatorname{tr}\left(\widetilde{\mathbf F}_{X1}^\top\widetilde P_S\widetilde{\mathbf F}_{X1}\right)=O_p\left(\frac1{n_c}\right),
\end{align*}
where $\operatorname{tr}(\widetilde{\mathbf F}_{X1}^\top\widetilde P_S\widetilde{\mathbf F}_{X1})=O_p(1)$ by Lemma~\ref{lem:tracebound}. Thus, the centered part is also $O_p(n_c^{-1/2})$. Now, combining the mean and centered parts, $\|L_{4,2}\|=O_p(n_c^{-1/2})$. Together with the bound for $L_{4,1}$ and the $O_p(1)$ outside factor, we obtain
\begin{align*}
|L_4|=O_p\left(p^{-1/2}+n_c^{-1/2}\right)=o_p(1).
\end{align*}

\noindent\textbf{\underline{$L_5=o_p(1)$.}}\quad
We have
\begin{align*}
L_5=e_{\mathbf 1}^\top\mathbf H_{X1}^\top\widetilde M^{-1}\mathbf H_{X1}\,\tfrac1{\sqrt{n_c}}\mathbf F_{X1}^\top\mathcal R_n.
\end{align*}
By the control analogue of \eqref{eq:BI1-bound}, the left factor has operator norm $O_p(1)$, so it suffices to bound $\tfrac1{\sqrt{n_c}}\mathbf F_{X1}^\top\mathcal R_n$. We first split the remainder into its covariate and mediator parts, we get
\begin{align}\label{eq:rm}
\mathcal R_n=\mathbf E_{X0}\mathbf a_n+\mathcal R_n^M,\qquad
\mathbf a_n=(0 \quad (-(\mathbf H_X^\dagger)^\top\mathbf B_0^\top\Delta_{\gamma,n})^\top)^\top,\qquad
(\mathcal R_n^M)_i=\Delta_{\gamma,n}^\top\mathbf u_{XM,0,i}+\hat\gamma_1^\top\mathbf e_{M_i}.
\end{align}
By Proposition~\ref{prop:loading-cond} and Lemma~\ref{lemma:approxgamma1}, $\|\mathbf a_n\|=O_p(n_t^{-1/2})=O_p(1)$. Let $\mathcal G_t$ denote the training-fold and treated-sample information, as defined earlier. Conditional on $\mathcal G_t$, the coefficients are fixed and the control errors are independent and mean zero. We first consider the covariate part. Write $\mathbf F_{X1}=\mathbf 1\mu_X^\top+\mathbf F_X$, where $\mathbf F_X$ contains the centered factors. For the mean part, since the rows of $\mathbf E_{X0}$ are independent and mean zero conditional on $\mathcal G_t$, we have
\begin{align*}
\mathbb E\left[\left\|\tfrac1{\sqrt{n_c}}\mu_X\mathbf 1^\top\mathbf E_{X0}\mathbf a_n\right\|^2\;\middle|\;\mathcal G_t\right]
&\le\frac Cp\|\mu_X\|^2\|\mathbf a_n\|^2=O_p\left(\frac1p\right),
\end{align*}
using the bounded covariance of $\mathbf u_X$ and $\|\mu_X\|=O(1)$. For the fluctuation part, the centered factors are independent of the control proxy errors and have mean zero. Conditional on $(\mathcal G_t,\mathbf E_{X0})$, we have
\begin{align*}
\mathbb E\left[\left\|\tfrac1{\sqrt{n_c}}\mathbf F_X^\top\mathbf E_{X0}\mathbf a_n\right\|^2\;\middle|\;\mathcal G_t,\mathbf E_{X0}\right]
\le\frac C{n_c}\|\mathbf E_{X0}\mathbf a_n\|^2\le C\left\|\tfrac1{n_c}\mathbf E_{X0}^\top\mathbf E_{X0}\right\|_{\mathrm{op}}\|\mathbf a_n\|^2
=O_p\left(\frac1p\right),
\end{align*}
where the last equality follows from Lemma~\ref{lem:E0bounds}(iii). Now, for the mediator part, Assumption~\ref{Assumption:covariances}(ii) gives that $\mathcal R_n^M$ is mean zero and independent of the control factors, conditional on $\mathcal G_t$. Moreover, Lemmas~\ref{lemma:approxgamma1} and~\ref{lemma:nullspace}, together with Proposition~\ref{prop:loading-cond}, give $\|\hat\gamma_1\|=O_p(1+\sqrt{q/n_t})$. Using the bounded error covariances, we obtain
\begin{align*}
\mathbb E\left[((\mathcal R_n^M)_i)^2\mid\mathcal G_t\right]
&\le C\left(\|\Delta_{\gamma,n}\|^2+\frac1q\|\hat\gamma_1\|^2\right)
=O_p\left(\frac1{n_t}+\frac1q\right).
\end{align*}
Since the control observations are independent and $\mathbb E\|\mathbf f_{X1,i}\|^2\le C$, we have
\begin{align*}
\mathbb E\left[\left\|\tfrac1{\sqrt{n_c}}\mathbf F_{X1}^\top\mathcal R_n^M\right\|^2\;\middle|\;\mathcal G_t\right]
&\le C\max_{i:A_i=0}\mathbb E\left[((\mathcal R_n^M)_i)^2\mid\mathcal G_t\right]=O_p\left(\frac1q+\frac1{n_t}\right).
\end{align*}
Thus, the mediator contribution is $O_p(q^{-1/2}+n_t^{-1/2})$. Combining the covariate and mediator parts gives
\begin{align*}
\left\|\tfrac1{\sqrt{n_c}}\mathbf F_{X1}^\top\mathcal R_n\right\|
&=O_p\left(\frac1{\sqrt p}+\frac1{\sqrt q}+\frac1{\sqrt{n_t}}\right).
\end{align*}
Therefore, together with the $O_p(1)$, we get that
\begin{align*}
|L_5|=O_p\left(\frac1{\sqrt p}+\frac1{\sqrt q}+\frac1{\sqrt{n_t}}\right)=o_p(1).
\end{align*}

\noindent\textbf{\underline{$L_6=o_p(1)$.}}\quad
By definition, $L_6=e_{\mathbf 1}^\top\mathbf H_{X1}^\top\widetilde M^{-1}\tfrac1{\sqrt{n_c}}\mathbf E_{X0}^\top\mathcal R_n$. This has the same structure as $L_4$, with $\mathcal R_n$ in place of $\boldsymbol\eta$. Applying the control one-sided identity and Equation \eqref{eq:PS-fold} with $w=\mathcal R_n$, we obtain
\begin{align*}
L_6&=e_{\mathbf 1}^\top\widetilde R^\top(\widetilde D+\widetilde E_n)^{-1}\Big[\underbrace{\tfrac1{\sqrt{n_c}}\widetilde U^\top\mathbf E_{X0}^\top(I-\widetilde P_S)\mathcal R_n}_{=:L_{6,1}}-\underbrace{\tfrac1{\sqrt{n_c}}\widetilde R\mathbf F_{X1}^\top\widetilde P_S\mathcal R_n}_{=:L_{6,2}}\Big].
\end{align*}
The outside factor is $O_p(1)$ by Equation \eqref{eq:Etil-rate}, so it suffices to bound $L_{6,1}$ and $L_{6,2}$. In \eqref{eq:rm}, we have that by Lemma~\ref{lemma:approxgamma1} and Proposition~\ref{prop:loading-cond}, $\|\mathbf a_n\|=O_p(n_t^{-1/2})=O_p(1)$. Let $\mathcal G_t$ denote the training-fold and treated-sample information defined earlier. Conditional on $\mathcal G_t$, the coefficients are fixed, and $\mathcal R_n^M$ is independent of $\mathbf E_{X0}$ and has mean zero. Moreover, the bounds for $\hat\gamma_1$ and $\Delta_{\gamma,n}$ give
\begin{align*}
r_{M,n}^2:=\max_{i:A_i=0}\mathbb E[((\mathcal R_n^M)_i)^2\mid\mathcal G_t]=O_p\left(\frac1q+\frac1{n_t}\right).
\end{align*}
For $L_{6,1}$, we first consider the covariate part. Since $\|\widetilde U\|_{\mathrm{op}}=1$ and $\|I-\widetilde P_S\|_{\mathrm{op}}\le1$, we have
\begin{align*}
\Big\|\tfrac1{\sqrt{n_c}}\widetilde U^\top\mathbf E_{X0}^\top(I-\widetilde P_S)\mathbf E_{X0}\mathbf a_n\Big\|
&\le\tfrac1{\sqrt{n_c}}\|\mathbf E_{X0}\|_{\mathrm{op}}^2\|\mathbf a_n\|=O_p\left(\frac{\sqrt{n_c}}p\right),
\end{align*}
where $\|\mathbf E_{X0}\|_{\mathrm{op}}^2=O_p(n_c/p)$ by Lemma~\ref{lem:E0bounds}(iii). For the mediator part, conditional on $(\mathcal G_t,\mathbf E_{X0})$, the errors in $\mathcal R_n^M$ are independent and mean zero. Hence,
\begin{align*}
&\mathbb E\left[\Big\|\tfrac1{\sqrt{n_c}}\widetilde U^\top\mathbf E_{X0}^\top(I-\widetilde P_S)\mathcal R_n^M\Big\|^2\;\middle|\;\mathcal G_t,\mathbf E_{X0}\right]\le\tfrac C{n_c}\|\mathbf E_{X0}\|_{\mathrm{op}}^2r_{M,n}^2=O_p\left(\frac1p\right)O_p\left(\frac1q+\frac1{n_t}\right).
\end{align*}
Thus, combining the two parts, we get
\begin{align*}
\|L_{6,1}\|=O_p\left(\frac{\sqrt{n_c}}p+\frac1{\sqrt p}\left(\frac1{\sqrt q}+\frac1{\sqrt{n_t}}\right)\right)=o_p(1),
\end{align*}
under Assumption~\ref{assump:growth}(ii). For $L_{6,2}$, we again consider the covariate and mediator parts separately. Write $\mathbf F_{X1}=\mathbf 1\mu_X^\top+\mathbf F_X$, where $\mathbf F_X$ contains the centered factors. For the mean contribution, Cauchy--Schwarz gives
\begin{align*}
\Big\|\tfrac1{\sqrt{n_c}}\mu_X\mathbf 1^\top\widetilde P_S\mathbf E_{X0}\mathbf a_n\Big\|&\le\tfrac{\|\mu_X\|}{\sqrt{n_c}}\big(\mathbf 1^\top\widetilde P_S\mathbf 1\big)^{1/2}\|\mathbf E_{X0}\|_{\mathrm{op}}\|\mathbf a_n\|=O_p(p^{-1/2}),
\end{align*}
where we use $\mathbf 1^\top\widetilde P_S\mathbf 1=O_p(1)$ by Lemma~\ref{1PE1}, together with $\|\mathbf E_{X0}\|_{\mathrm{op}}=O_p(\sqrt{n_c/p})$. For the fluctuation contribution, conditioning on $(\mathcal G_t,\mathbf E_{X0})$ gives
\begin{align*}
&\mathbb E\left[\Big\|\tfrac1{\sqrt{n_c}}\mathbf F_X^\top\widetilde P_S\mathbf E_{X0}\mathbf a_n\Big\|^2\;\middle|\;\mathcal G_t,\mathbf E_{X0}\right]\le\tfrac C{n_c}\|\widetilde P_S\mathbf E_{X0}\mathbf a_n\|^2\le\tfrac C{n_c}\|\mathbf E_{X0}\|_{\mathrm{op}}^2\|\mathbf a_n\|^2=O_p(p^{-1}),
\end{align*}
since the centered factors are independent of the control proxy errors and have bounded covariance. Thus, the covariate contribution is $O_p(p^{-1/2})$. For the mediator part, conditional on $(\mathcal G_t,\mathbf F_{X1},\mathbf E_{X0})$, the vector $\mathcal R_n^M$ has mean zero and covariance bounded by $r_{M,n}^2I_{n_c}$. Therefore,
\begin{align*}
&\mathbb E\left[\Big\|\tfrac1{\sqrt{n_c}}\widetilde R\mathbf F_{X1}^\top\widetilde P_S\mathcal R_n^M\Big\|^2\;\middle|\;\mathcal G_t,\mathbf F_{X1},\mathbf E_{X0}\right]\le\tfrac C{n_c}\|\widetilde R\|_{\mathrm{op}}^2\operatorname{tr}(\mathbf F_{X1}^\top\widetilde P_S\mathbf F_{X1})r_{M,n}^2=O_p\left(\frac1q+\frac1{n_t}\right),
\end{align*}
using $\|\widetilde R\|_{\mathrm{op}}=O(1)$, $\|\widetilde P_S\|_{\mathrm{op}}\le1$, and $n_c^{-1}\|\mathbf F_{X1}\|_F^2=O_p(1)$. Hence, the mediator contribution is $O_p(q^{-1/2}+n_t^{-1/2})$. Thus, combining the covariate and mediator parts gives us
\begin{align*}
\|L_{6,2}\|=O_p\left(\frac1{\sqrt p}+\frac1{\sqrt q}+\frac1{\sqrt{n_t}}\right)=o_p(1).
\end{align*}
Now, combining the two bounds with the $O_p(1)$ outside factor, and noting that $\sqrt{n_c}/p=O(p^{-1/2})$ under Assumption~\ref{assump:growth}(ii), we obtain
\begin{align*}
|L_6|=O_p\left(\frac1{\sqrt{n_t}}+\frac1{\sqrt p}+\frac1{\sqrt q}\right)=o_p(1).
\end{align*}

\noindent\textbf{\underline{$L_7=\sqrt{{n_c}/{n_t}}\,\delta_0^\top J_3+o_p(1)$.}}\quad
Unlike the bias terms above, $L_7$ does not vanish, so we identify its first-order form. Recall that $L_7=\sqrt{n_c}\,\delta_0^\top(\mathbf H_M^\top\hat\gamma_1-\gamma_1)$. By Lemma~\ref{lemma:approxgamma1}, we have
\begin{align*}
\sqrt{n_t}(\mathbf H_M^\top\hat\gamma_1-\gamma_1)=-J_1-J_2+J_3+J_4,\qquad J_1,J_2,J_4=o_p(1),
\end{align*}
where the remaining term is
\begin{align*}
J_3&=\mathbf P_{\bar q}\mathbf H_1^\top M^{-1}\mathbf H_1\,\tfrac1{\sqrt{n_t}}\mathbf F_1^\top\epsilon_1=\mathbf P_{\bar q}\Omega_1^{-1}\tfrac1{\sqrt{n_t}}\sum_{i:A_i=1}\mathbf f_{1,i}\epsilon_{1,i}+o_p(1),
\end{align*}
using $\mathbf H_1^\top M^{-1}\mathbf H_1-\Omega_1^{-1}=o_p(1)$ from Lemma~\ref{lemma:approxgamma1}.

Since $L_7$ is on the control $\sqrt{n_c}$ scale, while the expansion above is on the treated $\sqrt{n_t}$ scale, we keep the ratio when substituting:
\begin{align*}
L_7&=\sqrt{\frac{n_c}{n_t}}\,\delta_0^\top(-J_1-J_2+J_3+J_4)=\sqrt{\frac{n_c}{n_t}}\,\delta_0^\top J_3+o_p(1).
\end{align*}
Here, $|\delta_0^\top J_j|\le\|\delta_0\|\|J_j\|=o_p(1)$ for $j=1,2,4$, since $\|\delta_0\|=O(1)$ by Assumption~\ref{Assumption:pervasiveness}(i). Also, $n_c/n_t=O_p(1)$ since the treated and control sample proportions converge to positive limits. In particular,
\begin{align*}
\delta_0^\top J_3=(\mathbf P_{\bar q}^\top\delta_0)^\top\Omega_1^{-1}\,\tfrac1{\sqrt{n_t}}\sum_{i:A_i=1}\mathbf f_{1,i}\epsilon_{1,i}+o_p(1).
\end{align*}
For regression fold $k$, let $n_k$ denote the fold size, with treated and control counts $n_k^t$ and $n_k^c$. Since
\begin{align*}
\frac{n_k^t}{n_k}\xrightarrow{p}\pi_1,\qquad\frac{n_k^c}{n_k}\xrightarrow{p}\pi_0,
\end{align*}
we place the treated and control expansions on the common $\sqrt{n_k}$ scale. Now, combining the bounds above, we have
\begin{align*}
G_1=G_2=G_4=o_p(1),\quad L_1=L_2=L_4=L_5=L_6=o_p(1),\quad L_7=\sqrt{\tfrac{n_c}{n_t}}\,\delta_0^\top J_3+o_p(1).
\end{align*}
Combining the bounds above, we get that
\begin{align}
\sqrt{n_k}(\hat\alpha_1-\alpha_1)&=\sqrt{\frac{n_k}{n_k^t}}\,G_3+o_p(1),\notag\\
\sqrt{n_k}\left(\widehat{\gamma_1^\top\delta_0}-\gamma_1^\top\delta_0\right)&=\sqrt{\frac{n_k}{n_k^c}}\,L_3+\sqrt{\frac{n_k}{n_k^t}}\,\delta_0^\top J_3+o_p(1).
\label{eq:survivors}
\end{align}
Therefore, we have that
\begin{align*}
\sqrt{n_k}(\hat\theta_{0,k}-\theta_0)=\sqrt{\frac{n_k}{n_k^t}}(G_3+\delta_0^\top J_3)+\sqrt{\frac{n_k}{n_k^c}}L_3+o_p(1).
\end{align*}
The remaining terms are $G_3$, $L_3$, and $\delta_0^\top J_3$. We next derive their joint asymptotic distribution.

\medskip\noindent\textbf{Part 3: Asymptotic Normality of $\hat\theta_0$.}\label{part3}

\noindent The remaining terms involve the sandwich matrices $\mathbf H_1^\top M^{-1}\mathbf H_1$ for $G_3$ and $\delta_0^\top J_3$, and $\mathbf H_{X1}^\top\widetilde M^{-1}\mathbf H_{X1}$ for $L_3$. We proceed in six steps. Step~1 shows that these matrices converge to $\Omega_1^{-1}$ and $\Omega_{X0}^{-1}$. Steps~2--3 express the remaining terms as sums of i.i.d.\ mean-zero variables. Step~4 combines these terms into a single influence function and derives its variance. Step~5 establishes asymptotic normality for one fold, while Step~6 extends the result to the cross-fitted estimator. We then show that $\sigma^2$ is bounded above and below.

Throughout this part, we work on one regression fold $\mathcal D$ of size $m=|\mathcal D|$, with treated and control counts $n_t=|\{i\in\mathcal D:A_i=1\}|$ and $n_c=|\{i\in\mathcal D:A_i=0\}|$. Under an approximately equal split, we have that $n_t/m\to_p\pi_1$ and $n_c/m\to_p\pi_0$. We consider the full-sample estimator in Step~6.

\medskip\noindent\textbf{\underline{Step 1: Convergence of sandwich matrices.}}
By Lemma~\ref{lem:masterbias}, $\mathbf H_1=UR$, where $U^\top U=I$ and $R$ is square and invertible, with $\nu_{\min}(R)\ge c_R>0$. Moreover, we have
\begin{align*}
\mathbf H_1^\top M^{-1}\mathbf H_1=R^\top(D+E_n)^{-1}R,\qquad D=\tfrac1{n_t}R\mathbf F_1^\top\mathbf F_1R^\top.
\end{align*}
Now, conditional on the training fold, define $D_\infty=R\Omega_1R^\top$. Since $R$ is invertible and $\Omega_1\succ0$, we have $\lambda_{\min}(D_\infty)\ge\nu_{\min}^2(R)\lambda_{\min}(\Omega_1)\ge c\,c_R^2>0$. Furthermore, we have
\begin{align}\label{eq:S1-identity}
\mathbf S_1:=R^\top D_\infty^{-1}R=R^\top(R\Omega_1R^\top)^{-1}R=R^\top R^{-\top}\Omega_1^{-1}R^{-1}R=\Omega_1^{-1}.
\end{align}
This is for square invertible $R$, and we now show that the random sandwich converges to this.

\medskip\noindent\textbf{(a) $\|D-D_\infty\|_{\mathrm{op}}=o_p(1)$.}
The treated outer products $\mathbf f_{1,i}\mathbf f_{1,i}^\top$ are i.i.d.\ with conditional mean $\Omega_1$ and finite fourth moments by Assumption~\ref{assump:subgaussian}(i). For each fixed pair $(j,k)$, applying the central limit theorem entrywise gives
\begin{align*}
\sqrt{n_t}\left\{\left(\tfrac1{n_t}\mathbf F_1^\top\mathbf F_1\right)_{jk}-(\Omega_1)_{jk}\right\}=O_p(1).
\end{align*}
Since the factor dimension is fixed, entrywise convergence also gives $\|\tfrac1{n_t}\mathbf F_1^\top\mathbf F_1-\Omega_1\|_{\mathrm{op}}=O_p(n_t^{-1/2})$. Therefore,
\begin{align*}
\|D-D_\infty\|_{\mathrm{op}}&=\Big\|R\Big(\tfrac1{n_t}\mathbf F_1^\top\mathbf F_1-\Omega_1\Big)R^\top\Big\|_{\mathrm{op}}\le\|R\|_{\mathrm{op}}^2\,O_p(n_t^{-1/2})=O_p(n_t^{-1/2}),
\end{align*}
using $\|R\|_{\mathrm{op}}=O_p(1)$ from Lemma~\ref{lem:masterbias}.

\medskip\noindent\textbf{(b) $\|(D+E_n)^{-1}-D_\infty^{-1}\|_{\mathrm{op}}=o_p(1)$.}
Using the identity $A^{-1}-B^{-1}=A^{-1}(B-A)B^{-1}$ with $A=D+E_n$ and $B=D_\infty$, we have $(D+E_n)^{-1}-D_\infty^{-1}=(D+E_n)^{-1}\big[(D_\infty-D)-E_n\big]D_\infty^{-1}$. Taking operator norms and using sub-multiplicativity, we obtain
\begin{align}\label{eq:resolvent-rate}
\|(D+E_n)^{-1}-D_\infty^{-1}\|_{\mathrm{op}}&\le\|(D+E_n)^{-1}\|_{\mathrm{op}}\big(\|D-D_\infty\|_{\mathrm{op}}+\|E_n\|_{\mathrm{op}}\big)\|D_\infty^{-1}\|_{\mathrm{op}}=o_p(1),
\end{align}
since $\|(D+E_n)^{-1}\|_{\mathrm{op}}=O_p(1)$ by Lemma~\ref{lem:masterbias}, $\|D_\infty^{-1}\|_{\mathrm{op}}=O_p(1)$, $\|D-D_\infty\|_{\mathrm{op}}=O_p(n_t^{-1/2})$, and $\|E_n\|_{\mathrm{op}}=o_p(1)$ by \eqref{eqEn_rate}. Therefore, combining this with Equation \eqref{eq:S1-identity}, and using $\|R\|_{\mathrm{op}}=O_p(1)$, we get that
\begin{align*}
\big\|\mathbf H_1^\top M^{-1}\mathbf H_1-\Omega_1^{-1}\big\|_{\mathrm{op}}&=\big\|R^\top\big[(D+E_n)^{-1}-D_\infty^{-1}\big]R\big\|_{\mathrm{op}}\le\|R\|_{\mathrm{op}}^2\,o_p(1)=o_p(1).
\end{align*}
The same argument applies to the control side with $\mathbf H_{X1}=\widetilde U\widetilde R$ and $\widetilde D_\infty=\widetilde R\Omega_{X0}\widetilde R^\top$, for which $\widetilde R^\top\widetilde D_\infty^{-1}\widetilde R=\Omega_{X0}^{-1}$. Thus,
\begin{align}\label{eq:sandwich-conv}
\big\|\mathbf H_1^\top M^{-1}\mathbf H_1-\Omega_1^{-1}\big\|_{\mathrm{op}}&=o_p(1),\qquad \big\|\mathbf H_{X1}^\top\widetilde M^{-1}\mathbf H_{X1}-\Omega_{X0}^{-1}\big\|_{\mathrm{op}}=o_p(1).
\end{align}

\medskip\noindent\textbf{\underline{Step 2: $G_3+\delta_0^\top J_3$ as a treated i.i.d.\ average.}}
Recall that $\delta_0^\top J_3=(\mathbf P_{\bar q}^\top\delta_0)^\top\mathbf H_1^\top M^{-1}\mathbf H_1\tfrac1{\sqrt{n_t}}\mathbf F_1^\top\epsilon_1$ has the same structure as $G_3=e_{\mathbf 1}^\top\mathbf H_1^\top M^{-1}\mathbf H_1\tfrac1{\sqrt{n_t}}\mathbf F_1^\top\epsilon_1$. The two terms can therefore be combined using $a_1=e_{\mathbf 1}+\mathbf P_{\bar q}^\top\delta_0$, as defined before Theorem~\ref{thm:clt}. Since $\mathbf P_{\bar q}^\top\delta_0$ lies in the mediator block, its intercept coordinate is zero, so
\begin{equation}\label{eq:a1}
\|a_1\|^2=1+\|\mathbf P_{\bar q}^\top\delta_0\|^2,\qquad 1\le\|a_1\|=O(1),
\end{equation}
where $\|\delta_0\|=O(1)$ by Assumption~\ref{Assumption:pervasiveness}(i). Combining the two terms, we have
\begin{align*}
G_3+\delta_0^\top J_3&=a_1^\top\mathbf H_1^\top M^{-1}\mathbf H_1\,\tfrac1{\sqrt{n_t}}\mathbf F_1^\top\epsilon_1\\
&=a_1^\top\Omega_1^{-1}\tfrac1{\sqrt{n_t}}\mathbf F_1^\top\epsilon_1+a_1^\top\big(\mathbf H_1^\top M^{-1}\mathbf H_1-\Omega_1^{-1}\big)\tfrac1{\sqrt{n_t}}\mathbf F_1^\top\epsilon_1.
\end{align*}
The second term is $o_p(1)$ by \eqref{eq:sandwich-conv}, provided $\|n_t^{-1/2}\mathbf F_1^\top\epsilon_1\|=O_p(1)$. By Assumption~\ref{Assumption:covariances}(ii), $\epsilon_1$ is mean zero and independent of $\mathbf F_1$, while Assumption~\ref{assump:subgaussian}(i) gives uniformly bounded second moments. Therefore,
\begin{equation}\label{eq:F1eps-Op1}
\begin{aligned}
\mathbb E\Big[\Big\|\tfrac1{\sqrt{n_t}}\mathbf F_1^\top\epsilon_1\Big\|^2\,\Big|\,\mathbf F_1\Big]&=\operatorname{tr}\Big(\tfrac1{n_t}\mathbf F_1^\top\mathbb E(\epsilon_1\epsilon_1^\top\mid\mathbf F_1)\mathbf F_1\Big)\le C\operatorname{tr}\Big(\tfrac1{n_t}\mathbf F_1^\top\mathbf F_1\Big)=O_p(1).
\end{aligned}
\end{equation}
Hence $\|n_t^{-1/2}\mathbf F_1^\top\epsilon_1\|=O_p(1)$ by the second-moment bound. Together with $\|a_1\|=O(1)$, this shows that the second term above is $o_p(1)$. We therefore obtain
\begin{equation}\label{eq:xi1}
G_3+\delta_0^\top J_3=\frac1{\sqrt{n_t}}\sum_{i:A_i=1}\xi_{1,i}+o_p(1),\qquad \xi_{1,i}=\big(\Omega_1^{-1}a_1\big)^\top\mathbf f_{1,i}\epsilon_{1,i}.
\end{equation}
By Assumption~\ref{Assumption:covariances}(i)--(ii), the $\xi_{1,i}$ are i.i.d.\ across treated observations and have mean zero. In particular, $\mathbb E[\epsilon_{1,i}\mid\mathbf f_{1,i},A_i=1]=0$ implies $\mathbb E[\xi_{1,i}\mid A_i=1]=0$. Their variance is
\begin{equation}\label{eq:sigma1}
\begin{aligned}
\sigma_1^2&:=\operatorname{Var}(\xi_{1,i}\mid A_i=1)=\mathbb E\big[\epsilon_{1,i}^2\{(\Omega_1^{-1}a_1)^\top\mathbf f_{1,i}\}^2\mid A_i=1\big]=\sigma_{\epsilon_1,1}^2\,a_1^\top\Omega_1^{-1}a_1,
\end{aligned}
\end{equation}
where $\sigma_{\epsilon_1,1}^2=\mathbb E[\epsilon_{1,i}^2\mid A_i=1]$. The last equality follows from the independence of $\epsilon_{1,i}$ and $\mathbf f_{1,i}$ under Assumption~\ref{Assumption:covariances}(ii). The variance is finite because $\|\Omega_1^{-1}\|_{\mathrm{op}}=O(1)$, $\|a_1\|=O(1)$, and the relevant second moments are bounded by Assumption~\ref{assump:subgaussian}(i).

\medskip\noindent\textbf{\underline{Step 3: $L_3$ as a control i.i.d.\ average.}}
Repeating Step~2 with $(\widetilde M,\mathbf H_{X1},\mathbf F_{X1},\boldsymbol\eta,\Omega_{X0})$ in place of $(M,\mathbf H_1,\mathbf F_1,\epsilon_1,\Omega_1)$, and using the second bound in \eqref{eq:sandwich-conv} together with the control analogue of \eqref{eq:F1eps-Op1}, we obtain
\begin{equation}\label{eq:xi2}
L_3=\frac1{\sqrt{n_c}}\sum_{i:A_i=0}\xi_{2,i}+o_p(1),
\end{equation}
where $\xi_{2,i}=\big(\Omega_{X0}^{-1}e_{\mathbf 1}\big)^\top\mathbf f_{X1,i}\eta_i$ and $\eta_i=\gamma_1^\top\mathbf u_{XM,0,i}$. Here, $e_{\mathbf 1}$ denotes the intercept coordinate vector $e_{1,1+\bar p}$ defined before Theorem~\ref{thm:clt}. By Assumption~\ref{Assumption:covariances}(ii), $\mathbf u_{XM,0,i}$ is mean zero and independent of $(\mathbf f_{X,i},A_i)$, so $\mathbb E[\eta_i\mid\mathbf f_{X1,i},A_i=0]=0$. Thus, the $\xi_{2,i}$ are i.i.d.\ mean-zero across control observations with conditional variance
\begin{equation}\label{eq:sigma2}
\sigma_2^2:=\operatorname{Var}(\xi_{2,i}\mid A_i=0)=\mathbb E\big[\eta_i^2\{(\Omega_{X0}^{-1}e_{\mathbf 1})^\top\mathbf f_{X1,i}\}^2\mid A_i=0\big]<\infty.
\end{equation}
To see that this variance is finite, we have
\begin{align*}
\mathbb E[\eta_i^2\mid A_i=0]&=\gamma_1^\top\Sigma_{\mathbf u_{XM,0}}\gamma_1\le\|\gamma_1\|_2^2\|\Sigma_{\mathbf u_{XM,0}}\|_{\mathrm{op}}\le C,
\end{align*}
using Assumption~\ref{Assumption:pervasiveness}(i) for $\|\gamma_1\|=O(1)$ and Assumption~\ref{Assumption:covariances}(v) for $\|\Sigma_{\mathbf u_{XM,0}}\|_{\mathrm{op}}\le C$. Together with the bounded second moments of $\mathbf f_{X1,i}$ and $\|\Omega_{X0}^{-1}\|_{\mathrm{op}}=O(1)$, this establishes the claimed finiteness.

\medskip\noindent\textbf{\underline{Step 4: Influence function and vanishing cross term.}}
Substituting \eqref{eq:survivors}, \eqref{eq:xi1}, and \eqref{eq:xi2} into the fold decomposition \eqref{eqtheta_decomp_start}, and placing the terms on the common fold scale $\sqrt m$, we have
\begin{align*}
\sqrt m\,(\hat\theta_{\mathcal D}-\theta_0)&=\sqrt{\tfrac{m}{n_t}}\Big[\tfrac1{\sqrt{n_t}}\sum_{i:A_i=1}\xi_{1,i}\Big]+\sqrt{\tfrac{m}{n_c}}\Big[\tfrac1{\sqrt{n_c}}\sum_{i:A_i=0}\xi_{2,i}\Big]+o_p(1).
\end{align*}
For the treated group, regrouping the constants gives
\begin{align*}
\sqrt{\tfrac{m}{n_t}}\cdot\tfrac1{\sqrt{n_t}}\sum_{i:A_i=1}\xi_{1,i}&=\tfrac1{\sqrt m}\cdot\tfrac{m}{n_t}\sum_{i:A_i=1}\xi_{1,i}\\
&=\tfrac1{\sqrt m}\sum_{i\in\mathcal D}\mathbf 1(A_i=1)\,\tfrac{m}{n_t}\,\xi_{1,i}.
\end{align*}
Since $m/n_t\to_p1/\pi_1$ and $m^{-1/2}\sum_{i\in\mathcal D}\mathbf 1(A_i=1)\xi_{1,i}=O_p(1)$, replacing $m/n_t$ by $1/\pi_1$ introduces an error of $(m/n_t-1/\pi_1)\cdot O_p(1)=o_p(1)$ by Slutsky's theorem. The control group is analogous, with $m/n_c\to_p1/\pi_0$. Hence,
\begin{equation}\label{eq:IF}
\sqrt m\,(\hat\theta_{\mathcal D}-\theta_0)=\frac1{\sqrt m}\sum_{i\in\mathcal D}\psi_i+o_p(1),
\end{equation}
where
\begin{equation}\label{eq:true_influence}
\psi_i=\frac{\mathbf 1(A_i=1)}{\pi_1}\big(\Omega_1^{-1}a_1\big)^\top\mathbf f_{1,i}\epsilon_{1,i}+\frac{\mathbf 1(A_i=0)}{\pi_0}\big(\Omega_{X0}^{-1}e_{\mathbf 1}\big)^\top\mathbf f_{X1,i}\eta_i.
\end{equation}
By the conditional mean-zero properties established in Steps~2--3, we have $\mathbb E[\psi_i]=0$.

Now, write $\psi_i=\psi_i^{(1)}+\psi_i^{(0)}$ for the treated and control contributions. Their product contains the factor $\mathbf 1(A_i=1)\mathbf 1(A_i=0)=0$, since each observation belongs to exactly one treatment group. Thus, $\psi_i^{(1)}\psi_i^{(0)}=0$ for every $i$, and
\begin{align*}
\sigma^2:=\operatorname{Var}(\psi_i)=\mathbb E\big[(\psi_i^{(1)})^2\big]+\mathbb E\big[(\psi_i^{(0)})^2\big].
\end{align*}
Now, evaluating each term using $\mathbb E[\mathbf 1(A_i=1)]=\pi_1$ and the conditional variances in \eqref{eq:sigma1} and \eqref{eq:sigma2}, we obtain
\begin{align*}
\mathbb E\big[(\psi_i^{(1)})^2\big]=\frac1{\pi_1^2}\,\mathbb E[\mathbf 1(A_i=1)]\,\mathbb E\big[\{(\Omega_1^{-1}a_1)^\top\mathbf f_{1,i}\epsilon_{1,i}\}^2\mid A_i=1\big]=\frac{\sigma_1^2}{\pi_1},\quad
\mathbb E\big[(\psi_i^{(0)})^2\big]=\frac{\sigma_2^2}{\pi_0}.
\end{align*}
Therefore, we have
\begin{equation}\label{eq:var-final}
\sigma^2=\frac{\sigma_1^2}{\pi_1}+\frac{\sigma_2^2}{\pi_0}.
\end{equation}

\medskip\noindent\textbf{\underline{Step 5: Lyapunov central limit theorem.}}
The $\{\psi_i\}_{i\in\mathcal D}$ are i.i.d.\ and mean zero. Fix $\omega\in(0,1]$. Each part of $\psi_i$ is a fixed scalar ($1/\pi_1$ or $1/\pi_0$) times the product of a bounded linear form in a sub-Gaussian factor vector and a sub-Gaussian noise ($\epsilon_{1,i}$ or $\eta_i=\gamma_1^\top\mathbf u_{XM,0,i}$). By the bound $|x+y|^{2+\omega}\le2^{1+\omega}(|x|^{2+\omega}+|y|^{2+\omega})$, the Cauchy--Schwarz inequality, and the finiteness of all moments of sub-Gaussian variables under Assumption~\ref{assump:subgaussian}(i), we have
\begin{align*}
\mathbb E|\psi_i|^{2+\omega}&\le 2^{1+\omega}\Big(\tfrac1{\pi_1^{2+\omega}}\mathbb E\big|(\Omega_1^{-1}a_1)^\top\mathbf f_{1,i}\epsilon_{1,i}\big|^{2+\omega}+\tfrac1{\pi_0^{2+\omega}}\mathbb E\big|(\Omega_{X0}^{-1}e_{\mathbf 1})^\top\mathbf f_{X1,i}\eta_i\big|^{2+\omega}\Big)\le C<\infty,
\end{align*}
uniformly in $i$. Using the lower bound $\sigma^2\ge c>0$, established below, the Lyapunov ratio satisfies
\begin{align*}
\frac{\sum_{i\in\mathcal D}\mathbb E|\psi_i|^{2+\omega}}{\big(\sum_{i\in\mathcal D}\operatorname{Var}(\psi_i)\big)^{1+\omega/2}}&=\frac{m\,\mathbb E|\psi_1|^{2+\omega}}{(m\sigma^2)^{1+\omega/2}}\le\frac{Cm}{(mc)^{1+\omega/2}}=O(m^{-\omega/2})\to0.
\end{align*}
This verifies the Lyapunov condition. Therefore, we have by Lyapunov central limit theorem that
\begin{align*}
\frac1{\sqrt m\,\sigma}\sum_{i\in\mathcal D}\psi_i\xrightarrow{d}N(0,1).
\end{align*}
Now, combining this with \eqref{eq:IF} and Slutsky's theorem gives us
\begin{align*}
\frac{\sqrt m\,(\hat\theta_{\mathcal D}-\theta_0)}{\sigma}\xrightarrow{d}N(0,1).
\end{align*}

\medskip\noindent\textbf{\underline{Step 6: Cross-fitting.}}
The expansion above is for one regression fold. Let $\mathcal D_1$ and $\mathcal D_2$ be the two folds, with $|\mathcal D_1|=|\mathcal D_2|=n/2$, and recall that the projection directions for each fold are estimated using its complement. The cross-fitted estimator is $\hat\theta_0^{\mathrm{cf}}=\tfrac12(\hat\theta_{\mathcal D_1}+\hat\theta_{\mathcal D_2})$. Applying Steps~1 - 5 to each fold $\mathcal D_k$ gives
\begin{equation}\label{eq:fold-expansion}
\hat\theta_{\mathcal D_k}-\theta_0=\frac1{|\mathcal D_k|}\sum_{i\in\mathcal D_k}\psi_i+o_p\big(|\mathcal D_k|^{-1/2}\big),
\end{equation}
with the same influence function $\psi_i$ from \eqref{eq:IF}. Although the training-fold projection directions differ across folds, Step~1 shows that the corresponding sandwich matrices converge to the same population limits $\Omega_1^{-1}$ and $\Omega_{X0}^{-1}$. Together with the negligible factor estimation errors established in Part~2, this gives the same leading influence function for both folds, with the differences absorbed into the $o_p$ remainder.

The two fold sums are over disjoint sets $\mathcal D_1$ and $\mathcal D_2$. Since the leading summands $\psi_i$ depend only on the individual observations and fixed population quantities, they are i.i.d.\ across the full sample. The dependence introduced by estimating the projection directions on the opposite fold enters through the remainders in \eqref{eq:fold-expansion}. Therefore, using $2|\mathcal D_k|=n$, we have
\begin{align*}
\sqrt n(\hat\theta_0^{\mathrm{cf}}-\theta_0)&=\frac{\sqrt n}{2}\sum_{k=1}^2(\hat\theta_{\mathcal D_k}-\theta_0)\\
&=\frac{\sqrt n}{2}\sum_{k=1}^2\frac1{|\mathcal D_k|}\sum_{i\in\mathcal D_k}\psi_i+o_p(1)\\
&=\frac1{\sqrt n}\sum_{i=1}^n\psi_i+o_p(1).
\end{align*}
The $\{\psi_i\}_{i=1}^n$ are i.i.d.\ mean zero with variance $\sigma^2$. Thus, the Lyapunov argument in Step~5 applies with $m$ replaced by $n$ and gives us
\begin{align*}
\frac{\sqrt n(\hat\theta_0^{\mathrm{cf}}-\theta_0)}{\sigma}\xrightarrow{d}N(0,1),
\end{align*}
It remains to show $0<c\le\sigma^2\le C<\infty$. The upper bound follows from Equation \eqref{eq:var-final}, since $\sigma_1^2,\sigma_2^2=O(1)$ by Steps~2--3 and $\pi_1,\pi_0$ are fixed positive probabilities. For the lower bound, $\sigma^2$ is a sum of two nonnegative terms, so it suffices to bound one of them below. By Assumption~\ref{Assumption:covariances}(ii), $\epsilon_{1,i}$ is independent of $\mathbf f_{1,i}$ and $A_i$. Thus, Assumption~\ref{assump:subgaussian}(ii) gives $\operatorname{Var}(\epsilon_{1,i}\mid\mathbf f_{1,i},A_i=1)=\operatorname{Var}(\epsilon_{1,i})\ge\underline\sigma^2>0$. Thus, conditioning on $\mathbf f_{1,i}$ inside \eqref{eq:sigma1}, we get
\begin{align*}
\sigma_1^2&=\mathbb E\big[\operatorname{Var}(\epsilon_{1,i}\mid\mathbf f_{1,i},A_i=1)\{(\Omega_1^{-1}a_1)^\top\mathbf f_{1,i}\}^2\mid A_i=1\big]\ge\underline\sigma^2\,\mathbb E\big[\{(\Omega_1^{-1}a_1)^\top\mathbf f_{1,i}\}^2\mid A_i=1\big].
\end{align*}
The inner expectation is the $\Omega_1$-quadratic form, we get
\begin{align*}
\mathbb E\big[\{(\Omega_1^{-1}a_1)^\top\mathbf f_{1,i}\}^2\mid A_i=1\big]&=(\Omega_1^{-1}a_1)^\top\Omega_1(\Omega_1^{-1}a_1)=a_1^\top\Omega_1^{-1}a_1\ge\lambda_{\min}(\Omega_1^{-1})\|a_1\|^2.
\end{align*}
Since $\|a_1\|\ge1$ by \eqref{eq:a1} and $\lambda_{\min}(\Omega_1^{-1})=\lambda_{\max}(\Omega_1)^{-1}>0$, we have
\begin{align*}
\sigma^2\ge\frac{\sigma_1^2}{\pi_1}\ge\frac{\underline\sigma^2\,\lambda_{\max}(\Omega_1)^{-1}}{\pi_1}=c>0.
\end{align*}
Now, combining the upper and lower bounds gives $0<c\le\sigma^2\le C<\infty$, which completes the proof of asymptotic normality of the cross-fitted estimator $\hat\theta_0^{\mathrm{cf}}$.
\end{proof}

\subsection{Proof of Lemma \ref{lemma:ve}}\label{sec:appendix-variance}

\begin{proof}
We write $\hat{\sigma}^2$ as the empirical second moment of an estimated influence function. Then, we show that this function is close to the influence function as defined in Equation \eqref{eq:true_influence}. Here, the weights from the projected factors must be close to the weights from true factors and the regression residuals to the structural errors. When that would hold, the sample averages in $\hat{\sigma}_1^2$ and $\hat{\sigma}_2^2$ can be compared directly with the true averages, and this gives the result. Recall that the expansion in \eqref{eq:IF} from the proof of Theorem~\ref{thm:clt} uses
\begin{align}\label{eq:ve_true}
\psi_i=\frac{\mathbf 1(A_i=1)}{\pi_1}\bigl(\Omega_1^{-1}a_1\bigr)^\top\mathbf f_{1,i}\epsilon_{1,i}+\frac{\mathbf 1(A_i=0)}{\pi_0}\bigl(\Omega_{X0}^{-1}e_{1,1+\bar p}\bigr)^\top\mathbf f_{X1,i}\eta_i.
\end{align}
By the mean-zero properties of the structural errors and Equation \eqref{eq:var-final}, $\mathbb E(\psi_i)=0$ and $\mathbb E(\psi_i^2)=\sigma^2$. These terms depend on the unobserved factors and structural errors, so we replace them with the projected regression quantities.

Fix a regression fold $\mathcal D_k$. Write $m=n_k$, $n_t=n_k^t$, and $n_c=n_k^c$. The projections used on this fold were estimated on the other fold. We condition on that training fold whenever we work with the projected errors. For the control bridge we also condition on the treated observations, because $\hat\gamma_1$ is fitted on them. We have that $n_t\asymp n_c\asymp m\asymp n$ in probability. Let $\hat{\mathbf Z}_i=(1,\hat{\mathbf f}_{X,i}^\top,\hat{\mathbf f}_{M,i}^\top)^\top$ and $\hat{\mathbf R}_i=(1,\hat{\mathbf f}_{X,i}^\top)^\top$ denote the projected treated and control design vectors, respectively. By Equation \eqref{equation:zhat}, they satisfy
\begin{align*}
\hat{\mathbf Z}_i&=\mathbf H_1\mathbf f_{1,i}+\mathbf e_{0i}, \quad \hat{\mathbf R}_i=\mathbf H_{X1}\mathbf f_{X1,i}+\mathbf e_{X0,i}.
\end{align*}
Since $\mathbf H_{X1}=\operatorname{diag}(1,\mathbf H_X)$, we have $e_{\mathbf 1}=\mathbf H_{X1}e_{1,1+\bar p}$. Since $\mathbf H_1=\operatorname{diag}(1,\mathbf H_X,\mathbf H_M)$, we have $\mathbf H_1a_1=e_{\mathbf 1}+\mathbf S_{\tilde q}^\top\mathbf H_M\delta_0$. We know that $M=n_t^{-1}\hat{\mathbf Z}^\top\hat{\mathbf Z}$ and $\widetilde M=n_c^{-1}\hat{\mathbf R}^\top\hat{\mathbf R}$ denote the treated and control gram matrices on this fold. 

Let $\hat\epsilon_{1,i}$ and $\hat\eta_i$ be the least-squares residuals from the treated outcome regression and the control bridge regression of Algorithm~\ref{algo:FARMA}. Since $a_1$ contains the unknown mediator intercept $\delta_0$, we estimate $\mathbf H_M\delta_0$ by the intercept from a least-squares regression of $\hat{\mathbf f}_{M,i}$ on $\hat{\mathbf R}_i$ among the controls as we did in the proof of Theorem \ref{thm:clt}. On each regression fold, define $\widehat{\mathbf H_M\delta_0}=\{(1/{n_k^c})\sum_{i\in\mathcal D_k:A_i=0}\hat{\mathbf f}_{M,i}\hat{\mathbf R}_i^\top\}\widetilde M^{-1}e_{\mathbf 1}$ and $\widehat{\mathbf H_1a_1}=e_{\mathbf 1}+\mathbf S_{\tilde q}^\top\widehat{\mathbf H_M\delta_0}$. With these fold specific quantities, the terms in \ref{eq:ve-estimator}
are $\hat\pi_1=n^{-1}\sum_{i=1}^n\mathbf 1(A_i=1)$, $\hat\pi_0=1-\hat\pi_1$, and
\begin{align}
\hat\sigma_1^2=\frac1{n\hat\pi_1}\sum_{k=1}^2\sum_{i\in\mathcal D_k:A_i=1}\big(\widehat{\mathbf H_1a_1}^\top M^{-1}\hat{\mathbf Z}_i\,\hat\epsilon_{1,i}\big)^2,\quad \hat\sigma_2^2=\frac1{n\hat\pi_0}\sum_{k=1}^2\sum_{i\in\mathcal D_k:A_i=0}\big(e_{\mathbf 1}^\top\widetilde M^{-1}\hat{\mathbf R}_i\,\hat\eta_i\big)^2,\label{eq:ve-sigma12}
\end{align}
where for $i\in\mathcal D_k$, the matrices $M$ and $\widetilde M$, the residuals, and $\widehat{\mathbf H_1a_1}$ are those computed on $\mathcal D_k$. For $i\in\mathcal D_k$, define
\begin{align}\displaystyle
\hat\psi_{k,i}=\frac{\mathbf1(A_i=1)}{ \hat{\pi}_1}(\widehat{\mathbf H_1a_1})^\top M^{-1}\hat{\mathbf Z}_i\hat\epsilon_{1,i}+\frac{\mathbf1(A_i=0)}{ \hat{\pi}_0}e_{\mathbf 1}^\top\widetilde M^{-1}\hat{\mathbf R}_i\hat\eta_i,\label{eq:ve-estimator}
\end{align}
This is the sample version of $\psi_i$. The two indicators cannot both equal one, so the cross term in $\hat\psi_{k,i}^2$ vanishes. We sum over all observations and use Equation \eqref{eq:ve-sigma12}. This gives
\begin{align}
\frac1n\sum_{k=1}^2\sum_{i\in\mathcal D_k}\hat\psi_{k,i}^2
&=\frac1{n\hat\pi_1^2}\sum_{k=1}^2\sum_{i\in\mathcal D_k:A_i=1}\big(\widehat{\mathbf H_1a_1}^\top M^{-1}\hat{\mathbf Z}_i\,\hat\epsilon_{1,i}\big)^2+\frac1{n\hat\pi_0^2}\sum_{k=1}^2\sum_{i\in\mathcal D_k:A_i=0}\big(e_{\mathbf 1}^\top\widetilde M^{-1}\hat{\mathbf R}_i\,\hat\eta_i\big)^2\notag\\
&=\frac{\hat\sigma_1^2}{\hat\pi_1}+\frac{\hat\sigma_2^2}{\hat\pi_0}\notag\\
&=\hat\sigma^2.\label{eq:ve-sigma-psi}
\end{align}
It therefore suffices to show that $n^{-1}\sum_{k}\sum_{i\in\mathcal D_k}\hat\psi_{k,i}^2\xrightarrow{p}\sigma^2$.

Now, if we use the coordinates of Lemma~\ref{lemma:nullspace}, where $\mathbf H_1=UR$, $J=[\,U\ \ S\,]$ is orthogonal, and $\Sigma_S=\operatorname{diag}(p^{-1}\Omega_X,q^{-1}\Omega_M)$. Also let $\widetilde{\mathbf E}_S=\mathbf E_0S\Sigma_S^{-1/2}$ and $\bar M=n_t^{-1}\bar{\mathbf Z}^\top\bar{\mathbf Z}$, where $\bar{\mathbf Z}$ from Step~2 of the lemma. There, $\|\bar M^{-1}\|_{\mathrm{op}}=O_p(1)$, and its decomposition gives us that
\begin{align*}
M^{-1}&=J\begin{pmatrix}I&0\\0&\Sigma_S^{-1/2}\end{pmatrix}\bar M^{-1}\begin{pmatrix}I&0\\0&\Sigma_S^{-1/2}\end{pmatrix}J^\top.
\end{align*}
Hence, for every vector $v\in\mathbb R^{1+\tilde p+\tilde q}$, we have
\begin{align}\label{eq:ve-quad}
v^\top M^{-1}v&\le\|\bar M^{-1}\|_{\mathrm{op}}\big(\|U^\top v\|^2+\|\Sigma_S^{-1/2}S^\top v\|^2\big).
\end{align}
For the control part, use $\mathbf H_{X1}=\widetilde U\widetilde R$ and the $X$-only version of the same rescaling as we  have in the proof of Theorem~\ref{thm:clt}. Its null directions are spanned by $S_X$ with covariance scale $p^{-1}\Omega_X$. Thus, for every $v=(v_0,v_X^\top)^\top\in\mathbb R^{1+\tilde p}$, we have
\begin{align}\label{eq:ve-quad-control}
v^\top\widetilde M^{-1}v&\le O_p(1)\big(\|\widetilde U^\top v\|^2+p\,\|\Omega_X^{-1/2}S_X^\top v_X\|^2\big).
\end{align}
As $e_{\mathbf 1}$ has zero $X$-block and $\|\widetilde U^\top e_{\mathbf 1}\|\le1$, \eqref{eq:ve-quad-control} gives us that
\begin{align}\label{eq:ve-e1}
e_{\mathbf 1}^\top\widetilde M^{-1}e_{\mathbf 1}&=O_p(1).
\end{align}

For a control observation, the mediator model gives $\mathbf f_{M,i}=\delta_0+\mathbf B_0\mathbf f_{X,i}+\mathbf u_{XM,0,i}$. We also have $\mathbf f_{X,i}=\mathbf H_X^\dagger(\hat{\mathbf f}_{X_i}-\mathbf e_{X_i})$. We substitute both into $\hat{\mathbf f}_{M_i}=\mathbf H_M\mathbf f_{M,i}+\mathbf e_{M_i}$ and get that
\begin{align*}
\hat{\mathbf f}_{M_i}&=\mathbf H_M\big(\delta_0\ \ \mathbf B_0\mathbf H_X^\dagger\big)\hat{\mathbf R}_i+\mathbf H_M\mathbf u_{XM,0,i}-\mathbf H_M\mathbf B_0\mathbf H_X^\dagger\mathbf e_{X_i}+\mathbf e_{M_i}.
\end{align*}
The first term is linear in $\hat{\mathbf R}_i$. Its least-squares intercept is $\mathbf H_M(\delta_0\ \ \mathbf B_0\mathbf H_X^\dagger)e_{\mathbf 1}=\mathbf H_M\delta_0$. Hence
\begin{align*}
\widehat{\mathbf H_M\delta_0}-\mathbf H_M\delta_0&=\frac1{n_c}\sum_{i:A_i=0}\big(\mathbf H_M\mathbf u_{XM,0,i}-\mathbf H_M\mathbf B_0\mathbf H_X^\dagger\mathbf e_{X_i}+\mathbf e_{M_i}\big)\hat{\mathbf R}_i^\top\widetilde M^{-1}e_{\mathbf 1}.
\end{align*}
The scalars $\hat{\mathbf R}_i^\top\widetilde M^{-1}e_{\mathbf 1}$ satisfy that
\begin{align*}
\frac1{n_c}\sum_{i:A_i=0}\big(\hat{\mathbf R}_i^\top\widetilde M^{-1}e_{\mathbf 1}\big)^2=e_{\mathbf 1}^\top\widetilde M^{-1}\widetilde M\widetilde M^{-1}e_{\mathbf 1}=e_{\mathbf 1}^\top\widetilde M^{-1}e_{\mathbf 1}=O_p(1)
\end{align*}
using the Equation \eqref{eq:ve-e1}. By Assumption~\ref{Assumption:covariances}(ii), $\mathbf u_{XM,0,i}$ and $\mathbf e_{M_i}$ are mean zero and independent of $\hat{\mathbf R}_i$, conditional on the training fold. So, we have that the conditional second moment of the first term is at most $Cn_c^{-1}e_{\mathbf 1}^\top\widetilde M^{-1}e_{\mathbf 1}$, and that of the third term is at most $C(n_cq)^{-1}e_{\mathbf 1}^\top\widetilde M^{-1}e_{\mathbf 1}$. For the second term, we have using the Cauchy--Schwarz and the $X$-only version of Lemma~\ref{lem:E0bounds}(i) give a bound of order $O_p(p^{-1/2})$, since $\|\mathbf H_M\mathbf B_0\mathbf H_X^\dagger\|_{\mathrm{op}}=O_p(1)$. Therefore, we get
\begin{align}\label{eq:ve-d-rate}
\|\widehat{\mathbf H_M\delta_0}-\mathbf H_M\delta_0\|&=O_p\left(\frac1{\sqrt{m}}+\frac1{\sqrt{p}}\right).
\end{align}
Since $S_M^\top\mathbf H_M=0$, we have
\begin{align*}
S_M^\top\widehat{\mathbf H_M\delta_0}&=\frac1{n_c}\sum_{i:A_i=0}S_M^\top\mathbf e_{M_i}\hat{\mathbf R}_i^\top\widetilde M^{-1}e_{\mathbf 1}.
\end{align*}
Using the Step~2 in the proof of Lemma~\ref{lemma:nullspace}, we have that $\mathbb E\|S_M^\top\mathbf e_{M_i}\|^2=q^{-1}\operatorname{tr}(\Omega_M)$. The conditional second moment is therefore at most $C(n_cq)^{-1}e_{\mathbf 1}^\top\widetilde M^{-1}e_{\mathbf 1}$. Using Equation \eqref{eq:ve-e1},
\begin{align}\label{eq:ve-d-null}
\,\|S_M^\top\widehat{\mathbf H_M\delta_0}\|&=O_p\left(\frac1{\sqrt{qm}}\right).
\end{align}
Now, for treated observations, let $\hat h_{1,i}=\widehat{\mathbf H_1a_1}^\top M^{-1}\hat{\mathbf Z}_i$ and $h_{1,i}=a_1^\top\Omega_1^{-1}\mathbf f_{1,i}$. For control part, let $\hat h_{0,i}=e_{\mathbf 1}^\top\widetilde M^{-1}\hat{\mathbf R}_i$ and $h_{0,i}=e_{1,1+\bar p}^\top\Omega_{X0}^{-1}\mathbf f_{X1,i}$. Using Equation \eqref{eq:xi1} and \eqref{eq:xi2}, we have that $\xi_{1,i}=h_{1,i}\epsilon_{1,i}$ and $\xi_{2,i}=h_{0,i}\eta_i$.

Now, we start with the treated weights. Since $\mathbf H_1$ has full column rank, $\mathbf H_1^\top(\mathbf H_1^\dagger)^\top=I$. So, we have
\begin{align*}
\hat{\mathbf Z}(\mathbf H_1^\dagger)^\top\Omega_1^{-1}a_1
&=\mathbf F_1\mathbf H_1^\top(\mathbf H_1^\dagger)^\top\Omega_1^{-1}a_1+\mathbf E_0(\mathbf H_1^\dagger)^\top\Omega_1^{-1}a_1\\&=\mathbf F_1\Omega_1^{-1}a_1+\mathbf E_0(\mathbf H_1^\dagger)^\top\Omega_1^{-1}a_1.
\end{align*}
The vector $\mathbf F_1\Omega_1^{-1}a_1$ has entries $h_{1,i}$. So, we subtract and get that
\begin{align*}
\big(\hat h_{1,i}-h_{1,i}\big)_{i:A_i=1}
&=\hat{\mathbf Z}M^{-1}\widehat{\mathbf H_1a_1}-\hat{\mathbf Z}(\mathbf H_1^\dagger)^\top\Omega_1^{-1}a_1+\mathbf E_0(\mathbf H_1^\dagger)^\top\Omega_1^{-1}a_1\\&=\hat{\mathbf Z}M^{-1}v+\mathbf E_0(\mathbf H_1^\dagger)^\top\Omega_1^{-1}a_1,
\end{align*}
where $v=\widehat{\mathbf H_1a_1}-M(\mathbf H_1^\dagger)^\top\Omega_1^{-1}a_1$.
Since $n_t^{-1}\|\hat{\mathbf Z}M^{-1}v\|^2=v^\top M^{-1}v$, we have that
\begin{align}\label{eq:ve-h1-split}
\frac1{n_t}\sum_{i:A_i=1}|\hat h_{1,i}-h_{1,i}|^2
&\le2v^\top M^{-1}v+2\Big\|\frac1{n_t}\mathbf E_0^\top\mathbf E_0\Big\|_{\mathrm{op}}\|(\mathbf H_1^\dagger)^\top\Omega_1^{-1}a_1\|^2.
\end{align}
The second term is $O_p((p\wedge q)^{-1})$ using the Lemma~\ref{lem:E0bounds}(i), and also since $\|\mathbf H_1^\dagger\|_{\mathrm{op}}$, $\|\Omega_1^{-1}\|_{\mathrm{op}}$, and $\|a_1\|$ are $O_p(1)$. Now, we expand $v$. We substitute $\hat{\mathbf Z}=\mathbf F_1\mathbf H_1^\top+\mathbf E_0$ into $M=n_t^{-1}\hat{\mathbf Z}^\top\hat{\mathbf Z}$ and use $\widehat{\mathbf H_1a_1}=\mathbf H_1a_1+\mathbf S_{\tilde q}^\top(\widehat{\mathbf H_M\delta_0}-\mathbf H_M\delta_0)$. This gives
\begin{align*}
v
&=\mathbf H_1\Big(\Omega_1-\frac1{n_t}\mathbf F_1^\top\mathbf F_1\Big)\Omega_1^{-1}a_1-\frac1{n_t}\mathbf H_1\mathbf F_1^\top\mathbf E_0(\mathbf H_1^\dagger)^\top\Omega_1^{-1}a_1\\&\quad-\frac1{n_t}\mathbf E_0^\top\mathbf F_1\Omega_1^{-1}a_1-\frac1{n_t}\mathbf E_0^\top\mathbf E_0(\mathbf H_1^\dagger)^\top\Omega_1^{-1}a_1+\mathbf S_{\tilde q}^\top\big(\widehat{\mathbf H_M\delta_0}-\mathbf H_M\delta_0\big).
\end{align*}
For the component $U^\top v$, we use $U^\top\mathbf H_1=R$ and $\|R\|_{\mathrm{op}}=O_p(1)$. Here, the sample factor term is $O_p(n_t^{-1/2})$ by Part~3 of the proof of Theorem~\ref{thm:clt}, the next three terms are $O_p(\{n_t(p\wedge q)\}^{-1/2}+(p\wedge q)^{-1})$ by Lemma~\ref{lem:E0bounds}(i)-(ii), and the last term is controlled by \eqref{eq:ve-d-rate}. Hence, we have that
\begin{align*}
\|U^\top v\|&=O_p\left(\frac{1}{\sqrt{m}}+\frac{1}{\sqrt{p}}+\frac{1}{\sqrt{q}}\right).
\end{align*}
For the null-space component, $S^\top\mathbf H_1=0$ removes the first two terms. The third term is as $n_t^{-1}\widetilde{\mathbf E}_S^\top\mathbf F_1\Omega_1^{-1}a_1=O_p(n_t^{-1/2})$ by Step~2 of Lemma~\ref{lemma:nullspace}. The fourth term is $n_t^{-1}\widetilde{\mathbf E}_S^\top\mathbf E_0(\mathbf H_1^\dagger)^\top\Omega_1^{-1}a_1=O_p(p^{-1/2}+q^{-1/2})$ by the $K_2$ bound in that lemma with $\Omega_1^{-1}a_1$ in place of our $\Delta$. For the last term, the block form of $S$ gives us that
\begin{align*}
\Sigma_S^{-1/2}S^\top\mathbf S_{\tilde q}^\top\big(\widehat{\mathbf H_M\delta_0}-\mathbf H_M\delta_0\big)&=\begin{pmatrix}0\\\sqrt q\,\Omega_M^{-1/2}S_M^\top\widehat{\mathbf H_M\delta_0}\end{pmatrix},
\end{align*}
which is $O_p(m^{-1/2})$ using Equation \eqref{eq:ve-d-null}. So, we have that
\begin{align*}
\|\Sigma_S^{-1/2}S^\top v\|&=O_p\left(\frac{1}{\sqrt{m}}+\frac{1}{\sqrt{p}}+\frac{1}{\sqrt{q}}\right).
\end{align*}
We combine the two parts with Equations \eqref{eq:ve-quad} and \eqref{eq:ve-h1-split}. This gives us that
\begin{align}
\frac1{n_t}\sum_{i:A_i=1}|\hat h_{1,i}-h_{1,i}|^2&\le O_p(1)\big(\|U^\top v\|^2+\|\Sigma_S^{-1/2}S^\top v\|^2\big)+O_p\left(\frac1{p\wedge q}\right)=O_p\left(\frac1m+\frac1p+\frac1q\right).\label{eq:ve-weight-treated}
\end{align}
For control part, there is no estimated mediator intercept. Now, by applying the same decomposition with $e_{\mathbf 1}=\mathbf H_{X1}e_{1,1+\bar p}$, $(\mathbf H_{X1}^\dagger)^\top\Omega_{X0}^{-1}e_{1,1+\bar p}$, and the $X$-only versions of Equation \eqref{eq:ve-quad} and Lemma~\ref{lem:E0bounds} gives us that
\begin{align}
\frac1{n_c}\sum_{i:A_i=0}|\hat h_{0,i}-h_{0,i}|^2&=O_p\left(\frac1m+\frac1p\right).\label{eq:ve-weight-control}
\end{align}
To bound the treated weights, consider the transformed
vector
\begin{align*}
b=\begin{pmatrix}
U^\top\widehat{H_1a_1}\\\Sigma_S^{-1/2}S^\top\widehat{H_1a_1}\end{pmatrix}.
\end{align*}
By Equation~\eqref{eq:ve-d-rate} and the boundedness of
$\|H_1a_1\|$, the first block is $O_p(1)$. Since $S^\top H_1=0$, Equation~\eqref{eq:ve-d-null}, and the block structure of $\Sigma_S$ give $\|\Sigma_S^{-1/2}S^\top\widehat{\mathbf H_1a_1}\|=O_p(m^{-1/2})$.
Thus, $\|b\|=O_p(1)$.
Together with $\|\bar M^{-1}\|_{\mathrm{op}}=O_p(1)$, this gives
\[|\hat h_{1,i}|\le O_p(1)\bigl(\|\hat{\mathbf Z}_i\|+\|\tilde{\mathbf e}_{S,i}\|\bigr),\]
uniformly in $i$. Using Assumption~\ref{assump:subgaussian}(i), the true factors and the structural errors have sub-Gaussian projections with parameter at most $C$. The same holds for $\mathbf f_{M,i}$ which is a linear combination of $\mathbf f_{X,i}$ and $\mathbf u_{XM,a,i}$. Since $\mathbf W_X=\sqrt p\,\mathbf V_X$, each coordinate of $\mathbf e_{X_i}$ is $p^{-1/2}$ times a unit projection of $\mathbf u_{X,i}$. The $X$-block of $\tilde e_{S,i}$ is $\Omega_X^{-1/2}(\mathbf V_XS_X)^\top\mathbf u_{X,i}$, and $\mathbf V_XS_X$ has orthonormal columns. So, the coordinates of $\mathbf e_{0i}$ and $\tilde e_{S,i}$ are sub-Gaussian with parameters bounded by a constant. The mediator part is also handled in the same way. The number of coordinates is fixed, so a union bound over the $m$ observations gives us
\begin{align}
\max_{i\in\mathcal D_k:A_i=1}\bigl(|\hat h_{1,i}|^2+|\epsilon_{1,i}|^2\bigr)=O_p(\log m),\quad
\max_{i\in\mathcal D_k:A_i=0}\bigl(|\hat h_{0,i}|^2+|\eta_i|^2\bigr)=O_p(\log m).\label{eq:ve-max}
\end{align}
The control bound uses Equation \eqref{eq:ve-quad-control} in the same way that together with $\eta_i=\gamma_1^\top\mathbf u_{XM,0,i}$ and $\|\gamma_1\|=O(1)$. 

We compare the least-squares residuals with the structural errors, beginning with the treated observations. Using Equations \eqref{equation:yhat} and \eqref{equation:zhat}, the outcome satisfies the exact identity $\mathbf Y=\hat{\mathbf Z}\delta+\boldsymbol\epsilon_1-\mathbf E_0\delta$. Proposition~\ref{prop:loading-cond} and Assumption~\ref{Assumption:pervasiveness}(i) give $\|\delta\|=O_p(1)$. We know that the least-squares residuals are $\hat{\boldsymbol\epsilon}_1=(I-P_{\hat{\mathbf Z}})\mathbf Y$. Since $(I-P_{\hat{\mathbf Z}})\hat{\mathbf Z}=0$, we have that
\begin{align*}
\hat{\boldsymbol\epsilon}_1-\boldsymbol\epsilon_1&=(I-P_{\hat{\mathbf Z}})(\boldsymbol\epsilon_1-\mathbf E_0\delta)-\boldsymbol\epsilon_1=-(I-P_{\hat{\mathbf Z}})\mathbf E_0\delta-P_{\hat{\mathbf Z}}\boldsymbol\epsilon_1.
\end{align*}
The first term comes from the use of projected factors. By Lemma~\ref{lem:E0bounds}(i), $\|\mathbf E_0\delta\|^2\le\|\mathbf E_0\|_{\mathrm{op}}^2\|\delta\|^2=O_p(n_t(p\wedge q)^{-1})$. For the second term, the structural outcome errors are independent of the projected design, conditional on the training fold, and have bounded variance. Let $\mathcal T$ denote information from the training fol. The rank of $P_{\hat{\mathbf Z}}$ is at most $1+\tilde p+\tilde q$. Hence $\mathbb E(\|P_{\hat{\mathbf Z}}\boldsymbol\epsilon_1\|^2\mid\hat{\mathbf Z},\mathcal{T})\le C(1+\tilde p+\tilde q)$, and so $\|P_{\hat{\mathbf Z}}\boldsymbol\epsilon_1\|^2=O_p(1)$. An orthogonal projector does not increase the Euclidean norm. Therefore, we have that
\begin{align}
\frac1{n_t}\sum_{i:A_i=1}|\hat\epsilon_{1,i}-\epsilon_{1,i}|^2&\le\frac2{n_t}\|\mathbf E_0\delta\|^2+\frac2{n_t}\|P_{\hat{\mathbf Z}}\boldsymbol\epsilon_1\|^2\notag\\
&=O_p\left(\frac1{p\wedge q}\right)+O_p\left(\frac{1}{n_t}\right)\notag\\
&=O_p\left(\frac{1}m+\frac{1}p+\frac1q\right).\label{eq:ve-treated-residual}
\end{align}
Now, we turn to the control bridge. In the proof of Theorem~\ref{thm:clt}, Equation \eqref{eqbridge} and the expansion of $\widehat{\gamma_1^\top\delta_0}$ that follows it write the bridge response as $\mathbf b=\mathbf R\phi+\Delta_{\gamma,n}^\top\delta_0\cdot\mathbf 1_{n_c}+\hat{\mathbf R}\tilde c+\boldsymbol\eta+\mathcal R_n$. Since $\mathbf R=\hat{\mathbf R}-\mathbf E_{X0}$ and $\mathbf 1_{n_c}=\hat{\mathbf R}e_{\mathbf 1}$, then we have
\begin{align*}
\mathbf b&=\hat{\mathbf R}\big(\phi+\tilde c+\Delta_{\gamma,n}^\top\delta_0\,e_{\mathbf 1}\big)+\boldsymbol\eta+\mathcal R_n-\mathbf E_{X0}\phi.
\end{align*}
Since $(I-P_{\hat{\mathbf R}})\hat{\mathbf R}=0$, the control least-squares residuals satisfy
\begin{align*}
\hat{\boldsymbol\eta}-\boldsymbol\eta
&=(I-P_{\hat{\mathbf R}})\mathbf b-\boldsymbol\eta\\
&=(I-P_{\hat{\mathbf R}})(\boldsymbol\eta+\mathcal R_n-\mathbf E_{X0}\phi)-\boldsymbol\eta\\
&=(I-P_{\hat{\mathbf R}})(\mathcal R_n-\mathbf E_{X0}\phi)-P_{\hat{\mathbf R}}\boldsymbol\eta.
\end{align*}
We bound the three pieces. Here, Part~2 of the proof of Theorem~\ref{thm:clt} gives us that $\|\phi\|=O_p(1)$, and the $X$-only version of Lemma~\ref{lem:E0bounds}(i) gives us $\|\mathbf E_{X0}\|_{\mathrm{op}}^2=O_p(n_c/p)$. So, we have that $n_c^{-1}\|\mathbf E_{X0}\phi\|^2=O_p(p^{-1})$.

The remainder has entries $(\mathcal R_n)_i=\Delta_{\gamma,n}^\top\mathbf u_{XM,0,i}-\Delta_{\gamma,n}^\top\mathbf B_0\mathbf H_X^\dagger\mathbf e_{X_i}+\hat\gamma_1^\top\mathbf e_{M_i}$. Now, conditional on the training fold and the treated observations, $\Delta_{\gamma,n}$ and $\hat\gamma_1$ are fixed and the control observations are independent. Lemma~\ref{lemma:approxgamma1} gives $\|\Delta_{\gamma,n}\|=O_p(n_t^{-1/2})$, so the first two entries have empirical second moments of order $O_p(n_t^{-1})$. For the third entry, we split $\hat\gamma_1$ as in \eqref{eq:whitening-split}. The row-space part has second moment $O_p(q^{-1})$, since $\|\mathbf H_M^\dagger\|_{\mathrm{op}}=O_p(1)$ and $\mathbb E\|\mathbf e_{M_i}\|^2=O(q^{-1})$. The null-space part has conditional variance $O_p(n_t^{-1})$ by \eqref{eq:null-whitening-var}.  Hence, we have that
\begin{align*}
\frac1{n_c}\|\mathcal R_n\|^2
&=O_p\left(\frac1{n_t}\right)+O_p\left(\frac1q\right)+O_p\left(\frac{1}{n_t}\right)=O_p\left(\frac1{n_t}+\frac{1}{q}\right).
\end{align*}
Using Assumption~\ref{Assumption:covariances}(ii), the $\eta_i$ are also independent of the control projected design as mentioned earlier, conditional on the training fold and the treated observations, and have bounded variance. The rank of $P_{\hat{\mathbf R}}$ is at most $1+\tilde p$. So, we have that $\|P_{\hat{\mathbf R}}\boldsymbol\eta\|^2=O_p(1)$. We combine the three bounds with the contraction property of an orthogonal projector. This gives us that
\begin{align}
\frac1{n_c}\sum_{i:A_i=0}|\hat\eta_i-\eta_i|^2
&\le\frac4{n_c}\|\mathcal R_n\|^2+\frac4{n_c}\|\mathbf E_{X0}\phi\|^2+\frac2{n_c}\|P_{\hat{\mathbf R}}\boldsymbol\eta\|^2\notag\\
&=O_p\left(\frac1{n_t}+\frac1{q}\right)+O_p\left(\frac1p\right)+O_p\left(\frac1{n_c}\right)\notag\\
&=O_p\left(\frac1m+\frac1p+\frac1q\right).\label{eq:ve-control-residual}
\end{align}
Now, to the estimated influence function. For treated observations,
\begin{align*}
\hat h_{1,i}\hat\epsilon_{1,i}-h_{1,i}\epsilon_{1,i}
&=\hat h_{1,i}(\hat\epsilon_{1,i}-\epsilon_{1,i})+(\hat h_{1,i}-h_{1,i})\epsilon_{1,i}.
\end{align*}
We apply $(x+y)^2\le2x^2+2y^2$ and then we have from the Equations \eqref{eq:ve-weight-treated}, \eqref{eq:ve-max}, and \eqref{eq:ve-treated-residual} that
\begin{align*}
\frac1{n_t}\sum_{i:A_i=1}|\hat h_{1,i}\hat\epsilon_{1,i}-h_{1,i}\epsilon_{1,i}|^2
&\le\frac2{n_t}\sum_{i:A_i=1}|\hat h_{1,i}|^2|\hat\epsilon_{1,i}-\epsilon_{1,i}|^2+\frac2{n_t}\sum_{i:A_i=1}|\hat h_{1,i}-h_{1,i}|^2|\epsilon_{1,i}|^2\\
&\le2\max_{i:A_i=1}|\hat h_{1,i}|^2\frac1{n_t}\sum_{i:A_i=1}|\hat\epsilon_{1,i}-\epsilon_{1,i}|^2+2\max_{i:A_i=1}|\epsilon_{1,i}|^2\frac1{n_t}\sum_{i:A_i=1}|\hat h_{1,i}-h_{1,i}|^2\\&=O_p\left(\log m\,\left(\frac1m+\frac1p+\frac1q\right)\right).
\end{align*}
Now, using the same argument for the control part with Equations \eqref{eq:ve-weight-control} and \eqref{eq:ve-control-residual}, gives us that
\begin{align*}
\frac1{n_c}\sum_{i:A_i=0}|\hat h_{0,i}\hat\eta_i-h_{0,i}\eta_i|^2
&=O_p\left(\log m\,\left(\frac1m+\frac1p+\frac1q\right)\right).
\end{align*}
The last equality uses Assumption~\ref{assump:growth} and $m\asymp n$. Since $\hat\pi_1\xrightarrow{p}\pi_1$ and $\hat\pi_0\xrightarrow{p}\pi_0$ where $\pi_1,\pi_0>0$ using Assumption~\ref{assump:subgaussian}(iii), and the empirical second moments of $\xi_{1,i}$ and $\xi_{2,i}$ are bounded in probability, replacing $\pi_1$ and $\pi_0$ by $\hat\pi_1$ and $\hat\pi_0$, we get that
\begin{align}\label{eq:ve-IF-L2}
\frac1m\sum_{i\in\mathcal D_k}(\hat\psi_{k,i}-\psi_i)^2&=o_p(1).
\end{align}
Also, the same argument applies to the other fold. There are only two folds, so we have that
\begin{align}\label{eq:ve-IF-L2-full}
\frac1n\sum_{k=1}^2\sum_{i\in\mathcal D_k}(\hat\psi_{k,i}-\psi_i)^2&=o_p(1).
\end{align}
Now, when we compare the empirical second moments by adding and subtracting $n^{-1}\sum_{i=1}^n\psi_i^2$, we get that
\begin{align*}
\hat\sigma^2-\sigma^2
&=\frac1n\sum_{k=1}^2\sum_{i\in\mathcal D_k}(\hat\psi_{k,i}^2-\psi_i^2)+\Big(\frac1n\sum_{i=1}^n\psi_i^2-\sigma^2\Big).
\end{align*}
Using the Steps~4 and~6 of Part~3 in the proof of Theorem~\ref{thm:clt}, the $\psi_i$ are i.i.d.\ with mean zero, and Equation \eqref{eq:var-final} gives $\mathbb E(\psi_i^2)=\sigma^2$. Therefore, the law of large numbers gives $n^{-1}\sum_{i=1}^n\psi_i^2-\sigma^2=o_p(1)$. For the first term, we simply use $x^2-y^2=(x-y)(x+y)$ and using the Cauchy-Schwarz inequality. This gives us that
\begin{align*}
\left|\frac1n\sum_{k=1}^2\sum_{i\in\mathcal D_k}(\hat\psi_{k,i}^2-\psi_i^2)\right|&\le\left\{\frac1n\sum_{k=1}^2\sum_{i\in\mathcal D_k}(\hat\psi_{k,i}-\psi_i)^2\right\}^{1/2}\left\{\frac1n\sum_{k=1}^2\sum_{i\in\mathcal D_k}(\hat\psi_{k,i}+\psi_i)^2\right\}^{1/2}.
\end{align*}
the first term is $o_p(1)$ using Equation \eqref{eq:ve-IF-L2-full}. For the second factor, we write $x+y=(x-y)+2y$, so that $(x+y)^2\le2(x-y)^2+8y^2$. Then
\begin{align*}
\frac1n\sum_{k=1}^2\sum_{i\in\mathcal D_k}(\hat\psi_{k,i}+\psi_i)^2&\le\frac2n\sum_{k=1}^2\sum_{i\in\mathcal D_k}(\hat\psi_{k,i}-\psi_i)^2+\frac8n\sum_{i=1}^n\psi_i^2=O_p(1).
\end{align*}
Therefore, the first term is $o_p(1)$, and hence, we have $\hat\sigma^2-\sigma^2=o_p(1)$. Finally, Theorem~\ref{thm:clt} gives us that $\sqrt n(\hat\theta_0^{\mathrm{cf}}-\theta_0)\overset{d}{\to} N(0,\sigma^2)$, and its proof shows $\sigma^2\ge c>0$. So, the consistency of $\hat\sigma^2$ implies $\hat\sigma\overset{p}{\to}\sigma$. Using Slutsky's theorem, we get that
\begin{align*}
\frac{\sqrt n(\hat\theta_0^{\mathrm{cf}}-\theta_0)}{\hat\sigma}&=\frac{\sqrt n(\hat\theta_0^{\mathrm{cf}}-\theta_0)}{\sigma}\frac{\sigma}{\hat\sigma}\overset{d}{\to} N(0,1).
\end{align*}
\end{proof}

\section{Auxiliary Material}

\subsection{Technical Lemmas}\label{sec:technical-lemmas}

\begin{lemma}[Inverse of an augmented second-moment matrix]\label{lemma:augmented-inverse}
Suppose Assumptions \ref{Assumption:covariances} and \ref{Assumption:pervasiveness} hold. Let $\mathbf H = \text{diag}(\mathbf{H}_X, \mathbf{H}_M)$ where $\mathbf{H}_X$ and $\mathbf{H}_M$ be defined as in Definition \ref{definition:H}, and $\mathbf e = (\mathbf{e}_X^\top \quad \mathbf{e}_M^\top)^\top$ be a vector with mean zero, covariance $\Sigma_{\mathbf e}\succ0$ as defined in Section \ref{sec:prelim-notation}. Set $L=(1,(\mathbf H\mathbf f+\mathbf e)^\top)^\top$, then $M:=\mathbb E[LL^\top\mid A=1]$ is invertible, with
\begin{align*}
M^{-1}&=\begin{pmatrix}1+\mu_{\mathbf f\mid1}^\top\mathbf H^\top(\mathbf H\Sigma_{\mathbf f\mid1}\mathbf H^\top+\Sigma_{\mathbf e})^{-1}\mathbf H\mu_{\mathbf f\mid1} & -\mu_{\mathbf f\mid1}^\top\mathbf H^\top(\mathbf H\Sigma_{\mathbf f\mid1}\mathbf H^\top+\Sigma_{\mathbf e})^{-1} \\
-(\mathbf H\Sigma_{\mathbf f\mid1}\mathbf H^\top+\Sigma_{\mathbf e})^{-1}\mathbf H\mu_{\mathbf f\mid1} & (\mathbf H\Sigma_{\mathbf f\mid1}\mathbf H^\top+\Sigma_{\mathbf e})^{-1}\end{pmatrix}.
\end{align*}
\end{lemma}

\begin{proof}
Using Assumption \ref{Assumption:covariances}(ii), we know that $\mathbf{e}$ do not depend on $A$, so $\mathbb E[\mathbf e\mid A=1]=\mathbb E[\mathbf e]=0$, $\mathbb E[\mathbf e\mathbf e^\top\mid A=1]=\mathbb E[\mathbf e\mathbf e^\top]=\Sigma_{\mathbf e}$ and $\mathbb E[\mathbf f\mathbf e^\top\mid A=1]=0$.  From $L=(1,(\mathbf H\mathbf f+\mathbf e)^\top)^\top$, we have
\begin{align*}
LL^\top&=\begin{pmatrix}1 & (\mathbf H\mathbf f+\mathbf e)^\top \\
\mathbf H\mathbf f+\mathbf e & (\mathbf H\mathbf f+\mathbf e)(\mathbf H\mathbf f+\mathbf e)^\top\end{pmatrix}.
\end{align*}
Now, expanding the bottom-right block,
\begin{align*}
&(\mathbf H\mathbf f+\mathbf e)(\mathbf H\mathbf f+\mathbf e)^\top=\mathbf H\mathbf f \ \mathbf f^\top\mathbf H^\top+\mathbf H\mathbf f\mathbf e^\top+\mathbf e\mathbf f^\top\mathbf H^\top+\mathbf e\mathbf e^\top.
\end{align*}
We take the conditional expectation given $A=1$ entrywise. In the linear blocks, we have that
\begin{align*}
&\mathbb E[\mathbf H\mathbf f+\mathbf e\mid A=1]=\mathbf H\,\mathbb E[\mathbf f\mid A=1]+\mathbb E[\mathbf e]=\mathbf H\mu_{\mathbf f\mid1},
\end{align*}
using $\mathbb E[\mathbf e]=0$. Also, the two cross terms satisfy
\begin{align*}
&\mathbb E[\mathbf H\mathbf f\mathbf e^\top\mid A=1]=\mathbf H\,\mathbb E[\mathbf f\mathbf e^\top\mid A=1]=0, \qquad \mathbb E[\mathbf e\mathbf f^\top\mathbf H^\top\mid A=1]=\mathbb E[\mathbf f\mathbf e^\top\mid A=1]^\top\mathbf H^\top=0,
\end{align*}
as $\mathbb E[\mathbf f\mathbf e^\top\mid A=1]=0$, and the remaining noise term is $\mathbb E[\mathbf e\mathbf e^\top\mid A=1]=\mathbb E[\mathbf e\mathbf e^\top]=\Sigma_{\mathbf e}$. So, collecting these gives us
\begin{align*}
M&=\begin{pmatrix}1 & \mu_{\mathbf f\mid1}^\top\mathbf H^\top \\
\mathbf H\mu_{\mathbf f\mid1} & \mathbf H\,\mathbb E[\mathbf f\mathbf f^\top\mid A=1]\,\mathbf H^\top+\Sigma_{\mathbf e}\end{pmatrix}.
\end{align*}
We write $A_{11}=1$, $A_{12}=\mu_{\mathbf f\mid1}^\top\mathbf H^\top$, and $A_{22}=\mathbf H\,\mathbb E[\mathbf f \   \mathbf f^\top\mid A=1]\,\mathbf H^\top+\Sigma_{\mathbf e}$.
The Schur complement of $A_{11}$ is
\begin{align*}
A_{22}-A_{12}^\top A_{11}^{-1}A_{12}
&=\mathbf H\,\mathbb E[\mathbf f \   \mathbf f^\top\mid A=1]\,\mathbf H^\top+\Sigma_{\mathbf e} -\mathbf H\mu_{\mathbf f\mid1}\mu_{\mathbf f\mid1}^\top\mathbf H^\top \\
&=\mathbf H\big(\mathbb E[\mathbf f \  \mathbf f^\top\mid A=1]-\mu_{\mathbf f\mid1}\mu_{\mathbf f\mid1}^\top\big)\mathbf H^\top+\Sigma_{\mathbf e} \\
&=\mathbf H\Sigma_{\mathbf f\mid1}\mathbf H^\top+\Sigma_{\mathbf e},
\end{align*}
using $\Sigma_{\mathbf f\mid1}=\mathbb E[\mathbf f\  \mathbf f^\top\mid A=1]-\mu_{\mathbf f\mid1}\mu_{\mathbf f\mid1}^\top$ in the last line. This is positive definite, so $M$ is invertible. Applying
block-inverse formula with these blocks, we get
\begin{align*}
B_{22}&=(\mathbf H\Sigma_{\mathbf f\mid1}\mathbf H^\top+\Sigma_{\mathbf e})^{-1}, \\
B_{12}&=-\mu_{\mathbf f\mid1}^\top\mathbf H^\top(\mathbf H\Sigma_{\mathbf f\mid1}\mathbf H^\top+\Sigma_{\mathbf e})^{-1}, \\
B_{11}&=1+\mu_{\mathbf f\mid1}^\top\mathbf H^\top(\mathbf H\Sigma_{\mathbf f\mid1}\mathbf H^\top+\Sigma_{\mathbf e})^{-1}\mathbf H\mu_{\mathbf f\mid1},
\end{align*}
which is the formula.
\end{proof}

\begin{lemma}[Woodbury matrix identity]\label{lemma:woodbury}
Let $A\in\mathbb R^{n\times n}$ and $C\in\mathbb R^{k\times k}$ be invertible, and let $U\in\mathbb R^{n\times k}$, $V\in\mathbb R^{k\times n}$. If $C^{-1}+VA^{-1}U$ is invertible, then $A+UCV$ is invertible and
\begin{align*}
(A+UCV)^{-1}&=A^{-1}-A^{-1}U\big(C^{-1}+VA^{-1}U\big)^{-1}VA^{-1}.
\end{align*}
\end{lemma}

\begin{proof}
We verify the right-hand side is a right inverse of $A+UCV$. Since both matrices are square, a right inverse is a two-sided inverse. We have
\begin{align*}
&(A+UCV)\big[A^{-1}-A^{-1}U(C^{-1}+VA^{-1}U)^{-1}VA^{-1}\big] \\*
&=I+UCVA^{-1}-U(C^{-1}+VA^{-1}U)^{-1}VA^{-1}
-UCVA^{-1}U(C^{-1}+VA^{-1}U)^{-1}VA^{-1} \\
&=I+UCVA^{-1}-U\big[I+CVA^{-1}U\big](C^{-1}+VA^{-1}U)^{-1}VA^{-1} \\
&\overset{(1)}{=}I+UCVA^{-1}-UC\big[C^{-1}+VA^{-1}U\big](C^{-1}+VA^{-1}U)^{-1}VA^{-1} \\*
&=I+UCVA^{-1}-UCVA^{-1} =I,
\end{align*}
Here, $(1)$ follows simply using $C[C^{-1}+VA^{-1}U](C^{-1}+VA^{-1}U)^{-1}=C$.
\end{proof}

\subsection{Calculation of Effects \label{effects}}

Throughout this subsection, we center the latent factors so that $\mathbb E[\mathbf f_X]=\mathbb E[\mathbf f_M]=0$. The intercepts $\alpha_a$ and $\delta_a$ therefore account for the corresponding mean levels. Note that the arm-specific conditional means $\mu_{\mathbf f\mid1}$ and $\mu_{\mathbf f_X\mid0}$ used elsewhere in the Appendix can still be nonzero.

By the regression model in \eqref{eq:factor}, for $a\in\{0,1\}$,
\begin{align*}\mathbb E[Y\mid\mathbf f_X,\mathbf f_M,A=a]=\alpha_a+\beta_a^\top\mathbf f_X+\gamma_a^\top\mathbf f_M,\qquad \mathbb E[\mathbf f_M\mid\mathbf f_X,A=a]=\delta_a+B_a\mathbf f_X.\end{align*}
For $Y^{(a,
\mathbf{f}_M^{(a')})}$, we first average over the mediator distribution under $A=a'$ conditional on $\mathbf f_X$, and then over the marginal distribution of $\mathbf f_X$. Therefore,
\begin{align*}\begin{aligned}
\mathbb E\big[Y^{(a,\mathbf{f}_M^{(a')})}\big]
&=\alpha_a+\mathbb E[\mathbf f_X]^\top\beta_a+\mathbb E_{\mathbf f_X}[\mathbb E[\mathbf f_M\mid\mathbf f_X,A=a']]^\top\gamma_a\\
&=(\alpha_a+\delta_{a'}^\top\gamma_a)+\mathbb E[\mathbf f_X]^\top(\beta_a+B_{a'}^\top\gamma_a)\\
&=\alpha_a+\delta_{a'}^\top\gamma_a,
\end{aligned}\end{align*}
where the last equality follows from $\mathbb E[\mathbf f_X]=0$.

For the four choices of $(a,a')$, we have
\begin{align*}
\mathbb E[Y^{(1,\mathbf{f}_M^{(1)})}]  &= \alpha_1 + \mathbb E[\mathbf f_X]^\top \beta_1 + \mathbb E_{\mathbf f_X}[\mathbb E [\mathbf  f_M \mid \mathbf f_X, A=1]]^\top \gamma_1\\&= (\alpha_1 + \delta_1^\top \gamma_1) + \mathbb E[\mathbf f_X]^\top (\beta_1 + B_1^\top \gamma_1)\\
&= \alpha_1 + \delta_1^\top \gamma_1\,,
\\
\mathbb E[Y^{(0,\mathbf{f}_M^{(0)})}]  &= \alpha_0 + \mathbb E[\mathbf f_X]^\top \beta_0 + \mathbb E_{\mathbf f_X}[\mathbb E[\mathbf f_M \mid \mathbf f_X, A = 0]]^\top \gamma_0\\&= (\alpha_0 + \delta_0^\top \gamma_0) + \mathbb E[\mathbf f_X]^\top (\beta_0 + B_0^\top \gamma_0)\\&= \alpha_0 + \delta_0^\top \gamma_0\,,
\\
\mathbb E[Y^{(1,\mathbf{f}_M^{(0)})}]  &= \alpha_1 + \mathbb E[\mathbf f_X]^\top \beta_1 + \mathbb E_{\mathbf f_X}[\mathbb E[\mathbf f_M \mid \mathbf f_X, A = 0]]^\top \gamma_1\\&= (\alpha_1 + \delta_0^\top \gamma_1) + \mathbb E[\mathbf f_X]^\top (\beta_1 + B_0^\top \gamma_1)\\&= \alpha_1 + \delta_0^\top \gamma_1\,,
\\
\mathbb E[Y^{(0,\mathbf{f}_M^{(1)})}] &= \alpha_0 + \mathbb E[\mathbf f_X]^\top \beta_0 + \mathbb E_{\mathbf f_X}[\mathbb E[\mathbf f_M \mid \mathbf f_X, A = 1]]^\top \gamma_0\\&= (\alpha_0 + \delta_1^\top \gamma_0) + \mathbb E[\mathbf f_X]^\top (\beta_0 + B_1^\top \gamma_0)\\&= \alpha_0 + \delta_1^\top \gamma_0\,.
\end{align*}
Note that $\mathbb E[Y^{(1,\mathbf{f}_M^{(0)})}]=\alpha_1+\delta_0^\top\gamma_1$ is exactly the estimand $\theta_0=\alpha_1+\gamma_1^\top\delta_0$ studied in Theorem~\ref{thm:id} and throughout the rest of this Appendix.

The natural indirect effect holds exposure fixed at treatment and varies the mediator between its two natural values
\begin{align*}
\text{NIE} &= \mathbb E[Y^{(1,\mathbf{f}_M^{(1)})}] - \mathbb E[Y^{(1,\mathbf{f}_M^{(0)})}] = (\alpha_1 + \delta_1^\top \gamma_1) - (\alpha_1 + \delta_0^\top \gamma_1) = (\delta_1 - \delta_0)^\top \gamma_1 \,.
\end{align*}
The natural direct effect holds the mediator fixed at its control-arm natural value and varies exposure
\begin{align*}
\text{NDE} & = \mathbb E[Y^{(1,\mathbf{f}_M^{(0)})}] - \mathbb E[Y^{(0,\mathbf{f}_M^{(0)})}]  = (\alpha_1 + \delta_0^\top \gamma_1) - (\alpha_0 + \delta_0^\top \gamma_0)= \alpha_1 - \alpha_0 + \delta_0^\top (\gamma_1 - \gamma_0)\,.
\end{align*}
The remaining potential outcome, $\mathbb E[Y^{(0,\mathbf{f}_M^{(1)})}]=\alpha_0+\delta_1^\top\gamma_0$, is not needed for NIE/NDE above but gives the complementary decomposition, $\text{TDE}=\mathbb E[Y^{(1,\mathbf{f}_M^{(1)})}]-\mathbb E[Y^{(0,\mathbf{f}_M^{(1)})}]$ and $\text{TIE}=\mathbb E[Y^{(0,\mathbf{f}_M^{(1)})}]-\mathbb E[Y^{(0,\mathbf{f}_M^{(0)})}]$, which satisfies the same identity $\text{TDE}+\text{TIE}=\text{NDE}+\text{NIE}$.

\subsection{More on Simulation Study and Sensitivity Analysis \label{sensitivity}}

\begin{table}[ht!]
\centering
\small
\begin{tabular}{cc cccc cccc}
\toprule
 &  & \multicolumn{4}{c}{$\sigma = 0.1$} & \multicolumn{4}{c}{$\sigma = 1$} \\
\cmidrule(lr){3-6} \cmidrule(lr){7-10}
$p = q$ & $n$ & RMSE & SD & CI Length & Coverage & RMSE & SD & CI Length & Coverage \\
\midrule
\multirow{5}{*}{500} & 500 & 0.1988 & 0.1990 & 0.7799 & 0.956 & 0.2209 & 0.2211 & 0.8665 & 0.944 \\
 & 1000 & 0.1493 & 0.1494 & 0.5855 & 0.942 & 0.1496 & 0.1497 & 0.5867 & 0.954 \\
 & 1500 & 0.1156 & 0.1157 & 0.4534 & 0.948 & 0.1207 & 0.1206 & 0.4729 & 0.956 \\
 & 2000 & 0.1046 & 0.1046 & 0.4098 & 0.936 & 0.1060 & 0.1043 & 0.4089 & 0.948 \\
 & 2500 & 0.0851 & 0.0849 & 0.3330 & 0.954 & 0.0939 & 0.0937 & 0.3674 & 0.954 \\
\midrule
\multirow{5}{*}{1000} & 500 & 0.1645 & 0.1646 & 0.6452 & 0.950 & 0.1355 & 0.1352 & 0.5299 & 0.950 \\
 & 1000 & 0.1078 & 0.1075 & 0.4214 & 0.952 & 0.0937 & 0.0938 & 0.3678 & 0.952 \\
 & 1500 & 0.0923 & 0.0924 & 0.3621 & 0.948 & 0.0747 & 0.0745 & 0.2920 & 0.944 \\
 & 2000 & 0.0813 & 0.0814 & 0.3189 & 0.950 & 0.0651 & 0.0652 & 0.2556 & 0.942 \\
 & 2500 & 0.0746 & 0.0746 & 0.2923 & 0.954 & 0.0557 & 0.0557 & 0.2185 & 0.948 \\
\midrule
\multirow{5}{*}{2500} & 500 & 0.1737 & 0.1738 & 0.6814 & 0.954 & 0.2758 & 0.2760 & 1.0818 & 0.964 \\
 & 1000 & 0.1266 & 0.1267 & 0.4968 & 0.948 & 0.1900 & 0.1897 & 0.7436 & 0.956 \\
 & 1500 & 0.0951 & 0.0952 & 0.3731 & 0.930 & 0.1575 & 0.1574 & 0.6169 & 0.936 \\
 & 2000 & 0.0887 & 0.0888 & 0.3480 & 0.948 & 0.1265 & 0.1266 & 0.4961 & 0.956 \\
 & 2500 & 0.0763 & 0.0763 & 0.2993 & 0.956 & 0.1205 & 0.1206 & 0.4726 & 0.952 \\
\midrule
\multirow{5}{*}{5000} & 500 & 0.1754 & 0.1755 & 0.6880 & 0.952 & 0.2106 & 0.2108 & 0.8264 & 0.952 \\
 & 1000 & 0.1190 & 0.1191 & 0.4668 & 0.958 & 0.1323 & 0.1324 & 0.5188 & 0.940 \\
 & 1500 & 0.1012 & 0.1013 & 0.3970 & 0.948 & 0.1077 & 0.1078 & 0.4227 & 0.942 \\
 & 2000 & 0.0880 & 0.0872 & 0.3417 & 0.942 & 0.0976 & 0.0977 & 0.3830 & 0.956 \\
 & 2500 & 0.0791 & 0.0792 & 0.3103 & 0.950 & 0.0852 & 0.0850 & 0.3332 & 0.968 \\
\bottomrule
\end{tabular}
\caption{Performance of $\hat\theta_0$ under variance-matched $t_6$ errors, over $500$ replications.}
\label{tab:simu_heavy}
\end{table}

\footnotesize
\setlength{\tabcolsep}{3pt}
\renewcommand{\arraystretch}{1.0}
\setlength{\LTleft}{\fill}
\setlength{\LTright}{\fill}
\captionsetup[longtable]{font=small}

\begin{longtable}{@{}ccc*{8}{r}@{}}
\caption{Performance of $\hat\theta_0$ as the strength of the mediated pathway varies, $\gamma_1 = \gamma_0 = c\,g$, over $500$ replications.}
\label{tab:simu_signal}\\
\toprule
 &  &  & \multicolumn{4}{c}{$\sigma = 0.1$}
 & \multicolumn{4}{c}{$\sigma = 1$} \\
\cmidrule(lr){4-7}
\cmidrule(lr){8-11}
$p = q$ & $c$ & $n$
& RMSE & SD & CI Length & Coverage
& RMSE & SD & CI Length & Coverage \\
\midrule
\endfirsthead

\multicolumn{11}{c}{\tablename~\thetable\ --- continued}\\
\toprule
 &  &  & \multicolumn{4}{c}{$\sigma = 0.1$}
 & \multicolumn{4}{c}{$\sigma = 1$} \\
\cmidrule(lr){4-7}
\cmidrule(lr){8-11}
$p = q$ & $c$ & $n$
& RMSE & SD & CI Length & Coverage
& RMSE & SD & CI Length & Coverage \\
\midrule
\endhead

\midrule
\multicolumn{11}{r}{Continued on next page}\\
\endfoot

\bottomrule
\endlastfoot

\multirow{5}{*}{1000} & \multirow{5}{*}{0}
& 500  & 0.0085 & 0.0085 & 0.0333 & 0.948 & 0.0859 & 0.0859 & 0.3365 & 0.950 \\*
& & 1000 & 0.0056 & 0.0056 & 0.0219 & 0.956 & 0.0585 & 0.0585 & 0.2292 & 0.950 \\*
& & 1500 & 0.0047 & 0.0047 & 0.0185 & 0.950 & 0.0469 & 0.0468 & 0.1836 & 0.952 \\*
& & 2000 & 0.0039 & 0.0039 & 0.0152 & 0.956 & 0.0393 & 0.0384 & 0.1506 & 0.952 \\*
& & 2500 & 0.0036 & 0.0036 & 0.0142 & 0.950 & 0.0371 & 0.0372 & 0.1456 & 0.954 \\*
\cmidrule(lr){2-3}

\multirow{5}{*}{1000} & \multirow{5}{*}{0.1}
& 500  & 0.0204 & 0.0203 & 0.0797 & 0.958 & 0.0840 & 0.0840 & 0.3294 & 0.958 \\*
& & 1000 & 0.0145 & 0.0145 & 0.0570 & 0.938 & 0.0590 & 0.0590 & 0.2311 & 0.948 \\*
& & 1500 & 0.0112 & 0.0112 & 0.0438 & 0.946 & 0.0465 & 0.0465 & 0.1823 & 0.946 \\*
& & 2000 & 0.0101 & 0.0101 & 0.0395 & 0.942 & 0.0395 & 0.0394 & 0.1544 & 0.950 \\*
& & 2500 & 0.0085 & 0.0085 & 0.0333 & 0.948 & 0.0366 & 0.0366 & 0.1433 & 0.958 \\*
\cmidrule(lr){2-3}

\multirow{5}{*}{1000} & \multirow{5}{*}{0.25}
& 500  & 0.0433 & 0.0432 & 0.1692 & 0.950 & 0.0938 & 0.0939 & 0.3681 & 0.948 \\*
& & 1000 & 0.0317 & 0.0317 & 0.1242 & 0.946 & 0.0635 & 0.0635 & 0.2488 & 0.958 \\*
& & 1500 & 0.0250 & 0.0251 & 0.0982 & 0.948 & 0.0569 & 0.0570 & 0.2234 & 0.952 \\*
& & 2000 & 0.0225 & 0.0225 & 0.0881 & 0.948 & 0.0452 & 0.0453 & 0.1774 & 0.944 \\*
& & 2500 & 0.0200 & 0.0200 & 0.0784 & 0.954 & 0.0399 & 0.0398 & 0.1561 & 0.940 \\*
\cmidrule(lr){2-3}

\multirow{5}{*}{1000} & \multirow{5}{*}{0.5}
& 500  & 0.0908 & 0.0908 & 0.3561 & 0.952 & 0.1164 & 0.1165 & 0.4567 & 0.944 \\*
& & 1000 & 0.0632 & 0.0632 & 0.2479 & 0.948 & 0.0879 & 0.0879 & 0.3447 & 0.944 \\*
& & 1500 & 0.0531 & 0.0531 & 0.2083 & 0.942 & 0.0699 & 0.0700 & 0.2743 & 0.954 \\*
& & 2000 & 0.0463 & 0.0463 & 0.1817 & 0.952 & 0.0611 & 0.0610 & 0.2392 & 0.940 \\*
& & 2500 & 0.0382 & 0.0382 & 0.1499 & 0.946 & 0.0535 & 0.0533 & 0.2089 & 0.960 \\*
\cmidrule(lr){2-3}

\multirow{5}{*}{1000} & \multirow{5}{*}{1}
& 500  & 0.1885 & 0.1886 & 0.7394 & 0.958 & 0.1942 & 0.1941 & 0.7610 & 0.952 \\*
& & 1000 & 0.1263 & 0.1263 & 0.4950 & 0.950 & 0.1442 & 0.1443 & 0.5658 & 0.950 \\*
& & 1500 & 0.0999 & 0.1000 & 0.3920 & 0.944 & 0.1120 & 0.1121 & 0.4393 & 0.940 \\*
& & 2000 & 0.0886 & 0.0884 & 0.3464 & 0.950 & 0.0966 & 0.0963 & 0.3776 & 0.954 \\*
& & 2500 & 0.0858 & 0.0857 & 0.3361 & 0.960 & 0.0835 & 0.0835 & 0.3275 & 0.956 \\
\midrule

\multirow{5}{*}{2500} & \multirow{5}{*}{0}
& 500  & 0.0093 & 0.0093 & 0.0364 & 0.950 & 0.0965 & 0.0966 & 0.3785 & 0.954 \\*
& & 1000 & 0.0066 & 0.0066 & 0.0257 & 0.942 & 0.0658 & 0.0659 & 0.2583 & 0.956 \\*
& & 1500 & 0.0050 & 0.0050 & 0.0195 & 0.962 & 0.0508 & 0.0504 & 0.1977 & 0.948 \\*
& & 2000 & 0.0044 & 0.0044 & 0.0173 & 0.946 & 0.0429 & 0.0429 & 0.1681 & 0.958 \\*
& & 2500 & 0.0040 & 0.0039 & 0.0154 & 0.952 & 0.0390 & 0.0390 & 0.1529 & 0.944 \\*
\cmidrule(lr){2-3}

\multirow{5}{*}{2500} & \multirow{5}{*}{0.1}
& 500  & 0.0185 & 0.0185 & 0.0726 & 0.948 & 0.0936 & 0.0937 & 0.3672 & 0.944 \\*
& & 1000 & 0.0132 & 0.0132 & 0.0518 & 0.948 & 0.0655 & 0.0655 & 0.2568 & 0.942 \\*
& & 1500 & 0.0103 & 0.0103 & 0.0405 & 0.962 & 0.0528 & 0.0527 & 0.2066 & 0.948 \\*
& & 2000 & 0.0092 & 0.0092 & 0.0361 & 0.950 & 0.0467 & 0.0467 & 0.1832 & 0.960 \\*
& & 2500 & 0.0085 & 0.0085 & 0.0334 & 0.960 & 0.0408 & 0.0408 & 0.1599 & 0.946 \\*
\cmidrule(lr){2-3}

\multirow{5}{*}{2500} & \multirow{5}{*}{0.25}
& 500  & 0.0410 & 0.0410 & 0.1606 & 0.940 & 0.1009 & 0.1010 & 0.3959 & 0.948 \\*
& & 1000 & 0.0321 & 0.0321 & 0.1258 & 0.958 & 0.0741 & 0.0740 & 0.2900 & 0.938 \\*
& & 1500 & 0.0236 & 0.0237 & 0.0927 & 0.952 & 0.0579 & 0.0580 & 0.2272 & 0.952 \\*
& & 2000 & 0.0195 & 0.0196 & 0.0767 & 0.952 & 0.0468 & 0.0468 & 0.1835 & 0.952 \\*
& & 2500 & 0.0186 & 0.0187 & 0.0731 & 0.942 & 0.0403 & 0.0403 & 0.1579 & 0.952 \\*
\cmidrule(lr){2-3}

\multirow{5}{*}{2500} & \multirow{5}{*}{0.5}
& 500  & 0.0831 & 0.0830 & 0.3254 & 0.948 & 0.1314 & 0.1315 & 0.5155 & 0.952 \\*
& & 1000 & 0.0597 & 0.0598 & 0.2342 & 0.950 & 0.0808 & 0.0808 & 0.3169 & 0.942 \\*
& & 1500 & 0.0467 & 0.0467 & 0.1832 & 0.950 & 0.0684 & 0.0684 & 0.2682 & 0.958 \\*
& & 2000 & 0.0384 & 0.0384 & 0.1505 & 0.948 & 0.0606 & 0.0603 & 0.2363 & 0.950 \\*
& & 2500 & 0.0350 & 0.0350 & 0.1371 & 0.956 & 0.0514 & 0.0514 & 0.2014 & 0.952 \\*
\cmidrule(lr){2-3}

\multirow{5}{*}{2500} & \multirow{5}{*}{1}
& 500  & 0.1615 & 0.1615 & 0.6331 & 0.948 & 0.1838 & 0.1838 & 0.7205 & 0.948 \\*
& & 1000 & 0.1153 & 0.1154 & 0.4522 & 0.946 & 0.1348 & 0.1349 & 0.5287 & 0.958 \\*
& & 1500 & 0.0884 & 0.0885 & 0.3467 & 0.954 & 0.1045 & 0.1045 & 0.4095 & 0.944 \\*
& & 2000 & 0.0826 & 0.0825 & 0.3232 & 0.952 & 0.0891 & 0.0887 & 0.3478 & 0.950 \\*
& & 2500 & 0.0734 & 0.0733 & 0.2874 & 0.952 & 0.0811 & 0.0812 & 0.3183 & 0.946 \\
\midrule

\multirow{5}{*}{5000} & \multirow{5}{*}{0}
& 500  & 0.0099 & 0.0099 & 0.0387 & 0.948 & 0.1024 & 0.1024 & 0.4014 & 0.952 \\*
& & 1000 & 0.0065 & 0.0065 & 0.0255 & 0.948 & 0.0650 & 0.0651 & 0.2551 & 0.940 \\*
& & 1500 & 0.0055 & 0.0055 & 0.0216 & 0.952 & 0.0495 & 0.0495 & 0.1940 & 0.944 \\*
& & 2000 & 0.0045 & 0.0045 & 0.0176 & 0.956 & 0.0452 & 0.0453 & 0.1775 & 0.958 \\*
& & 2500 & 0.0042 & 0.0042 & 0.0163 & 0.956 & 0.0420 & 0.0420 & 0.1645 & 0.934 \\*
\cmidrule(lr){2-3}

\multirow{5}{*}{5000} & \multirow{5}{*}{0.1}
& 500  & 0.0210 & 0.0210 & 0.0823 & 0.956 & 0.0998 & 0.0999 & 0.3914 & 0.942 \\*
& & 1000 & 0.0159 & 0.0159 & 0.0625 & 0.950 & 0.0647 & 0.0647 & 0.2536 & 0.954 \\*
& & 1500 & 0.0119 & 0.0119 & 0.0465 & 0.948 & 0.0551 & 0.0550 & 0.2155 & 0.946 \\*
& & 2000 & 0.0100 & 0.0100 & 0.0391 & 0.952 & 0.0480 & 0.0480 & 0.1883 & 0.948 \\*
& & 2500 & 0.0094 & 0.0094 & 0.0367 & 0.946 & 0.0434 & 0.0435 & 0.1704 & 0.952 \\*
\cmidrule(lr){2-3}

\multirow{5}{*}{5000} & \multirow{5}{*}{0.25}
& 500  & 0.0495 & 0.0495 & 0.1940 & 0.960 & 0.1117 & 0.1116 & 0.4376 & 0.946 \\*
& & 1000 & 0.0364 & 0.0364 & 0.1426 & 0.960 & 0.0730 & 0.0730 & 0.2862 & 0.960 \\*
& & 1500 & 0.0274 & 0.0275 & 0.1076 & 0.948 & 0.0593 & 0.0593 & 0.2325 & 0.954 \\*
& & 2000 & 0.0259 & 0.0259 & 0.1015 & 0.942 & 0.0527 & 0.0528 & 0.2069 & 0.952 \\*
& & 2500 & 0.0220 & 0.0221 & 0.0865 & 0.956 & 0.0462 & 0.0463 & 0.1814 & 0.954 \\*
\cmidrule(lr){2-3}

\multirow{5}{*}{5000} & \multirow{5}{*}{0.5}
& 500  & 0.0960 & 0.0960 & 0.3764 & 0.950 & 0.1483 & 0.1483 & 0.5815 & 0.942 \\*
& & 1000 & 0.0660 & 0.0661 & 0.2590 & 0.950 & 0.0942 & 0.0943 & 0.3696 & 0.962 \\*
& & 1500 & 0.0559 & 0.0560 & 0.2194 & 0.934 & 0.0797 & 0.0797 & 0.3126 & 0.946 \\*
& & 2000 & 0.0477 & 0.0477 & 0.1871 & 0.958 & 0.0684 & 0.0685 & 0.2683 & 0.964 \\*
& & 2500 & 0.0435 & 0.0435 & 0.1707 & 0.952 & 0.0623 & 0.0623 & 0.2444 & 0.948 \\*
\cmidrule(lr){2-3}

\multirow{5}{*}{5000} & \multirow{5}{*}{1}
& 500  & 0.1888 & 0.1890 & 0.7409 & 0.944 & 0.2238 & 0.2240 & 0.8781 & 0.950 \\*
& & 1000 & 0.1433 & 0.1432 & 0.5613 & 0.958 & 0.1534 & 0.1535 & 0.6017 & 0.940 \\*
& & 1500 & 0.1126 & 0.1126 & 0.4415 & 0.940 & 0.1216 & 0.1216 & 0.4766 & 0.946 \\*
& & 2000 & 0.0983 & 0.0983 & 0.3854 & 0.954 & 0.1069 & 0.1070 & 0.4194 & 0.946 \\*
& & 2500 & 0.0875 & 0.0876 & 0.3432 & 0.946 & 0.0916 & 0.0916 & 0.3592 & 0.956 \\

\end{longtable}

\footnotesize
\setlength{\tabcolsep}{3pt}
\renewcommand{\arraystretch}{1.0}
\setlength{\LTleft}{\fill}
\setlength{\LTright}{\fill}
\captionsetup[longtable]{font=small}

\begin{longtable}{@{}ccc*{8}{r}@{}}
\caption{Sensitivity of $\hat\theta_0$ to the common working dimension $\kappa = \tilde p = \tilde q$, with $\bar p = \bar q = 3$, over $500$ replications.}
\label{tab:simu_kappa}\\
\toprule
 &  &  & \multicolumn{4}{c}{$\sigma = 0.1$}
 & \multicolumn{4}{c}{$\sigma = 1$} \\
\cmidrule(lr){4-7}
\cmidrule(lr){8-11}
$p = q$ & $\kappa$ & $n$
& RMSE & SD & CI Length & Coverage
& RMSE & SD & CI Length & Coverage \\
\midrule
\endfirsthead

\multicolumn{11}{c}{\tablename~\thetable\ --- continued}\\
\toprule
 &  &  & \multicolumn{4}{c}{$\sigma = 0.1$}
 & \multicolumn{4}{c}{$\sigma = 1$} \\
\cmidrule(lr){4-7}
\cmidrule(lr){8-11}
$p = q$ & $\kappa$ & $n$
& RMSE & SD & CI Length & Coverage
& RMSE & SD & CI Length & Coverage \\
\midrule
\endhead

\midrule
\multicolumn{11}{r}{Continued on next page}\\
\endfoot

\bottomrule
\multicolumn{11}{@{}l}{\footnotesize All $500$ replications were usable in every cell.}\\
\endlastfoot

\multirow{5}{*}{1000} & \multirow{5}{*}{1} & 500 & 0.2470 & 0.1902 & 0.7454 & 0.880 & 0.2609 & 0.2072 & 0.8122 & 0.894 \\*
& & 1000 & 0.2044 & 0.1355 & 0.5311 & 0.796 & 0.2140 & 0.1518 & 0.5952 & 0.840 \\*
& & 1500 & 0.1925 & 0.1076 & 0.4218 & 0.684 & 0.1975 & 0.1200 & 0.4704 & 0.770 \\*
& & 2000 & 0.1846 & 0.0994 & 0.3895 & 0.654 & 0.1916 & 0.1000 & 0.3919 & 0.630 \\*
& & 2500 & 0.1726 & 0.0875 & 0.3432 & 0.610 & 0.1764 & 0.0909 & 0.3564 & 0.610 \\
\cmidrule(lr){2-3}

\multirow{5}{*}{1000} & \multirow{5}{*}{2} & 500 & 0.1986 & 0.1983 & 0.7772 & 0.960 & 0.2207 & 0.2209 & 0.8659 & 0.944 \\*
& & 1000 & 0.1461 & 0.1451 & 0.5690 & 0.944 & 0.1609 & 0.1610 & 0.6311 & 0.954 \\*
& & 1500 & 0.1183 & 0.1179 & 0.4620 & 0.944 & 0.1321 & 0.1302 & 0.5102 & 0.944 \\*
& & 2000 & 0.1067 & 0.1060 & 0.4154 & 0.950 & 0.1122 & 0.1104 & 0.4328 & 0.942 \\*
& & 2500 & 0.0878 & 0.0870 & 0.3411 & 0.942 & 0.1008 & 0.0996 & 0.3903 & 0.946 \\
\cmidrule(lr){2-3}

\multirow{5}{*}{1000} & \multirow{5}{*}{3} & 500 & 0.1956 & 0.1951 & 0.7647 & 0.952 & 0.2240 & 0.2242 & 0.8790 & 0.954 \\*
& & 1000 & 0.1484 & 0.1484 & 0.5818 & 0.946 & 0.1485 & 0.1483 & 0.5815 & 0.944 \\*
& & 1500 & 0.1151 & 0.1152 & 0.4515 & 0.944 & 0.1269 & 0.1269 & 0.4975 & 0.952 \\*
& & 2000 & 0.1053 & 0.1054 & 0.4130 & 0.944 & 0.1106 & 0.1107 & 0.4339 & 0.952 \\*
& & 2500 & 0.0931 & 0.0932 & 0.3653 & 0.954 & 0.1016 & 0.1015 & 0.3980 & 0.952 \\
\cmidrule(lr){2-3}

\multirow{5}{*}{1000} & \multirow{5}{*}{5} & 500 & 0.2138 & 0.2140 & 0.8390 & 0.948 & 0.2285 & 0.2287 & 0.8965 & 0.950 \\*
& & 1000 & 0.1492 & 0.1493 & 0.5852 & 0.946 & 0.1700 & 0.1697 & 0.6653 & 0.948 \\*
& & 1500 & 0.1252 & 0.1253 & 0.4913 & 0.936 & 0.1306 & 0.1307 & 0.5123 & 0.948 \\*
& & 2000 & 0.1098 & 0.1098 & 0.4306 & 0.954 & 0.1104 & 0.1104 & 0.4329 & 0.938 \\*
& & 2500 & 0.0900 & 0.0901 & 0.3531 & 0.944 & 0.1035 & 0.1032 & 0.4047 & 0.942 \\
\cmidrule(lr){2-3}

\multirow{5}{*}{1000} & \multirow{5}{*}{7} & 500 & 0.2235 & 0.2236 & 0.8767 & 0.960 & 0.2349 & 0.2351 & 0.9216 & 0.950 \\*
& & 1000 & 0.1500 & 0.1500 & 0.5881 & 0.948 & 0.1608 & 0.1609 & 0.6308 & 0.962 \\*
& & 1500 & 0.1190 & 0.1191 & 0.4670 & 0.944 & 0.1297 & 0.1297 & 0.5084 & 0.944 \\*
& & 2000 & 0.1052 & 0.1049 & 0.4112 & 0.950 & 0.1110 & 0.1111 & 0.4353 & 0.956 \\*
& & 2500 & 0.1012 & 0.1012 & 0.3966 & 0.960 & 0.1019 & 0.1019 & 0.3996 & 0.948 \\
\cmidrule(lr){2-3}

\multirow{5}{*}{1000} & \multirow{5}{*}{10} & 500 & 0.2242 & 0.2244 & 0.8795 & 0.956 & 0.2392 & 0.2394 & 0.9384 & 0.942 \\*
& & 1000 & 0.1536 & 0.1535 & 0.6017 & 0.942 & 0.1643 & 0.1643 & 0.6440 & 0.942 \\*
& & 1500 & 0.1211 & 0.1212 & 0.4752 & 0.956 & 0.1269 & 0.1270 & 0.4979 & 0.950 \\*
& & 2000 & 0.1035 & 0.1035 & 0.4058 & 0.948 & 0.1210 & 0.1210 & 0.4744 & 0.944 \\*
& & 2500 & 0.0925 & 0.0926 & 0.3630 & 0.950 & 0.0965 & 0.0962 & 0.3772 & 0.948 \\
\cmidrule(lr){2-3}

\multirow{5}{*}{1000} & \multirow{5}{*}{15} & 500 & 0.2176 & 0.2175 & 0.8525 & 0.954 & 0.2529 & 0.2531 & 0.9922 & 0.948 \\*
& & 1000 & 0.1501 & 0.1503 & 0.5891 & 0.948 & 0.1674 & 0.1675 & 0.6565 & 0.942 \\*
& & 1500 & 0.1233 & 0.1234 & 0.4837 & 0.960 & 0.1265 & 0.1265 & 0.4961 & 0.950 \\*
& & 2000 & 0.1067 & 0.1067 & 0.4182 & 0.954 & 0.1111 & 0.1112 & 0.4361 & 0.954 \\*
& & 2500 & 0.0949 & 0.0950 & 0.3723 & 0.960 & 0.1047 & 0.1048 & 0.4109 & 0.962 \\
\cmidrule(lr){2-3}

\multirow{5}{*}{1000} & \multirow{5}{*}{20} & 500 & 0.2290 & 0.2293 & 0.8987 & 0.948 & 0.2499 & 0.2493 & 0.9774 & 0.958 \\*
& & 1000 & 0.1577 & 0.1577 & 0.6180 & 0.952 & 0.1678 & 0.1677 & 0.6575 & 0.948 \\*
& & 1500 & 0.1244 & 0.1243 & 0.4873 & 0.948 & 0.1350 & 0.1350 & 0.5291 & 0.948 \\*
& & 2000 & 0.1036 & 0.1037 & 0.4065 & 0.950 & 0.1176 & 0.1176 & 0.4611 & 0.944 \\*
& & 2500 & 0.0970 & 0.0971 & 0.3807 & 0.950 & 0.1026 & 0.1027 & 0.4027 & 0.950 \\
\cmidrule(lr){2-3}

\multirow{5}{*}{1000} & \multirow{5}{*}{30} & 500 & 0.2370 & 0.2371 & 0.9293 & 0.950 & 0.2665 & 0.2665 & 1.0445 & 0.952 \\*
& & 1000 & 0.1634 & 0.1635 & 0.6409 & 0.948 & 0.1680 & 0.1682 & 0.6592 & 0.964 \\*
& & 1500 & 0.1222 & 0.1222 & 0.4792 & 0.954 & 0.1342 & 0.1342 & 0.5262 & 0.946 \\*
& & 2000 & 0.1080 & 0.1078 & 0.4225 & 0.934 & 0.1099 & 0.1099 & 0.4309 & 0.946 \\*
& & 2500 & 0.0961 & 0.0959 & 0.3758 & 0.954 & 0.1094 & 0.1095 & 0.4291 & 0.944 \\
\midrule

\multirow{5}{*}{2500} & \multirow{5}{*}{1} & 500 & 0.3131 & 0.1519 & 0.5956 & 0.566 & 0.3236 & 0.1722 & 0.6749 & 0.656 \\*
& & 1000 & 0.2947 & 0.1080 & 0.4233 & 0.274 & 0.2878 & 0.1262 & 0.4947 & 0.464 \\*
& & 1500 & 0.2747 & 0.0889 & 0.3485 & 0.154 & 0.2923 & 0.1004 & 0.3935 & 0.234 \\*
& & 2000 & 0.2747 & 0.0743 & 0.2912 & 0.052 & 0.2855 & 0.0861 & 0.3375 & 0.118 \\*
& & 2500 & 0.2692 & 0.0660 & 0.2586 & 0.028 & 0.2704 & 0.0759 & 0.2976 & 0.066 \\
\cmidrule(lr){2-3}

\multirow{5}{*}{2500} & \multirow{5}{*}{2} & 500 & 0.3131 & 0.1993 & 0.7814 & 0.770 & 0.3550 & 0.2184 & 0.8563 & 0.744 \\*
& & 1000 & 0.3093 & 0.1662 & 0.6516 & 0.626 & 0.3039 & 0.1807 & 0.7083 & 0.728 \\*
& & 1500 & 0.3041 & 0.1402 & 0.5496 & 0.532 & 0.3036 & 0.1581 & 0.6196 & 0.616 \\*
& & 2000 & 0.2897 & 0.1437 & 0.5632 & 0.590 & 0.2920 & 0.1383 & 0.5421 & 0.564 \\*
& & 2500 & 0.2938 & 0.1370 & 0.5370 & 0.536 & 0.3044 & 0.1463 & 0.5734 & 0.530 \\
\cmidrule(lr){2-3}

\multirow{5}{*}{2500} & \multirow{5}{*}{3} & 500 & 0.1532 & 0.1533 & 0.6010 & 0.944 & 0.1749 & 0.1749 & 0.6858 & 0.946 \\*
& & 1000 & 0.1159 & 0.1159 & 0.4542 & 0.952 & 0.1249 & 0.1251 & 0.4903 & 0.956 \\*
& & 1500 & 0.0859 & 0.0860 & 0.3372 & 0.948 & 0.1016 & 0.1017 & 0.3987 & 0.964 \\*
& & 2000 & 0.0738 & 0.0739 & 0.2897 & 0.962 & 0.0905 & 0.0903 & 0.3540 & 0.958 \\*
& & 2500 & 0.0692 & 0.0692 & 0.2713 & 0.934 & 0.0766 & 0.0767 & 0.3006 & 0.962 \\
\cmidrule(lr){2-3}

\multirow{5}{*}{2500} & \multirow{5}{*}{5} & 500 & 0.1584 & 0.1582 & 0.6202 & 0.944 & 0.1870 & 0.1869 & 0.7328 & 0.956 \\*
& & 1000 & 0.1127 & 0.1128 & 0.4421 & 0.948 & 0.1254 & 0.1254 & 0.4917 & 0.958 \\*
& & 1500 & 0.0899 & 0.0899 & 0.3523 & 0.944 & 0.1037 & 0.1038 & 0.4067 & 0.950 \\*
& & 2000 & 0.0745 & 0.0745 & 0.2922 & 0.944 & 0.0891 & 0.0891 & 0.3493 & 0.960 \\*
& & 2500 & 0.0672 & 0.0673 & 0.2637 & 0.958 & 0.0787 & 0.0788 & 0.3089 & 0.958 \\
\cmidrule(lr){2-3}

\multirow{5}{*}{2500} & \multirow{5}{*}{7} & 500 & 0.1592 & 0.1590 & 0.6234 & 0.944 & 0.1846 & 0.1845 & 0.7233 & 0.944 \\*
& & 1000 & 0.1108 & 0.1108 & 0.4345 & 0.942 & 0.1239 & 0.1235 & 0.4842 & 0.946 \\*
& & 1500 & 0.0843 & 0.0843 & 0.3306 & 0.950 & 0.0983 & 0.0984 & 0.3857 & 0.946 \\*
& & 2000 & 0.0794 & 0.0793 & 0.3108 & 0.950 & 0.0877 & 0.0877 & 0.3438 & 0.946 \\*
& & 2500 & 0.0701 & 0.0701 & 0.2748 & 0.958 & 0.0783 & 0.0784 & 0.3073 & 0.950 \\
\cmidrule(lr){2-3}

\multirow{5}{*}{2500} & \multirow{5}{*}{10} & 500 & 0.1494 & 0.1496 & 0.5863 & 0.946 & 0.1835 & 0.1834 & 0.7189 & 0.954 \\*
& & 1000 & 0.1073 & 0.1074 & 0.4210 & 0.944 & 0.1272 & 0.1273 & 0.4991 & 0.954 \\*
& & 1500 & 0.0939 & 0.0938 & 0.3677 & 0.950 & 0.0986 & 0.0987 & 0.3868 & 0.940 \\*
& & 2000 & 0.0776 & 0.0776 & 0.3042 & 0.946 & 0.0869 & 0.0870 & 0.3410 & 0.948 \\*
& & 2500 & 0.0681 & 0.0681 & 0.2671 & 0.950 & 0.0793 & 0.0790 & 0.3098 & 0.936 \\
\cmidrule(lr){2-3}

\multirow{5}{*}{2500} & \multirow{5}{*}{15} & 500 & 0.1678 & 0.1679 & 0.6580 & 0.944 & 0.1960 & 0.1962 & 0.7689 & 0.946 \\*
& & 1000 & 0.1054 & 0.1054 & 0.4131 & 0.948 & 0.1305 & 0.1306 & 0.5120 & 0.952 \\*
& & 1500 & 0.0847 & 0.0846 & 0.3314 & 0.960 & 0.1096 & 0.1096 & 0.4298 & 0.952 \\*
& & 2000 & 0.0704 & 0.0703 & 0.2755 & 0.934 & 0.0862 & 0.0862 & 0.3380 & 0.940 \\*
& & 2500 & 0.0648 & 0.0647 & 0.2537 & 0.954 & 0.0764 & 0.0765 & 0.2998 & 0.950 \\
\cmidrule(lr){2-3}

\multirow{5}{*}{2500} & \multirow{5}{*}{20} & 500 & 0.1651 & 0.1653 & 0.6479 & 0.954 & 0.2014 & 0.2015 & 0.7900 & 0.948 \\*
& & 1000 & 0.1108 & 0.1109 & 0.4346 & 0.948 & 0.1374 & 0.1375 & 0.5390 & 0.948 \\*
& & 1500 & 0.0853 & 0.0854 & 0.3346 & 0.940 & 0.1043 & 0.1043 & 0.4089 & 0.938 \\*
& & 2000 & 0.0779 & 0.0777 & 0.3046 & 0.954 & 0.0874 & 0.0875 & 0.3428 & 0.948 \\*
& & 2500 & 0.0685 & 0.0685 & 0.2685 & 0.942 & 0.0747 & 0.0745 & 0.2919 & 0.954 \\
\cmidrule(lr){2-3}

\multirow{5}{*}{2500} & \multirow{5}{*}{30} & 500 & 0.1851 & 0.1851 & 0.7254 & 0.942 & 0.2226 & 0.2225 & 0.8724 & 0.958 \\*
& & 1000 & 0.1134 & 0.1135 & 0.4448 & 0.954 & 0.1374 & 0.1373 & 0.5382 & 0.950 \\*
& & 1500 & 0.0913 & 0.0914 & 0.3581 & 0.960 & 0.1072 & 0.1073 & 0.4206 & 0.948 \\*
& & 2000 & 0.0771 & 0.0772 & 0.3024 & 0.952 & 0.0923 & 0.0924 & 0.3621 & 0.956 \\*
& & 2500 & 0.0675 & 0.0675 & 0.2648 & 0.948 & 0.0809 & 0.0810 & 0.3174 & 0.944 \\
\midrule

\multirow{5}{*}{5000} & \multirow{5}{*}{1} & 500 & 0.3050 & 0.2377 & 0.9318 & 0.872 & 0.3197 & 0.2629 & 1.0304 & 0.908 \\*
& & 1000 & 0.2479 & 0.1642 & 0.6437 & 0.790 & 0.2586 & 0.1761 & 0.6903 & 0.818 \\*
& & 1500 & 0.2335 & 0.1285 & 0.5038 & 0.670 & 0.2356 & 0.1440 & 0.5645 & 0.750 \\*
& & 2000 & 0.2201 & 0.1151 & 0.4513 & 0.630 & 0.2325 & 0.1264 & 0.4956 & 0.632 \\*
& & 2500 & 0.2148 & 0.1017 & 0.3988 & 0.540 & 0.2205 & 0.1063 & 0.4167 & 0.556 \\
\cmidrule(lr){2-3}

\multirow{5}{*}{5000} & \multirow{5}{*}{2} & 500 & 0.3087 & 0.2478 & 0.9712 & 0.888 & 0.3139 & 0.2486 & 0.9745 & 0.884 \\*
& & 1000 & 0.2540 & 0.1662 & 0.6515 & 0.794 & 0.2758 & 0.1974 & 0.7738 & 0.840 \\*
& & 1500 & 0.2442 & 0.1371 & 0.5375 & 0.696 & 0.2420 & 0.1427 & 0.5593 & 0.726 \\*
& & 2000 & 0.2297 & 0.1208 & 0.4736 & 0.640 & 0.2282 & 0.1220 & 0.4782 & 0.646 \\*
& & 2500 & 0.2235 & 0.1031 & 0.4042 & 0.528 & 0.2349 & 0.1136 & 0.4451 & 0.560 \\
\cmidrule(lr){2-3}

\multirow{5}{*}{5000} & \multirow{5}{*}{3} & 500 & 0.2370 & 0.2370 & 0.9292 & 0.958 & 0.2530 & 0.2519 & 0.9872 & 0.950 \\*
& & 1000 & 0.1759 & 0.1760 & 0.6900 & 0.962 & 0.1818 & 0.1815 & 0.7116 & 0.946 \\*
& & 1500 & 0.1357 & 0.1358 & 0.5325 & 0.958 & 0.1457 & 0.1457 & 0.5713 & 0.948 \\*
& & 2000 & 0.1261 & 0.1262 & 0.4946 & 0.944 & 0.1384 & 0.1383 & 0.5420 & 0.948 \\*
& & 2500 & 0.1079 & 0.1080 & 0.4235 & 0.956 & 0.1098 & 0.1099 & 0.4307 & 0.954 \\
\cmidrule(lr){2-3}

\multirow{5}{*}{5000} & \multirow{5}{*}{5} & 500 & 0.2346 & 0.2347 & 0.9201 & 0.942 & 0.2498 & 0.2500 & 0.9801 & 0.950 \\*
& & 1000 & 0.1640 & 0.1642 & 0.6436 & 0.946 & 0.1834 & 0.1835 & 0.7195 & 0.948 \\*
& & 1500 & 0.1378 & 0.1379 & 0.5405 & 0.936 & 0.1538 & 0.1539 & 0.6034 & 0.954 \\*
& & 2000 & 0.1180 & 0.1180 & 0.4626 & 0.950 & 0.1223 & 0.1225 & 0.4800 & 0.946 \\*
& & 2500 & 0.1077 & 0.1078 & 0.4226 & 0.948 & 0.1091 & 0.1092 & 0.4279 & 0.948 \\
\cmidrule(lr){2-3}

\multirow{5}{*}{5000} & \multirow{5}{*}{7} & 500 & 0.2342 & 0.2345 & 0.9191 & 0.944 & 0.2677 & 0.2678 & 1.0497 & 0.952 \\*
& & 1000 & 0.1777 & 0.1776 & 0.6963 & 0.958 & 0.1751 & 0.1750 & 0.6861 & 0.960 \\*
& & 1500 & 0.1399 & 0.1400 & 0.5489 & 0.942 & 0.1463 & 0.1460 & 0.5724 & 0.952 \\*
& & 2000 & 0.1226 & 0.1227 & 0.4810 & 0.956 & 0.1236 & 0.1231 & 0.4823 & 0.946 \\*
& & 2500 & 0.1097 & 0.1098 & 0.4304 & 0.942 & 0.1094 & 0.1090 & 0.4273 & 0.948 \\
\cmidrule(lr){2-3}

\multirow{5}{*}{5000} & \multirow{5}{*}{10} & 500 & 0.2441 & 0.2444 & 0.9579 & 0.954 & 0.2755 & 0.2751 & 1.0785 & 0.956 \\*
& & 1000 & 0.1640 & 0.1641 & 0.6434 & 0.954 & 0.1807 & 0.1809 & 0.7090 & 0.946 \\*
& & 1500 & 0.1423 & 0.1425 & 0.5585 & 0.936 & 0.1546 & 0.1545 & 0.6058 & 0.942 \\*
& & 2000 & 0.1246 & 0.1247 & 0.4887 & 0.956 & 0.1346 & 0.1347 & 0.5281 & 0.950 \\*
& & 2500 & 0.1061 & 0.1057 & 0.4142 & 0.938 & 0.1269 & 0.1268 & 0.4970 & 0.956 \\
\cmidrule(lr){2-3}

\multirow{5}{*}{5000} & \multirow{5}{*}{15} & 500 & 0.2487 & 0.2489 & 0.9758 & 0.954 & 0.2773 & 0.2775 & 1.0879 & 0.950 \\*
& & 1000 & 0.1818 & 0.1820 & 0.7134 & 0.952 & 0.1875 & 0.1870 & 0.7331 & 0.956 \\*
& & 1500 & 0.1401 & 0.1398 & 0.5481 & 0.944 & 0.1476 & 0.1476 & 0.5787 & 0.958 \\*
& & 2000 & 0.1174 & 0.1174 & 0.4601 & 0.958 & 0.1242 & 0.1244 & 0.4875 & 0.936 \\*
& & 2500 & 0.1073 & 0.1074 & 0.4212 & 0.946 & 0.1143 & 0.1145 & 0.4487 & 0.954 \\
\cmidrule(lr){2-3}

\multirow{5}{*}{5000} & \multirow{5}{*}{20} & 500 & 0.2577 & 0.2576 & 1.0099 & 0.942 & 0.2903 & 0.2905 & 1.1389 & 0.960 \\*
& & 1000 & 0.1664 & 0.1665 & 0.6525 & 0.954 & 0.1900 & 0.1901 & 0.7453 & 0.962 \\*
& & 1500 & 0.1416 & 0.1417 & 0.5556 & 0.954 & 0.1469 & 0.1470 & 0.5764 & 0.942 \\*
& & 2000 & 0.1189 & 0.1190 & 0.4663 & 0.956 & 0.1227 & 0.1226 & 0.4805 & 0.944 \\*
& & 2500 & 0.1118 & 0.1120 & 0.4389 & 0.948 & 0.1157 & 0.1158 & 0.4540 & 0.944 \\
\cmidrule(lr){2-3}

\multirow{5}{*}{5000} & \multirow{5}{*}{30} & 500 & 0.2774 & 0.2775 & 1.0878 & 0.950 & 0.3107 & 0.3110 & 1.2193 & 0.950 \\*
& & 1000 & 0.1789 & 0.1791 & 0.7021 & 0.940 & 0.1963 & 0.1965 & 0.7702 & 0.944 \\*
& & 1500 & 0.1445 & 0.1446 & 0.5668 & 0.954 & 0.1548 & 0.1549 & 0.6071 & 0.948 \\*
& & 2000 & 0.1202 & 0.1203 & 0.4717 & 0.950 & 0.1252 & 0.1253 & 0.4912 & 0.946 \\*
& & 2500 & 0.1117 & 0.1118 & 0.4381 & 0.948 & 0.1167 & 0.1168 & 0.4577 & 0.948 \\

\end{longtable}

\begin{table}[ht!]
\centering
\small
\setlength{\tabcolsep}{4pt}
\begin{tabular}{cc ccc ccc}
\toprule
 & & \multicolumn{3}{c}{$\sigma = 0.1$} & \multicolumn{3}{c}{$\sigma = 1$} \\
\cmidrule(lr){3-5} \cmidrule(lr){6-8}
$p=q$ & $\tilde p$ & $\tilde q = 3$ & $\tilde q = 7$ & $\tilde q = 20$ & $\tilde q = 3$ & $\tilde q = 7$ & $\tilde q = 20$ \\
\midrule
\multirow{3}{*}{1000} & 3 & 0.0739 (0.940) & 0.0702 (0.968) & 0.0722 (0.956) & 0.0699 (0.946) & 0.0690 (0.946) & 0.0720 (0.940) \\
 & 7 & 0.0690 (0.946) & 0.0688 (0.934) & 0.0706 (0.954) & 0.0733 (0.944) & 0.0749 (0.952) & 0.0711 (0.938) \\
 & 20 & 0.0726 (0.956) & 0.0750 (0.942) & 0.0714 (0.944) & 0.0724 (0.946) & 0.0746 (0.942) & 0.0764 (0.946) \\
\midrule
\multirow{3}{*}{2500} & 3 & 0.0581 (0.948) & 0.0550 (0.952) & 0.0568 (0.950) & 0.0725 (0.954) & 0.0687 (0.954) & 0.0694 (0.948) \\
 & 7 & 0.0536 (0.948) & 0.0581 (0.962) & 0.0578 (0.956) & 0.0721 (0.948) & 0.0714 (0.952) & 0.0669 (0.962) \\
 & 20 & 0.0594 (0.952) & 0.0571 (0.950) & 0.0534 (0.940) & 0.0704 (0.956) & 0.0707 (0.946) & 0.0701 (0.944) \\
\midrule
\multirow{3}{*}{5000} & 3 & 0.0932 (0.942) & 0.0972 (0.946) & 0.0938 (0.952) & 0.0829 (0.944) & 0.0816 (0.946) & 0.0820 (0.954) \\
 & 7 & 0.0966 (0.938) & 0.0933 (0.956) & 0.1036 (0.954) & 0.0832 (0.944) & 0.0829 (0.964) & 0.0827 (0.954) \\
 & 20 & 0.1002 (0.940) & 0.0973 (0.944) & 0.1014 (0.956) & 0.0828 (0.946) & 0.0862 (0.948) & 0.0842 (0.956) \\
\bottomrule
\end{tabular}
\caption{RMSE of $\hat\theta_0$ (coverage in parentheses) for $\tilde p$ and $\tilde q$ varied separately, with $\bar p = \bar q = 3$ and $n = 2500$, over $500$ replications.}
\label{tab:simu_pq_cross}
\end{table}

\subsection{Covariates \label{covariates}}
\begin{table}[H]
\centering
\begin{tabular}{ll}
\hline
\textbf{Variable Name} & \textbf{Description} \\
\hline
\texttt{PTDOBYY} & Date of Birth (in year) \\
\texttt{PTGENDER} & Gender of the subject (e.g., Male or Female) \\
\texttt{PTMARRY}  & Marital status (e.g., Married, Single, Divorced) \\
\texttt{PTEDUCAT} & Years of formal education completed \\
\texttt{PTNOTRT}  & Indicator for treatment group or non-treatment \\
\texttt{VSWEIGHT} & Body weight (in pounds) \\
\texttt{VSPULSE}  & Pulse rate (beats per minute) \\
\texttt{VSRESP}   & Respiratory rate (breaths per minute) \\
\texttt{VSTEMP}   & Body temperature (in Fahrenheit) \\
\texttt{VSHEIGHT} & Height (in inches) \\
\hline
\end{tabular}
\caption{Descriptions of patient background variables used as covariates in the model.}
\label{tabpatient_variables}
\end{table}

\section{Comparison to Related Literature}

\subsection{Comparison to \texorpdfstring{\cite{zhang2026statistical}}{Zhang et al. (2026)}}
The method of \cite{zhang2026statistical} is related to FARMA because both use latent factor models in high-dimensional mediation settings. However, the high-dimensional variables and latent factors play different roles in the two methods, and the causal targets are also different.
Using the notation of \cite{zhang2026statistical}, let $\boldsymbol x\in\mathbb R^q$ be a high-dimensional exposure vector, $\boldsymbol m\in\mathbb R^p$ a high-dimensional mediator vector, and $y\in\mathbb R$ the outcome. Their model is
\begin{align*}
y&=\boldsymbol\alpha_0^\top\boldsymbol m+\boldsymbol\alpha_1^\top\boldsymbol x+\varepsilon_1,\qquad
\boldsymbol m=\boldsymbol\Gamma^\top\boldsymbol x+\boldsymbol\varepsilon.
\end{align*}
The direct, indirect, and total effects are $\balpha_1$, $\boldsymbol\beta=\boldsymbol\Gamma\boldsymbol\alpha_0$ and $\boldsymbol\gamma=\boldsymbol\alpha_1+\boldsymbol\beta$ respectively. 
In order to account for strong dependence among the high-dimensional variables, PRIME uses the factor models
\begin{align*}\boldsymbol x=B_1\boldsymbol f_1+\boldsymbol v_1,\qquad \begin{pmatrix}\boldsymbol m\\ \boldsymbol x\end{pmatrix}=B_2\boldsymbol f_2+\boldsymbol v_2.\end{align*}
For each prespecified exposure group $g$, PRIME estimates $\boldsymbol\alpha_{1,g}$ and $\boldsymbol\gamma_g$, and estimates the indirect effect as
\begin{align*}\widehat{\boldsymbol\beta}_g=\widehat{\boldsymbol\gamma}_g-\widehat{\boldsymbol\alpha}_{1,g}.\end{align*}
The main differences from FARMA are given below.

\begin{enumerate}

\item \textbf{Observed variables and surrogate measurements.} In \cite{zhang2026statistical}, $\boldsymbol x$ and $\boldsymbol m$ are observed exposures and mediators, and the direct and indirect effects are defined for their original coordinates. The factors $\boldsymbol f_1$ and $\boldsymbol f_2$ are introduced to account for the dependence among these observed variables.
In FARMA, $X$ and $M$ are not the covariates and mediators themselves. They are high-dimensional surrogate measurements of the latent factors $\boldsymbol f_X$ and $\boldsymbol f_M$:
\begin{align*}X=\Gamma_X\boldsymbol f_X+\boldsymbol u_X,\quad M=\Gamma_M\boldsymbol f_M+\boldsymbol u_M.\end{align*}
Therefore, the individual coordinates of $M$ are not treated as separate mediators. Instead, they provide information about the underlying latent mediator $\boldsymbol f_M$.

\item \textbf{Exposure and causal target.} \cite{zhang2026statistical} considers a high-dimensional exposure vector and studies direct and indirect effects for individual exposures or prespecified groups:
\begin{align*}H_{0,\mathrm{dir}}^{(g)}:\boldsymbol\alpha_{1,g}=0,\qquad H_{0,\mathrm{ind}}^{(g)}:\boldsymbol\beta_g=0.\end{align*}
The corresponding inference problem is therefore high-dimensional and involves multiple testing.
FARMA considers a binary treatment $A\in\{0,1\}$. The main target is the population-level counterfactual mean
\begin{align*}\theta_0=E\left\{Y^{(1,\boldsymbol f_M^{(0)})}\right\},\end{align*}
which is then used together with the treatment-specific counterfactual means to obtain the natural direct, natural indirect, and total effects. The goal is therefore different from identifying which coordinate of a high-dimensional exposure has a direct or indirect effect.

\item \textbf{Role of the factor structure.} In \cite{zhang2026statistical}, the exposures and mediators are observed, and the factors are used to handle the strong dependence among the high-dimensional variables.

In FARMA, the factors are themselves the latent covariates and mediators of interest. As a result, the factor structure also plays an important role in identification. With fixed $p$ and $q$, the latent mediation functional is generally not point identified from $(X,M)$ alone. We show that the observed-data functional approaches the latent target as the numbers of surrogate measurements increase. We also allow the working dimensions to be larger than the true factor dimensions, $\widetilde p\geq\bar p$ and $\widetilde q\geq\bar q.$ Thus, the exact latent factor dimensions do not need to be known.

\item \textbf{Model assumptions.} PRIME uses a high-dimensional linear structural equation model with sparsity assumptions on the regression coefficients. The main model does not include unmeasured confounding, although observed confounders can be added. The mediator-outcome coefficient is also common across exposure levels.

FARMA includes latent pre-treatment covariates $\boldsymbol f_X$, which adjust for confounding among the treatment, mediator, and outcome. Since the regressions involving the latent factors are low-dimensional, our theory does not require sparsity. We also allow the coefficients $(B_0,B_1), (\boldsymbol\beta_0,\boldsymbol\beta_1),$ and $(\boldsymbol\gamma_0,\boldsymbol\gamma_1)$ to vary across treatment levels.

\item \textbf{Estimation and inference.} PRIME estimates the factor and idiosyncratic components and then uses group-wise partially penalized least squares. Inference is based on tests for direct and indirect effects together with false discovery rate control.

FARMA uses projected surrogate variables followed by two cross-fitted low-dimensional regression steps. The estimator targets a scalar mediation functional. We show that the estimator is $\sqrt n$-consistent and asymptotically normal, which allows inference for the population natural direct and indirect effects.

\end{enumerate}

The difference can also be seen in the ADNI applications. \cite{zhang2026statistical} use SNPs as high-dimensional exposures and regional PET measurements as observed mediators. In our application, depressive status is the treatment, while high-dimensional DNA methylation measurements are used as surrogates for an underlying latent mediating process.

\noindent
\textbf{Summary of differences:}
\begin{center}
\small
\resizebox{\textwidth}{!}{%
\begin{tabular}{lcc}
\toprule
&\textbf{Zhang et al.\ (2026a): PRIME}&\textbf{FARMA}\\
\midrule
High-dimensional variables&Observed exposures and mediators&Surrogate measurements $X$ and $M$\\
Role of factors&Account for dependence&Latent covariates and mediators\\
Exposure&High-dimensional $\boldsymbol x$&Binary treatment $A$\\
Target&Group-wise direct and indirect effects&Population NDE, NIE, and TE\\
Covariates&Not included in the core model&Latent $\boldsymbol f_X$ measured through $X$\\
Sparsity&Required for high-dimensional coefficients&Not required\\
Treatment-specific coefficients&No&Yes\\
Estimation&Factor-adjusted penalized regression&Cross-fitted projected regressions\\
Inference&Group-wise tests and FDR control&$\sqrt n$-based inference\\
\bottomrule
\end{tabular}%
}
\end{center}

\end{document}